\documentclass[3p,authoryear,a4paper]{elsarticle}
\usepackage{graphicx}
\usepackage{xcolor}
\usepackage{subcaption}
\usepackage{dsfont}
\usepackage{amsmath}
\usepackage{amssymb}
\usepackage{amsfonts}
\usepackage{amsbsy}
\usepackage{siunitx}
\usepackage{amsthm}
\usepackage{array}
\usepackage{booktabs}
\usepackage{verbatim}
\usepackage[english]{babel}
\usepackage{float}
\usepackage{epsfig}
\usepackage{multirow}
\usepackage{natbib}
\usepackage{hyperref}
\usepackage{float}
\usepackage[font=small,skip=3pt, labelfont=bf]{caption}
\usepackage{graphicx}
\usepackage{caption}
\usepackage{subcaption}
\usepackage{float}
\usepackage{wrapfig}
\usepackage{babel}
\usepackage{comment}
\usepackage{algorithm}
\usepackage{algpseudocode}
\usepackage{graphbox,graphicx}
\usepackage{xurl}
\usepackage{sidecap}
\usepackage{bm}
\usepackage[table]{xcolor}
\usepackage{colortbl}
\usepackage{xfp}
\definecolor{excelblue}{RGB}{0,112,192}
\definecolor{excelwhite}{RGB}{255,255,255}
\newcommand{\grad}[2]{%
  \cellcolor{excelblue!#1!excelwhite}{#2}%
}

\newcolumntype{C}[1]{>{\centering\let\newline\\\arraybackslash\hspace{0pt}}m{#1}}
\renewcommand\appendix{\par
\setcounter{section}{0}%
\setcounter{subsection}{0}%
\setcounter{table}{0}
\setcounter{figure}{0}
\gdef\thetable{\Alph{table}}
\gdef\thefigure{\Alph{figure}}
\gdef\thesection{\Alph{section}}
\setcounter{section}{0}}

\newcolumntype{L}[1]{>{\raggedright\let\newline\\\arraybackslash\hspace{0pt}}m{#1}}
\newcolumntype{R}[1]{>{\raggedleft\let\newline\\\arraybackslash\hspace{0pt}}m{#1}}

\newcommand{\rev}[1]{\textcolor{black}{#1}}

\newtheorem{remark}{Remark}[section]
\newtheorem{limitation}{Limitation}[section]
\numberwithin{equation}{section}
\numberwithin{figure}{section}
\numberwithin{table}{section}

\newcounter{arclist}

\newcounter{arcenum}
\begin{document}

\begin{frontmatter}

\title{Dynamic Financial Analysis (DFA) of General Insurers under Climate Change}

\author[UM]{Benjamin Avanzi}
\ead{b.avanzi@unimelb.edu.au}

\author[UNSW]{Yanfeng Li \corref{cor}}
\ead{yanfeng.li@student.unsw.edu.au}

\cortext[cor]{Corresponding author. }

\author[UNSW]{Greg Taylor}
\ead{gregory.taylor@unsw.edu.au}

\author[UNSW]{Bernard Wong}
\ead{bernard.wong@unsw.edu.au}

\address[UM]{Centre for Actuarial Studies, Department of Economics, University of Melbourne VIC 3010, Australia}

\address[UNSW]{School of Risk and Actuarial Studies, UNSW Business School, UNSW Sydney NSW 2052, Australia}

\begin{abstract}

Climate change is expected to significantly affect the physical, financial, and economic environments over the long term, posing risks to the financial health of general insurers. While general insurers typically use Dynamic Financial Analysis (DFA) for a comprehensive view of financial impacts, traditional DFA as presented in the literature does not consider the impact of climate change. To address this gap, we \rev{extend the stationary DFA framework to integrate climate risk}, \rev{enabling} a holistic assessment of the long-term impact of climate change on the general insurance industry \rev{and offering a foundational architecture for the DFA of individual insurers.}

\rev{Our framework} captures the long-term impact of climate change on the assets and liabilities of general insurers by considering both physical and economic dimensions across different climate scenarios within an interconnected structure. \rev{Furthermore}, it addresses the uncertainty of climate change impacts using stochastic simulations within climate scenario analysis that are useful for actuarial applications. \rev{Our extensions are} tailored to the general insurance sector \rev{and address} its unique characteristics.

To demonstrate the practical application of our model, we conduct an extensive empirical study using Australian data \rev{and} assess the long-term financial impact of climate change on the general insurance market under various climate scenarios. \rev{The results are benchmarked against those of a stationary DFA framework and}  show that the interaction between economic growth and physical risk plays a key role in shaping general insurers' risk-return profiles. \rev{They highlight the advantages of the climate-dependent DFA over the stationary DFA in generating financial projections under climate change impacts.} Limitations of our framework are thoroughly discussed.

\end{abstract}


\end{frontmatter}

\section{Introduction} \label{intro}

\subsection{Background} \label{background}

\rev{Dynamic Financial Analysis (DFA) is a widely used framework in the general insurance industry for evaluating the financial performance and resilience of insurers under uncertainty. Early precursors of DFA include the generalised cash flow model proposed by \cite{CoDe86}, the Solvency Working Party (SWP) model \citep{DaBeCo87a, DaBeCo87b}, the cash-flow simulation model of \cite{PaDi89}, and the management model introduced in \cite{DaHe90}. The term DFA was formally introduced in actuarial practice through the CAS Dynamic Financial Analysis Handbook \citep{CAS95}, after which a growing body of literature introduced firm-specific DFA systems \citep[see, e.g.,][]{LoSt97, CaZi98, BeMa99}. A systematic academic introduction to DFA is provided by \cite{KaGaKl01}, which presents a general modelling framework that integrates key components common to many DFA models proposed in earlier studies.}



\rev{DFA provides a holistic modelling environment that projects the distribution of potential future financial outcomes of insurers by jointly simulating assets, liabilities, and capital \citep{ElTo09}. Typical applications include economic capital modelling, solvency monitoring, and strategy testing. As a virtual representation of an insurance enterprise, DFA allows management to assess the financial implications of strategic decisions in a simulated environment before implementing them in practice \citep{KaGaKl01, ElPa07}. It can also be used to analyse market behaviour within the general insurance sector \citep[see, e.g.,][]{Ta12}. Through these capabilities, DFA has become an important tool for supporting risk management and long-term strategic planning in insurance.}

\rev{Conventional DFA frameworks are primarily designed around historical economic and insurance relationships and generally assume that future risk dynamics evolve in a stationary manner. As a result,  DFA models in the existing literature typically do not explicitly incorporate climate dynamics or the long-term structural changes associated with climate change. This limitation is increasingly problematic as climate change poses multifaceted risks to general insurers. On the liabilities side, changing weather patterns are expected to alter the frequency and severity of future claims \citep[see, e.g.,][]{HaDiVrAlMe11, LyuNeGhMaGe19}, potentially increasing claims costs and affecting underwriting profitability. On the assets side, climate change may influence key macroeconomic variables such as inflation \citep[see, e.g.,][]{Pa18, EcXe18, KoKuLiNi24}, interest rates \citep[see, e.g.,][]{ByJo20, MoPo22}, and equity returns \citep[see, e.g.,][]{kaXe22,Ve22, Ba23}, thereby affecting insurers' investment performance. The combined effects of these asset and liability channels may ultimately affect insurers' capital positions and the financial health of the broader insurance market. The growing importance of climate-related financial risks has also been recognised by regulators, as evidenced by the growing number of climate-related disclosure requirements, such as IFRS S2 \citep{IFRSS2}. These developments further highlight the need for analytical frameworks capable of evaluating climate risks in an integrated financial context.}

\rev{Outside the DFA literature, a growing body of literature has examined specific aspects of climate-related risks faced by insurers. However, most of those studies focus on individual components of insurers' balance sheets rather than analysing their joint financial implications. On the liabilities side, numerous studies have investigated the impacts of climate change on natural hazards, including floods \citep[see, e.g.,][]{SeZhAdBa21, BoBoCaBoRa24}, bushfires \citep[see, e.g.,][]{QuBaRiPa22}, tropical cyclones and storms \citep[see, e.g.,][]{JaElSa08, JaElBu11, MeVoBlCiLe22}. In addition, weather-related non-catastrophe claims have also been analysed \citep[see, e.g.,][]{HaDiVrAlMe11, LyuHaYuRi17}. Outside the general insurance domain, climate impacts on mortality rates \citep{MiliNaLi25, GuPiPl25} and life insurance reserves \citep{ArSh25} have also been investigated. On the asset side, climate change impacts on interest rates, inflation, and equity returns have been explored both theoretically using general equilibrium models \citep{EcXe18, kaXe22, Ba23} and empirically using regression and factor models \citep[see, e.g.,][]{Pa18, Ve22}.}

\rev{Beyond these useful contributions, a comprehensive framework that captures the interconnected financial impacts of climate change across both assets and liabilities for general insurers remains largely unexplored in the literature. In particular, existing studies often focus on individual risk channels rather than providing an integrated view of how climate change may affect insurers' overall financial position. This may misconstrue the overall financial impact due to asset–liability interdependencies. Moreover, widely used climate scenarios--such as the Shared Socioeconomic Pathways (SSPs) developed by the IPCC \citep{OnKrEbKeRiRo17}--combine both physical and economic narratives, further highlighting the need for modelling frameworks capable of jointly capturing these interconnected dimensions.} 


\rev{An initial attempt to jointly assess the climate impacts on both assets and liabilities of general insurers is presented in \cite{GaOz24} within a stress-testing framework, although it is not embedded within the modern DFA framework introduced by \cite{KaGaKl01} and subsequent studies. Furthermore, the analysis is limited to an instantaneous climate shock over a one-year horizon and does not capture the evolution of climate dynamics over time. Given the long-term nature of climate change and recent regulatory requirements to disclose its long-term financial implications \citep{IFRSS2}, a longer-term perspective is necessary. Such a perspective is also essential for strategic decisions by insurers and policymakers, including relocation planning \citep{BoWe21} and reinsurance planning \citep{MeOuFr06, MeOu10}. These considerations highlight the need for a multi-year modelling framework capable of evaluating the long-term financial impacts of climate change.}

\rev{In light of the above considerations, we develop, in this paper,} a comprehensive yet tractable ``climate-dependent DFA" framework for examining the multifaceted impacts of climate change on the general insurance market, as detailed in the following section.

\subsection{Statement of contributions} \label{Section:StatementofContributions}

In this paper, we extend the traditional DFA framework \rev{to include major climate change future drivers.} \rev{This differs from the traditional DFA approach, which largely relies on historical experiences to generate stochastic financial outcomes. It represents a paradigm shift toward a forward-looking framework that recognises future economic and physical environments may differ fundamentally from historical conditions under the impact of climate change.} \rev{While the framework is developed at the macro level, focusing on national-level projections for the general insurance sector, it provides a foundational structure that can be extended to individual insurers.} This framework \rev{therefore} serves as an initial step \rev{toward supporting} decision-making by insurers and regulators facing climate-related challenges. Specifically, \rev{our key contributions are as follows.}

\begin{itemize}
    \item \rev{\textit{Translation of climate scenarios into the balance sheets of general insurers based on tractable models:} The proposed climate-dependent DFA framework enables the translation of climate change narratives into the financial figures on general insurers' balance sheets under different emission pathways. When designing the component models within the climate-dependent DFA framework, we adopt parsimonious and tractable modelling structures to avoid unnecessary complexity that could obscure key insights. This is a desirable feature of DFA models \citep{ElPa07}, which are intended to provide transparent and interpretable simulations to support strategic decision-making. At the same time, the proposed models capture the core assumptions of climate scenarios and the main attributes of climate change impacts, such as their long-term nature and high uncertainty. To this end, we draw inspiration from the physical climate science and climate-economy literature when developing the climate, hazard, macroeconomic, and asset modules, while introducing necessary simplifications to ensure tractability and compatibility with the DFA framework. This design keeps the framework computationally manageable while capturing the main channels through which climate change affects the financial performance of the general insurance sector.}

    \item \rev{\textit{Holistic assessment of climate change impact under the DFA framework:} By systematically integrating climate change impacts into each component of the DFA model, the proposed framework captures the interdependencies between assets and liabilities and enables a holistic assessment of the financial impacts of climate change on general insurers. This integration is non-trivial, as it requires linking scenario-dependent socioeconomic narratives, climate projections, hazard dynamics, macroeconomic variables, and financial outcomes within a unified stochastic simulation framework. Specifically, climate variables under each SSP scenario are translated into hazard risks to generate catastrophe losses and insurers’ claims liabilities, while socioeconomic projections and climate damage estimates jointly inform the simulation of investment returns. Combining these asset and liability projections allows us to derive insurers’ surplus and overall market-wide financial performance. By extending the traditional DFA structure to incorporate climate dynamics across multiple channels, the climate-dependent DFA provides an integrated and forward-looking assessment framework of the financial impacts of climate change on general insurers.}

\end{itemize}

Our proposed framework is tailored to the general insurance sector by incorporating its unique features: it captures asset–liability interdependence through an interconnected structure and models the high variability of liability cash flows via DFA's stochastic simulations. It also includes catastrophe reinsurance programs and accounts for the sensitivity of reinsurance premiums to capital constraints, which are also key factors affecting insurer profitability and solvency \citep[see, e.g.,][]{MeOuFr06}. This differs from studies such as \cite{MiliNaLi25} and \cite{GuPiPl25}, which analyse climate change impacts on mortality risk, a risk type more pertinent to life insurance.

\rev{The contributions of the proposed framework are illustrated by comparing the climate-dependent DFA with a conventional, stationary DFA in the Australian context.} Australia's geography and climate make it highly exposed to hazards such as bushfires, floods, and tropical cyclones, with risks expected to intensify under high-emission scenarios \citep[Chapter 12 of][pp. 1805--1812]{IPCC2021Ch12}. These pressures may challenge insurers' ability to underwrite and remain solvent. \rev{By calibrating the framework} using Australian data on insurance losses, macroeconomic indicators, and financial markets, \rev{the results demonstrate the advantages of the climate-dependent DFA over stationary DFA approaches in capturing long-term climate trends, incorporating alternative emission pathways, and accounting for the multiple sources of uncertainty associated with future climate impacts. The case study also provides insights into how climate change may affect the Australian general insurance market under different emission scenarios, offering useful implications for insurers and regulators in managing climate-related risks.}

\subsection{Outline of the paper}

In Section \ref{Section:ModelFramework}, we introduce the modelling framework for the proposed climate-dependent DFA. Section \ref{Section:ModelOverview} then provides an overview of the structure of this framework \rev{and highlights the main extensions relative to the conventional, stationary DFA}. The design of its component modules is discussed from Section \ref{Section:ClimateHazards} through Section \ref{Section:Surplus}. \rev{Section \ref{Section:StationaryDFA} presents the benchmark stationary DFA model used for comparison.} 
In Section \ref{Section:Results}, we \rev{illustrate our framework by calibrating it to Australia. Numerical simulation outcomes are presented, analysed, and compared with those obtained from the benchmark stationary DFA model. } 
\rev{Limitations and potential directions for future research are discussed in Section \ref{Section:LimitationsandExtensions}. Section \ref{Section:Conclusions} concludes.}

\section{Modelling framework for climate-dependent DFA: A DFA framework underpinned by climate inputs} \label{Section:ModelFramework}

In this section, we begin by providing an overview of our proposed framework’s structure in Section \ref{Section:ModelOverview},  \rev{beginning with a conceptual comparison with traditional DFA frameworks}. We then present the design of the component modules within the climate-dependent DFA framework, which collectively enable users to capture the comprehensive impacts of climate change on general insurers. In particular, Section \ref{Section:ClimateHazards} introduces the climate and hazard modules, \rev{whose} outputs serve as key inputs to \rev{our proposed climate-dependent DFA model}. Subsequently, Sections \ref{Section:Assets} and \ref{Section:Liabilities} describe the assets and liabilities modules, respectively, which project future investment returns and underwriting results under each climate scenario, based on outputs from the climate and hazards modules. Section \ref{Section:Surplus} introduces the surplus module, which combines outputs from both the assets and liabilities modules, and presents key measures of general insurance financial performance. \rev{Finally, Section \ref{Section:StationaryDFA} introduces a benchmark stationary DFA model and highlights key differences with our extended frameworks. The differences between the stationary and climate-dependent DFA results are further illustrated in Section \ref{Section:Results}. To avoid breaking flow, limitations of the proposed framework are discussed throughout this section in separate remarks labelled Limitation \ref{Remark:ScenariosLimitations} to Limitation \ref{Remark:ReinsuranceMarketLimitations}.}


\begin{figure}[H]
    \centering
    \includegraphics[width= 0.7\textwidth]{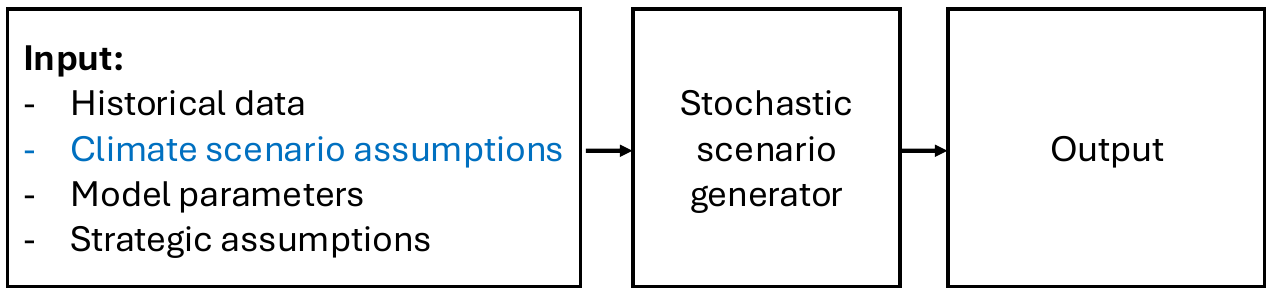}
    \caption{\rev{Main structure of the climate-dependent DFA framework. Elements consistent with traditional DFA frameworks \citep[see, e.g.,][]{KaGaKl01} are shown in black, while the key extensions introduced by the climate-dependent DFA are highlighted in blue.}}
    \label{fig:DFAMainStructureComparison}
\end{figure}

\begin{figure}[H]
    \hspace{-2cm}
    \includegraphics[width=1.2\textwidth]{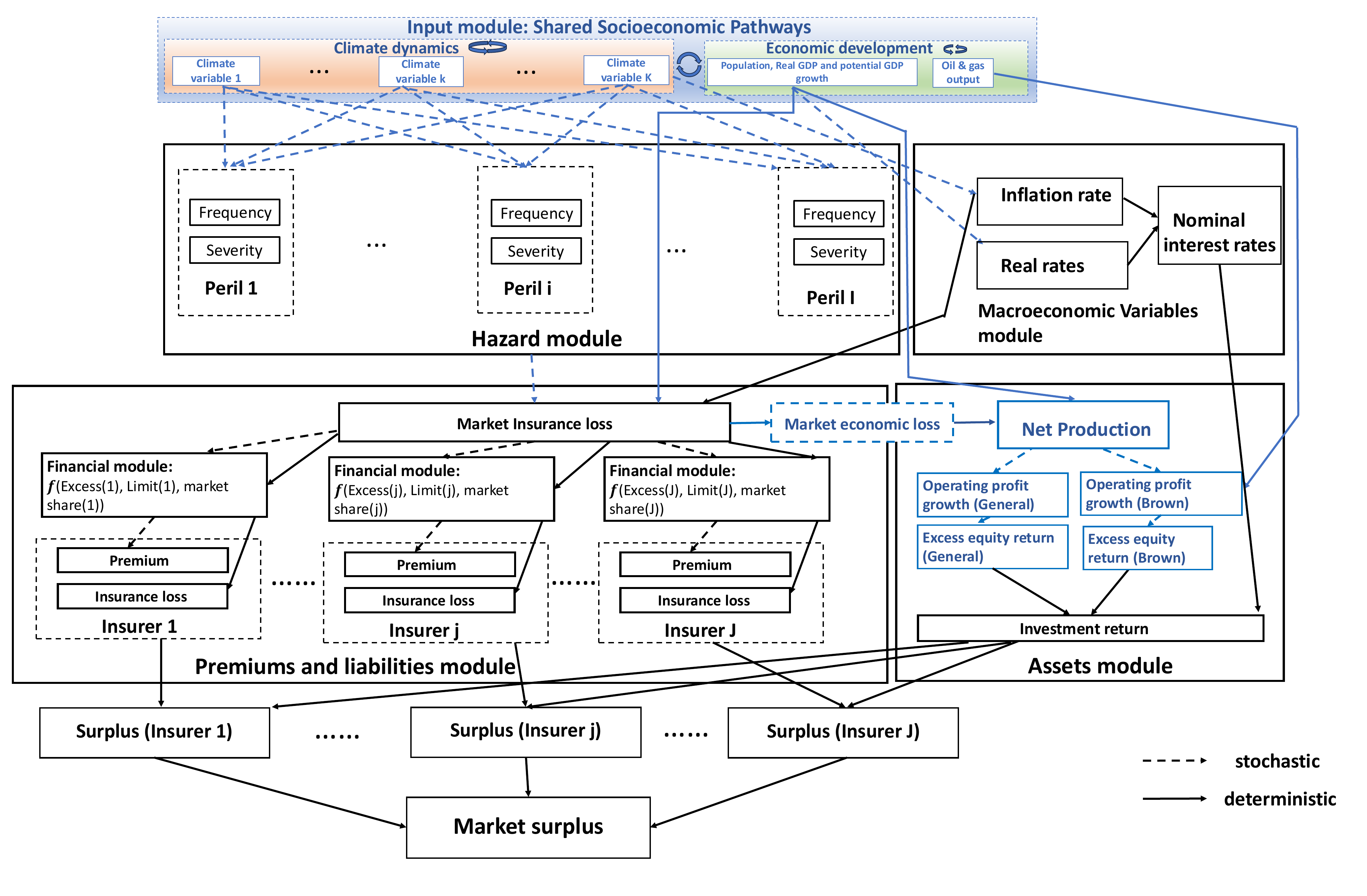}
    \caption{Modelling framework of climate-dependent DFA. \rev{Components consistent with the conventional DFA frameworks are shown in black, while elements added or modified under the climate-dependent DFA framework are highlighted in blue.}}
    \label{fig:DFAStructure}
\end{figure}

\subsection{Model overview} \label{Section:ModelOverview}

\rev{A high-level overview of the climate-dependent DFA structure, compared with the traditional DFA framework, is shown in Figure \ref{fig:DFAMainStructureComparison}. Under the traditional DFA approach, historical data (e.g., past series of interest and inflation rates), model parameters (e.g., distribution parameters of the catastrophe loss), and strategic assumptions (e.g., asset allocations) are used as inputs to a stochastic scenario generator to simulate future financial outcomes \citep{KaGaKl01}. The key extension in the climate-dependent DFA is the incorporation of climate scenario assumptions into the stochastic scenario generator, allowing the effects of climate change to be reflected in the simulated financial outcomes. More broadly, this represents a paradigm shift towards a forward-looking analytical framework, recognising that future economic and physical environments may differ fundamentally from historical conditions and may not exhibit mean-reverting behaviour.}

\rev{Building on this conceptual extension, Figure \ref{fig:DFAStructure} presents the full proposed climate-dependent DFA framework. The framework illustrates how climate scenario assumptions can be incorporated into a traditional DFA structure\footnote{\footnotesize{The underlying DFA structure largely follows that introduced in \cite{KaGaKl01}, but is adapted here to operate at the industry level.}} and highlights the key modelling components that require modification to account for climate change. Detailed specifications of these components are provided in the subsequent sections.}


\rev{A distinct feature of the climate-dependent DFA framework, compared with the traditional DFA framework, is that it begins with a set of climate scenarios as a key input. For the selection of climate scenarios,} we adopt the Shared Socioeconomic Pathways (SSPs). These SSPs form a widely adopted framework in climate research and they are central to the IPCC's climate risk assessments \citep{OnKrEbKeRiRo17}. It should be noted, however, that the SSPs represent only one of several possible scenario frameworks. In particular, it is acknowledged that the SSPs may be succeeded by the Representative Emission Pathways (REPs) in the IPCC's Seventh Assessment Report \citep{MeScBaBoBo24}, which aim to provide more policy-relevant, flexible, and up-to-date scenario representations. Such alternative scenarios \rev{could be integrated by analogy.}

Each SSP scenario is associated with a narrative, from which the economic growth rate at the technological frontier is derived \citep{DeChLa12}. Starting from historical values, country-specific GDP projections are generated under the assumption that individual economies gradually converge toward this frontier. The convergence speed is determined by the degree of trade openness, as inferred from the scenario narratives \citep{DeChLa12}. The emissions pathways consistent with the economic and environmental assumptions underlying each scenario are then used as inputs to climate models to produce projections of future climate at a much finer spatial resolution, typically at the level of gridded cells \citep{EyBoMeSeStStTa16}. 

The narratives of the selected representative climate scenarios are outlined below \citep{OnKrEbKeRiRo17}:
    \begin{itemize}
        \item SSP 2.6 (``Sustainability"): Envisions a world characterised by progressive economic development and improving environmental conditions. The combination of low physical risk and sustainable economic growth results in low challenges for both mitigation and adaptation.
        \item SSP 4.5 (``Middle of the Road") \footnote{\footnotesize{We follow the naming convention for the SSP 4.5 scenario as used in the literature \citep{OnKrEbKeRiRo17}. However, it should be noted that projections under SSP 4.5 might underestimate both emissions and temperature increases under current policies. For instance, while the best estimate of global temperature increase under SSP~4.5 is about 1.5 degree by 2030, this level of warming had already been reached by 2023 \citep{GoRaJu25}.}}: Represents a development pathway aligned roughly with typical historical trends observed over the past century, leading to moderate mitigation and adaptation challenges. 
        \item SSP 7.0 (``Regional rivalry"): Describes a world characterised by slowing economic growth and environmental degradation due to regional rivalries. Here, the combination of weak economic growth and elevated physical risk gives rise to high mitigation and adaptation challenges. 
        \item SSP 8.5 (``Taking the highway"): Describes a world with rapid economic growth driven by competitive markets and innovation. Heavy reliance on fossil fuels, however, contributes to high physical risk and consequently high mitigation challenges, though strong economic growth leads to relatively low adaptation challenges. 
    \end{itemize}

\rev{The future climate and socioeconomic projections under each SSP scenario are then used as inputs for modelling other variables within the DFA framework, following the cascading structure illustrated in Figure \ref{fig:DFAStructure}. Given the deterministic nature of physical climate models and the fact that they provide simplified representations of complex climate processes, the raw outputs from these models need to be adjusted to account for prediction bias and uncertainty before being used to simulate other variables. This step represents an additional modelling stage in the climate-dependent DFA framework compared with traditional DFA models.}


The projections of climate variables underlying each SSP scenario are used to estimate the frequency and severity of major natural hazard events in a selected country, thereby generating the market-level catastrophe insurance losses. \rev{This requires calibrating the relationships between the frequency and severity of each peril and their underlying climate drivers, as indicated by the blue dashed links in Figure \ref{fig:DFAStructure}. Such linkages are not required under the stationary setting.} On the liabilities side, these catastrophe loss estimates translate into insurance claims liabilities for general insurers, taking into account their reinsurance structures and market shares. On the assets side, socioeconomic projections under each SSP scenario, along with climate damage estimates from the hazards module, inform the simulation of investment returns. \rev{Specifically, simulated climate damages and projected economic growth under the SSP scenarios are linked to equity returns through the channel of operating profits, with the detailed modelling process discussed in Section \ref{Section:Equity}. This differs from asset models commonly used in traditional DFA or Economic Scenario Generators (ESGs) in the literature, which typically treat asset returns as a self-contained stochastic process \citep[see, e.g.,][]{ChKoWaHaLaToZh21, AlDaGo05}.} Finally, combining the resulting asset and liability forecasts allows us to derive the surplus of general insurers, representing an overall measure of market-wide financial performance.

\rev{Overall, by extending the interconnected structure of traditional DFA frameworks to incorporate climate dynamics, the proposed framework provides a holistic perspective on how climate change may affect the general insurance sector.}

\begin{limitation} \label{Remark:ScenariosLimitations}

Note that caution is warranted when interpreting the results derived from these scenarios, given the limitations of their underlying assumptions, particularly under high-emission pathways such as SSP 8.5. This scenario assumes continued economic growth without accounting for the risk of economic collapse \rev{under the impact of climate tipping points \citep{KeLeGoYi21, NeNePi25}.} Moreover, labour productivity is treated as exogenous in SSP scenarios, overlooking potential negative impacts of climate change \citep{KeLeGoYi21}. In addition, recent studies suggest that the increasing duration and intensity of extreme heat under high-emission scenarios tend to reduce life expectancy \citep{GuPiPl25, MiliNaLi25}, which could, in turn, negatively affect labour force growth. However, this potential impact is likewise not incorporated into the labour force growth assumptions in the current SSP frameworks. While SSP 8.5 also assumes strong investments in health, education, and highly engineered infrastructure \citep{OnKrEbKeRiRo17}, which may mitigate climate-related productivity and labour force losses, the extent of such mitigation remains uncertain. In addition, the SSP scenarios do not capture abrupt or disorderly transition pathways and are therefore less suitable for stress tests involving severe transition shocks to the economy. Assessing or refining these assumptions lies beyond the scope of this paper and is left for future research.
    
\end{limitation}

\begin{limitation} \label{Remark:OtherRisks}
    This paper focuses on the direct financial impacts of climate change on general insurers, as outlined in the modelling framework presented in Figure \ref{fig:DFAStructure}. While indirect effects, such as shifts in customer preference toward ``greener" insurers or the potential rise in liability risks  \citep[e.g., lawsuits against commercial policyholders for environmental damage][]{Al03Nature, Bu22} that represent transition risks on the liability side are not included here, these are important areas for future research. In addition, we do not consider unidentified risks stemming from unforeseen responses of environmental and social systems to climate change, as these currently remain entirely unknown and unquantified \citep{RiTePiSt22}. As quantitative and qualitative methodologies for assessing such risks continue to evolve, their integration into DFA frameworks may become more feasible.
\end{limitation}

\begin{remark} \label{Remark:MacrovsIndividual}
While \cite{KaGaKl01} and related studies develop DFA frameworks at the level of individual insurers, the framework introduced in this paper adopts a macro-level representation of the general insurance sector. Consequently, several insurer-specific components in conventional DFA frameworks are modified or omitted. For example, the payment patterns module in \cite{KaGaKl01} is not included in the proposed climate-dependent framework. Nevertheless, such elements could be incorporated when extending the framework to the analysis of individual insurers. Further discussion of the extensions is provided in Online Appendix \ref{Appendix:ModelExtension}.
\end{remark}

\subsection{Climate inputs} \label{Section:ClimateHazards}


To incorporate the physical risk narratives embedded in various climate scenarios, we adapt climate projections from global climate models (subject to necessary modifications) to simulate the future frequency and severity of catastrophe events via the proposed climate and hazard modules. These outputs are then translated into the financial impacts on general insurers through assets and liabilities modules in the DFA framework based on its cascading structure. This approach serves as a major addition to traditional DFA models \citep{KaGaKl01}, which typically rely on stationary hazard loss distributions derived from historical data. By integrating both historical information and scenario-based climate outlooks, our framework aims to produce forward-looking simulations of catastrophe insurance losses under the influence of climate change.

\subsubsection{Climate module} \label{Section:Climate}

The raw forecasts of climate variables (e.g., temperature, precipitation, and sea-level pressure) are derived from outputs of the Coupled Model Intercomparison Project Phase 6 (CMIP6). The findings based on the CMIP6 models, which play a crucial role in informing the IPCC Sixth Assessment Report \citep{IPCC2023}. CMIP6 comprises a set of global climate model experiments that simulate historical, present and future climate conditions under IPCC's SSP scenarios \citep{EyBoMeSeStStTa16}. CMIP6 model outputs are typically provided as gridded datasets, representing climate variables across latitude–longitude grids over time, with spatial resolutions ranging from $1^{\circ}$ to $2.5^{\circ}$. These outputs are aggregated by averaging across grid cells within defined regions, with the selection of regions for each hazard type discussed in detail in Section \ref{Section:WeatherCovariatesSelection}.

One limitation of the raw outputs from CMIP6 models is that they are deterministic in nature. As highlighted in Section~\ref{Section:StatementofContributions}, it is essential for actuarial applications, especially capital modelling, to incorporate the stochastic variability (i.e., aleatoric uncertainty) of climate forecasts. Additionally, model outputs can exhibit biases relative to observations (often due to resolution discrepancies) \citep{HaDiVrAlMe11, Ma13}. Furthermore, uncertainties can also arise from limitations in the climate models used (i.e., model uncertainty) \citep{LiRa21}. 

To address these biases and both aleatoric and model uncertainty, we adopt the following procedure for simulating future climate variables, building on the approach of \cite{LiRa21}:

\begin{enumerate} 
\item \textit{Model uncertainty:} We use an ensemble of CMIP6 models to capture differences among future climate forecasts \citep{LiRa21}. This ensemble approach acknowledges that distinct models can yield varying projections.

\item \textit{Bias correction:}  For each ensemble member, we correct bias by comparing model backcasts to historical observations via the quantile mapping method, a simple but effective technique frequently used in the literature \citep{HaDiVrAlMe11,Ma13,SaQiLiCe22}. Specifically, using a quantile mapping approach with a linear transformation function \citep{PiWeBeGoViHaHa10, QiCh21}, we estimate
\begin{equation}\label{Equation:QM}
    \hat{\theta}_{q} \;=\; \hat{\beta}_0^{(m)} \;+\; \hat{\beta}_1^{(m)} \,\hat{\theta}_{q}^{(m)},
\end{equation}
where $\hat{\theta}_{q}$ and $\hat{\theta}_{q}^{(m)}$ are the $q^\text{th}$ quantiles of the historical observations and model~$m$ backcasts, respectively, over the same reference period.

\item \textit{Aleatoric uncertainty:}  
To incorporate inherent randomness, we collect residuals $\displaystyle z^{(m)}_t = \theta_t - \hat{\theta}_{t}^{(m),*}$ by comparing the bias-corrected model backcasts $\hat{\theta}_{t}^{(m),*}$ with actual historical data $\theta_t$. We then calibrate a Normal distribution on the residuals (i.e., $z^{(m)}_t \sim \text{N}(0,\sigma^2_{(m)})$). We also acknowledge that, although this assumption is supported by normality tests  \citep[e.g., Shapiro–Wilk][]{YaYo07} for most ensemble members in our calibration data (see Online Appendix \ref{Appendix:NormalityTest}), alternative error distributions may better fit other regions. We therefore recommend validating the residual distributional assumptions when applying the method to new data, as indicated in Step 3 of the proposed flow diagram.

\item \textit{Future projections and simulations:} For each simulation path in the future projection period, we randomly select a CMIP6 model $m$ to generate a deterministic forecast $\hat{\theta}_{t}^{(m)}$. We apply the bias correction as $\hat{\theta}_{t}^{(m),*} \;=\; \hat{\beta}_0^{(m)} + \hat{\beta}_1^{(m)} \,\hat{\theta}_{t}^{(m)}$, and then draw one trajectory of residuals $\tilde{z}^{(m)}_t$ to account for aleatoric uncertainty. The final simulated climate variable is thus:
\begin{equation}\label{Equation:SimulatedVariables}
    \tilde{\theta}_{t} = \hat{\theta}_{t}^{(m),*} \;+\; \tilde{z}^{(m)}_t.
\end{equation}
\end{enumerate}

\begin{figure}[htb]
    \centering
    \includegraphics[width= 0.95\textwidth]{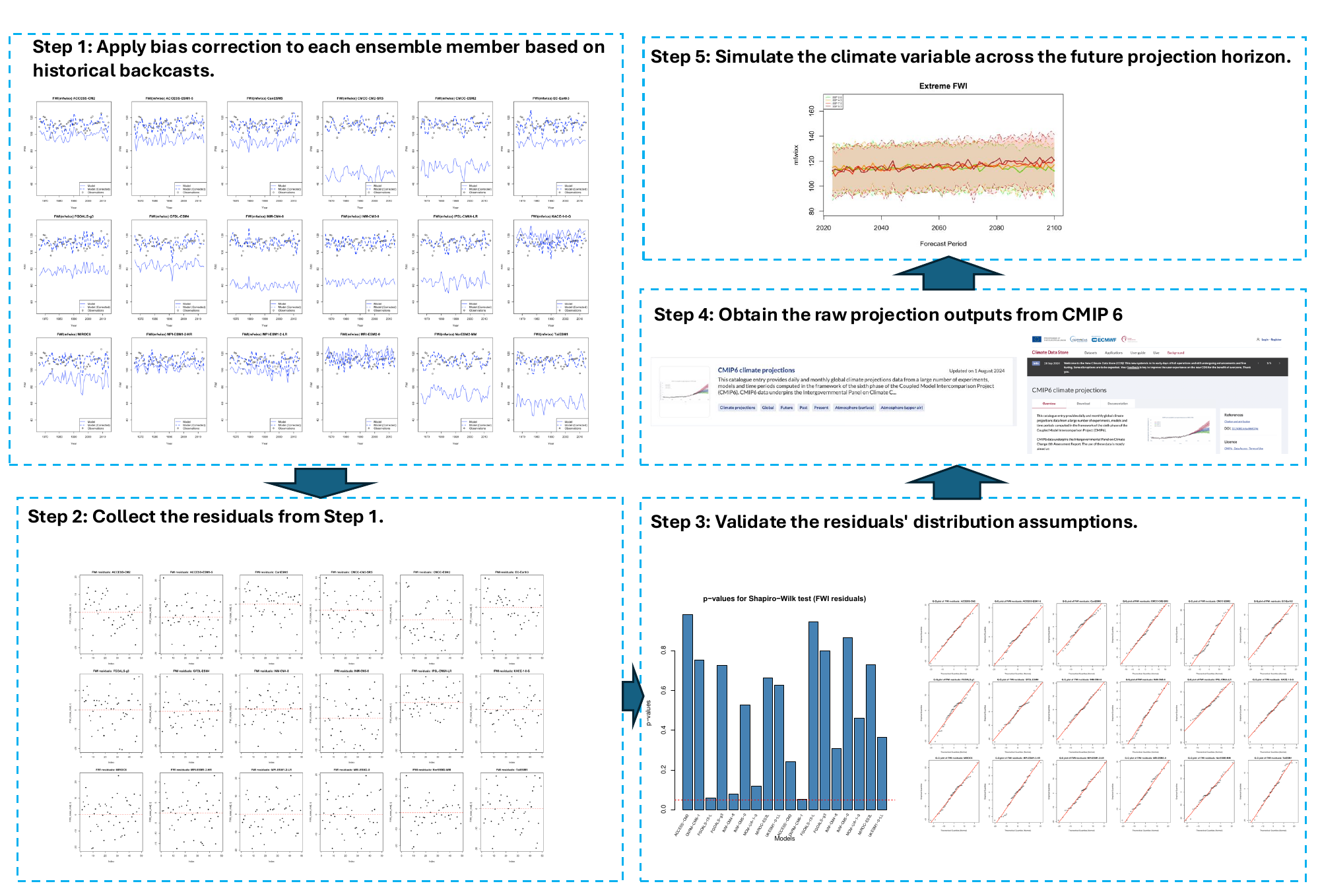}
    \caption{An illustrative diagram of climate variable simulations}
    \label{fig:IllustrativeDiagramofDFAbiasCorrection}
\end{figure}

A schematic illustration of the process described above is shown in Figure \ref{fig:IllustrativeDiagramofDFAbiasCorrection}. 

\subsubsection{Hazards module} \label{Section:Hazards}

Based on the projected climate variables from the previous module, this section forecasts the frequency and severity of natural hazards. This constitutes a critical component of the DFA model, as the resulting hazard forecasts will be employed to model the general insurance assets and liabilities in subsequent sections. Numerous approaches exist for hazard modelling in the literature; however, as discussed in Section \ref{Section:StatementofContributions}, balancing model interpretability and comprehensiveness is essential. 

At one extreme, traditional Collective Risk Models (CRMs) \citep{KlPaWi12} offer a simplistic, intuitive means of modelling aggregate insurance losses, and it is also often used in traditional DFA applications \citep{KaGaKl01}. Yet, their static assumption regarding insurance loss distributions neglects the dynamics introduced by climate change. At the other extreme, CAT models are sophisticated models that are usually capable of capturing the complex environmental process affected by climate change to generate hazard events based on advanced physical and mathematical models \citep{MiJoHiFo17}. However, these proprietary models usually have complex structures with modelling details usually not accessible by general insurers, making them less comprehensible for insurers \citep{WeJepi17}, leading to challenges in interpretability.

In light of the above considerations, we have opted for the weather-dependent CRMs \citep{HaDiVrAlMe11} for modelling insurance losses. These models combine the high interpretability of traditional CRMs with the capability to incorporate climate effects by integrating meteorological variables in the modelling of insurance loss frequency and severity. In essence, the aggregate catastrophe loss ($\tilde{X}_t$) is modelled as: 
\begin{equation} \label{Eq:AggregatLoss}
    \tilde{X}_t = \sum_{i=1}^{I} \sum_{m=1}^{M_t^{(i)}} \tilde{X}_t^{(i), m},
\end{equation}
where $M_t^{(i)}$ is the number of event of hazard type $i$ in year $t$, and $\tilde{X}_t^{(i), m}$ is the insurance loss associated with the $m^{th}$ event of hazard type $i$ in year $t$. We further assume:
\rev{
\begin{equation} \label{Eq:HazardDistclimateDFA}
    M_t^{(i)} \sim F(\kappa_1 (\bm{\Theta}^{(i)}_t), \dots, \kappa_p(\bm{\Theta}^{(i)}_t)), 
    \quad 
    X_t^{(i), m} \sim G(\vartheta_1(\bm{\Theta}^{(i)}_t), \dots, \vartheta_q(\bm{\Theta}^{(i)}_t)),
\end{equation}
}\rev{where $F$ denotes the frequency distribution with parameters $\kappa_1, \dots, \kappa_p$, and $G$ denotes the severity distribution with parameters $\vartheta_1, \dots, \vartheta_q$. In both cases, the parameters are functions of the weather covariate vector $\bm{\Theta}_t^{(i)}$ associated with hazard type $i$.} The variable $X_t^{(i),m}$ represents the normalised catastrophe loss, adjusted for both inflation and wealth exposure
\footnote{\footnotesize{The per-event normalised loss is defined as: $X_t^{(i),m} = \tilde{X}_t^{(i),m} \cdot \frac{\text{CPI}_s}{\text{CPI}_t} \cdot \frac{\text{Real GDP}_s}{\text{Real GDP}_t}$, which adjusts the nominal loss ($\tilde{X}_t^{(i)}$) for both the price level and total wealth in reference year $s$. Although our normalization technique is relatively simple, it aligns with conventional approaches in the literature \citep[see, e.g.,][]{PiLa08,VrPi09}. While more granular factors could be incorporated into normalization \citep[see, e.g.,][]{CrMc08,Pi21}, our focus on future projections rather than historical trends -- and on national-level rather than granular losses -- supports the adoption of a more parsimonious approach.}}. \rev{Typical choices for $F$ include the Poisson or Negative Binomial distributions, while $G$ is often specified as a heavy-tailed distribution (e.g., Log-Normal, Pareto, or Weibull) to capture catastrophic loss behaviour \citep{KaGaKl01}.}

We further specify:
\rev{
\begin{equation} \label{Eq:HazardDistParamsfunction}
   f(\kappa_{1,t}^{(i)}) = \bm{\beta}_{1,(i)}^{\prime}\bm{\Theta}_t^{(i)},..., f(\kappa_{p,t}^{(i)}) = \bm{\beta}_{p,(i)}^{\prime}\bm{\Theta}_t^{(i)}; \;\; g(\vartheta^{(i)}_{1, t}) = \bm{\alpha}_{1,(i)}^{\prime}\bm{\Theta}_t^{(i)},...,g(\vartheta^{(i)}_{q, t}) = \bm{\alpha}_{q,(i)}^{\prime}\bm{\Theta}_t^{(i)};
\end{equation}
}
where the set of coefficients \rev{$\{\bm{\beta}_{1,(i)},..,\bm{\beta}_{p,(i)}\}$ and $\{\bm{\alpha}_{1,(i)},...,\bm{\alpha}_{q,(i)}\}$} are estimated via regression, \rev{and $f$ and $g$ denote the link functions.}

We select the weather covariates based on the physical mechanisms driving each hazard type $i$ and validate them statistically. \rev{A detailed illustrative example of the selection of weather covariates is provided in Section~\ref{Section:WeatherCovariatesSelection}.} 

The hazard model presented in this section is designed to capture industry-level trends consistent with the scope of this paper. For more granular decision-making at the level of individual insurers (e.g., portfolio management), a higher-resolution modelling framework may be required. A discussion of how the current hazard model can be extended to regional projections and integrated with CAT model outputs is provided in Online Appendix \ref{Appendix:ModelExtension}. Nonetheless, incorporating such extensions should be carefully balanced against the trade-off between precision and interpretability depending on the intended business applications, as discussed in Section \ref{Section:StatementofContributions}.

\begin{limitation} \label{Remark:RegulatoryControl}

The hazard loss modelling in this paper does not explicitly incorporate potential government interventions. The primary aim of this work is to develop a general framework, rather than a complete predictive analysis for any specific country. As a baseline model, it can serve as a foundation for future studies to explore the potential impacts of various policy interventions.
    
\end{limitation}

\begin{limitation} \label{Remark:HistoricalRelationship}

Forecasts of hazard-related losses are typically derived from historically calibrated relationships between climate variables and observed insurance losses. However, these relationships may change, especially under the impact of tipping points \citep{NeNePi25}. Future research could improve hazard modelling by incorporating tipping point effects and conducting sensitivity analyses to account for the high degree of uncertainty in their timing, triggers, and impact magnitude \citep{Le08, No13}. 

\end{limitation}

\begin{limitation} \label{Remark:Uncertanties}

While our climate input modules account for scenario uncertainty, model uncertainty, and aleatoric uncertainty as discussed in previous sections, they do not incorporate parameter uncertainty and model inadequacy (i.e., the inherent limitations of models in fully representing real-world systems) as highlighted in \cite{RiTePiSt22}. These two sources of uncertainty are left for future research.

\end{limitation}

\subsection{Assets and macro-economic variables} \label{Section:Assets}


Beyond the modelling of hazard losses, DFA frameworks also commonly generate simulations of future macroeconomic variables and asset returns to capture the effects of changing economic and financial environments on the financial performance and position of general insurers \citep{CoDe89, KaGaKl01}. To incorporate the influence of climate change on these factors, as discussed in Section \ref{background}, we introduce our assets and macro-economic variables modules. These modules aim to account for both the physical risks and the broader economic dimensions of climate change under different scenarios. This is achieved by extending traditional modelling approaches, drawing on relevant literature to reflect the long-term impacts of climate change on financial markets and economic conditions.

\subsubsection{Inflation rates} \label{Section:InflationRates}

General inflation rates can influence both the liabilities and assets of general insurers by affecting claims inflation and nominal interest rates \citep{KaGaKl01}. Baseline inflation rates are modelled following the common approaches in literature by using the mean-reverting AR(1) process \citep[see, e.g.,][]{ChKoWaHaLaToZh21,Be22}:
\begin{equation} \label{Eq:BaseCPI}
    i_t=\mu_i+a_i(i_{t-1}-\mu_i)+\sigma_i \epsilon_{i,t},
\end{equation}
where $i_t$ denotes the inflation rate at time $t$, $\mu_i$ is the long-run mean inflation, $a_i$ is the autoregressive parameter, $\sigma_i$ is the volatility, and $\epsilon_{i,t}$ represents a standard error term.

Studies have shown that historical fluctuations in weather conditions -- such as temperature shocks and increased temperature variability -- can exert inflationary pressures on food, energy, and service prices \citep{FaPaSt21, MuOu21, CiKuHe23}. This inflationary effect ultimately contributes to general inflation. Since climate change is expected to exacerbate weather fluctuations, it is crucial to account for its impact in modelling inflation rates \citep{KoKuLiNi24}. To incorporate the influence of climate on inflation, we apply a climate overlay to the baseline inflation rates, following the methodology proposed by \cite{KoKuLiNi24}. To the best of our knowledge, the study by \cite{KoKuLiNi24} is the first to quantitatively assess and project the effects of future climate change on both food and general inflation. Specifically, the climate-adjusted inflation rate is given by:
\begin{equation} \label{Eq:ClimateOverlay}
    i_t^{\text{Clim}} = i_t + i_t^{\text{Clim-Impact}},
\end{equation}
where $i_t^{\text{Clim-Impact}}$ captures the additional inflationary effects from climate change. Following the approach in \cite{KoKuLiNi24}, the monthly climate impact on inflation is modelled as:
\begin{equation} \label{Eq:ClimateImpactonCPI}
       i_{m}^{\text{Clim-Impact}} = \sum_{L=0}^{11} (\alpha_{1+L} \Delta \bar{T}_{m-L}^{\text{NS}}+\beta_{1+L}\bar{T}_{m-L}^{\text{NS}} \cdot \Delta \bar{T}_{m-L}^{\text{NS}}),
\end{equation}
where $\bar{T}_m^{\text{NS}}$ denotes the monthly average near-surface temperature over the selected country and $\Delta \bar{T}_m^{\text{NS}}$ represents the deviation of future monthly averages from the 1990--2021 baseline. This formulation assumes a one-year lag effect. The term $\beta_{1+L},\bar{T}_{m-L}^{\text{NS}} \cdot \Delta \bar{T}_{m-L}^{\text{NS}}$ is introduced to capture the interaction effect whereby higher temperatures during hotter months lead to larger inflationary impacts \citep{FaPaSt21, KoKuLiNi24}. For future projections, the monthly average near-surface temperature will be sourced from the outputs of the climate module described in Section \ref{Section:Climate}. The annual climate impact on inflation rates for year $t$ is then obtained by summing the monthly impacts \citep{KoKuLiNi24}: $i_t^{\text{Clim-Impact}} = \sum_{m \in t} i_{m}^{\text{Clim-Impact}}$.

\subsubsection{Risk-free interest rates} \label{Section:RiskfreeRates}

Drawing inspiration from \cite{LaThWi03} and \cite{HoLaWi17}, we model the real risk-free short-term interest rate as:
\begin{equation} \label{Eq:RealRates}
    r_t = \beta_0 + \beta_1 g_t + z_t,
\end{equation}
which is closely related to the Ramsey's equation \citep{Ramsey28}, given by $r^* = \rho+\gamma g$, where $g$ denotes the growth rate of potential output, $\rho$ represents the rate of time preference, and $r^*$ denotes the natural rate of interest. A positive relationship between $r^*$ and $g$ is expected, as higher potential growth enhances future income prospects, reducing households’ incentives to save today and thereby placing upward pressure on natural rate of interest \citep{MoPo22}.

In our model, $g_t$ denotes the growth rate of potential real GDP. For calibration, these growth rates will be obtained from the \href{https://www.worldbank.org/en/research/brief/potential-growth-database}{World Bank Potential Growth Database} \citep{KiKoOhRu23}. For future projections, $g_t$ will be derived from the GDP forecasts under each SSP scenario provided in the SSP database \citep{RiVaKr17}. The residual term $z_t$ is assumed to follow an AR(1) process: 
\begin{equation} \label{Eq:ResidRealRates}
z_t = \mu_r + \phi_r (z_{t-1} - \mu_r) + \epsilon_r(t),    
\end{equation}
which captures residual factors not explained by the growth rate. The nominal risk-free rate is then derived by incorporating inflationary effects using Fisher's equation: $\tilde{r}_t = r_t + i_t^{\text{Clim}}$. 

In summary, the key inputs for this model are the real GDP growth rate $g_t$ and the climate-adjusted inflation rate $i_t^{\text{Clim}}$ (as output from the inflation model; see Section \ref{Section:InflationRates}). These inputs yield the nominal risk-free rate, $\tilde{r}_t$, as the final output.

\begin{remark}
     Choice of $g_t$: To mitigate the potential endogeneity issue, here $g_t$ is chosen as the growth rate of potential (full-capacity) GDP in the historical calibration; for future forecasts, $g_t$ will be derived from the potential real GDP forecasts underlying each SSP scenario \citep{DeChLaMa17}. Therefore, the impact of monetary policy (through manipulation of $i_t$) on $g_t$ is limited, as it mainly affects short-term output gaps.
\end{remark}

\subsubsection{Equity module} \label{Section:Equity}

We begin by considering the benchmark equity return model proposed by \cite{AlDaGo05}, which is given by:
\begin{equation} \label{Eq:ExcessEquityReturns}
    r_t^{(S)} = \tilde{r}_t+x_t,
\end{equation}
where $\tilde{r}_t$ is the nominal risk-free rates, and $x_t$ is the excess equity return. 

Under traditional DFA or ESG frameworks \citep[see, e.g.,][]{Wil95, AlDaGo05, ChKoWaHaLaToZh21, Be22}, excess equity returns are often modelled as independent stochastic processes, which is a self-contained approach that can enhance reliability in light of the considerable uncertainty surrounding exogenous variables over long-term horizons \citep{Wil95}. However, relying solely on historical data limits the capacity to capture the forward-looking climate impacts and the evolving socio-economic conditions under different scenarios. 

By contrast, factor models leverage a wide range of climate proxies -- often at a granular level -- to assess their influence on equity returns \citep[see, e.g.,][]{BaKiOc19, HoLiXu19, Go20, Ve22}. While these models are effective for empirical, in-sample analyses of individual or portfolio assets, their extensive data requirements and focus on asset-specific rather than market-level returns pose challenges for long-term projections, whereas ESG or DFA typically focuses on market-level returns.

\begin{figure}[H]
    \includegraphics[width=0.9\linewidth]{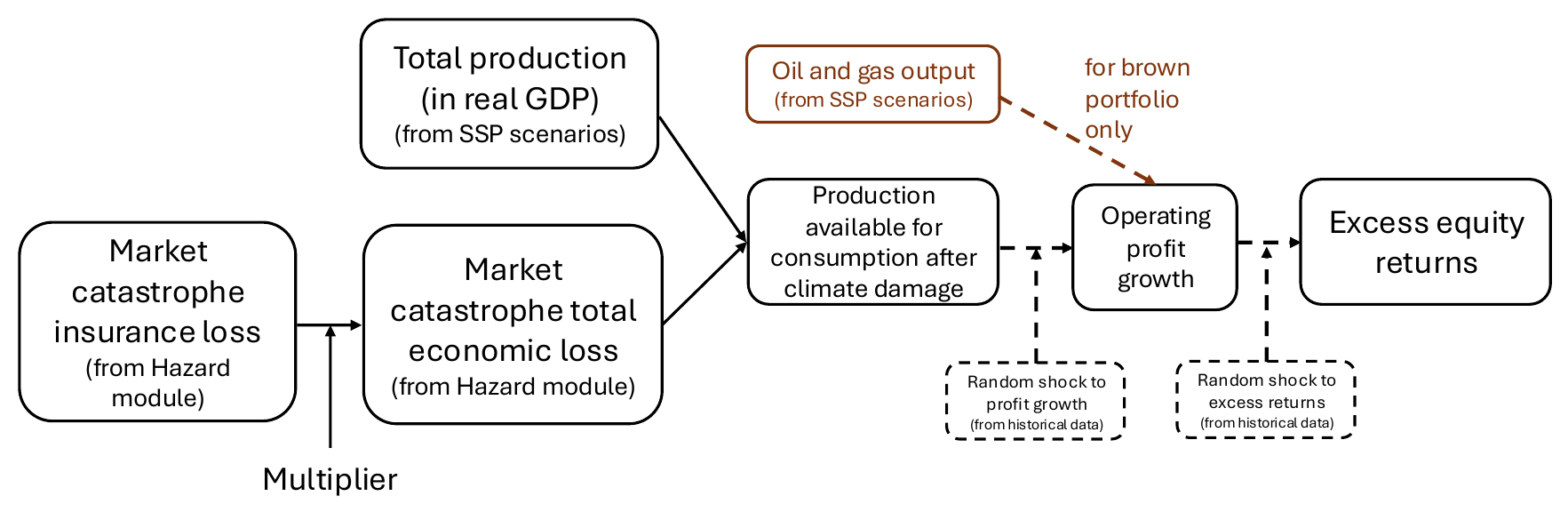}%
    \caption{Illustrative diagram showing the simulation flow for equity excess returns}
    \label{fig:DiagramforGeneralReturns}
\end{figure}

To strike a balance between these two approaches, we propose a partial self-contained framework that incorporates forward-looking climate considerations without relying on overly granular external data. Inspired by the climate-economic literature \citep[see, e.g.,][]{kaXe22, Ba23}, our method channels climate's influence on equity returns through climate-damaged consumption. A flow diagram illustrating the simulation of excess equity returns is presented in Figure \ref{fig:DiagramforGeneralReturns}.
In the first step, we obtain the simulated nominal market insurance catastrophe loss $\tilde{X_t}$ from the hazard module. Next, these insurance losses are scaled to represent uninsured economic damage, yielding $\eta \tilde{X}_t$, where $\eta$ is the multiplier. The production available for consumption after climate damage is then computed as:
\begin{equation} \label{Eq:ConsumptionSimulation}
    C_t = Y_t - \eta \tilde{X}_t,
\end{equation}
where $Y_t$ is the real GDP projection under each SSP scenario.

To transmit the effects of economic growth and climate damage to equity returns, excess total equity returns are modelled as a function of corporate earnings (based on operating profits) growth, $\Delta \text{OP}_t$, which is itself modelled as a function of consumption growth, $\Delta C_t$. Specifically, we have: 
\begin{equation} \label{Eq:EquityReturns}
        \Delta \text{OP}_t = \alpha_0 + \alpha_1 \Delta C_t + \epsilon_t^O,   \;\;\; x_t = \beta_0 + \beta_1 \Delta \text{OP}_t + \epsilon_t^{\text{x}}
\end{equation}
where the parameters are estimated through regression using historical data. The random shocks to operating profit growth and excess equity returns are represented by $\epsilon_t^{\text{x}} \sim N(0, \sigma^2_{\text{x}})$ and $\epsilon_t^O \sim N(0, \sigma^2_O)$, with their volatilities calibrated from historical observations.

For the brown sector, we further apply a transition stress overlay factor \citep{GrMa20} on its operating profit growth:
\begin{equation} \label{Eq:TransitionStress}
    \Delta \text{OP}_t^B = \Delta \text{OP}_t + \beta \Delta Y_t^B,
\end{equation}
where $\Delta Y_t^B$ represents the change in brown energy production, and $\beta$ is the sensitivity of brown firms' corporate profits to these changes. The adjusted operating profit growth for the brown portfolio is then used to simulate its excess equity returns based on \eqref{Eq:EquityReturns}.

Based on the outputs from the interest rate (Section \ref{Section:RiskfreeRates}) and equity modules, investment returns are calculated as: $r_t^I = w_{f} \tilde{r}_t + (1 - w_f)r_t^{(S)}$, where $w_{f}$ is the proportion of the portfolio allocated to risk-free assets.

In summary, the key inputs to this model are the nominal risk-free rate $\tilde{r}_t$, aggregate catastrophe losses $\tilde{X}_t$, real GDP projections, and brown energy production for each SSP scenario ($Y_t$ and $Y_t^B$), with $\tilde{r}_t$ and $\tilde{X}_t$ obtained from the respective interest rate and hazards modules. The model outputs are the equity returns for the general portfolio, $r_t^{(S,G)}$, and for the brown portfolio, $r_t^{(S,B)}$. By preserving the simplicity of traditional methods while integrating forward-looking climate damage projections, this partial self-contained approach captures key trends in evolving climate and socio-economic conditions without using high-dimensional external factors. 

\begin{limitation} \label{Remark:AssetModuleLimitations}

The equity model presented here focuses solely on domestic market investments. In practice, however, insurers often hold foreign asset exposures. This limitation is less concerning over longer horizons, as the SSP framework assumes convergence in global economic growth \citep{DeChLa12}, and market return differentials are expected to narrow over time due to arbitrage. The asset model is also intentionally simplified to align with the overall framework.

Another key assumption in the asset module is that short-term government bonds are considered free of default risk. This approach aligns with the risk-free treatment commonly adopted in conventional DFA studies \citep[see, e.g.,][]{KaGaKl01, DaGo04, CoMoViMe18} and is consistent (in the Australian context) with the Australian Prudential Regulation Authority's (APRA) Prescribed Capital Amount (PCA) framework for Australian sovereign bonds \citep{apra_gps114}. However, this assumption may warrant reconsideration in light of potential climate-induced sovereign downgrades as climate-related damages intensify \citep{KlAgBuKrMo23}. The extent and significance of climate impacts on sovereign risk, however, vary across the literature and depend on the specific countries examined \citep{CeJa22,Ma22, KlAgBuKrMo23}. Future research could incorporate default risk into the modelling of government bond returns within DFA frameworks under climate change scenarios, leveraging existing findings to assess the potential material impact on the financial performance of general insurers.

In addition, when deriving net production after accounting for climate damage, we do not explicitly incorporate the reconstruction effects of natural disasters on GDP. The direction and magnitude of these effects remain largely inconclusive in the empirical literature. \cite{AlJo93} found that, for most countries in their sample, short-term GDP growth increased following natural disasters, primarily due to the replacement of destroyed capital with more efficient ones, while the long-term growth rate remained unaffected. \cite{SkTo02} reported that climatic disasters can also exert a positive influence on long-run economic growth in addition to short-term effects. In contrast, based on cyclone impacts, \cite{HsJi14} rejected the hypothesis that disasters stimulate growth, showing that national income tends to decline and fails to recover in the short term. Similarly, \cite{No09} identified statistically significant short-term GDP losses, but with smaller magnitudes in developed economies due to their greater capacity to mobilize reconstruction resources. The theoretical framework of \cite{HaDu09} further suggests that while disasters can affect the level of production, they do not alter the long-run growth rate, and the production response depends on the quality of reconstruction. Our assumption regarding economic damage shares some slight (and limited) similarity with \cite{HaDu09} and \cite{No09}, implying a short-run decline in production but no long-term effect on economic growth. However, the explicit incorporation and testing of alternative assumptions regarding the reconstruction effects on GDP are left for future research.

Finally, similar to other bottom-up approaches \citep[e.g.,][]{WaShPu21}, the derivation of economic damage in \eqref{Eq:ConsumptionSimulation} accounts only for direct losses from acute climate risks. Future work could also integrate chronic risks into the climate-adjusted economic growth used in equity modelling, providing a more comprehensive representation of climate-related impacts.

\end{limitation}

\subsection{Liabilities and premiums} \label{Section:Liabilities}


Building on the outputs from the hazard and macroeconomic variable modules, the liabilities and premiums module introduced below aims to translate the impacts of climate change into the financial statements of general insurers.

\subsubsection{Insurance costs} \label{Section:InsuranceCost}

Drawing on the hazard module outputs and assuming an aggregate excess-of-loss reinsurance contract, the net catastrophe loss allocated to insurer $j$ is determined by:
\begin{equation} \label{Eq:NetCATLoss}
    \tilde{X}_{t,(j)}^{\text{net}} = w_j \tilde{X}_t - \text{min}((w_j \tilde{X}_{t} - d_{(j)})_+, L_{(j)}),
\end{equation}
where $\tilde{X}_t$ represents the gross CAT losses adjusted for CPI and GDP growth, $w_j$ denotes the market share of insurer $j$, and $d{(j)}$ and $L_{(j)}$ are the inflation and GDP-adjusted reinsurance excess and limit levels for insurer $j$. The second term in \eqref{Eq:NetCATLoss} represents the recoverables from reinsurers based on the excess-of-loss contract.

In addition to catastrophe losses, another key component of DFA liability modelling is non-catastrophe losses \citep{KaGaKl01}. We model non-catastrophe losses with a Tweedie distribution \citep{JoPa94}, which is a commonly used distribution assumption for modelling non-catastrophe loss. Specifically,
\begin{equation} \label{Eq:NC}
X_t^{\text{NC}} \sim \text{Tweedie}(\mu^{\text{NC}}, \phi),    
\end{equation}
where $\mu^{\text{NC}}$ is the location parameter and $\phi$ is the dispersion parameter, both calibrated using historical data. 
Using a Tweedie distribution implies that claim frequency follows a Poisson distribution, while claim severity follows a Gamma distribution, reflecting the typically high-frequency, low-severity nature of non-catastrophe losses. Some studies have investigated the influence of weather on non-catastrophe claims \citep[e.g.,][]{Mc08,HaDiVrAlMe11,SchFeFrHaHiMe13,ReGuPeReAg23}, but these typically require high-resolution daily or monthly municipal-level data, exceeding the usual granularity of DFA. Moreover, the long-term effect of weather on non-catastrophe claims is uncertain. For instance, \cite{Mc08} found that same-day precipitation increases motor claims frequency, whereas lagged precipitation decreases it possibly due to cleaner road conditions \citep{Ei04,Mc08}. Consequently, we do not explicitly incorporate climate impacts in our modelling of non-catastrophe losses, and the integration of climate covariates is left for future research as more data become available.

For projections, the aggregate non-catastrophe loss is computed as
\begin{equation} \label{Eq:nonCATloss} 
\tilde{X}_{t, (j)}^{\text{NC}} = X_t^{\text{NC}} \cdot \omega_t \cdot w_j \cdot \frac{\text{CPI}_t}{\text{CPI}_{s}}, 
\end{equation}
where $\omega_t$ is the total number of risks, $s$ denotes the reference year, and $\text{CPI}_t$ is the climate-adjusted CPI from the inflation module (see Section \ref{Section:InflationRates}). The total number of risks is modeled as a linear function of population:
\begin{equation} \label{Eq:ExposureUnits} 
\hat{\omega}_t = \hat{\omega}_0 + \hat{\omega}_1 \text{Pop}_t, 
\end{equation}
where $\text{Pop}_t$ is the projected population for each climate scenario, and the parameters $\hat{\omega}_0$ and  $\hat{\omega}_1$ are calibrated on historical data via linear regression. Under these assumptions, climate change does not directly affect non-catastrophe losses; however, it still influences them indirectly through population growth and inflation.

\begin{limitation} \label{Remark:BusinessVolume}
    The exposure growth modelling presented here assumes that business volume grows in line with population growth and does not consider the potential loss of business volume due to premium increases driven by climate change. This simplification relies on the assumption that household income growth will generally keep pace with rising premiums. While this may be less concerning under scenarios such as SSP 8.5—where both economic growth and climate risk are high—or SSP 2.6—where climate risk is low—the issue of affordability may become more significant under scenarios with weak income growth but elevated climate risk (e.g., SSP 7.0). As a potential extension, one could incorporate the influence of GDP-per-capita growth relative to premium growth on premium affordability, and its implications on insurance demand.
    
    In Australia, an initial effort to assess premium affordability in the context of climate change is being undertaken through the Insurance Climate Vulnerability Assessment project initiated by the Australian Prudential Regulation Authority (APRA) \citep{APRA23}, with results expected by the end of 2025. Future research could build on these findings (or similar such findings in other jurisdictions) to incorporate the impact of premium affordability on business volume.
\end{limitation}

\subsubsection{Insurance premiums}

Based on the distribution assumptions of catastrophe and non-catastrophe losses, the insurance premium is then calculated using the standard deviations loading principle  \citep{PaBoAe13, PaBoAeDi15, TeBoAe20}:
\begin{equation} \label{Eq:InsurersPremiumsCalc}
    \pi_{t,(j)} = \underbrace{\text{E}(\tilde{X}_{t, (j)}) + \rho \sqrt{\text{Var}(\tilde{X}_{t, (j)})}}_{\text{CAT premium}} + \underbrace{\text{E}(\tilde{X}_{t, (j)}^{\text{NC}}) + \rho \sqrt{\text{Var}(\tilde{X}_{t, (j)}^{\text{NC}})}}_{\text{Non-CAT premium}},
\end{equation}
where $\rho$ is the risk aversion parameter that reflects the level of insurer's risk aversion towards the extreme nature of the
risk \citep{PaBoAe13}. The risk aversion parameter could be selected empirically. We adopt the assumed risk aversion parameter of 0.55 in \cite{KuMiRa11} and \cite{PaBoAe13}, which is based on an empirical survey analysis conducted by \cite{KuMi11}. The implication of this assumption on the projected premiums growth will also be examined in Section \ref{Section: ResultsPremiumsUL}. 

\subsubsection{Reinsurance premiums} \label{Section:ReinsurancePremiums}

Based on the reinsurance structure specified in Section \ref{Section:InsuranceCost}, the reinsurance premium is derived as: 
\begin{equation} \label{Eq:ReinsurnacePremiumBase}
 \pi_{t,(j)}^{RI} = \text{E}[ \min((\tilde{X}_{t, (j)} - d_{(j)})_+, L_{(j)})]+ \rho \sqrt{\text{Var}(\min((\tilde{X}_{t, (j)} - d_{(j)})_+, L_{(j)}))}.   
\end{equation}
Similarly, the second term in \eqref{Eq:ReinsurnacePremiumBase} represents the surcharge on the premium above the expected value of the loss, which is dependent on the variability of the reinsurance losses. 

Reinsurers typically cover the extreme tail of insurers' risk portfolios through excess-of-loss coverage, making them particularly vulnerable to large natural catastrophes. Such events can strain reinsurance capital and trigger hard markets with higher premiums, which is a trend expected to intensify as climate change increases catastrophe losses \citep{TeBoAe20}. Capital constraint theory offers a common explanation, suggesting firms prefer to accumulate surplus through higher premiums rather than raise costly external capital \citep{Wi88, Wi94, DiGa22}.

To capture this relationship, we model reinsurance premiums as a function of reinsurance capital using a negative exponential form, inspired by \cite{Ta12}:
\begin{equation} \label{Eq:ReinsurancePremiums}
    \pi_{t,(j)}^{RI,*} = \max\left(\pi_{t,(j)}^{RI}, \pi_{t,(j)}^{RI} e^{-k_1 \cdot(S_{t-1}-S_0)}\right),
\end{equation}
where $S_{t-1}$ denotes the solvency ratio at the end of period $t-1$, $S_0$ represents the reference (or steady-state) solvency ratio, and $k_1$ is the premium-to-solvency sensitivity parameter. The solvency ratio is defined as the ratio of reinsurance capital to premium (i.e., $S_{t-1} = K_{t-1}^{\text{Re}}/\pi_{t-1}^{RI,*}$), where the reinsurance capital is derived as:
\begin{equation} \label{Eq:ReinsuranceCapital}
     K_t^{\text{Re}} =  (1 + \tilde{r}_t^{(I)})\left(K_{t-1}^{\text{Re}} + \sum_{j=1}^J \pi_{t,(j)}^{RI,*}\right) - \sum_{j=1}^J \min\left[(\tilde{X}_{t, (j)} - d_{(j)})_+, L_{(j)}\right].
\end{equation}
Here, we assume that reinsurers earn the same investment returns as direct insurers. The specification in \eqref{Eq:ReinsurancePremiums} ensures that premiums increase as the solvency ratio decreases. Furthermore, due to the concave nature of the function, a decline in capital levels results in a more pronounced increase in premiums compared to the decrease in premiums when capital levels rise. This asymmetry aligns with the assumptions in \cite{Wi94}, where insurers are assumed to be averse to the risk of bankruptcy.

\begin{limitation} \label{Remark:ReinsuranceMarketLimitations}

In modelling insurance premiums, we assume that insurance contracts are repriced annually to reflect changes in risk. While this assumption is appropriate for the scope of this paper, which is to provide a baseline framework, it should be noted that, in practice, premium adjustments may lag behind changes in risk due to underwriting cycles \citep[see, e.g.,][]{Wi94}, market competition \citep[see, e.g.,][]{Ta12}, and rate regulation \citep[see, e.g.,][]{WeTeRe10, Ta12}. Such pricing delays may further increase insolvency risk through periods of underpricing. Future research could incorporate these factors into the modelling of premium dynamics and assess their implications for the risk–return profile of general insurers under various SSP scenarios, relative to the baseline results presented in this paper.

We acknowledge that the reinsurance model presented here is a simplified representation, as it assumes a single reinsurer exclusively covering catastrophe losses in Australia and does not allow reinsurance prices to fall below their risk-based premium levels. In practice, the reinsurance market typically involves multiple participants, with reinsurers operating globally \citep{BeInsureMedia25}, offering both catastrophe and non-catastrophe coverage, and prices that can occasionally fall below their risk-based premium levels \citep{Enz02,MeOu10}. Despite its highly stylised nature, the model generally reflects the prudence that reinsurance prices are likely to rise above risk-based premiums as reinsurance capital tightens. However, it should not be interpreted as a comprehensive modelling of reinsurance market cycles due to its simplified assumption mentioned above. The potential impacts of climate change on the reinsurance cycle itself constitute an interesting topic for future research.

\end{limitation}

\subsection{Surplus and performance measures} \label{Section:Surplus}

Based on the simulated quantities from the previous module, the market surplus process is derived as \citep{KaGaKl01}: 
\begin{equation} \label{SurplusFormula}
    K_t = \sum_{j= 1}^J K_t^{(j)} = \sum_{j=1}^J (1 + \tilde{r}_t^{(I)})(K_{t-1}^{(j)} + \tilde{\pi}_t^{(j)}) - (\tilde{X}_{t,(j)}^{\text{net}} + \tilde{X}_{t, (j)}^{\text{NC}}),
\end{equation}
where $K_t^{(j)}$ represents the surplus for entity $j$ at time $t$. Market insolvency is defined as $K_t < 0$.

To determine the initial capital level $K_0^{(j)}$, the base capital requirement $\tilde{K}_0^{(j)}$ is calibrated to satisfy $\text{Pr}(K_1^{(j)} \leq 0) = 0.5\%$, consistent with the solvency standards under Solvency II \citep{ChNi14}. Recognizing that insurers typically maintain capital buffers above the minimum required capital to mitigate insolvency risk, we scale this base requirement using a target capital ratio $\tau$. Therefore, the final starting capital is thus computed as $K_0^{(j)} = \tau \tilde{K}_0^{(j)}$. 

The financial performance of general insurers is typically evaluated using both returns and risk measures. For returns, we consider the median surplus $\text{med}(K_t)$ -- the median of the surplus distribution at time $t$ -- alongside the expected surplus:
\begin{equation} \label{Eq:ExpectedSurplus}
    \text{E}(K_t) = \frac{1}{N} \sum_{n=1}^N K_t^{(n)},
\end{equation}
where $N$ is the total number of simulations. These two measures are commonly used returns metrics in DFA studies \citep[see, e.g.,][]{KaGaKl01}.

For the risk measures, we use both insolvency probability and the deficit-given-insolvency ratio. The insolvency probability, a common DFA risk metric \citep{KaGaKl01}, is calculated as the proportion of simulations yielding zero or negative capital:
\begin{equation} \label{Eq:InsolvencyProb}
    \text{P}(K_t < 0) = \frac{1}{N} \sum_{n=1}^N \mathbb{I}(K_t^{(n)} \le 0).
\end{equation}
The second measure we consider is the deficit-given-insolvency ratio, given by:
\begin{equation}
    \text{E}\left[\frac{-K_t}{L_t} \mid K_t < 0\right],
\end{equation}
where $L_t = \sum_{j= 1}^J \tilde{X}_{t,(j)}^{\text{net}} + \tilde{X}_{t, (j)}^{\text{NC}}$ represents the total claims liabilities at time $t$. This ratio measures the severity of the market deficit conditional on insolvency and is analogous to the Loss-Given-Default (LGD) metric used in reinsurance credit risk \citep{ChCuSuWe20}. Together, these risk measures provide a comprehensive assessment of both the likelihood and severity of adverse outcomes in the surplus process.

\subsection{Benchmark Model: Stationary DFA} \label{Section:StationaryDFA}

For benchmarking purposes, we consider a stationary DFA model based on the main structure of the conventional DFA framework \citep[see, e.g.,][]{KaGaKl01}. This model can be viewed as a special case of the climate-dependent DFA obtained by suppressing all climate-related components. It is calibrated using the same data sources as the climate-dependent DFA, and its simulation results are compared with those from the climate-dependent DFA in Section \ref{Section:ComparisontoStationaryResults}.

\subsubsection{Hazards Model} \label{Section:StationaryDFAHazards}

Under the stationary specification, event frequency and severity follow time-invariant distributions \citep{KaGaKl01}:
\begin{equation}
    M_t^{(i)} \sim F(\kappa_1, \dots, \kappa_p), 
    \quad 
    X_t^{(i), m} \sim G(\vartheta_1, \dots, \vartheta_q),
\end{equation}
where the parameters $\kappa_1, \dots, \kappa_p$ and $\vartheta_1, \dots, \vartheta_q$ are constant over time. 
This corresponds to the climate-dependent hazard model in Section~\ref{Section:Hazards} with all coefficients on climate variables set to zero.

\subsubsection{Inflation Rate} \label{Section:StationaryDFAInflation}

Inflation is modelled as a mean-reverting AR(1) process consistent with the baseline specification in \eqref{Eq:BaseCPI}. 
This is a special case of the climate-dependent inflation model obtained by setting the climate overlay component (i.e., $i_{m}^{\text{Clim-Impact}}$) in \eqref{Eq:ClimateOverlay} to zero.

\subsubsection{Interest Rates} \label{Section:StationaryDFAInterest}

Interest rates follow the generalised one-factor CIR model in \cite{KaGaKl01}:
\begin{equation}
    r_t = m_r + \varphi_r (r_{t-1} - m_r) + (r_{t-1})^g \epsilon_r(t),
    \quad \epsilon_r(t) \sim N(0, \sigma_r^2).
\end{equation}
For comparability, we set $g = 0$, yielding a mean-reverting AR(1) process. 
This corresponds to the climate-dependent specification in Section~\ref{Section:RiskfreeRates} when the coefficient on scenario-dependent economic growth is set to zero.

\subsubsection{Equity Model} \label{Section:EquityModel}

Excess equity returns are modelled as a stationary Normal distribution:
\begin{equation}
    x_t \sim N(\mu_{\text{x}}, \sigma_{\text{x}}^2).
\end{equation}
This is a special case of the climate-dependent equity model in Section~\ref{Section:Equity}, obtained by muting sensitivity to scenario-dependent production growth and energy variables. Implicitly, this assumes that future economic conditions follow historical patterns.

\subsubsection{Liabilities, Premiums, and Surplus} \label{Section:LiabPremiums}

Adopting a similar approach to \cite{KaGaKl01}, the exposure growth under the stationary setting is modelled as:
\begin{equation}
   \delta_t = \mu_{\omega} + \phi_{\omega} (\delta_{t-1} - \mu_{\omega}) + \epsilon_t^{\omega},
\end{equation}
where $\delta_t = \Delta \omega_t / \omega_{t-1}$ denotes the growth rate of exposure units, with its long-run mean denoted as $\hat{\mu}_{\omega}$.

Gross premiums and surplus are computed in the same manner as in the climate-dependent DFA, except that hazard distributions are assumed stationary and independent of climate variables. While the calculation of base reinsurance premiums is similar in both frameworks, the climate-dependent DFA incorporates the impact of climate change on reinsurance capital when deriving the solvency uplift applied to reinsurance premiums, whereas the stationary DFA does not. As a result, the difference in final reinsurance premiums between the two frameworks reflects not only changes in the underlying base premiums due to evolving climate risk, but also the additional solvency uplift arising from climate-related impacts on reinsurance capital.

\section{Application to the Australian general insurance market} \label{Section:Results}

After introducing the modelling framework for the climate-dependent DFA, this section presents numerical examples to demonstrate its application within the Australian general insurance market. \rev{Section \ref{Section:WeatherCovariatesSelection} discusses the selection of weather covariates and distributional assumptions for modelling hazard losses, tailored to the region under study.} Section \ref{Section:DataCalibrations} outlines the data sources and parameter calibrations used in the proposed framework; \rev{detailed descriptions of the data sources and calibration results are provided in the Online Appendix \ref{Appendix:DataSourcesandMainCalibration}.} Section \ref{Section:KeyResultsfromIndividualModels} analyses key simulation results from individual modules and discusses their potential financial implications for general insurers. Section \ref{Section:Returns} analyses and compares general insurers' financial performance under different climate scenarios using both risk and return measures. \rev{The simulation results from the benchmark stationary DFA model are incorporated into the figures presented in Sections \ref{Section:KeyResultsfromIndividualModels} and \ref{Section:Returns} to avoid duplication. A detailed comparison between the stationary and climate-dependent DFA simulation results is provided separately in Section \ref{Section:ComparisontoStationaryResults}.}

\subsection{\rev{Selection of weather covariates and distribution assumptions}}\label{Section:WeatherCovariatesSelection}

While Section \ref{Section:Hazards} presents a general framework for modelling hazard losses under the impact of climate change, the specific choice of distributions and candidate weather covariates for each hazard type must be tailored to the region under study.

Following the framework outlined in Section \ref{Section:Hazards}, we assume that the frequency and severity of catastrophe events follow:
\begin{equation}
    M_t^{(i)} \sim \text{Poi}(\lambda(\bm{\Theta}_t^{(i)})); \;\; X_t^{(i), m} \sim \text{LN}(\mu(\bm{\Theta}_t^{(i)}), \sigma^2),
\end{equation}
where $\bm{\Theta}^{(i)}_t$ is a set of weather covariates for hazard type $i$. The Poisson distribution aligns with common practice in modelling the frequency of hazard events \citep[see, e.g.,][]{JaElSa08, JaElBu11}. Similarly, Log-Normal distribution is frequently used for modelling catastrophe losses due to its heavy-tailed nature \citep{KaGaKl01, McRuEm15}, and it is also selected from a class of heavy-tailed candidates based on our data. The detailed discussion of the selected distributions based on the Australian data is provided in Online Appendix \ref{Section:CompareDistFit}. However, alternative distributions may provide a better fit for other datasets; hence, it is recommended to validate the distributional assumptions proposed here when applying the model to new data.

The distribution parameters are then modelled as: 
\begin{equation}
    \log(\lambda^{(i)}_t) = \bm{\beta}_{(i)}^{\prime}\bm{\Theta}_t^{(i)}; \;\; \mu^{(i)}_t = \bm{\alpha}_{(i)}^{\prime}\bm{\Theta}_t^{(i)},
\end{equation}
where the set of coefficients $\bm{\beta}_{(i)}$ and $\bm{\alpha}_{(i)}$ are estimated via regression. We do not explicitly adjust the scale parameter of the severity distribution in this study. However, it could be modelled as a function of climate covariates using GAMLSS regression \citep{StRi08}. This is left for future research when stronger statistical and climate science evidence becomes available on the influence of climate covariates on the scale parameter.

Here, we focus on major Australian hazards: flood, bushfire, tropical cyclones, storms, hailstorm, and East Coast Lows \citep{ICA_data}. Below, we outline the candidate weather covariates for each hazard type, drawn from relevant literature on the associated physical processes. Although the data primarily cover Australia, the selection of candidate climate variables can be generalised to other regions, except for East Coast Lows, which is a unique hazard type occurring near the eastern coast of Australia.

\paragraph{Bushfire}

Bushfire risk depends on temperature, relative humidity, drought conditions, and wind speed \citep{ShCaFoMoEvFl16, Do18, QuBaRiPa22}. Generally, fire danger increases with higher temperatures, lower humidity, stronger winds, and more severe drought. The Fire Weather Index (FWI), derived from these factors, is commonly used to assess bushfire risk \citep{Do18, QuBaRiPa22}. In particular, bushfire occurrence responds most strongly to extremes of the FWI rather than average conditions \citep{QuBaRiPa22}. Hence, we choose: 
$$\bm{\Theta}_t^{\text{(BF)}} = \{\text{fwixx}_t, \text{fwixd}_t\},$$ where $\text{fwixx}_t$ is the annual maxima of fire weather index, and $\text{fwixd}_t$ is the number of days with extreme fire weather, which are two crucial statistics capturing FWI extremes \citep{QuBaRiPa22}. The FWI data is subsequently averaged over the land surface area of the selected country to derive a national-level indicator.

\paragraph{Flood}

Precipitation, particularly extreme precipitation, is a key driver of pluvial and river floods \citep{KoBhChChGa20}. Accordingly, we choose: 
$$\bm{\Theta}_t^{\text{(FL)}} = \{R^{\text{x5}}_t, R^{\text{x1}}_t, R_t\},$$  
where $R_t$ represents the annual total precipitation, $R^{\text{x1}}_t$ is the largest one-day precipitation (usually denoted as \texttt{rx1day}), and $R^{\text{x5}}_t$ is the largest five-day cumulative precipitation (usually denoted as \texttt{rx5day}). Both $R^{\text{x1}}_t$ and $R^{\text{x5}}_t$ are commonly used as proxies for extreme precipitation in the IPCC report \citep[Chapter 11 of][pp. 1557--1561]{IPCC2021Ch11}. Similarly, the precipitation data is then averaged across the land surface area of the selected country.

\paragraph{Tropical Cyclone and storms}

The critical climate drivers for the formation and intensity of cyclones include sea surface temperature (SST) and sea level pressure \citep{MeVoBlCiLe22}. Warm ocean water is an essential condition for storm formation; additionally, low sea level pressure contributes to storm development by causing warm, moist air to rise \citep{BoM23}. Therefore, storms typically form under conditions of high sea surface temperature and low sea level pressure. When the wind speed exceeds 119\,km/h, the storm is classified as a cyclone \citep{BoM23}.

To capture the influence of sea surface temperature and sea level pressure on tropical cyclones and storms, we choose:
$$\bm{\Theta}_m^{\text{(TC)}} = \{\overline{\text{SST}}_m, \overline{\text{MSLP}}_m \},$$
where $\overline{\text{SST}}_m$ denotes the monthly average sea surface temperature over the tropical cyclone basin near the selected country (e.g., the Australian Tropical Cyclone Basin for Australian applications) \footnote{\footnotesize{Note: The cyclone basin refers to the area of tropical cyclone formation.}}, and $\overline{\text{MSLP}}_m$ is the monthly average mean sea level pressure over the same region.  The use of monthly averages for sea surface temperature and mean sea level pressure is also consistent with common practices in the literature, such as in the CAT model STORM \citep{BlHaMoMu20}.

\paragraph{Hailstorm} \label{Section:HailstormFormulation}

Atmospheric and near-surface temperatures are key drivers of hailstorm formation and intensity \citep{RaMaAlKuLa21}. Rising near-surface temperatures increase low-level moisture and convective instability, potentially boosting hail frequency \citep{AlKawa14, RaMaAlKuLa21}. Higher atmospheric temperatures contribute to greater water vapor, intensifying hailstorms, but also raise the melting level, which can reduce hail frequency by melting smaller hailstones \citep{RaMaAlKuLa21}. Thus, the overall impact varies locally. For instance, an increase in hail frequency is projected in Australia \citep{LeLeBu08, AlKawa14, RaMaAlKuLa21}.

Accordingly, we choose:
$$\bm{\Theta}_m^{\text{(H)}} = \{ \bar{T}_m^{\text{NS}}, \bar{T}_m^{\text{MT}} \},$$
where $\bar{T}_m^{\text{NS}}$ is the monthly average near-surface temperature, and $\bar{T}_m^{\text{MT}}$ is the monthly average mid-tropospheric temperature.

\paragraph{East Coast Lows}

An East Coast Low (ECL) is a type of mid-latitude cyclone that forms near the east coast of Australia, commonly referred to as the ECL identification region \citep{PeAlEvSh16,PeDiJiAlEvSh16,SpLeJoSh21}. However, the formation mechanisms of ECLs differ from those of tropical cyclones. While tropical cyclones develop over warm ocean waters, ECL formation is primarily driven by sea-surface temperature gradients \citep{PeAlEvSh16}. The ECL module presented here is specifically designed for Australian applications; for use in other countries, this module may require modification or removal.

Based on the above mechanism, we choose: 
$$\bm{\Theta}_m^{\text{(ECL)}} = \{ \Delta \overline{\text{SST}}_m, \overline{\text{SST}}_m \},$$
where $\Delta \overline{\mathrm{SST}}_m$ represents the SST gradient near the east coast of Australia, defined as the average SST anomaly in the region $24^\circ$--$41^\circ\mathrm{S}$ and $148^\circ$--$155^\circ\mathrm{E}$ relative to the SST at the same latitudes and time between $160^\circ$--$165^\circ\mathrm{E}$ \citep{PeAlEvSh16}.

\subsection{Data and calibration} \label{Section:DataCalibrations}

\rev{The climate and hazard modules are calibrated using historical weather data from ERA 5 reanalysis \citep{ERA5_data} and future projections from CMIP 6 models \citep{CMIP6_data} under various emissions scenarios.}

\rev{Catastrophe frequency and severity are calibrated using the Insurance Council of Australia (ICA) dataset \citep{ICA_data}, with insured losses normalised for the Consumer Price Index (CPI) and exposure. Hazard models are selected based on both physical relevance (see Section \ref{Section:WeatherCovariatesSelection}) and statistical performance.}

\rev{Macroeconomic models are calibrated using the Reserve Bank of Australia (RBA) cash rate data \citep{RBACashRates_data}, World Bank estimates of potential GDP growth \citep{KiKoOhRu23}, and Consumer Price Index (CPI) data from the Australian Bureau of Statistics (ABS) \citep{ABSCPI_data}. The equity models are calibrated using corporate operating profits data from the Australian Bureau of Statistics \citep{ABSBI_data} and All Ordinaries total return data \citep{FactSet}.}

\rev{Insurance market assumptions and non-catastrophe loss calibrations are informed by the Australian Prudential Regulation Authority's General Insurance Performance and Institution-Level Statistics databases \citep{APRAQuarterlyStats, APRAGIIL_data}.}

\rev{Detailed data descriptions, calibration procedures and results are provided in Online Appendix \ref{Appendix:DataSourcesandMainCalibration}. The calibration results for the stationary DFA model described in Section~\ref{Section:StationaryDFA} are provided in Online Appendix~\ref{Appendix:CalibrationStationaryDFA}, using the same data sources listed above.}

\subsection{Key simulation results from individual modules} \label{Section:KeyResultsfromIndividualModels}

\subsubsection{Climate and hazards} \label{Section: ResultsClimateHazards}

The simulated climate variables derived from the CMIP6 model outputs \citep{CMIP6_data} are presented in Appendix \ref{Appendix:SimulationResultsClimate}. Most variables show an upward trend, especially under high-emission scenarios, indicating a general increase in climate risk. The projections also exhibit notable inter-annual variability and a high degree of uncertainty, both of important drivers for the results shown in the downstream modules. 

\begin{figure}[htb]
    \centering
    \includegraphics[width= 0.8\textwidth]{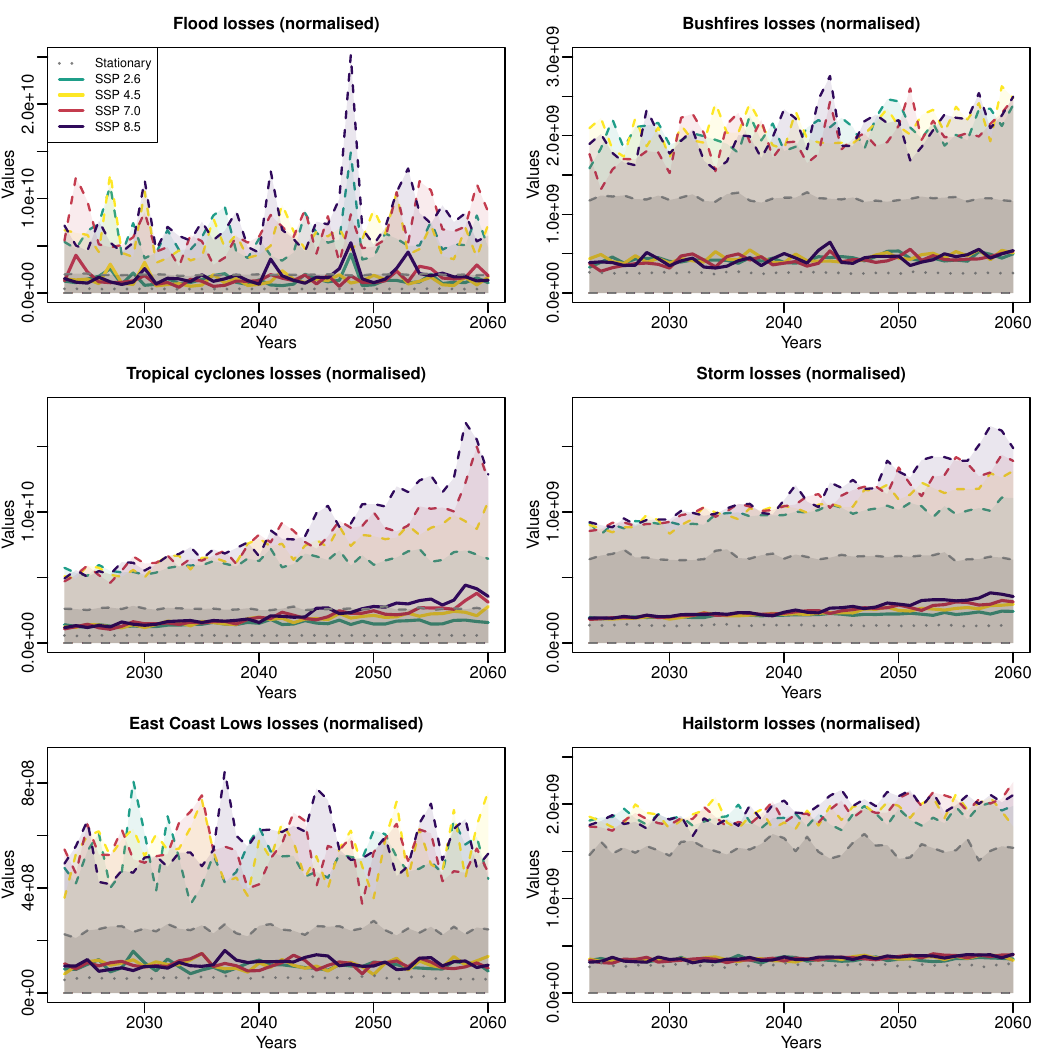}
    \caption{\textbf{Simulation results of normalised hazard losses.} Solid lines \rev{(and dotted lines under the stationary assumption)} represent the average simulation paths under different climate scenarios, while dashed lines denote the $5^{\text{th}}$ and $95^{\text{th}}$ percentiles. Results are derived from simulated climate variables and calibrated hazard models. The simulations reveal an increasing trend in both the mean and volatility of hazard losses for most hazard types under high-emission scenarios.}
    \label{fig:HazardSimulations}
\end{figure}

Based on the simulated climate variables, catastrophe losses for each hazard type are generated using the calibrated relationships between climate variables and hazard frequency and severity. Overall, the projected normalised losses exhibit an increasing trend for most hazards, particularly under high-emission scenarios, with notable increases in both the mean and the upper tails of the loss distributions. In addition, the projected losses exhibit substantial interannual variability. For example, a pronounced peak in simulated flood losses is observed around 2047, particularly at the $95^{\text{th}}$ percentile. This peak is primarily driven by the corresponding increase in extreme precipitation (see Online Appendix \ref{Appendix:FloodLossInvestigation}), which may be explained by the projected rise in the frequency of extreme ENSO events under high-emission scenarios reported in the literature \citep[see, e.g.,][]{CaGuMc15, HeFe23}. Moreover, the results reveal considerable uncertainty, driven by both the variability in simulation of climate variables as shown in Appendix \ref{Appendix:SimulationResultsClimate} and the heavy-tailed nature of catastrophe losses. The rising mean and volatility are expected to place upward pressure on both insurance and reinsurance premiums, as well as shocks to capital over time.

\begin{remark}
    
Despite the simplified structure of our hazard models, the simulation results generally align with physical risk trends reported in the literature. Projected increases in bushfire (with the upward trend more clearly illustrated in the extended simulation results presented in Online Appendix \ref{Appendix:BushfireLossInvestigation}) and flood losses are consistent with previous findings \citep[Chapter 12 of][pp. 1805--1811]{IPCC2021Ch12}, reflecting intensified extreme fire weather and increased extreme precipitation across much of Australia. 

For tropical cyclones, existing literature suggests a decline in total cyclone numbers but a rise in high-intensity events \citep[Chapter 12 of][pp. 1809--1810]{IPCC2021Ch12}. As our frequency models are based on the ICA dataset, which only includes catastrophe-level hazards, the projected increase reflects trends in severe tropical cyclones and is broadly consistent with the literature.

Regarding hailstorms, a net increase in losses is projected. This is primarily driven by the influence of near-surface temperatures on storm frequency (see Table \ref{Table:CalibrationResultsHazard}), and aligns with expected increases in hailstorm frequency \citep{LeLeBu08, AlKawa14, RaMaAlKuLa21}.

For East Coast Lows, simulations show a slight decreasing trend, driven by changes in sea-surface temperature gradients. This result is consistent with studies projecting reduced East Coast Low frequency under high-emission scenarios \citep{PeAlEvSh16, PeDiJiAlEvSh16, SpLeJoSh21}.

Our focus is on broad trends in hazard losses rather than precise magnitudes, as the analysis aims to capture industry-wide patterns. For granular decision-making—such as capital allocation or portfolio steering—insurers can use proprietary claims data to tailor hazard models to their specific exposures.

\end{remark}

\subsubsection{Investment returns} \label{Section: ResultsInvestmentReturns}

\begin{figure}[h!]
  \centering
  \subfloat[]{\includegraphics[width=0.49\textwidth]{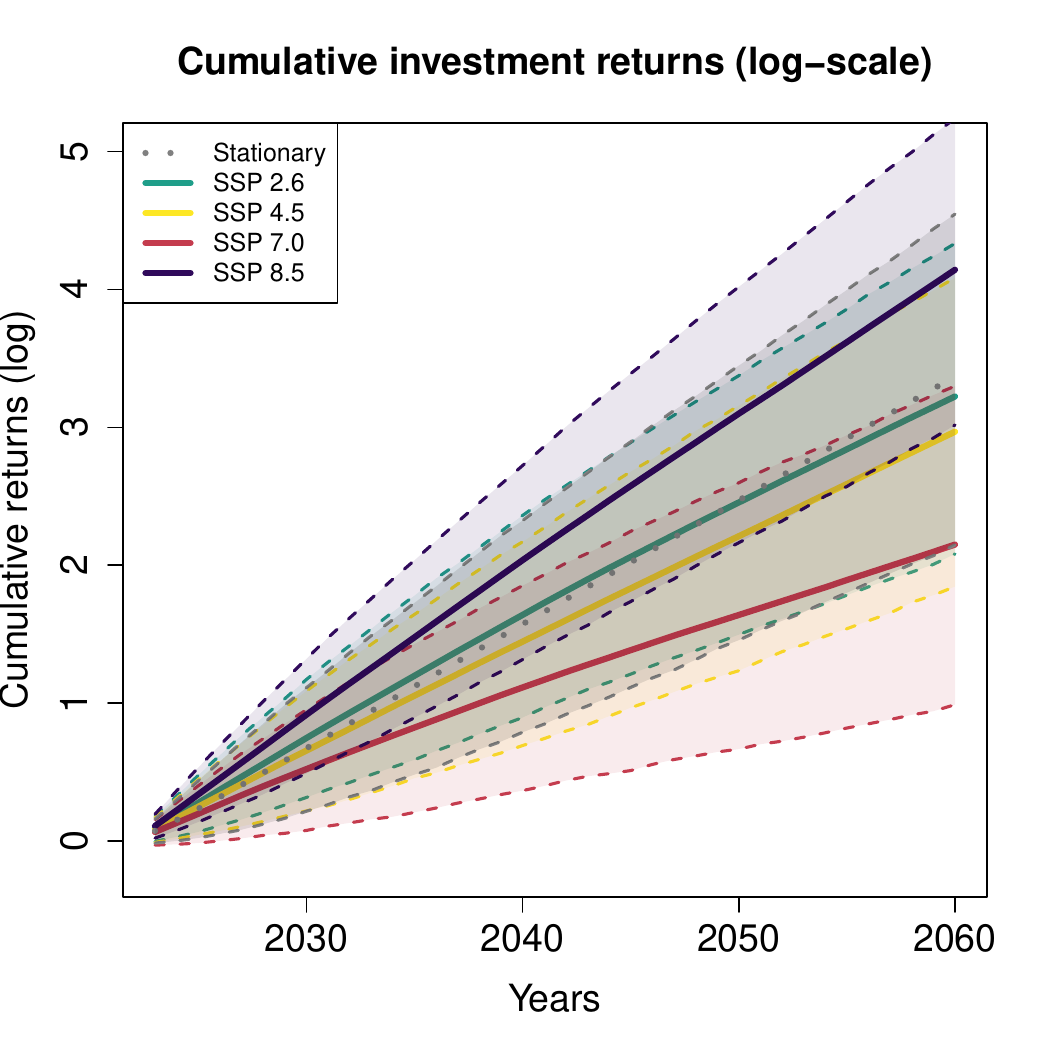}\label{fig:SimulatedCompInvReturns}}
  \subfloat[]{\includegraphics[width=0.49\textwidth]{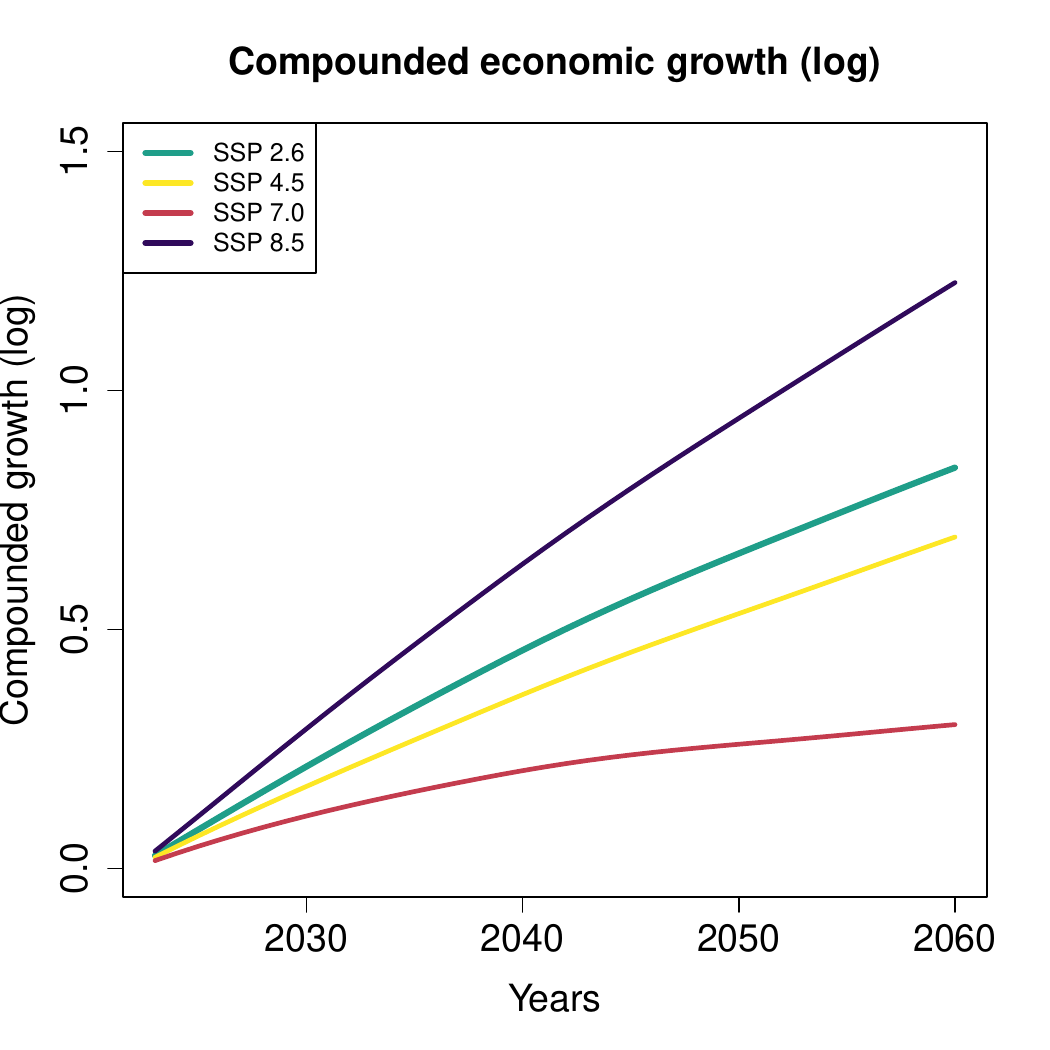}\label{fig:LogCompEconomicGrowth}}
 
  \caption{\textbf{Simulated (log) compounded investment returns \textbf{(a)} and projected compounded real GDP growth in Australia \textbf{(b)}.} Panel \textbf{(a)} shows the simulated compounded returns on the total investment portfolios generated from the economic growth assumptions and the simulated hazard losses underlying each scenario. Panel \textbf{(b)} presents compounded real GDP growth projections derived from the SSP database \citep{RiVaKr17}.}
\end{figure}

\begin{figure}[h!]
  \centering
  \subfloat[]{\includegraphics[width=0.49\textwidth]{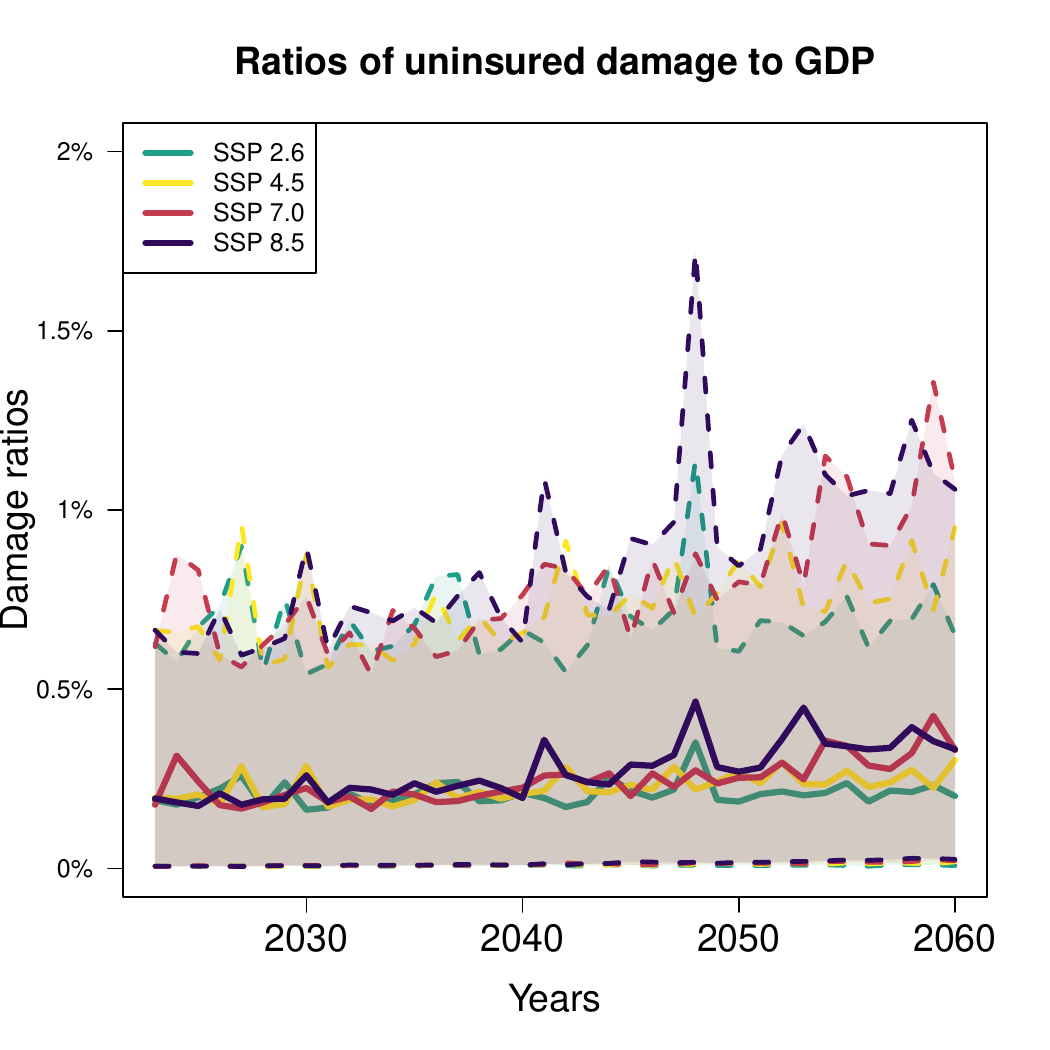}\label{fig:UninsuredDamageRatios}}
  \subfloat[]{\includegraphics[width=0.49\textwidth]{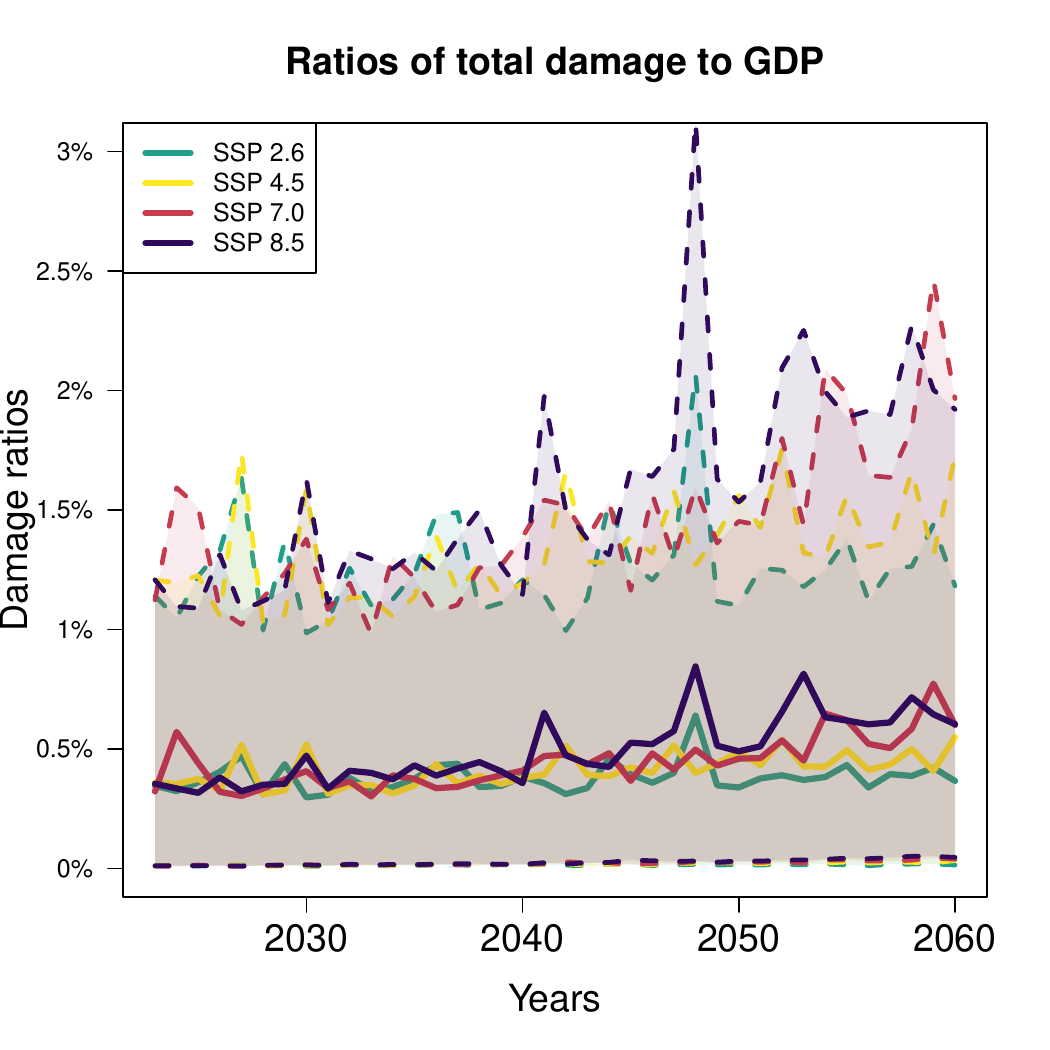}\label{fig:TotalDamageRatios}}
  \caption{\textbf{Ratios of uninsured (a) and total (b) catastrophe losses to GDP.} The results are obtained by dividing the scaled outputs from the hazard module (see Section \ref{Section:Equity}) by the corresponding GDP projections under each climate scenario.}
\end{figure}

The simulation results for the cumulative returns on the total investment portfolios, presented on a logarithmic scale, are shown in Figure \ref{fig:SimulatedCompInvReturns} \footnote{\footnotesize{Here, the log-scale cumulative investment return is defined as: $r^{\text{cum}}_t = \log\Bigl(\prod_{s=1}^t \bigl(1+r_s\bigr)\Bigr)$. This can also be written as: $r^{\text{cum}}_t = \log\Bigl(\prod_{s=1}^t \bigl(1+r_s\bigr)\Bigr)
= \log\Bigl(\frac{V_0 \prod_{s=1}^t (1+r_s)}{V_0}\Bigr)
= \log\Bigl(\frac{V_t}{V_0}\Bigr)$, where $V_t$ is the total investment value at time $t$.}}. The uncertainty bounds in the figure account for historical fluctuations in interest rates and equity returns, as well as the variability in climate-related damage to production available for consumption, as illustrated in Figure \ref{fig:UninsuredDamageRatios} and derived via \eqref{Eq:ConsumptionSimulation}.

The observed trends in cumulative investment returns are primarily driven by the economic growth assumptions underlying each SSP scenario (Figure \ref{fig:LogCompEconomicGrowth}). Under SSP 8.5, the highest cumulative investment return is projected, driven by its robust economic growth assumption \citep{OnKrEbKeRiRo17}, followed by SSP 2.6. Conversely, SSP 7.0 exhibits the weakest cumulative investment return, influenced by both its slow economic growth assumptions and high catastrophe damages, in line with its narrative \citep{OnKrEbKeRiRo17}. The relatively high investment returns under SSP 8.5 and SSP 2.6 are likely to accelerate capital accumulation. By contrast, the lower returns, coupled with high catastrophe losses in SSP 7.0 (see Figures~\ref{fig:TotalDamageRatios} and \ref{fig:HazardSimulations}), are expected to slow the pace of capital accumulation under this scenario. A sensitivity analysis using alternative economic damage assumptions has also been conducted (see Online Appendix~\ref{Appendix:SensitivityEconomicDamages}), indicates broadly consistent patterns with those reported in this section.

A breakdown analysis of investment returns for the general and brown portfolios is provided in Online Appendix~\ref{Appendix:BrownPortfolioResults}. Overall, the transition stress remains relatively mild, as the SSP scenarios considered do not assume abrupt or disorderly transitions, as noted in \rev{Limitation}~\ref{Remark:ScenariosLimitations}.

\begin{remark}
  The projected damage ratios from our climate-dependent DFA model are also compared with estimates from existing literature; a detailed discussion can be found in Online Appendix \ref{Appendix: ProjectedDamageRatiosComparison} and \ref{Appendix:SensitivityEconomicDamages}.
\end{remark}

\subsubsection{Premiums and underwriting losses} \label{Section: ResultsPremiumsUL}

\begin{figure}[H]
  \centering
  \subfloat[]{\includegraphics[width=0.49\textwidth]{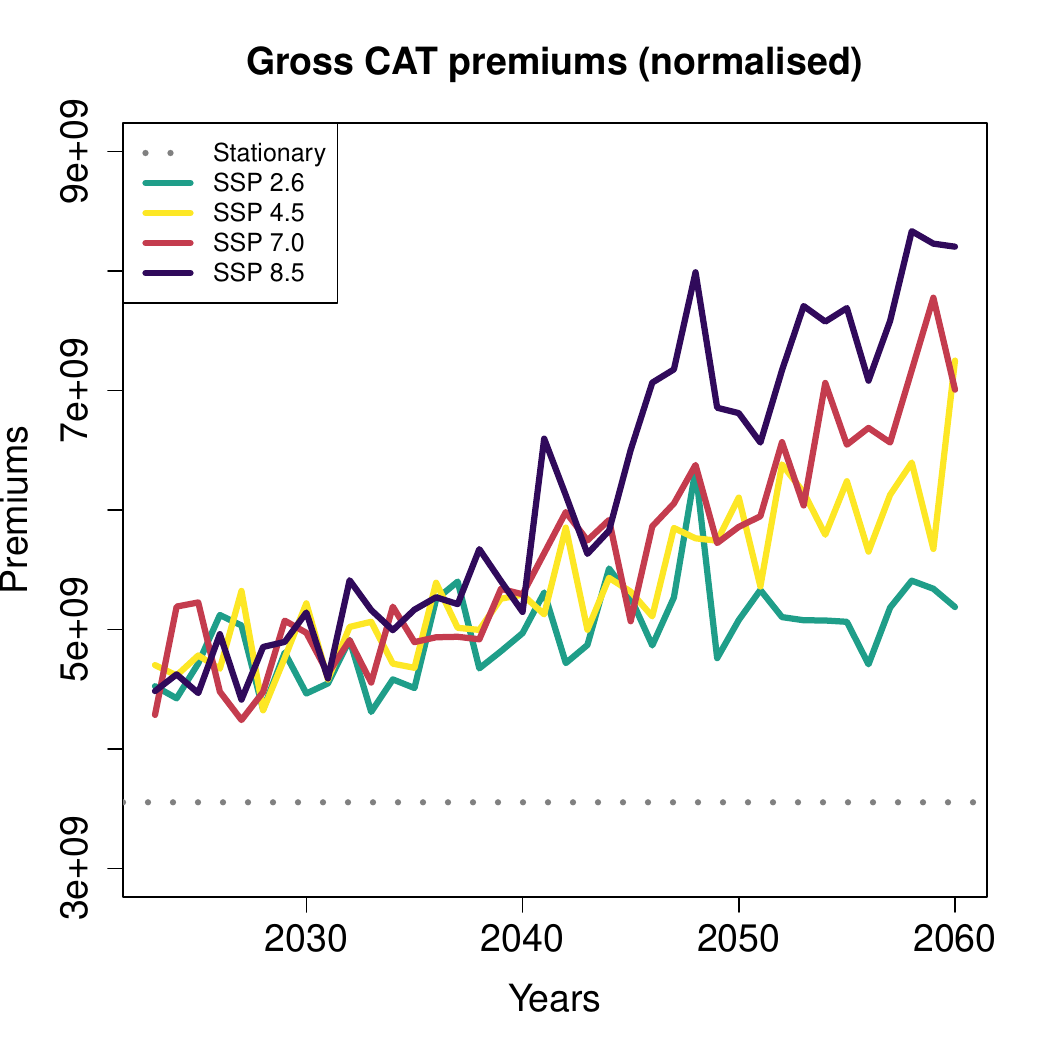}\label{fig:GrossPremiums}}
  \subfloat[]{\includegraphics[width=0.49\textwidth]{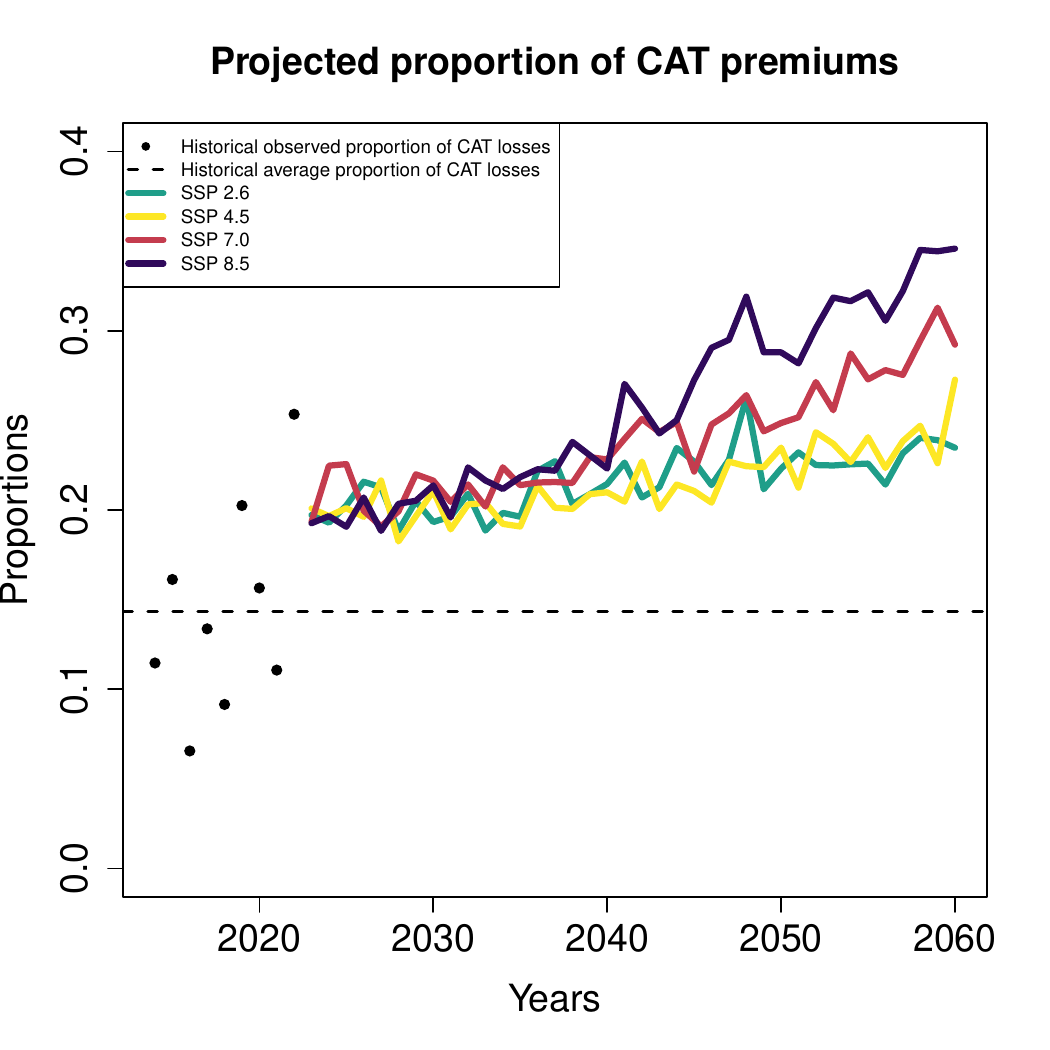}\label{fig:GrossPremiumsProp}}
  \caption{\textbf{Gross premiums (normalised) associated with catastrophe cover, shown as dollar values (a) and as a proportion of total gross premiums (b).} Panel \textbf{(a)} presents the normalised values of calculated gross catastrophe (CAT) premiums based on \eqref{Eq:InsurersPremiumsCalc}. Panel \textbf{(b)} shows the projected CAT premiums as a proportion of total premiums, compared with historically observed CAT losses as a share of total insurance losses, based on data from the APRA database and the ICA dataset. Both panels indicate an increasing trend in CAT premiums--both in dollar terms and as a share of total premiums--particularly under high-emission scenarios.}
\end{figure}

The normalized gross premiums associated with catastrophe (CAT) coverage are presented in Figure \ref{fig:GrossPremiums}. An overall increasing trend is found, driven by the rising risk of most hazards as shown in Figure \ref{fig:HazardSimulations}. Gross CAT premiums reach their highest levels under SSP 8.5, particularly in the later projection period, followed by SSP 7.0, SSP 4.5, and SSP 2.6, in line with the physical risk narratives underlying those scenarios as described in Section \ref{Section:ModelOverview}. The increase in premiums may also create affordability issues and, as noted in \rev{Limitation}~\ref{Remark:BusinessVolume}, affect insurance demand. This effect is noted as a limitation of the current paper but is encouraged for future research. In addition to this general upward trend, significant inter-annual variability is observed, reflecting the internal climate variability of the underlying climate variables as shown in Appendix\ref{Appendix:SimulationResultsClimate}. 

Moreover, the proportion of CAT premiums relative to total general insurance premiums is expected to increase relative to the historical levels, with a more pronounced increase under high-emission scenarios. This is illustrated in Figure \ref{fig:GrossPremiumsProp}, which compares projected CAT premium proportions with historically observed CAT loss proportions (used here as a proxy for historical CAT premium proportions) derived from the General Insurance Performance Database \citep{APRAGIperf_data} and the ICA dataset. These findings suggest that catastrophe losses will have an increasingly significant impact on the underwriting performance of general insurers under the influence of climate change. Consequently, managing CAT exposures will become a more critical component of portfolio management for general insurers.

\begin{figure}[h!]
  \centering
  \subfloat[]{\includegraphics[width=0.49\textwidth]{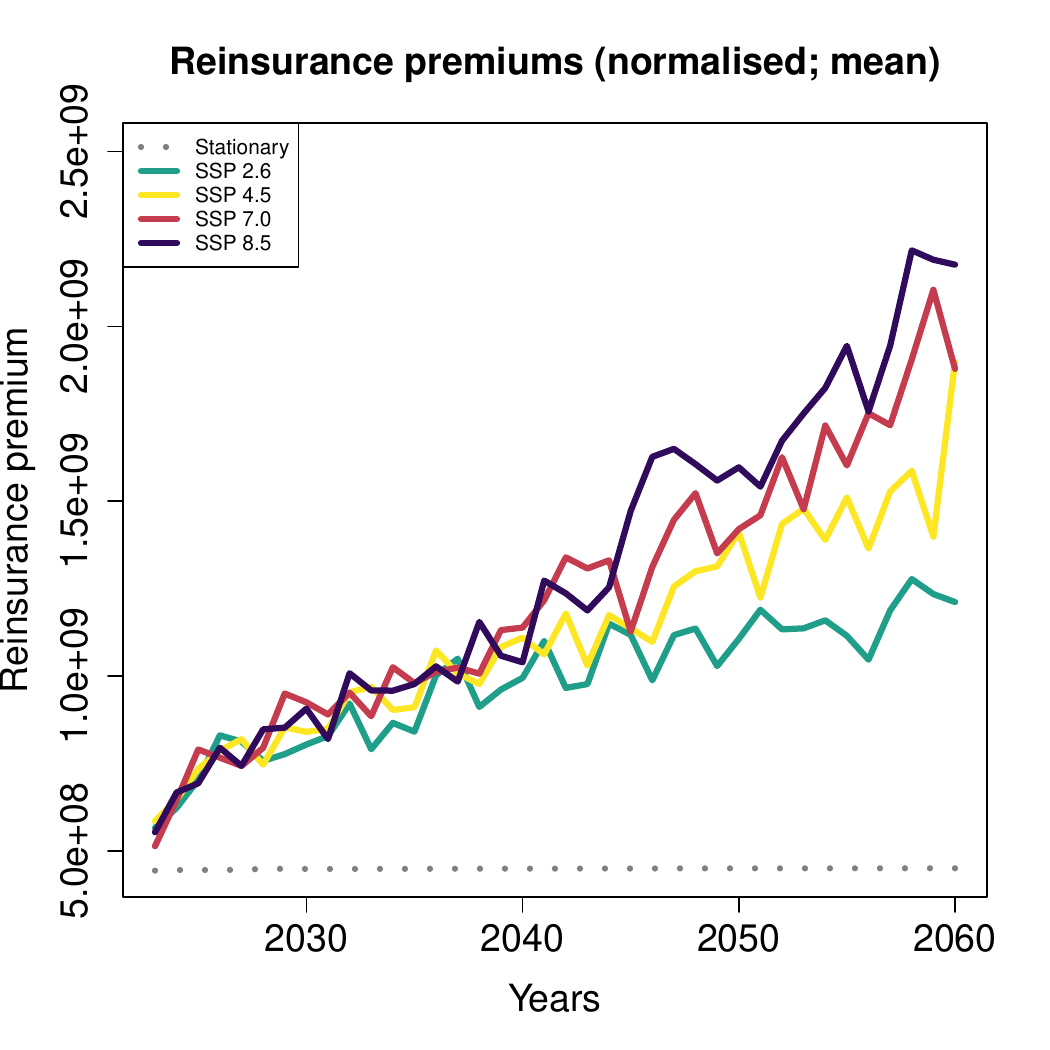}
  \label{fig:ReinsurancePremiums}}
  \subfloat[]{\includegraphics[width=0.49\textwidth]{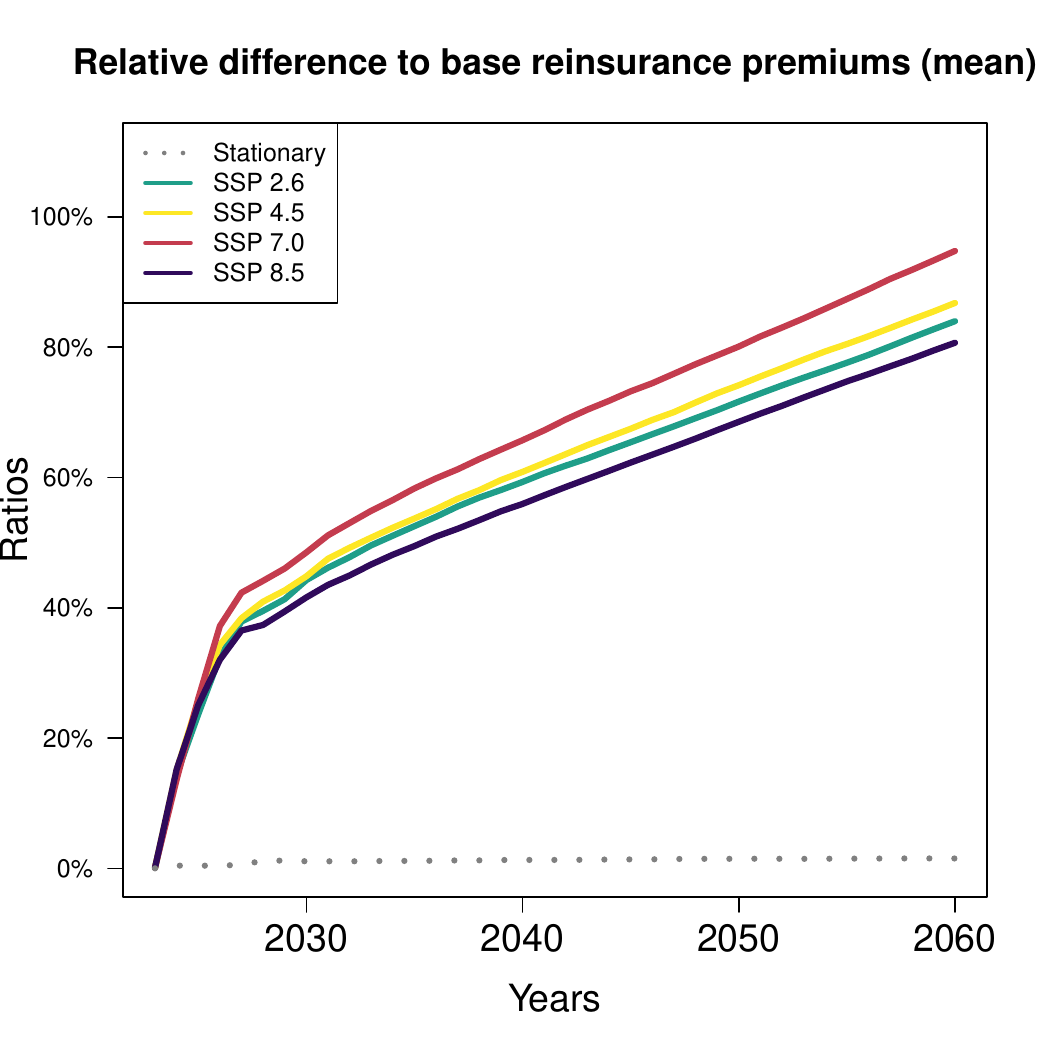}\label{fig:ReinsurancePremiumsChangeMean}}
  \caption{\textbf{Total reinsurance premiums (normalised) (a) and relative difference between total and base reinsurance premiums (b).} Panel \textbf{(a)} shows the normalised values of calculated total reinsurance premiums (including mark-up), averaged across simulation paths, based on \eqref{Eq:ReinsurancePremiums}. The results exhibit trends similar to those observed for gross CAT premiums. Panel \textbf{(b)} presents the relative difference between total reinsurance premiums and base reinsurance premiums, with the latter calculated using \eqref{Eq:ReinsurnacePremiumBase}. The largest premium uplift is projected under the SSP 7.0 scenario.}
\end{figure}

A similar pattern to the gross insurance premiums emerges in the reinsurance premiums (Figure \ref{fig:ReinsurancePremiums}), which show a general upward trend. The highest projected premium occurs under SSP 8.5, followed by SSP 7.0, while SSP 2.6 is expected to have the lowest premium. 

To illustrate the role of reinsurance capital constraints, Figure \ref{fig:ReinsurancePremiumsChangeMean} presents the average relative difference between the solvency-sensitive reinsurance premiums (in \eqref{Eq:ReinsurnacePremiumBase}) and the base premiums (in \eqref{Eq:ReinsurancePremiums}). The largest uplift appears under SSP 7.0, likely due to low investment returns coupled with high insurance losses (Sections \ref{Section: ResultsClimateHazards} and \ref{Section: ResultsInvestmentReturns}), resulting in slower capital accumulation and prolonged tight capacity. In contrast, SSP 8.5 exhibits the smallest uplift despite experiencing the highest catastrophe (CAT) losses, possibly because its stronger investment returns (see Section \ref{Section: ResultsInvestmentReturns}) accelerate capital accumulation and shorten tight-capacity periods. A similar pattern is also observed in the primary general insurance market (see Section \ref{Section:Returns}).


The results also suggest that the percentage growth in reinsurance premiums exceeds that of primary insurance, reflecting both rising physical risks and adjustments to tightening reinsurance capital. The modelled uplift relative to the base level is considered as conservative due to the simplified assumptions outlined in \rev{Limitation} \ref{Remark:ReinsuranceMarketLimitations}. Nevertheless, it is assumed that primary insurers will continue to purchase reinsurance despite higher prices; however, rising premiums may dampen demand, although this effect could be partly offset by the strengthening risk-transfer demand as climate risk intensifies \citep{Lloyds2025ClimateTrends}. The dynamics of these effects are left for future research.

\begin{figure}[h!]
\begin{minipage}{0.65\textwidth}
\includegraphics[width=0.85\linewidth]{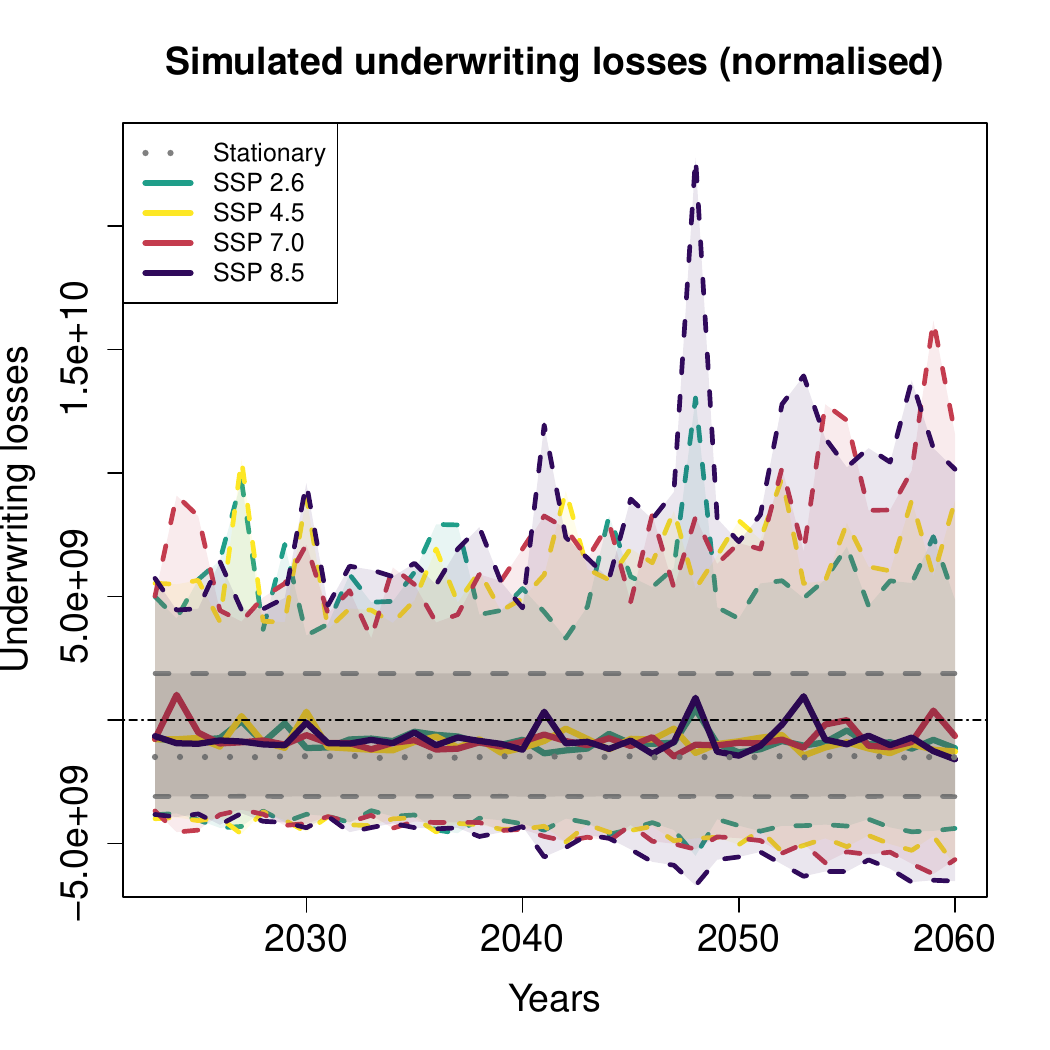} 
\end{minipage}
    \begin{minipage}{0.35\textwidth}
        \captionof{figure}{\textbf{Simulations of underwriting losses (normalised).} Underwriting losses are calculated by subtracting the net premium received (i.e., gross premium received minus reinsurance premiums paid) from the net simulated insurance losses (i.e., gross insurance losses net of reinsurance recoverables). Solid lines \rev{(and dotted lines under the stationary assumption)} represent the average simulation paths across climate scenarios, while dashed lines indicate the $5^{\text{th}}$ and $95^{\text{th}}$ percentiles. Across climate scenarios, the results are similar at the mean level but diverge at higher quantiles, with greater losses projected under high-emission scenarios.}
\label{fig:UnderwritingLossSimulations}
\end{minipage}
\end{figure}

Using the calculated premiums and simulated catastrophe losses from the hazard module (Section \ref{Section: ResultsClimateHazards}), we derive the simulated underwriting losses, with normalized results presented in Figure \ref{fig:UnderwritingLossSimulations}. On average, the simulated underwriting loss remains below zero (indicating a positive underwriting profit) and is relatively consistent across scenarios, which could be explained by the premium loadings. At higher quantiles, however, underwriting losses remain similar across scenarios initially but rise substantially in later periods under high-emission pathways (SSP 8.5 and SSP 7.0) due to escalating climate risks. This upward shift in the tail of underwriting losses can be attributed to the increasing volatility and extreme percentile of hazard losses observed under high-emission scenarios (see Section \ref{Section: ResultsClimateHazards}). Consequently, while downside liability impacts on financial performance appear moderate at first, they are expected to intensify over longer projection horizons in the high-emission scenarios.


\subsection{Risk and returns measures} \label{Section:Returns}

\begin{figure}[h!]
  \centering
  \subfloat[]{\includegraphics[width=0.49\textwidth]{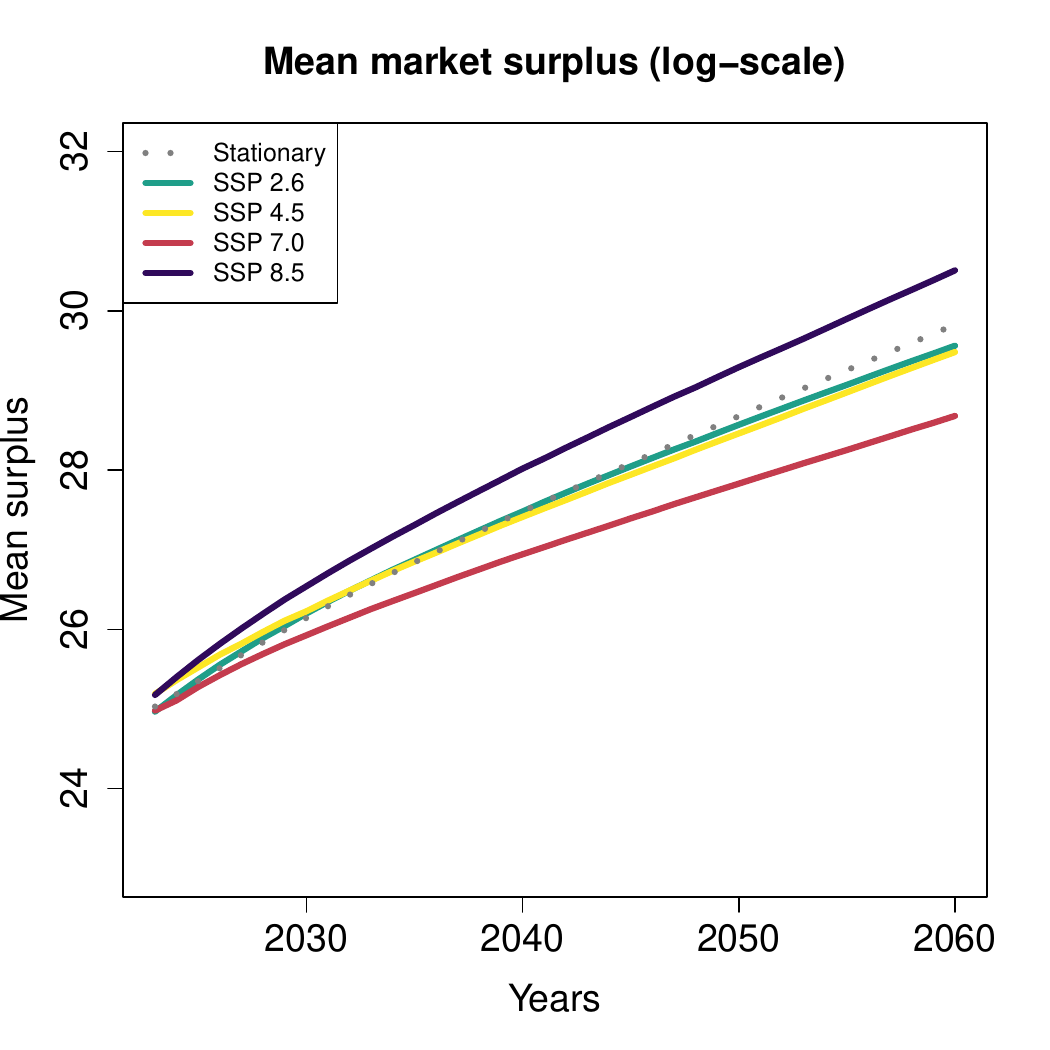}\label{fig:ExpectedMarketSurplus}}
  \subfloat[]{\includegraphics[width=0.49\textwidth]{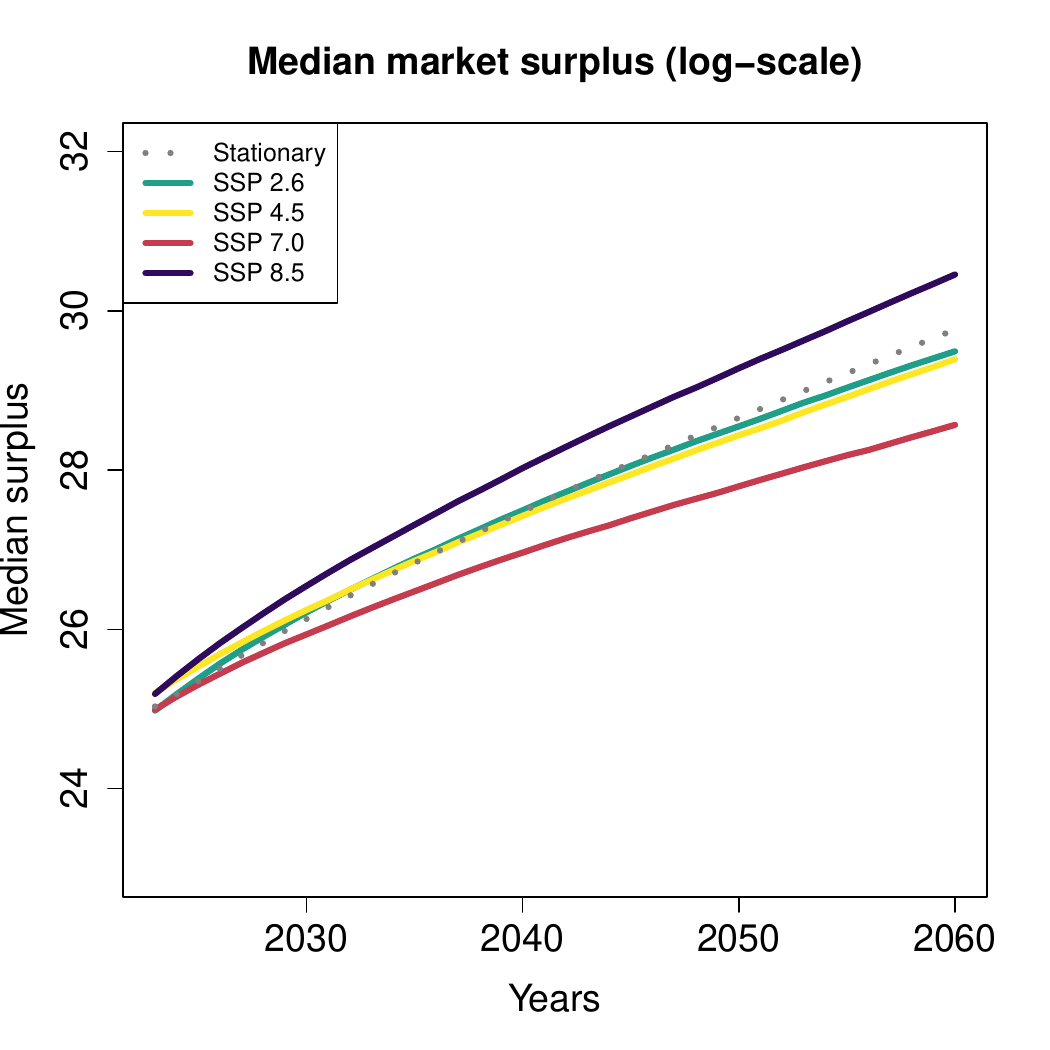}\label{fig:MedianMarketSurplus}}
  \caption{\textbf{Expected (a) and median (b) market surplus (log scale) derived from surplus simulations.} Both the expected and median surplus trajectories reflect the economic growth assumptions underlying each climate scenario.}
\end{figure}

Finally, the risk and return measures for the general insurance market can be derived from the surplus, which is calculated using outputs from the individual modules (see \eqref{SurplusFormula}). Figures \ref{fig:ExpectedMarketSurplus} and \ref{fig:MedianMarketSurplus} show the expected and median surplus, which are our return measures, across different SSP scenarios. Under SSP 8.5, surplus is highest, followed by SSP 2.6, whereas SSP 7.0 yields the lowest surplus.

To investigate the drivers of these surplus trends, we compare the expected and median surplus with cumulative investment returns at their average and median paths, revealing that differences in cumulative investment returns largely explain the observed surplus patterns (see Appendix \ref{Appendix:SurplusInvestment}). These findings align with earlier results showing that underwriting profits or losses are similar at mean levels across scenarios (Section \ref{Section: ResultsPremiumsUL}), making investment growth the primary driver of mean and median surplus trends. Since investment returns are predominantly affected by the economic growth assumptions underlying each scenario (Section \ref{Section: ResultsInvestmentReturns}), the divergent economic growth paths in different SSP scenarios are expected to be the key drivers of expected returns to the general insurance industry.

\begin{figure}[!h]
  \centering
  \subfloat[]{\includegraphics[width=0.49\textwidth]{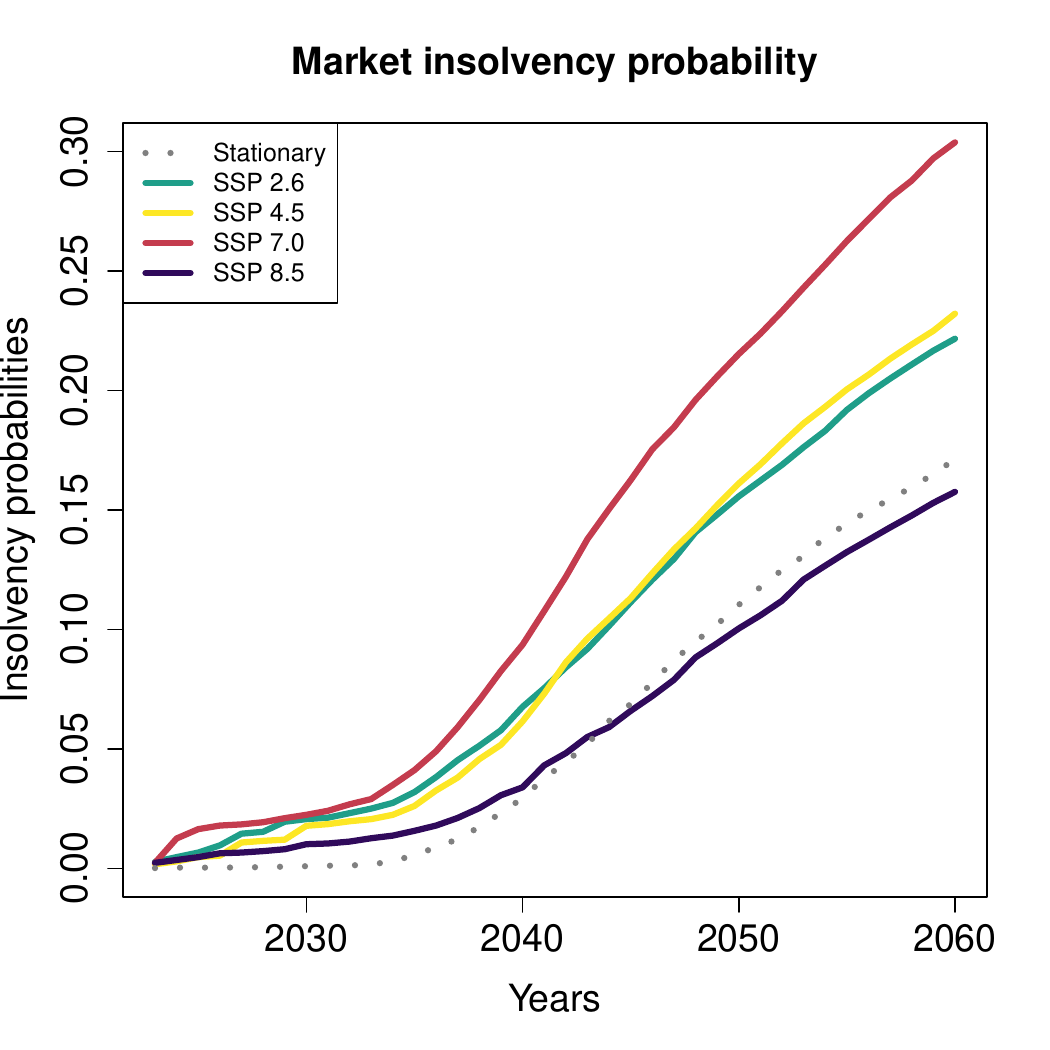}\label{fig:MarketInsolvencyProb}}
  \subfloat[]{\includegraphics[width=0.49\textwidth]{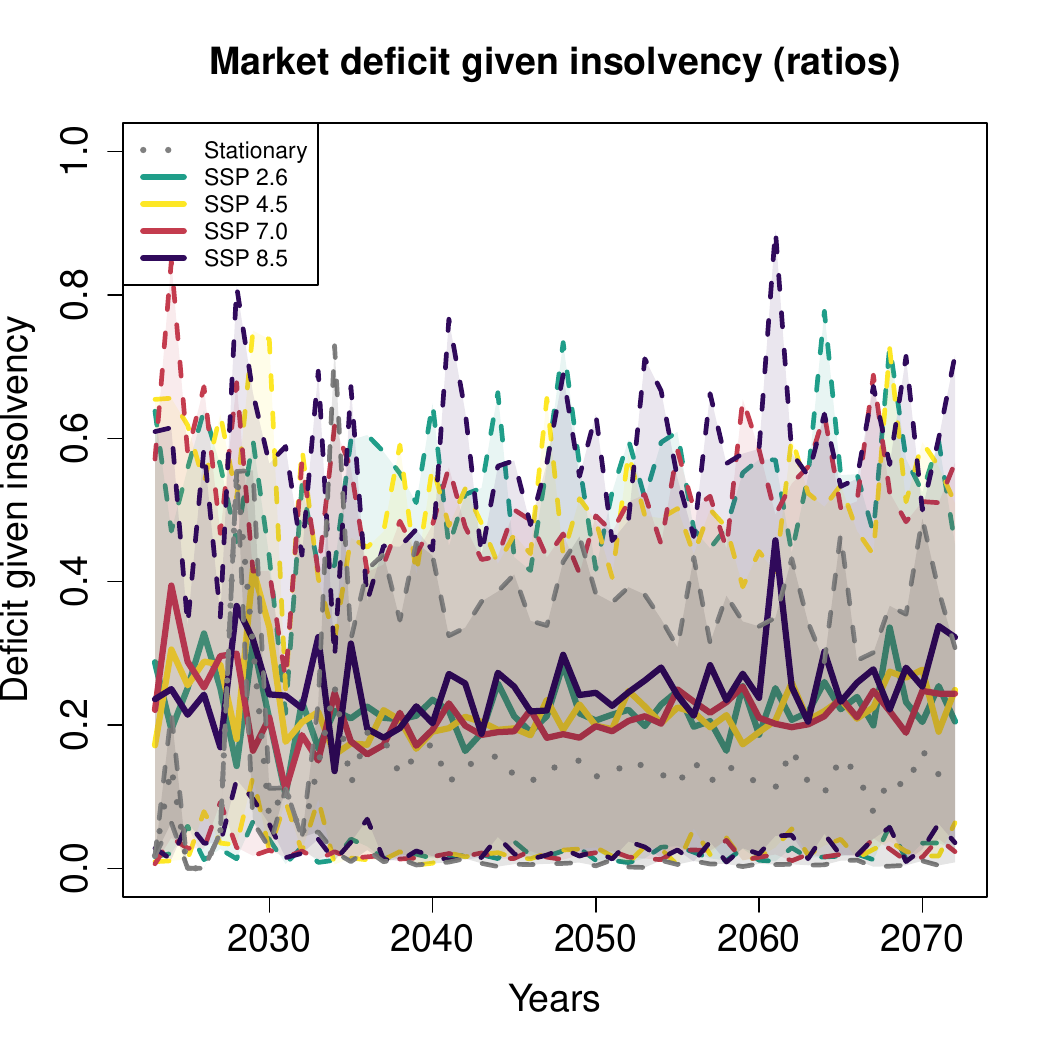}\label{fig:MarketLGD}}
  \caption{\textbf{Market insolvency probabilities (a) and market deficit-given-insolvency ratios \protect\footnotemark (b).} Panel \textbf{(a)} shows the proportion of simulations in which total market capital becomes negative, while Panel \textbf{(b)} presents the ratio of market deficit to total claims liabilities, conditional on market insolvency. Solid lines \rev{(and dotted lines under the stationary assumption)} represent the average simulation paths across climate scenarios, and dashed lines indicate the $5^{\text{th}}$ and $95^{\text{th}}$ percentiles. The SSP 8.5 scenario exhibits the lowest market insolvency probability but the highest market deficit-given-insolvency ratio.}
\end{figure}

Figure \ref{fig:MarketInsolvencyProb} shows market insolvency probabilities under various scenarios, a common risk metric in DFA studies \citep{KaGaKl01}. SSP 7.0 exhibits the highest insolvency rates, followed by SSP 4.5 and SSP 2.6. Although SSP 7.0 does not incur the highest hazard losses, its poor investment returns (Section \ref{Section: ResultsInvestmentReturns}), substantial reinsurance premium increases (Section \ref{Section: ResultsPremiumsUL}), and relatively high catastrophe (CAT) losses (Section \ref{Section: ResultsClimateHazards}) collectively erode profits and constrain capital accumulation, leading to high insolvency probabilities once both physical and economic aspects of the climate change are considered. These outcomes align with the  physical and economic narratives underlying the SSP 7.0 scenario (Section \ref{Section:ModelOverview}).
 
Interestingly, while SSP 8.5 is typically linked to the highest physical risks in climate science \citep[Chapter 11 of][pp. 1517--1605]{IPCC2021Ch11}, it yields the lowest projected insolvency probabilities when both physical and economic factors are modelled. This outcome stems from strong investment returns outpacing underwriting losses early in the projection horizon (see Figures \ref{fig:SimulatedCompInvReturns} and \ref{fig:UnderwritingLossSimulations}). As a result, capital accumulates rapidly, particularly in the early years. SSP 8.5 also shows the highest average Compound Annual Growth Rate (CAGR) of market surplus \footnote{\footnotesize{The Compound Annual Growth Rate (CAGR) of the insurance surplus over the projection horizon $T_h$ is given by: $\text{CAGR}_h = \left(\frac{K_{T_h}}{K_0}\right)^{\frac{1}{T_h}} - 1$. This measure, commonly used in finance to assess the average growth rate of an investment portfolio \citep{Gr23}, is employed here to evaluate the rate of capital accumulation across different projection horizons.}}(see Appendix \ref{Appendix:MeanGACR}), which could help insurers absorb potential losses at later stages. These findings align with the SSP 8.5 narrative of ``robust economic growth", which leads to ``low adaptation challenges" except in extreme case \citep{OnKrEbKeRiRo17}.

However, when insolvency occurs, SSP 8.5 results in the most severe impacts in later projection horizons, as shown by the market deficit-given-insolvency ratios (see Figure \ref{fig:MarketLGD}). These ratios, which indicate the share of unpaid claims, suggest policyholders tend to face the greatest losses under SSP 8.5 in tail events. This is likely due to higher underwriting losses in the distribution tails later in the projection (Section \ref{Section: ResultsPremiumsUL}; Figure \ref{fig:UnderwritingLossSimulations}), despite similar early-horizon losses across scenarios. Additionally, given the limitations of SSP 8.5 (Section \ref{Section:ModelOverview}), insurers should remain cautious about unmodeled risks such as economic collapse from climate tipping points.

SSP 2.6 offers a more balanced risk-return profile, with the second-highest expected surplus, second-lowest insolvency risk, and modest market deficits in insolvency. It also features the lowest projected catastrophe-related premiums, potentially enhancing affordability and insurance coverage against climate risks. These results align with SSP 2.6's narrative of ``sustainable economic development" and ``improving environmental conditions," leading to ``low mitigation and adaptation challenges" \citep{OnKrEbKeRiRo17}.

\footnotetext{\footnotesize{Note that the market deficit-given-insolvency ratio exhibits high volatility across all climate scenarios in the early projection horizon. This is likely due to the relatively small number of simulation paths that become insolvent at early stages (see Figure \ref{fig:MarketInsolvencyProb}), which leads to greater fluctuation in the calculated ratios.}}

\begin{remark} \label{Remark:DividendsPayouts}
    The results on surplus and insolvency probability presented in Section \ref{Section:Returns} are based on the assumption that no dividends are paid from the surplus, consistent with the paper's objective of establishing a baseline model. We also conducted a sensitivity analysis to examine the impact of different dividend policies on surplus and insolvency probability. Overall, although the magnitudes of surplus and insolvency probability vary under different dividend policies, the scenario rankings with respect to surplus and insolvency probability under varying dividend payout assumptions remain consistent with the baseline findings reported in this section. It is also found that the SSP 8.5 scenario shows greater sensitivity to higher dividend payouts, as larger dividends erode the capital buffer to absorb high catastrophe losses under this scenario. Detailed results and discussions are provided in Online Appendix \ref{Appendix:SensitivityDividendsPaids}.
\end{remark}

\subsection{Comparison with the stationary DFA model}
\label{Section:ComparisontoStationaryResults}

We also compare the simulation results from the climate-dependent DFA model with those from the benchmark stationary DFA model specified in Section \ref{Section:StationaryDFA}. The stationary DFA outcomes are shown as grey dotted lines in the figures presented in Sections \ref{Section:KeyResultsfromIndividualModels} and \ref{Section:Returns}. The main observations from this comparison are summarised below.

A key distinction between the two frameworks lies in the ability of the climate-dependent DFA to capture long-term climate trends embedded in the scenario assumptions. As illustrated by the simulations of hazard losses (Figure \ref{fig:HazardSimulations} in Section \ref{Section: ResultsClimateHazards}) and projected gross CAT premiums (Figure \ref{fig:GrossPremiums} in Section \ref{Section: ResultsPremiumsUL}), the stationary DFA produces relatively stable mean hazard losses and premiums over time. In contrast, the climate-dependent DFA captures the upward trend in hazard losses and associated premiums under climate change, particularly under higher-emission scenarios. Moreover, the stationary DFA does not reflect interannual variability in hazard losses arising from internal climate dynamics (e.g., ENSO cycles). By linking climate model inputs to the hazard models, as described in Section \ref{Section:Hazards}, the climate-dependent DFA captures both long-term trends and short-term variability in hazard losses.

Another distinguishing feature of the climate-dependent DFA is its ability to incorporate different scenario assumptions. As shown in the simulated cumulative investment returns (Figure \ref{fig:SimulatedCompInvReturns}) and the projected mean and median market surplus (Figures \ref{fig:MedianMarketSurplus} and \ref{fig:ExpectedMarketSurplus}), the stationary results are broadly similar to those under the SSP 2.6 and SSP 4.5 scenarios. However, the stationary DFA does not capture the trends observed under the higher-emission scenarios (SSP 7.0 and SSP 8.5). In particular, it tends to underestimate the insolvency probability if the SSP 7.0 scenario is realised (Figure \ref{fig:MarketInsolvencyProb}), as it does not incorporate the combination of weaker economic growth and elevated physical risks associated with this pathway. These factors lead to lower investment returns and higher underwriting losses, which jointly erode insurers' capital over the long term. This highlights the advantage of the climate-dependent DFA in generating long-term projections under climate change, where future economic and physical conditions may differ substantially from the present environment.

A further difference lies in the ability of the climate-dependent DFA to capture uncertainty in future climate impacts. For example, in the simulated underwriting losses (Figure \ref{fig:UnderwritingLossSimulations}), the mean losses are broadly similar across scenarios, but the stationary DFA exhibits lower volatility and smaller upper-tail percentiles compared with the climate-dependent DFA. A similar pattern is observed in the individual hazard losses (Figure \ref{fig:HazardSimulations}), where the stationary DFA produces consistently lower simulation volatility across hazard types. This difference reflects the additional model and stochastic uncertainty in future climate dynamics captured in the climate module of the climate-dependent DFA, which is subsequently propagated to the hazard and financial modules. Omitting this additional uncertainty may lead to an underestimation of tail risks when projecting future financial outcomes. Consistent with this observation, the market deficit given insolvency (Figure \ref{fig:MarketLGD}) is notably lower under the stationary DFA than under the climate-dependent DFA, indicating a lower estimated severity of insolvency in the stationary case.

\section{Discussion} \label{Section:LimitationsandExtensions}

The modelling framework necessarily adopts simplifying assumptions to manage the complexity of climate risk. While this offers a practical starting point, it introduces several key limitations that warrant attention. It also presents several opportunities for future extension, as discussed below.

\subsection{Reliances and limitations} \label{Section:Limitations}


The first limitation of our analysis is its reliance on economic and environmental assumptions in the SSP scenarios, which have inherent limitations (see \rev{Limitation} \ref{Remark:ScenariosLimitations}). In addition, post-disaster reconstruction effects have not been explicitly incorporated into adjusting the GDP assumptions for climate damage (see \rev{Limitation} \ref{Remark:AssetModuleLimitations}). 

Secondly, our hazard loss forecasts rely on historically calibrated relationships between climate variables and insurance losses, which may shift over time. Future research could adopt a more forward-looking approach to reflect these changes (see \rev{Limitation} \ref{Remark:HistoricalRelationship}).

Thirdly, the framework is limited to domestic disasters and excludes global spillover effects. However, climate-related economic damage can potentially spread via international trade \citep{NeNePi25}. Modelling these linkages would offer a more complete view of risks to local insurers.

Additionally, as noted in \rev{Limitation} \ref{Remark:BusinessVolume}, the framework does not consider potential business volume loss from rising premiums. Also, the analysis does not incorporate possible changes in future building standards or dwelling sizes that may result from adaptation measures when projecting hazard losses. 

Finally, while the framework incorporates several sources of climate-related uncertainty, it does not capture uncertainty arising from parameters and model inadequacy (see \rev{Limitation} \ref{Remark:Uncertanties}). The unidentified risks are also not quantified in our framework due to their unquantifiable and truly unknown nature (see \rev{Limitation} \ref{Remark:OtherRisks}). Users should therefore remain mindful of these additional risks and uncertainties when interpreting the model outputs.

\subsection{Extensions and future work} \label{Section:Extensions}

Our framework offers several avenues for extension. First, \rev{as discussed in Remark \ref{Remark:MacrovsIndividual},} while this paper assesses climate risks at an industry-wide level, it provides a foundation for individual insurers to develop firm-specific, climate-dependent DFA models aligned with their unique risk profiles. In addition, the current modelling framework does not differentiate between lines of business. Although this simplification is appropriate for the study's macro-level focus, future research could extend the framework to incorporate multiple business lines, which would be particularly useful for decision-making at the individual insurer level, such as capital allocation \citep{MyRe01}.

Second, as noted in \rev{Limitation} \ref{Remark:RegulatoryControl} and \rev{Limitation} \ref{Remark:ReinsuranceMarketLimitations}, the model serves as a baseline that does not incorporate the effects of government interventions, market competition, or regulatory actions. While certain reinsurance market dynamics have been considered in this study, it should not be regarded as a comprehensive modelling of reinsurance cycles due to the simplified assumptions outlined in \rev{Limitation}~\ref{Remark:ReinsuranceMarketLimitations}. Future research could incorporate these factors and compare the outcomes with this baseline to assess their impact on the financial performance of general insurers under different climate scenarios.

Third, on the liability side, the model focuses on climate impacts on catastrophe losses. Future research could extend this by incorporating climate effects on non-catastrophe losses as relevant data become available.

Moreover, as noted in \rev{Limitation} \ref{Remark:OtherRisks}, transition risks on the liability side are not yet included. Their integration could be explored in future work once quantification methods are more established in the literature.

Finally, the modelling approach is broad but simplified. Future studies may adopt more advanced component models for greater precision, such as the proprietary catastrophe (CAT) models. These enhancements, however, must be balanced against the trade-off between model complexity and parsimony, which is an essential consideration in DFA.

\section{Conclusions} \label{Section:Conclusions}

\rev{This paper extends the traditional Dynamic Financial Analysis (DFA) framework by incorporating climate change considerations. This extension represents a paradigm shift toward a forward-looking framework recognising that future economic and physical conditions may differ fundamentally from historical experience due to climate change. We show how climate-related impacts and scenario assumptions can systematically be integrated across the different modules of the DFA framework, drawing on insights from climate science, climate-economy literature, and the financial relationships embedded in conventional DFA models. This integrated structure allows the framework to capture the interdependencies between assets and liabilities under climate change, enabling a holistic assessment of climate-related impacts on general insurers that remains relatively under-explored in the literature. The proposed extension also translates climate scenario narratives into insurers’ balance-sheet outcomes in a tractable manner. By balancing tractability and comprehensiveness in the design of the component models, the framework captures the key assumptions of climate scenarios and the main characteristics of climate change impacts, while retaining the desirable features of conventional DFA frameworks, such as transparency and interpretability.}

\rev{To illustrate the contributions of the proposed framework, we apply it to the Australian general insurance market. Compared with the results from a conventional stationary DFA model, the climate-dependent DFA captures not only the long-term trends associated with climate change and the differences across emission pathways, but also the uncertainty surrounding future climate impacts. Ignoring these factors in financial projections may lead to a misestimation of insurers’ risk and return profiles under climate change and may consequently affect strategic decision-making.}

\rev{The empirical illustration of the proposed framework also provides several useful insights into the potential impact of climate change on the general insurance market.} In climate science, SSP 8.5 carries the highest physical risk, followed by SSP 7.0, SSP 4.5, and SSP 2.6 \citep[Chapter 11 of][pp. 1517--1605]{IPCC2021Ch11}, a ranking reflected in our hazard simulations and CAT premium projections. However, when both economic and physical factors are considered, SSP 7.0, which combines high catastrophe risk with poor economic growth, emerges as the most detrimental scenario for insurers, leading to low returns and high insolvency risk. In contrast, SSP 8.5 delivers the highest returns and lowest insolvency probability due to strong economic growth, but also results in the largest market deficits upon insolvency, increasing risks to policyholders.

These findings highlight the need for insurers to prepare for scenarios involving both high catastrophe risk and weak economic growth in strategic areas such as business planning, capital management, and reinsurance. Insurers should also account for the tail-end financial risks under high-emission scenarios like SSP 8.5. Regulators, in turn, should ensure adequate capital buffers and collaborate with governments on contingency plans, such as bail-out mechanisms, to manage systemic risks from potential insurer insolvencies. Further development of the proposed climate-dependent DFA framework might support these efforts. 
\rev{Indeed, the framework developed in this paper provides a useful foundation for assessing climate-related financial risks, which can be further developed in many useful directions, as discussed in Section \ref{Section:Extensions}.}




\section*{Data and Code}

The R code used in this study is available at \url{https://github.com/agi-lab/climate-dependent-DFA}, and the supporting datasets are hosted at \url{https://zenodo.org/records/18521235}. Instructions for data access and code execution are provided on the GitHub page.

Due to licensing restrictions, we cannot share the Total Returns series of the All-Ordinaries Shares Index; users should obtain it directly from FactSet. For those without access, pseudo data simulated from the calibrated equity model is provided for calibration.

Similarly, we cannot distribute Woodside Energy Limited’s 2014–2023 financial statements from FactSet. Users can retrieve them via FactSet or compile them manually from public reports at \url{https://www.woodside.com/investors/reports-investor-briefings}. For models using restricted datasets, the same parameters are applied in the code to ensure consistency with the results presented.

\section*{Acknowledgements}

This research was supported under Australian Research Council's Discovery Project (DP200101859) funding scheme \rev{and University International Postgraduate Award (UIPA)}. The views and opinions expressed in this paper are solely those of the authors and do not reflect those of their affiliated institutions.

The authors would like to extend their sincere gratitude to Sharanjit Paddam for his constructive feedback and insightful comments, which significantly improved the quality of this paper and provided valuable guidance for potential future research directions.

Earlier versions of this paper were presented at the 2024 Australasian Actuarial Education and Research Symposium (AAERS) in Melbourne (Australia), All Actuaries Summit 2025 in Sydney (Australia), the 2025 28th International Congress on Insurance: Mathematics and Economics held in Tartu (Estonia), \rev{the 2026 inaugural ASTIN Bulletin Conference}, as well as at research seminars hosted by the School of Risk and Actuarial Studies at UNSW. The authors are grateful for constructive comments received from colleagues who attended those events.

The authors also gratefully acknowledge the Actuaries Institute of Australia for awarding the Melville Prize and the Carol Dolan Prize to an earlier version of this paper, which was tailored for an industry audience and presented at the All Actuaries Summit 2025.

\section*{Declarations of interest}
None

\newpage

\section*{References}

\bibliographystyle{elsarticle-harv}
\bibliography{libraries}

@article{CaTr90,
  title={Regression-based tests for overdispersion in the Poisson model},
  author={Cameron, A Colin and Trivedi, Pravin K},
  journal={Journal of econometrics},
  volume={46},
  number={3},
  pages={347--364},
  year={1990},
  publisher={Elsevier}
}

@book{EmKlMi13,
  title={Modelling extremal events: for insurance and finance},
  author={Embrechts, Paul and Kl{\"u}ppelberg, Claudia and Mikosch, Thomas},
  volume={33},
  year={2013},
  publisher={Springer Science \& Business Media}
}

@article{JaElSa08,
  title={Forecasting US insured hurricane losses},
  author={Jagger, Thomas H and Elsner, James B and Saunders, Mark A},
  journal={Climate extremes and society},
  volume={189},
  pages={209},
  year={2008},
  publisher={Cambridge University Press Cambridge}
}

@article{JaElBu11,
  title={Climate and solar signals in property damage losses from hurricanes affecting the United States},
  author={Jagger, Thomas H and Elsner, James B and Burch, R King},
  journal={Natural Hazards},
  volume={58},
  pages={541--557},
  year={2011},
  publisher={Springer}
}

@article{Nord92,
  title={The DICE model: background and structure of a dynamic integrated climate-economy model of the economics of global warming},
  author={Nordhaus, William D},
  year={1992}
}

@article{No92,
  title={An optimal transition path for controlling greenhouse gases},
  author={Nordhaus, William D},
  journal={Science},
  volume={258},
  number={5086},
  pages={1315--1319},
  year={1992},
  publisher={American Association for the Advancement of Science}
}

@article{Al03Nature,
  title={Liability for climate change: Will it ever be possible to sue anyone for damaging the climate?},
  author={Alien, M},
  journal={Nature},
  volume={421},
  pages={891--892},
  year={2003}
}

@misc{APRA23,
  author       = {{APRA}},
  title        = {Insurance Climate Vulnerability Assessment},
  year         = {2023},
  howpublished          = {https://www.apra.gov.au/insurance-climate-vulnerability-assessment},
  note         = {Accessed: 2025-05-06}
}

@article{Nord18,
  title={Evolution of modeling of the economics of global warming: changes in the DICE model, 1992--2017},
  author={Nordhaus, William},
  journal={Climatic change},
  volume={148},
  number={4},
  pages={623--640},
  year={2018},
  publisher={Springer}
}

@article{HaDiVrAlMe11,
  title={Future building water loss projections posed by climate change},
  author={Haug, Ola and Dimakos, Xeni K and V{\aa}rdal, Jofrid F and Aldrin, Magne and Meze-Hausken, Elisabeth},
  journal={Scandinavian Actuarial Journal},
  volume={2011},
  number={1},
  pages={1--20},
  year={2011},
  publisher={Taylor \& Francis}
}

@article{SchFeFrHaHiMe13,
  title={A Bayesian hierarchical model with spatial variable selection: the effect of weather on insurance claims},
  author={Scheel, Ida and Ferkingstad, Egil and Frigessi, Arnoldo and Haug, Ola and Hinnerichsen, Mikkel and Meze-Hausken, Elisabeth},
  journal={Journal of the Royal Statistical Society Series C: Applied Statistics},
  volume={62},
  number={1},
  pages={85--100},
  year={2013},
  publisher={Oxford University Press}
}

@misc{apra_gps114,
  author       = {{APRA}},
  title        = {{GPS 114: Capital Adequacy – Asset Risk Charge}},
  year         = {2023},
  howpublished = {https://www.apra.gov.au/gps-114-capital-adequacy-asset-risk-charge},
  note         = {Accessed: 2025-05-06}
}

@article{LyuHaYuRi17,
  title={Where Home Insurance Meets Climate Change: Making Sense of Climate Risk, Data Uncertainty, and Projections},
  author={Lyubchich, Vyacheslav and Kilbourne, K Halimeda and Gel, Yulia R and Richardson, TX},
  journal={Variance},
  volume = {12},
  number = {2},
  year = {2017},
  pages={278--292}
}

@article{LyuNeGhMaGe19,
  title={Insurance risk assessment in the face of climate change: Integrating data science and statistics},
  author={Lyubchich, Vyacheslav and Newlands, Nathaniel K and Ghahari, Azar and Mahdi, Tahir and Gel, Yulia R},
  journal={Wiley Interdisciplinary Reviews: Computational Statistics},
  volume={11},
  number={4},
  pages={e1462},
  year={2019},
  publisher={Wiley Online Library}
}

@article{ReGuPeReAg23,
  title={Weather Conditions and Telematics Panel Data in Monthly Motor Insurance Claim Frequency Models},
  author={Reig Torra, Jan and Guillen, Montserrat and P{\'e}rez-Mar{\'\i}n, Ana M and Rey G{\'a}mez, Lorena and Aguer, Giselle},
  journal={Risks},
  volume={11},
  number={3},
  pages={57},
  year={2023},
  publisher={MDPI}
}

@article{KaGaKl01,
  title={Introduction to dynamic financial analysis},
  author={Kaufmann, Roger and Gadmer, Andreas and Klett, Ralf},
  journal={ASTIN Bulletin: The Journal of the IAA},
  volume={31},
  number={1},
  pages={213--249},
  year={2001},
  publisher={Cambridge University Press}
}

@book{KlPaWi12,
  title={Loss models: from data to decisions},
  author={Klugman, Stuart A and Panjer, Harry H and Willmot, Gordon E},
  volume={715},
  year={2012},
  publisher={John Wiley \& Sons}
}

@article{DaGo04,
  title={The use of dynamic financial analysis to determine whether an optimal growth rate exists for a property-liability insurer},
  author={D'Arcy, Stephen P and Gorvett, Richard W},
  journal={Journal of Risk and Insurance},
  volume={71},
  number={4},
  pages={583--615},
  year={2004},
  publisher={Wiley Online Library}
}

@article{ElPa07,
  title={Dynamic financial analysis: Classification, conception, and implementation},
  author={Eling, Martin and Parnitzke, Thomas},
  journal={Risk Management and Insurance Review},
  volume={10},
  number={1},
  pages={33--50},
  year={2007},
  publisher={Wiley Online Library}
}

@article{CoMoViMe18,
  title={Optimal insurance portfolios risk-adjusted performance through dynamic stochastic programming},
  author={Consigli, Giorgio and Moriggia, Vittorio and Vitali, Sebastiano and Mercuri, Lorenzo},
  journal={Computational Management Science},
  volume={15},
  pages={599--632},
  year={2018},
  publisher={Springer}
}

@article{kaXe22,
  title={Climate change financial risks: Implications for asset pricing and interest rates},
  author={Karydas, Christos and Xepapadeas, Anastasios},
  journal={Journal of Financial Stability},
  volume={63},
  pages={101061},
  year={2022},
  publisher={Elsevier}
}

@article{BoWe21,
  title={Leaving place, restoring home: enhancing the evidence base on planned relocation cases in the context of hazards, disasters, and climate change},
  author={Bower, Erica and Weerasinghe, Sanjula},
  journal={UNSW Sydney and Kaldor Centre for International Refugee Law},
  year={2021}
}

@article{MeOuFr06,
  title={Business cycles in insurance and reinsurance: the case of France, Germany and Switzerland},
  author={Meier, Ursina B and Outreville, J Fran{\c{c}}ois},
  journal={The Journal of Risk Finance},
  year={2006},
  publisher={Emerald Group Publishing Limited}
}

@article{WeJepi17,
  title={The truthiness about hurricane catastrophe models},
  author={Weinkle, Jessica and Pielke Jr, Roger},
  journal={Science, Technology, \& Human Values},
  volume={42},
  number={4},
  pages={547--576},
  year={2017},
  publisher={SAGE Publications Sage CA: Los Angeles, CA}
}

@article{ElTo09,
  title={Modeling and management of nonlinear dependencies--copulas in dynamic financial analysis},
  author={Eling, Martin and Toplek, Denis},
  journal={Journal of Risk and Insurance},
  volume={76},
  number={3},
  pages={651--681},
  year={2009},
  publisher={Wiley Online Library}
}

@article{Gr23,
  title={On volatile growth: Simple fitting of exponential functions taking into account values of every observation with any signs, applied to readily calculate a novel covariance-invariant CAGR},
  author={Grimm, Wolfgang M},
  journal={The Engineering Economist},
  volume={68},
  number={1},
  pages={34--58},
  year={2023},
  publisher={Taylor \& Francis}
}

@article{ChNi14,
  title={Fundamental definition of the solvency capital requirement in solvency II},
  author={Christiansen, Marcus C and Niemeyer, Andreas},
  journal={ASTIN Bulletin: The Journal of the IAA},
  volume={44},
  number={3},
  pages={501--533},
  year={2014},
  publisher={Cambridge University Press}
}

@misc{APRAQuarterlyStats,
  author = {{APRA}},
  title = {Quarterly General Insurance Performance Statistics},
  year = {2024},
  howpublished  = {https://www.apra.gov.au/quarterly-general-insurance-performance-statistics},
  note  = {Accessed: 2024-06-30}
}

@misc{IFRSS2,
	author = {{IFRS}},
	howpublished = {https://www.ifrs.org/issued-standards/ifrs-sustainability-standards-navigator/ifrs-s2-climate-related-disclosures/\#standard},
	title = {IFRS S2: Sustainability Disclosure Standard},
	year = {2023}}

@book{McRuEm15,
  title={Quantitative risk management: concepts, techniques and tools-revised edition},
  author={McNeil, Alexander J and Frey, R{\"u}diger and Embrechts, Paul},
  year={2015},
  publisher={Princeton university press}
}

@article{Pa18,
  title={The impact of disasters on inflation},
  author={Parker, Miles},
  journal={Economics of Disasters and Climate Change},
  volume={2},
  number={1},
  pages={21--48},
  year={2018},
  publisher={Springer}
}

@article{Ve22,
  title={Climate change, risk factors and stock returns: A review of the literature},
  author={Venturini, Alessio},
  journal={International Review of Financial Analysis},
  volume={79},
  pages={101934},
  year={2022},
  publisher={Elsevier}
}

@article{BaKiOc19,
  title={Climate change risk},
  author={Bansal, Ravi and Kiku, Dana and Ochoa, Marcelo},
  journal={Federal Reserve Bank of San Francisco Working Paper},
  year={2019}
}

@article{Go20,
  title={Carbon risk},
  author={G{\"o}rgen, Maximilian and Jacob, Andrea and Nerlinger, Martin and Riordan, Ryan and Rohleder, Martin and Wilkens, Marco},
  journal={Available at SSRN 2930897},
  year={2020}
}

@article{Ba23, 
  title={Climate change and uncertainty: An asset pricing perspective},
  author={Barnett, Michael},
  journal={Management Science},
  year={2023},
  publisher={INFORMS}
}

@article{Wi94,
  title={The dynamics of competitive insurance markets},
  author={Winter, Ralph A},
  journal={Journal of Financial Intermediation},
  volume={3},
  number={4},
  pages={379--415},
  year={1994},
  publisher={Elsevier}
}

@article{Wi88,
  title={The liability crisis and the dynamics of competitive insurance markets},
  author={Winter, Ralph A},
  journal={Yale J. on Reg.},
  volume={5},
  pages={455},
  year={1988},
  publisher={HeinOnline}
}

@article{DiGa22,
  title={Asymmetric information and insurance cycles},
  author={Dicks, David L and Garven, James R},
  journal={Journal of Risk and Insurance},
  volume={89},
  number={2},
  pages={449--474},
  year={2022},
  publisher={Wiley Online Library}
}

@article{EcXe18,
  title={Monetary policy under climate change},
  author={Economides, George and Xepapadeas, Anastasios},
  year={2018},
  publisher={CESifo Working Paper Series}
}

@article{Mc08,
  title={Does rainfall increase or decrease motor accidents?},
  author={McGuire, Grainne},
  year={2008},
  publisher={Taylor Fry Consulting}
}

@article{Ta12,
  title={A simple model of insurance market dynamics},
  author={Taylor, Greg},
  journal={North American Actuarial Journal},
  volume={12},
  number={3},
  pages={242--262},
  year={2008},
  publisher={Taylor \& Francis}
}

@book{Enz02,
  title={The insurance cycle as an entrepreneurial challenge},
  author={Enz, Rudolf},
  year={2002},
  publisher={Swiss Reinsurance Company}
}

@article{MeOu10,
  title={Business cycles in insurance and reinsurance: international diversification effects},
  author={Meier, Ursina B and Outreville, J Fran{\c{c}}ois},
  journal={Applied Financial Economics},
  volume={20},
  number={8},
  pages={659--668},
  year={2010},
  publisher={Taylor \& Francis}
}

@article{ByJo20,
  title={How does climate change affect the long-run real interest rate},
  author={Bylund, Emma and Jonsson, Magnus},
  journal={Economic Commentaries},
  volume={11},
  pages={90--108},
  year={2020}
}

@article{LeLu05,
  title={Expected returns and expected dividend growth},
  author={Lettau, Martin and Ludvigson, Sydney C},
  journal={Journal of Financial Economics},
  volume={76},
  number={3},
  pages={583--626},
  year={2005},
  publisher={Elsevier}
}

@article{MoPo22,
  title={The effects of climate change on the natural rate of interest: a critical survey},
  author={Mongelli, Francesco Paolo and Pointner, Wolfgang and van den End, Jan Willem},
  year={2022},
  publisher={ECB Working Paper}
}

@article{Wil95,
  title={More on a stochastic asset model for actuarial use},
  author={Wilkie, A David},
  journal={British Actuarial Journal},
  volume={1},
  number={5},
  pages={777--964},
  year={1995},
  publisher={Cambridge University Press}
}

@article{JoPa94,
  title={Fitting Tweedie's compound Poisson model to insurance claims data},
  author={J{\o}rgensen, Bent and Paes De Souza, Marta C},
  journal={Scandinavian Actuarial Journal},
  volume={1994},
  number={1},
  pages={69--93},
  year={1994},
  publisher={Taylor \& Francis}
}

@inproceedings{AlDaGo05,
  title={Modeling financial scenarios: A framework for the actuarial profession},
  author={Ahlgrim, Kevin C and D'Arcy, Stephen P and Gorvett, Richard W},
  booktitle={Proceedings of the Casualty Actuarial Society},
  pages={177--238},
  year={2005}
}

@article{ChKoWaHaLaToZh21,
  title={Using a stochastic economic scenario generator to analyse uncertain superannuation and retirement outcomes},
  author={Chen, Wen and Koo, Bonsoo and Wang, Yunxiao and O’Hare, Colin and Langren{\'e}, Nicolas and Toscas, Peter and Zhu, Zili},
  journal={Annals of Actuarial Science},
  volume={15},
  number={3},
  pages={549--566},
  year={2021},
  publisher={Cambridge University Press}
}

@misc{FaPaSt21,
  title={Feeling the heat: extreme temperatures and price stability},
  author={Faccia, Donata and Parker, Miles and Stracca, Livio},
  year={2021},
  publisher={ECB Working Paper}
}

@article{MuOu21,
  title={Climate and monetary policy: do temperature shocks lead to inflationary pressures?},
  author={Mukherjee, Krishnendu and Ouattara, Bakary},
  journal={Climatic change},
  volume={167},
  number={3},
  pages={32},
  year={2021},
  publisher={Springer}
}

@misc{CiKuHe23,
  title={The asymmetric effects of weather shocks on euro area inflation},
  author={Ciccarelli, Matteo and Kuik, Friderike and Hern{\'a}ndez, Catalina Mart{\'\i}nez},
  year={2023},
  publisher={ECB Working Paper}
}

@article{Be22,
  title={Ensemble Economic Scenario Generators: Unity Makes Strength},
  author={B{\'e}gin, Jean-Fran{\c{c}}ois},
  journal={North American Actuarial Journal},
  pages={1--28},
  year={2022},
  publisher={Taylor \& Francis}
}

@article{IPCC2023,
  title={Synthesis report of the IPCC Sixth Assessment Report (AR6), Longer report. IPCC.},
  author={Lee, Hoesung and Calvin, Katherine and Dasgupta, Dipak and Krinmer, Gerhard and Mukherji, Aditi and Thorne, Peter and Trisos, Christopher and Romero, Jose and Aldunce, Paulina and Barret, Ko and others},
  year={2023},
  publisher={Intergovernmental Panel on Climate Change (IPCC)}
}

@misc{BoM23,
  title = {The Southern Oscillation Index (SOI)},
  author = {{Bureau of Meteorology}},
  howpublished = {\url{http://www.bom.gov.au/climate/enso/history/ln-2010-12/SOI-what.shtml}},
  note = {Accessed: 2023-10-09}
}

@article{Ei04,
  title={The mixed effects of precipitation on traffic crashes},
  author={Eisenberg, Daniel},
  journal={Accident analysis \& prevention},
  volume={36},
  number={4},
  pages={637--647},
  year={2004},
  publisher={Elsevier}
}

@article{ArSh25,
  title={The Impact of Climate Change on Reserves in Life Insurance},
  author={Arandjelovic, Aleksandar and Shevchenko, Pavel V},
  journal={Available at SSRN 5870182},
  year={2025}
}

@article{StRi08,
  title={Generalized additive models for location scale and shape (GAMLSS) in R},
  author={Stasinopoulos, D Mikis and Rigby, Robert A},
  journal={Journal of Statistical Software},
  volume={23},
  pages={1--46},
  year={2008}
}

@article{CaGuMc15,
  title={Increased frequency of extreme La Ni{\~n}a events under greenhouse warming},
  author={Cai, Wenju and Wang, Guojian and Santoso, Agus and McPhaden, Michael J and Wu, Lixin and Jin, Fei-Fei and Timmermann, Axel and Collins, Mat and Vecchi, Gabriel and Lengaigne, Matthieu and others},
  journal={Nature Climate Change},
  volume={5},
  number={2},
  pages={132--137},
  year={2015},
  publisher={Nature Publishing Group UK London}
}

@article{HeFe23,
  title={Towards understanding the robust strengthening of ENSO and more frequent extreme El Ni{\~n}o events in CMIP6 global warming simulations},
  author={Heede, Ulla K and Fedorov, Alexey V},
  journal={Climate Dynamics},
  volume={61},
  number={5},
  pages={3047--3060},
  year={2023},
  publisher={Springer}
}

@article{BoBoCaBoRa24,
  title={A data science approach to climate change risk assessment applied to pluvial flood occurrences for the United States and Canada},
  author={Bourget, Mathilde and Boudreault, Mathieu and Carozza, David A and Boudreault, J{\'e}r{\'e}mie and Raymond, S{\'e}bastien},
  journal={ASTIN Bulletin: The Journal of the IAA},
  volume={54},
  number={3},
  pages={495--517},
  year={2024},
  publisher={Cambridge University Press}
}

@article{MiliNaLi25,
  title={Assessing Climate-Driven Mortality Risk: A Stochastic Approach with Distributed Lag Non-Linear Models},
  author={Min, Jiacheng and Li, Han and Nagler, Thomas and Li, Shuanming},
  journal={arXiv preprint arXiv:2506.00561},
  year={2025}
}

@misc{APRA2020,
  author       = {{APRA}},
  title        = {APRA updates guidance on capital management for banks and insurers},
  year         = {2020},
  howpublished = {\url{https://www.apra.gov.au/news-and-publications/apra-updates-guidance-on-capital-management-for-banks-and-insurers}}
}

@article{GuPiPl25,
  author    = {Guibert, Quentin and Pincemin, Guillaume and Planchet, Frédéric},
  title     = {Impact of Climate Change on Mortality: An Extrapolation of Temperature Effects Based on Time Series Data in France},
  journal   = {International Journal of Forecasting},
  year      = {2025},
  note      = {Forthcoming}
}

@article{Ramsey28,
  title={A mathematical theory of saving},
  author={Ramsey, Frank Plumpton},
  journal={The economic journal},
  volume={38},
  number={152},
  pages={543--559},
  year={1928},
  publisher={Oxford University Press Oxford, UK}
}

@article{LaThWi03,
  title={Measuring the natural rate of interest},
  author={Laubach, Thomas and Williams, John C},
  journal={Review of Economics and Statistics},
  volume={85},
  number={4},
  pages={1063--1070},
  year={2003},
  publisher={MIT Press 238 Main St., Suite 500, Cambridge, MA 02142-1046, USA journals~…}
}

@article{HoLaWi17,
  title={Measuring the natural rate of interest: International trends and determinants},
  author={Holston, Kathryn and Laubach, Thomas and Williams, John C},
  journal={Journal of International Economics},
  volume={108},
  pages={S59--S75},
  year={2017},
  publisher={Elsevier}
}

@article{KoKuLiNi24,
  title={Global warming and heat extremes to enhance inflationary pressures},
  author={Kotz, Maximilian and Kuik, Friderike and Lis, Eliza and Nickel, Christiane},
  journal={Nature Communications Earth \& Environment},
  volume={5},
  number={1},
  pages={1--13},
  year={2024},
  publisher={Nature Publishing Group}
}

@article{GaOz24,
  title={The impact of dependencies between climate risks on the asset and liability side of non-life insurers},
  author={Gatzert, Nadine and {\"O}zdil, Onur},
  journal={European Actuarial Journal},
  pages={1--19},
  year={2024},
  publisher={Springer}
}

@misc{IPCC2021Ch11,
  author       = {{IPCC}},
  title        = {Climate Change 2021: The Physical Science Basis. Contribution of Working Group I to the Sixth Assessment Report of the IPCC -- Chapter 11},
  year         = {2021},
  howpublished = {Cambridge University Press},
  url          = {https://www.ipcc.ch/report/ar6/wg1/}
}

@misc{IPCC2021Ch12,
  author       = {{IPCC}},
  title        = {Climate Change 2021: The Physical Science Basis. Contribution of Working Group I to the Sixth Assessment Report of the IPCC -- Chapter 12},
  year         = {2021},
  howpublished = {Cambridge University Press},
  url          = {https://www.ipcc.ch/report/ar6/wg1/}
}

@article{GoRaJu25,
  title={Recent global temperature surge intensified by record-low planetary albedo},
  author={Goessling, Helge F and Rackow, Thomas and Jung, Thomas},
  journal={Science},
  volume={387},
  number={6729},
  pages={68--73},
  year={2025},
  publisher={American Association for the Advancement of Science}
}

@article{CrMc08,
  title={Normalised Australian insured losses from meteorological hazards: 1967--2006},
  author={Crompton, Ryan P and McAneney, K John},
  journal={Environmental Science \& Policy},
  volume={11},
  number={5},
  pages={371--378},
  year={2008},
  publisher={Elsevier}
}

@article{Pi21,
  title={Economic ‘normalisation’of disaster losses 1998--2020: A literature review and assessment},
  author={Pielke, Roger},
  journal={Environmental Hazards},
  volume={20},
  number={2},
  pages={93--111},
  year={2021},
  publisher={Taylor \& Francis}
}

@article{PiLa08,
  title={Normalized hurricane damages in the United States: 1925--95},
  author={Pielke, Roger A and Landsea, Christopher W},
  journal={Weather and forecasting},
  volume={13},
  number={3},
  pages={621--631},
  year={1998},
  publisher={American Meteorological Society}
}

@article{VrPi09,
  title={Normalized earthquake damage and fatalities in the United States: 1900--2005},
  author={Vranes, Kevin and Pielke Jr, Roger},
  journal={Natural Hazards Review},
  volume={10},
  number={3},
  pages={84--101},
  year={2009},
  publisher={American Society of Civil Engineers}
}

@article{ShCaFoMoEvFl16,
  title={Natural hazards in Australia: extreme bushfire},
  author={Sharples, Jason J and Cary, Geoffrey J and Fox-Hughes, Paul and Mooney, Scott and Evans, Jason P and Fletcher, Michael-Shawn and Fromm, Mike and Grierson, Pauline F and McRae, Rick and Baker, Patrick},
  journal={Climatic Change},
  volume={139},
  pages={85--99},
  year={2016},
  publisher={Springer}
}

@article{Do18,
  title={Climatological variability of fire weather in Australia},
  author={Dowdy, Andrew J},
  journal={Journal of Applied Meteorology and Climatology},
  volume={57},
  number={2},
  pages={221--234},
  year={2018},
  publisher={American Meteorological Society}
}

@article{QuBaRiPa22,
  title={Fire weather index data under historical and SSP projections in CMIP6 from 1850 to 2100},
  author={Quilcaille, Yann and Batibeniz, Fulden and Ribeiro, Andreia FS and Padr{\'o}n, Ryan S and Seneviratne, Sonia I},
  journal={Earth System Science Data Discussions},
  volume={2022},
  pages={1--31},
  year={2022},
  publisher={G{\"o}ttingen, Germany}
}

@techreport{NCRAHazards25,
  author       = {{Australian Climate Service}},
  title        = {Australia's Future Climate and Hazards Report},
  institution  = {Australian Climate Service},
  year         = {2025},
  url          = {https://www.acs.gov.au/documents/fadb05a9fa254835b840db73383910d7},
  note         = {Accessed 2025-10-31}
}

@article{QiCh21,
  title={Projecting health impacts of future temperature: a comparison of quantile-mapping bias-correction methods},
  author={Qian, Weijia and Chang, Howard H},
  journal={International journal of environmental research and public health},
  volume={18},
  number={4},
  pages={1992},
  year={2021},
  publisher={MDPI}
}

@article{SeZhAdBa21,
  title={Weather and climate extreme events in a changing climate (Chapter 11)},
  author={Seneviratne, Sonia I and Zhang, Xuebin and Adnan, Muhammad and Badi, Wafae and Dereczynski, Claudine and Di Luca, Alejandro and Ghosh, Subimal and Iskander, I and Kossin, James and Lewis, Sophie and others},
  year={2021},
  publisher={Cambridge University Press}
}

@article{BlHaMoMu20,
  title={Generation of a global synthetic tropical cyclone hazard dataset using STORM},
  author={Bloemendaal, Nadia and Haigh, Ivan D and de Moel, Hans and Muis, Sanne and Haarsma, Reindert J and Aerts, Jeroen CJH},
  journal={Scientific data},
  volume={7},
  number={1},
  pages={40},
  year={2020},
  publisher={Nature Publishing Group UK London}
}

@article{MeVoBlCiLe22,
  title={Intercomparison of regional loss estimates from global synthetic tropical cyclone models},
  author={Meiler, Simona and Vogt, Thomas and Bloemendaal, Nadia and Ciullo, Alessio and Lee, Chia-Ying and Camargo, Suzana J and Emanuel, Kerry and Bresch, David N},
  journal={Nature Communications},
  volume={13},
  number={1},
  pages={6156},
  year={2022},
  publisher={Nature Publishing Group UK London}
}

@article{SpLeJoSh21,
  title={Changes in frequency and location of east coast low pressure systems affecting southeast Australia},
  author={Speer, Milton and Leslie, Lance and Hartigan, Joshua and MacNamara, Shev},
  journal={Climate},
  volume={9},
  number={3},
  pages={44},
  year={2021},
  publisher={MDPI}
}

@article{PeAlEvSh16,
  title={The influence of local sea surface temperatures on Australian east coast cyclones},
  author={Pepler, Acacia S and Alexander, Lisa V and Evans, Jason P and Sherwood, Steven C},
  journal={Journal of Geophysical Research: Atmospheres},
  volume={121},
  number={22},
  pages={13--352},
  year={2016},
  publisher={Wiley Online Library}
}

@article{PeDiJiAlEvSh16,
  title={Projected changes in east Australian midlatitude cyclones during the 21st century},
  author={Pepler, Acacia S and Di Luca, Alejandro and Ji, Fei and Alexander, Lisa V and Evans, Jason P and Sherwood, Steven C},
  journal={Geophysical Research Letters},
  volume={43},
  number={1},
  pages={334--340},
  year={2016},
  publisher={Wiley Online Library}
}

@article{RaMaAlKuLa21,
  title={The effects of climate change on hailstorms},
  author={Raupach, Timothy H and Martius, Olivia and Allen, John T and Kunz, Michael and Lasher-Trapp, Sonia and Mohr, Susanna and Rasmussen, Kristen L and Trapp, Robert J and Zhang, Qinghong},
  journal={Nature reviews earth \& environment},
  volume={2},
  number={3},
  pages={213--226},
  year={2021},
  publisher={Nature Publishing Group UK London}
}

@article{AlKawa14,
  title={Future Australian severe thunderstorm environments. Part II: The influence of a strongly warming climate on convective environments},
  author={Allen, John T and Karoly, David J and Walsh, Kevin J},
  journal={Journal of Climate},
  volume={27},
  number={10},
  pages={3848--3868},
  year={2014},
  publisher={American Meteorological Society}
}

@article{LeLeBu08,
  title={Estimating future trends in severe hailstorms over the Sydney Basin: A climate modelling study},
  author={Leslie, Lance M and Leplastrier, Mark and Buckley, Bruce W},
  journal={Atmospheric Research},
  volume={87},
  number={1},
  pages={37--51},
  year={2008},
  publisher={Elsevier}
}

@article{LiRa21,
  title={Country-based rate of emissions reductions should increase by 80\% beyond nationally determined contributions to meet the 2 degree target},
  author={Liu, Peiran R and Raftery, Adrian E},
  journal={Communications earth \& environment},
  volume={2},
  number={1},
  pages={29},
  year={2021},
  publisher={Nature Publishing Group UK London}
}

@article{KoBhChChGa20,
  title={Physics-guided probabilistic modeling of extreme precipitation under climate change},
  author={Kodra, Evan and Bhatia, Udit and Chatterjee, Snigdhansu and Chen, Stone and Ganguly, Auroop Ratan},
  journal={Scientific reports},
  volume={10},
  number={1},
  pages={10299},
  year={2020},
  publisher={Nature Publishing Group UK London}
}

@book{MiJoHiFo17,
  title={Natural catastrophe risk management and modelling: A practitioner's guide},
  author={Mitchell-Wallace, Kirsten and Jones, Matthew and Hillier, John and Foote, Matthew},
  year={2017},
  publisher={John Wiley \& Sons}
}

@article{GrMa20,
  title={Climate-Related Stress Testing: Transition Risks in Norway},
  author={Grippa, Pierpaolo and Mann, Samuel},
  journal={IMF Working Papers},
  volume={2020},
  number={232},
  year={2020},
  publisher={International Monetary Fund}
}

@article{HoLiXu19,
  title={Climate risks and market efficiency},
  author={Hong, Harrison and Li, Frank Weikai and Xu, Jiangmin},
  journal={Journal of econometrics},
  volume={208},
  number={1},
  pages={265--281},
  year={2019},
  publisher={Elsevier}
}

@article{MyRe01,
  title={Capital allocation for insurance companies},
  author={Myers, Stewart C and Read Jr, James A},
  journal={Journal of risk and insurance},
  pages={545--580},
  year={2001},
  publisher={JSTOR}
}

@article{BeCoGuMcMe23,
  title={The Impact of Climate Change Risk on Long-Term Asset Allocation.},
  author={Bertrand, Jean-Charles and Coqueret, Guillaume and McLoughlin, Nicholas and Mesnard, St{\'e}phane},
  journal={Journal of Portfolio Management},
  volume={50},
  number={5},
  year={2024}
}

@article{DeChLaMa17,
  title={Long-term economic growth projections in the Shared Socioeconomic Pathways},
  author={Dellink, Rob and Chateau, Jean and Lanzi, Elisa and Magn{\'e}, Bertrand},
  journal={Global Environmental Change},
  volume={42},
  pages={200--214},
  year={2017},
  publisher={Elsevier}
}

@misc{KiKoOhRu23,
  title={Potential Growth: A Global Database},
  author={Kilic Celik, Sinem and Kose, Ayhan M and Ohnsorge, Franziska and Ruch, Franz},
  year={2023}
}

@article{OnKrEbKeRiRo17,
  title={The roads ahead: Narratives for shared socioeconomic pathways describing world futures in the 21st century},
  author={O Neill, Brian C and Kriegler, Elmar and Ebi, Kristie L and Kemp-Benedict, Eric and Riahi, Keywan and Rothman, Dale S and Van Ruijven, Bas J and Van Vuuren, Detlef P and Birkmann, Joern and Kok, Kasper and others},
  journal={Global environmental change},
  volume={42},
  pages={169--180},
  year={2017},
  publisher={Elsevier}
}

@book{KuMi11,
  title={At war with the weather: Managing large-scale risks in a new era of catastrophes},
  author={Kunreuther, Howard C and Michel-Kerjan, Erwann O},
  year={2011},
  publisher={MIT Press}
}

@techreport{KuMiRa11,
  title={Insuring climate catastrophes in Florida: an analysis of insurance pricing and capacity under various scenarios of climate change and adaptation measures},
  author={Kunreuther, Howard and Michel-Kerjan, Erwann and Ranger, Nicola},
  year={2011},
  institution={Grantham Research Institute on Climate Change and the Environment}
}

@article{PaBoAe13,
  title={Estimation of insurance premiums for coverage against natural disaster risk: an application of Bayesian Inference},
  author={Paudel, Y and Botzen, WJW and Aerts, JCJH},
  journal={Natural Hazards and Earth System Sciences},
  volume={13},
  number={3},
  pages={737--754},
  year={2013},
  publisher={Copernicus Publications G{\"o}ttingen, Germany}
}

@article{PaBoAeDi15,
  title={Risk allocation in a public--private catastrophe insurance system: an actuarial analysis of deductibles, stop-loss, and premiums},
  author={Paudel, Y and Botzen, WJW and Aerts, JCJH and Dijkstra, TK},
  journal={Journal of Flood Risk Management},
  volume={8},
  number={2},
  pages={116--134},
  year={2015},
  publisher={Wiley Online Library}
}

@article{TeBoAe20,
  title={Impacts of climate change and remote natural catastrophes on EU flood insurance markets: An analysis of soft and hard reinsurance markets for flood coverage},
  author={Tesselaar, Max and Botzen, WJ Wouter and Aerts, Jeroen CJH},
  journal={Atmosphere},
  volume={11},
  number={2},
  pages={146},
  year={2020},
  publisher={MDPI}
}

@article{OECDInsurMarket23,
  title={Global Insurance Market Trends 2023},
  author={{OECD}},
  year={2023},
  journal={OECD Publishing},
  url={https://www.oecd.org/finance/insurance/global-insurance-market-trends.htm}
}

@article{EyBoMEsEsTsTtA16,
  title={Overview of the Coupled Model Intercomparison Project Phase 6 (CMIP6) experimental design and organization},
  author={Eyring, Veronika and Bony, Sandrine and Meehl, Gerald A and Senior, Catherine A and Stevens, Bjorn and Stouffer, Ronald J and Taylor, Karl E},
  journal={Geoscientific Model Development},
  volume={9},
  number={5},
  pages={1937--1958},
  year={2016},
  publisher={Copernicus GmbH}
}

@article{JaSc23,
  title={Importance of internal variability for climate model assessment},
  author={Jain, Shipra and Scaife, Adam A and Shepherd, Theodore G and Deser, Clara and Dunstone, Nick and Schmidt, Gavin A and Trenberth, Kevin E and Turkington, Thea},
  journal={npj Climate and Atmospheric Science},
  volume={6},
  number={1},
  pages={68},
  year={2023},
  publisher={Nature Publishing Group UK London}
}

@article{WuMiFaGoZhZh22,
  title={Quantifying the uncertainty sources of future climate projections and narrowing uncertainties with bias correction techniques},
  author={Wu, Yi and Miao, Chiyuan and Fan, Xuewei and Gou, Jiaojiao and Zhang, Qi and Zheng, Haiyan},
  journal={Earth's Future},
  volume={10},
  number={11},
  pages={e2022EF002963},
  year={2022},
  publisher={Wiley Online Library}
}

@article{ChCuSuWe20,
  title={The reinsurance network among US property--casualty insurers: Microstructure, insolvency risk, and contagion},
  author={Chen, Hua and Cummins, J David and Sun, Tao and Weiss, Mary A},
  journal={Journal of Risk and Insurance},
  volume={87},
  number={2},
  pages={253--284},
  year={2020},
  publisher={Wiley Online Library}
}

@article{Ma13,
  title={Bias correction, quantile mapping, and downscaling: Revisiting the inflation issue},
  author={Maraun, Douglas},
  journal={Journal of Climate},
  volume={26},
  number={6},
  pages={2137--2143},
  year={2013},
  publisher={American Meteorological Society}
}

@article{SaQiLiCe22,
  title={Bias correction of extreme values of high-resolution climate simulations for risk analysis},
  author={Sanabria, Luis Augusto and Qin, Xuerong and Li, Jin and Cechet, Robert Peter},
  journal={Theoretical and Applied Climatology},
  volume={150},
  number={3-4},
  pages={1015--1026},
  year={2022},
  publisher={Springer}
}

@article{YaYo07,
  title={A comparison of various tests of normality},
  author={Yazici, Berna and Yolacan, Senay},
  journal={Journal of statistical computation and simulation},
  volume={77},
  number={2},
  pages={175--183},
  year={2007},
  publisher={Taylor \& Francis}
}

@misc{ERA5_data,
  title        = {{ERA5-Land monthly averaged data from 1950 to present}},
  howpublished = {\url{https://cds.climate.copernicus.eu/cdsapp\#!/dataset/reanalysis-era5-land-monthly-means?tab=overview}},
  note         = {Accessed: 2024-06-30},
  year         = {2024},
  author       = {{Copernicus Climate Change Service}}
}

@misc{CMIP6_data,
  title        = {{CMIP6 climate projections}},
  howpublished = {\url{https://cds.climate.copernicus.eu/cdsapp!\#/dataset/projections-cmip6?tab=overview}},
  note         = {Accessed: 2024-06-30},
  year         = {2024},
  author       = {{Copernicus Climate Change Service}}
}

@misc{EMDAT,
  author = {{EM--DAT}},
  title = {The international disasters database},
  year = {2023},
  note = {data retrieved from \url{https://public.emdat.be/data}},
}

@misc{ICA_data,
  title        = {{Historical Normalised Catastrophe list - December 2024}},
  howpublished = {\url{https://insurancecouncil.com.au/wp-content/uploads/2025/01/ICA-Historical-Normalised-Catastrophe-Master-December-2024.xlsx}},
  note         = {Accessed: 2024-06-30},
  year         = {2024},
  author       = {{Insurance Council of Australia}}
}

@misc{RBACashRates_data,
  title        = {{Cash Rate Target}},
  howpublished = {\url{https://www.rba.gov.au/statistics/cash-rate/}},
  note         = {Accessed: 2024-06-30},
  year         = {2024},
  author       = {{Reserve Bank of Australia}}
}

@misc{ABSCPI_data,
  title        = {{Consumer Price Index, Australia}},
  howpublished = {\url{https://www.abs.gov.au/statistics/economy/price-indexes-and-inflation/consumer-price-index-australia/latest-release}},
  note         = {Accessed: 2024-06-30},
  year         = {2024},
  author       = {{Australian Bureau of Statistics}}
}

@misc{WorldBankIndicators_data,
  title        = {{Databank: World Development Indicators}},
  howpublished = {\url{https://databank.worldbank.org/source/world-development-indicators}},
  note         = {Accessed: 2024-06-30},
  year         = {2024},
  author       = {{World Bank Group}}
}

@article{RiVaKr17,
  title={The Shared Socioeconomic Pathways and their energy, land use, and greenhouse gas emissions implications: An overview},
  author={Riahi, Keywan and Van Vuuren, Detlef P and Kriegler, Elmar and Edmonds, Jae and O’neill, Brian C and Fujimori, Shinichiro and Bauer, Nico and Calvin, Katherine and Dellink, Rob and Fricko, Oliver and others},
  journal={Global environmental change},
  volume={42},
  pages={153--168},
  year={2017},
  publisher={Elsevier}
}

@article{BuHsMi15,
  title={Global non-linear effect of temperature on economic production},
  author={Burke, Marshall and Hsiang, Solomon M and Miguel, Edward},
  journal={Nature},
  volume={527},
  number={7577},
  pages={235--239},
  year={2015},
  publisher={Nature Publishing Group UK London}
}

@misc{ABSBI_data,
  title        = {{Business Indicators, Australia}},
  howpublished = {\url{https://www.abs.gov.au/statistics/economy/business-indicators/business-indicators-australia/latest-release}},
  note         = {Accessed: 2024-06-30},
  year         = {2024},
  author       = {{Australian Bureau of Statistics}}
}

@incollection{CoDe89,
  title={The assessment of the financial strength of insurance companies by a generalized cash flow model},
  author={Coutts, Stewart M and Devitt, Russell},
  booktitle={Financial models of insurance solvency},
  pages={1--36},
  year={1989},
  publisher={Springer}
}

@incollection{PaDi89,
  title={Cash flow simulation models for premium and surplus analysis},
  author={Paulson, Albert S and Dixit, R},
  booktitle={Financial Models of Insurance Solvency},
  pages={37--55},
  year={1989},
  publisher={Springer}
}

@article{KeLeGoYi21,
  title={Economists' erroneous estimates of damages from climate change},
  author={Keen, Stephen and Lenton, Timothy M and Godin, Antoine and Yilmaz, Devrim and Grasselli, Matheus and Garrett, Timothy J},
  journal={arXiv preprint arXiv:2108.07847},
  year={2021}
}

@book{No13,
  title={The climate casino: Risk, uncertainty, and economics for a warming world},
  author={Nordhaus, William},
  year={2013},
  publisher={Yale University Press}
}

@article{Le08,
  title={Tipping elements in the Earth's climate system},
  author={Lenton, Timothy M and Held, Hermann and Kriegler, Elmar and Hall, Jim W and Lucht, Wolfgang and Rahmstorf, Stefan and Schellnhuber, Hans Joachim},
  journal={Proceedings of the national Academy of Sciences},
  volume={105},
  number={6},
  pages={1786--1793},
  year={2008},
  publisher={National Academy of Sciences}
}

@misc{FactSet,
  author       = {{FactSet}},
  title        = {{FactSet Database}},
  howpublished = {{Available at: \url{https://www.factset.com/}}},
  year         = {2024},
  note         = "Accessed on 2024-06-30"
}

@misc{APRAGIperf_data,
  title        = {{Quarterly general insurance performance statistics}},
  howpublished = {https://www.apra.gov.au/quarterly-general-insurance-performance-statistics},
  note         = {Accessed: 2024-06-30},
  year         = {2024},
  author       = {{APRA}}
}

@misc{APRAGIIL_data,
  title        = {{Annual general insurance institution-level statistics}},
  howpublished = {https://www.apra.gov.au/annual-general-insurance-institution-level-statistics},
  note         = {Accessed: 2024-06-30},
  year         = {2024},
  author       = {{APRA}}
}

@misc{CorporateTaxRate_data,
  author       = {{Trading Economics}},
  title        = {{Australia Corporate Tax Rate}},
  howpublished = {\url{https://tradingeconomics.com/australia/corporate-tax-rate}},
  note         = {Accessed: 2024-06-30},
  year         = {2024}
}

@article{Er13,
  title={Spline interpolation techniques},
  author={Erdogan, KAYA},
  journal={Journal of Technical Science and Technologies},
  pages={47--52},
  year={2013}
}

@article{NeNePi25,
  title={Reconsidering the Macroeconomic Damage of Severe Warming},
  author={Neal, Timothy and Newell, Ben R and Pitman, Andy J},
  journal={Environmental Research Letters},
  year={2025}
}

@article{BaNo24,
  title={Policies, projections, and the social cost of carbon: Results from the DICE-2023 model},
  author={Barrage, Lint and Nordhaus, William},
  journal={Proceedings of the National Academy of Sciences},
  volume={121},
  number={13},
  pages={e2312030121},
  year={2024},
  publisher={National Acad Sciences}
}

@article{PiWeBeGoViHaHa10,
  title={Statistical bias correction of global simulated daily precipitation and temperature for the application of hydrological models},
  author={Piani, Claudio and Weedon, GP and Best, M and Gomes, SM and Viterbo, Pedro and Hagemann, Stefan and Haerter, JO},
  journal={Journal of hydrology},
  volume={395},
  number={3-4},
  pages={199--215},
  year={2010},
  publisher={Elsevier}
}

@article{Nord18DICE26R2,
  title={Projections and uncertainties about climate change in an era of minimal climate policies},
  author={Nordhaus, William},
  journal={American economic journal: economic policy},
  volume={10},
  number={3},
  pages={333--360},
  year={2018},
  publisher={American Economic Association}
}

@article{DeChLa12,
  title={Long-term economic growth projections in the Shared Socioeconomic Pathways},
  author={Dellink, Rob and Chateau, Jean and Lanzi, Elisa and Magn{\'e}, Bertrand},
  journal={Global Environmental Change},
  volume={42},
  pages={200--214},
  year={2017},
  publisher={Elsevier}
}

@article{Bu22,
  title={Public Nuisance and Climate Change: The Common Law's Solutions to the Plaintiff, Defendant and Causation Problems},
  author={Bullock, David},
  journal={The Modern Law Review},
  volume={85},
  number={5},
  pages={1136--1167},
  year={2022},
  publisher={Wiley Online Library}
}

@article{CeJa22,
  title={This changes everything: Climate shocks and sovereign bonds⁎},
  author={Cevik, Serhan and Jalles, Jo{\~a}o Tovar},
  journal={Energy Economics},
  volume={107},
  pages={105856},
  year={2022},
  publisher={Elsevier}
}

@article{Ma22,
  title={Natural disasters, climate change, and sovereign risk},
  author={Mallucci, Enrico},
  journal={Journal of International Economics},
  volume={139},
  pages={103672},
  year={2022},
  publisher={Elsevier}
}

@article{KlAgBuKrMo23,
  title={Rising temperatures, falling ratings: The effect of climate change on sovereign creditworthiness},
  author={Klusak, Patrycja and Agarwala, Matthew and Burke, Matt and Kraemer, Moritz and Mohaddes, Kamiar},
  journal={Management Science},
  volume={69},
  number={12},
  pages={7468--7491},
  year={2023},
  publisher={INFORMS}
}

@article{McSaCrMoMuPiGi19,
  title={Normalised insurance losses from Australian natural disasters: 1966--2017},
  author={McAneney, John and Sandercock, Benjamin and Crompton, Ryan and Mortlock, Thomas and Musulin, Rade and Pielke Jr, Roger and Gissing, Andrew},
  journal={Environmental Hazards},
  volume={18},
  number={5},
  pages={414--433},
  year={2019},
  publisher={Taylor \& Francis}
}

@article{WaShPu21,
  title={Linking temperature to catastrophe damages from hydrologic and meteorological extremes},
  author={Wasko, Conrad and Sharma, Ashish and Pui, Alexander},
  journal={Journal of Hydrology},
  volume={602},
  pages={126731},
  year={2021},
  publisher={Elsevier}
}

@article{RiTePiSt22,
  title={The missing risks of climate change},
  author={Rising, James and Tedesco, Marco and Piontek, Franziska and Stainforth, David A},
  journal={Nature},
  volume={610},
  number={7933},
  pages={643--651},
  year={2022},
  publisher={Nature Publishing Group UK London}
}

@techreport{HsJi14,
  title={The causal effect of environmental catastrophe on long-run economic growth: Evidence from 6,700 cyclones},
  author={Hsiang, Solomon M and Jina, Amir S},
  year={2014},
  institution={National Bureau of Economic Research}
}

@article{HaDu09,
  title={Can natural disasters have positive consequences? Investigating the role of embodied technical change},
  author={Hallegatte, St{\'e}phane and Dumas, Patrice},
  journal={Ecological economics},
  volume={68},
  number={3},
  pages={777--786},
  year={2009},
  publisher={Elsevier}
}

@book{AlJo93,
  title={Political economy of large natural disasters: with special reference to developing countries},
  author={Albala-Bertrand, Jose-Miguel},
  year={1993},
  publisher={Oxford University Press}
}

@article{SkTo02,
  title={Do natural disasters promote long-run growth?},
  author={Skidmore, Mark and Toya, Hideki},
  journal={Economic inquiry},
  volume={40},
  number={4},
  pages={664--687},
  year={2002},
  publisher={Wiley Online Library}
}

@article{No09,
  title={The macroeconomic consequences of disasters},
  author={Noy, Ilan},
  journal={Journal of Development economics},
  volume={88},
  number={2},
  pages={221--231},
  year={2009},
  publisher={Elsevier}
}

@article{WeTeRe10,
  title={The effects of regulated premium subsidies on insurance costs: An empirical analysis of automobile insurance},
  author={Weiss, Mary A and Tennyson, Sharon and Regan, Laureen},
  journal={Journal of Risk and Insurance},
  volume={77},
  number={3},
  pages={597--624},
  year={2010},
  publisher={Wiley Online Library}
}

@misc{BeInsureMedia25,
  author       = {{BeInsure Media}},
  title        = {IAIS Reinsurance Market Survey: Premiums, Assets, Retention \& Distribution},
  year         = {2025},
  howpublished = {\url{https://beinsure.com/iais-reinsurance-market-survey}},
  note         = {Accessed August 2025}
}

@article{MeScBaBoBo24,
  title={A perspective on the next generation of Earth system model scenarios: towards representative emission pathways (REPs)},
  author={Meinshausen, Malte and Schleussner, Carl-Friedrich and Beyer, Kathleen and Bodeker, Greg and Boucher, Olivier and Canadell, Josep G and Daniel, John S and Diongue-Niang, A{\"\i}da and Driouech, Fatima and Fischer, Erich and others},
  journal={Geoscientific Model Development},
  volume={17},
  number={11},
  pages={4533--4559},
  year={2024},
  publisher={Copernicus GmbH}
}

@article{MeLeRu24,
  title={Climate change and its impact on home insurance uptake in Australia},
  author={Melser, Daniel and Le, Trinh and Ruthbah, Ummul},
  journal={Ecological Economics},
  volume={222},
  pages={108195},
  year={2024},
  publisher={Elsevier}
}

@techreport{Lloyds2025ClimateTrends,
  author      = {{Lloyd's}},
  title       = {Climate Risk: Capital Modelling -- Market Trends},
  institution = {Lloyd's of London},
  year        = {2025},
  month       = {July},
  url         = {https://assets.lloyds.com/media/1cec7cba-a0b3-4744-b9e0-2a21f0558f57/Climate%20Risk%20-%20Market%20Trends%20-%20Published.pdf},
  note        = {Market Oversight report}
}

@misc{EMDAT2023Issues,
  author       = {{EM-DAT}},
  title        = {Known Issues and Limitations | EM-DAT Documentation},
  howpublished = {\url{https://doc.emdat.be/docs/known-issues-and-limitations/}},
  note         = {Accessed: 2025-10-27},
  year         = {2023}
}

@techreport{PAGE09,
  author       = {Hope, C. W.},
  year         = {2011},
  title        = {The PAGE09 Integrated Assessment Model: A Technical Description},
  institution  = {Cambridge Judge Business School, University of Cambridge},
  note         = {Working Paper No. 4/2011},
  url          = {https://www.jbs.cam.ac.uk/wp-content/uploads/2020/08/wp1104.pdf}
}

@techreport{FUND38,
  author       = {Anthoff, D. and Tol, R. S. J.},
  year         = {2014},
  title        = {The Climate Framework for Uncertainty, Negotiation and Distribution (FUND): Technical Description, Version 3.8},
  institution  = {Energy and Resources Group, University of California at Berkeley, and Department of Economics, University of Sussex},
  note         = {Technical Description of the FUND Integrated Assessment Model, Version 3.8},
  url          = {https://www.fund-model.org/files/documentation/Fund-3-8-Scientific-Documentation.pdf}
}

@misc{RMS2022,
  author       = {RMS},
  title        = {RMS launches new climate change models in U.S. and Japan},
  howpublished = {\url{https://www.moodys.com/web/en/us/insights/announcements/rms-launches-new-climate-change-models-in-us-and-japan.html}},
  year         = {2022}}

@article{DaBeCo87a,
  title={The solvency of a general insurance company in terms of emerging costs},
  author={Daykin, Chris D and Bernstein, GD and Coutts, SM and Devitt, ERF and Hey, GB and Reynolds, DIW and Smith, PD},
  journal={ASTIN Bulletin: The Journal of the IAA},
  volume={17},
  number={1},
  pages={85--132},
  year={1987},
  publisher={Cambridge University Press}
}

@article{DaBeCo87b,
  title={Assessing the solvency and financial strength of a general insurance company},
  author={Daykin, Christopher D and Bernstein, Geoffrey D and Coutts, Stewart M and Devitt, E Russell F and Hey, G Brian and Reynolds, D Ian W and Smith, Peter D},
  journal={Journal of the Institute of Actuaries},
  volume={114},
  number={2},
  pages={227--325},
  year={1987},
  publisher={Cambridge University Press}
}

@article{DaHe90,
  title={Managing uncertainty in a general insurance company},
  author={Daykin, Christopher D and Hey, G Brian},
  journal={Journal of the Institute of Actuaries},
  volume={117},
  number={2},
  pages={173--277},
  year={1990},
  publisher={Cambridge University Press}
}

@techreport{CAS95,
  author      = {{Casualty Actuarial Society}},
  title       = {CAS Dynamic Financial Analysis Handbook},
  year        = {1995},
  institution = {Casualty Actuarial Society},
  url         = {https://www.casact.org/sites/default/files/database/forum_96wforum_96wf001.pdf}
}

@article{LoSt97,
  title={An Integrated Dynamic Financial Analysis and Decision Support System for a Property Catastrophe Reinsurer},
  author={Lowe, Stephen P and Stanard, James N},
  journal={ASTIN Bulletin: The Journal of the IAA},
  volume={27},
  number={2},
  pages={339--371},
  year={1997},
  publisher={Cambridge University Press}
}

@inproceedings{BeMa99,
  title={A comprehensive system for selecting and evaluating DFA model parameters},
  author={Berger, Adam J and Madsen, C},
  booktitle={CAS Forum},
  pages={51},
  year={1999}
}

@article{CaZi98,
  title={Formulation of the Russell-Yasuda Kasai financial planning model},
  author={Carino, David R and Ziemba, William T},
  journal={Operations research},
  volume={46},
  number={4},
  pages={433--449},
  year={1998},
  publisher={INFORMS}
}

@incollection{CoDe86,
  title={The assessment of the financial strength of insurance companies by a generalized cash flow model},
  author={Coutts, Stewart M and Devitt, Russell},
  booktitle={Financial models of insurance solvency},
  pages={1--36},
  year={1986},
  publisher={Springer}
}

\newpage

\appendix

\section{Simulation results of climate variables} \label{Appendix:SimulationResultsClimate}

Figure \ref{fig:ClimateSimulations} presents the simulated climate variables derived from the CMIP6 model outputs \citep{CMIP6_data}. The results suggest that most climate variables, except for SST gradients, exhibit an upward trend, particularly under high-emission scenarios, indicating an overall increase in climate risk.

In addition to this long-term trend, the average simulation path also shows notable inter-annual variability. This variability, likely driven by internal climate processes (e.g., El Ni\~no cycles) captured by the CMIP6 models \citep{JaSc23}, may contribute to annual fluctuations in insurance premiums and costs as shown in the corresponding sections in the paper. 

Furthermore, the projections display a high degree of uncertainty. This uncertainty, also highlighted in other studies \citep[see, e.g.,][]{WuMiFaGoZhZh22}, arises from both the climate model uncertainty and the aleatoric uncertainty incorporated in our climate module (see Section \ref{Section:Climate}). Given that such uncertainty, especially at the upper percentiles, is crucial for actuarial applications and capital modelling, these findings underscore the importance of accounting for uncertainty in climate projections, as discussed in Section \ref{Section:StatementofContributions} and \ref{Section:Climate}.

\begin{figure}[H]
    \centering
    \includegraphics[width= 0.8\textwidth]{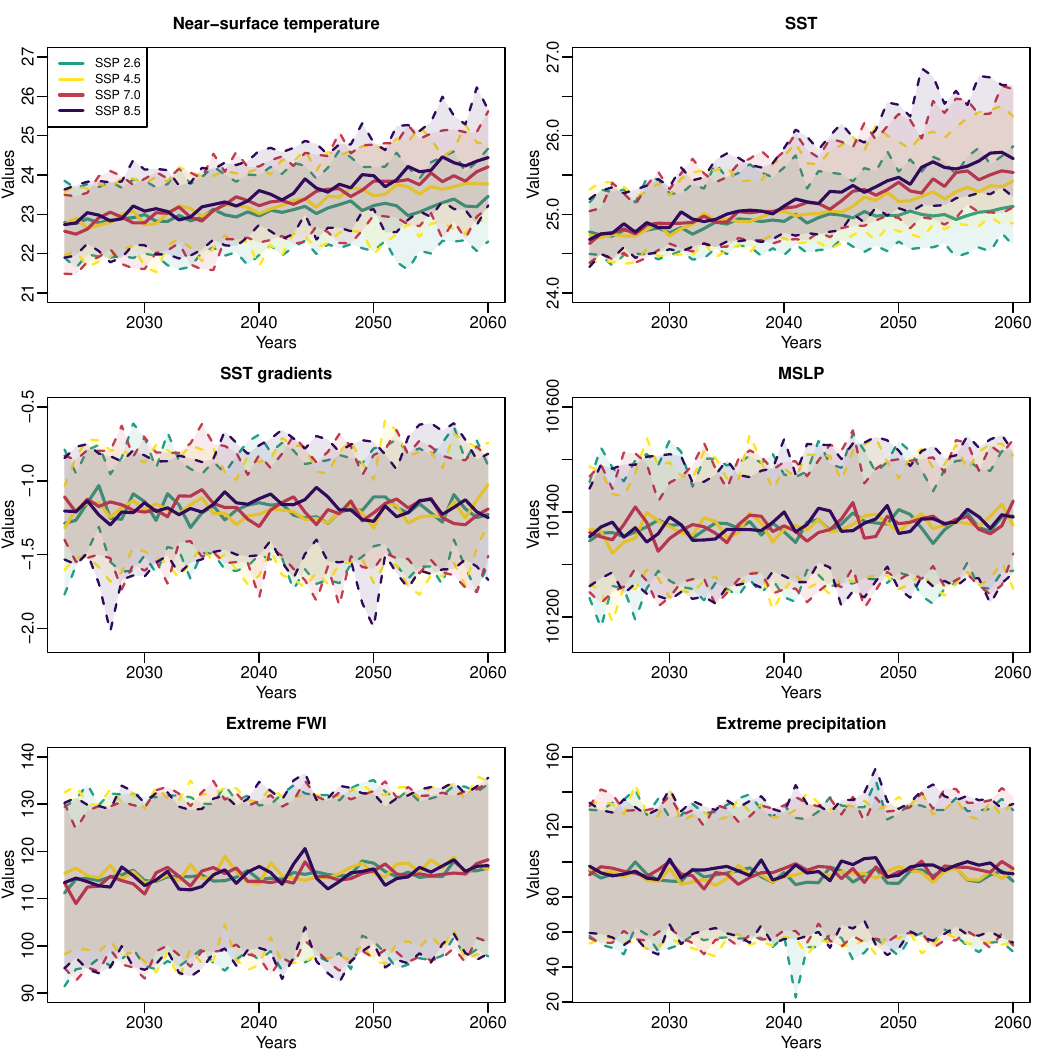}
    \caption{\textbf{Simulation results for climate variables.} Results are generated using CMIP6 model outputs and the simulation methodology described in Section \ref{Section:Climate}. Solid lines represent the average simulation paths under different climate scenarios, while dashed lines indicate the $5^{\text{th}}$ and $95^{\text{th}}$ percentiles. The simulations illustrate both long-term trends and inter-annual variability, as well as the uncertainty associated with the projections.}
    \label{fig:ClimateSimulations}
\end{figure}

\section{Relationships between market surplus and investment returns} \label{Appendix:SurplusInvestment}

\begin{figure}[H]
  \centering
  \subfloat[]{\includegraphics[width=0.49\textwidth]{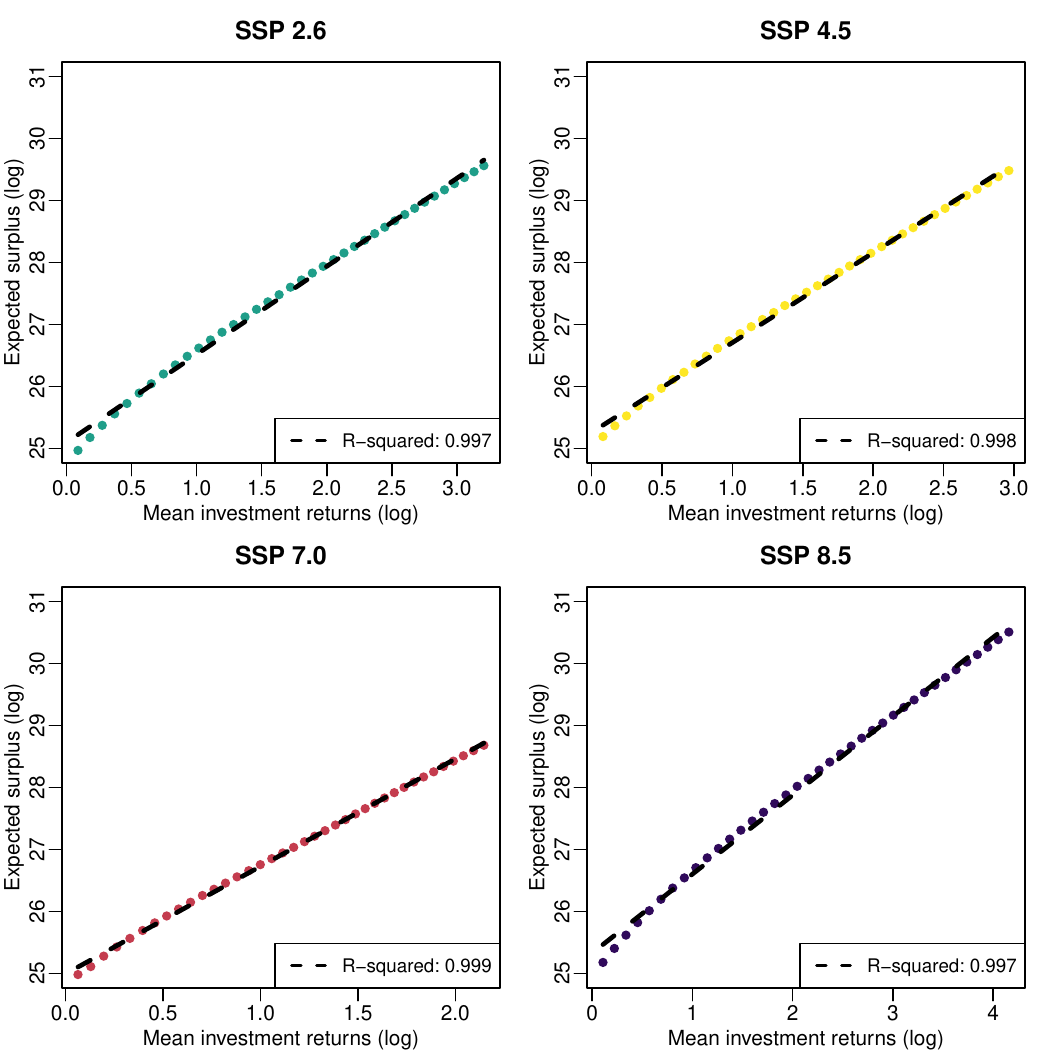}\label{fig:ExpectedMarketSurplusDrivers}}
  \subfloat[]{\includegraphics[width=0.49\textwidth]{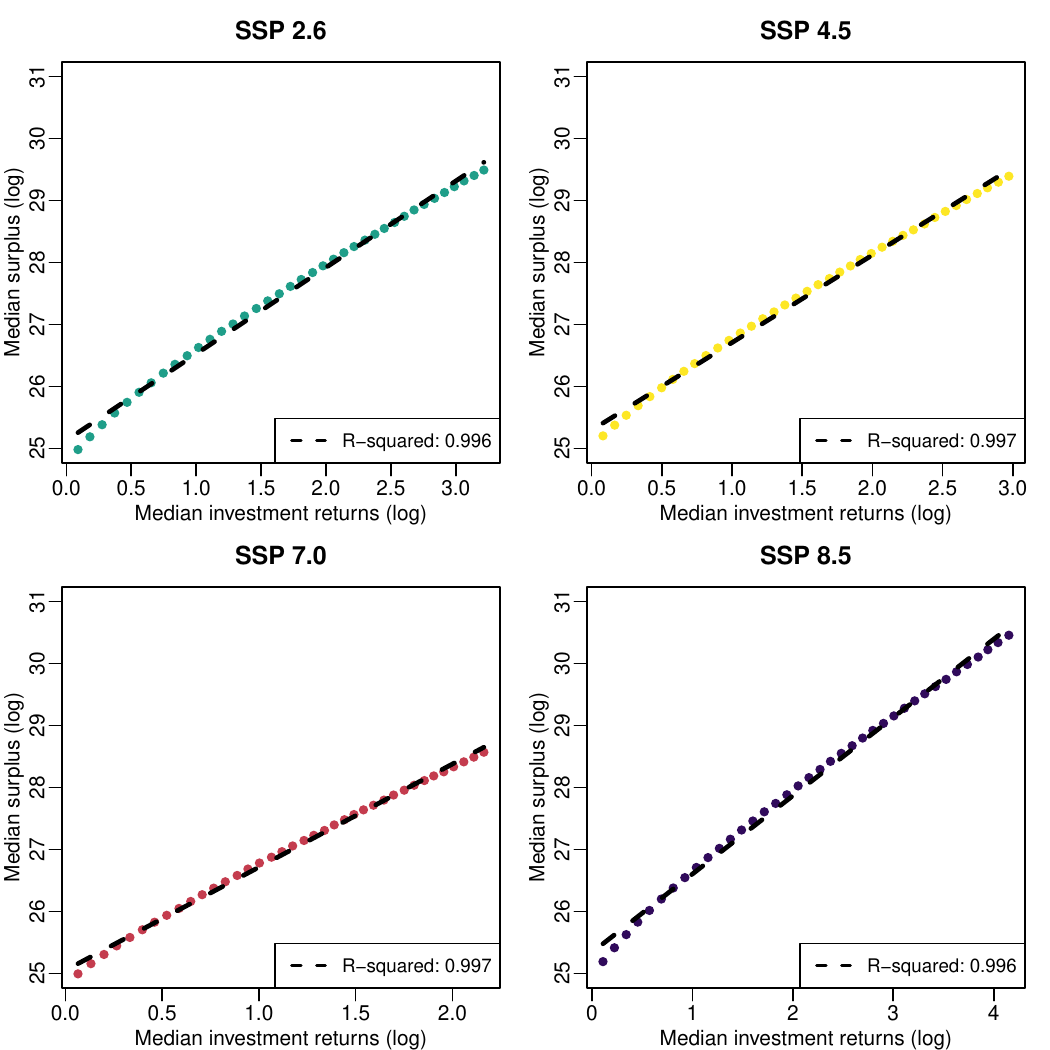}\label{fig:MedianMarketSurplusDrivers}}
  \caption{\textbf{Expected (a) and median (b) market surplus vs. average (a) and median (b) compounded investment returns (log scale).} The results indicate a strong correlation between market surplus and compounded investment returns at both the mean and median levels.}
\end{figure}

\section{Mean Compound Annual Growth Rate (CAGR) of market surplus} \label{Appendix:MeanGACR}

\begin{figure}[h!]
    \centering
    \begin{minipage}{0.65\textwidth}
        \includegraphics[width=0.8\linewidth]{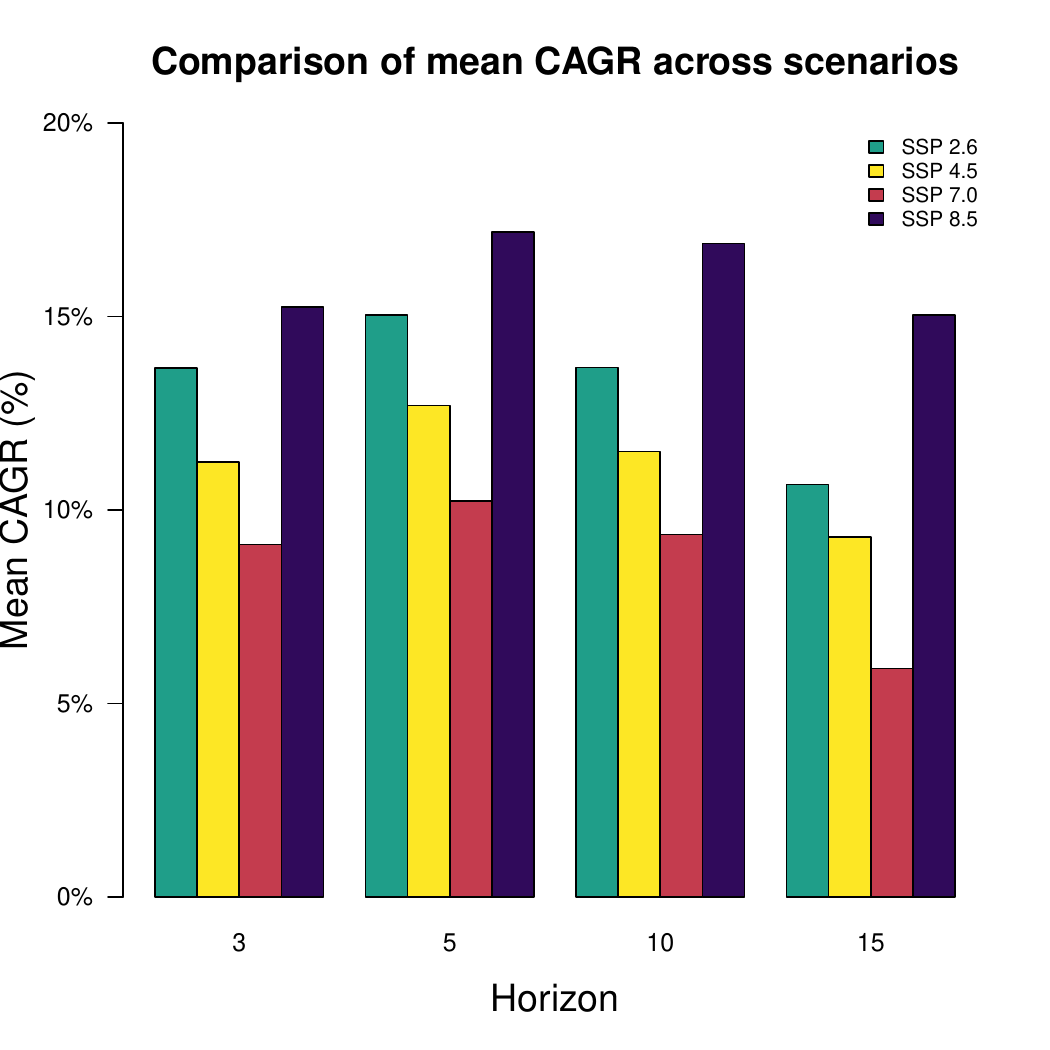} 
    \end{minipage}%
    \hfill
    \begin{minipage}{0.35\textwidth}
        \captionof{figure}{\textbf{Mean Compound Annual Growth Rate (CAGR) of market surplus, by projection horizon and climate scenario.} This metric reflects the rate of capital accumulation over different projection periods. The highest growth rate is observed under the SSP 8.5 scenario during the early projection horizon.}
    \end{minipage}
\end{figure}

\newpage
\pagenumbering{arabic}  
\setcounter{page}{1}
\renewcommand{\thesection}{\arabic{section}}
\section*{Online Appendix}

This online appendix accompanies the paper ``Dynamic Financial Analysis (DFA) of General Insurers under Climate Change" by Benjamin Avanzi, Yanfeng Li, Greg Taylor, and Bernard Wong.

Included here are \rev{detailed description of the data sources and a presentation of the main calibration results (see Section \ref{Appendix:DataSourcesandMainCalibration}); Calibration results for the benchmark stationary DFA model (see Section \ref{Appendix:CalibrationStationaryDFA});} supplementary details on model calibration results for the climate and hazard modules (see Section~\ref{Appendix:CalibrationResultsClimateandHazard}); supplementary simulation outputs from the hazard module (see Section~\ref{Appendix:SupplementaryModelSimulationResults}); supplementary details on the equity model (see Section~\ref{Appendix:EquitySupplementary}); a benchmarking analysis of projected damage ratios (see Section~\ref{Appendix: ProjectedDamageRatiosComparison}); discussion on potential extensions of the current model to multi-regional projections, portfolio management, and integration with CAT modelling outputs (see Section~\ref{Appendix:ModelExtension}); a sensitivity test on the impacts of dividend payout policies (see Section~\ref{Appendix:SensitivityDividendsPaids}); and a sensitivity test on the impacts of alternative economic damage assumptions on investment returns (see Section~\ref{Appendix:SensitivityEconomicDamages}).

\appendix

\section{\rev{Data sources, main calibration procedures and results}} \label{Appendix:DataSourcesandMainCalibration}

\subsection{\rev{Data sources}} \label{Appendix:DataSources}

\paragraph{Climate data}

The historical weather data used to calibrate the climate and hazard modules are sourced from ERA 5 reanalysis data \citep{ERA5_data}, which combines past observations with current weather computer models to provide consistent estimates of atmospheric, land and oceanic climate variables from 1950 to the present. For future projections, we use the outputs from the CMIP6 models \citep{CMIP6_data}, which offers projected meteorological variables under various emissions scenarios up to 2100 at both monthly and daily resolutions. Additionally, the CMIP6 models generate backcasts of climate variables from 1850 to 2014, which --together with historical data -- are used to calibrate the bias correction and aleatoric uncertainty models as detailed in Section \ref{Section:Climate}. As our analysis focuses on the frequency and severity of hazards at the national level, we use the average values of climate observations across all gridded cells for calibration and projection purposes.

\paragraph{Hazards data}

To calibrate the frequency and severity models for catastrophe insurance losses as detailed in Section \ref{Section:WeatherCovariatesSelection}, we draw on the ICA dataset \citep{ICA_data}, which is maintained by the Insurance Council of Australia. This dataset covers all recorded natural disasters in Australia from 1967 to 2024, including variables such as disaster locations, start and end dates, and total insured damages. The insured damages data has been normalised for CPI and exposure prior to model calibration.

\paragraph{Macro-economic data}

In Section \ref{Section:RiskfreeRates}, we calibrate the interest rate model using the RBA cash rates data \citep{RBACashRates_data} and potential GDP growth estimates from the World Bank Potential Growth Database \citep{KiKoOhRu23}. The RBA dataset covers cash rate targets and overnight cash rates from 1976 to 2023, while the World Bank Database offers annual Australian potential GDP growth from 1981 to 2021. Consequently, the calibration period spans 1981–2021.

In Section \ref{Section:InflationRates}, we calibrate the inflation rates model with data from \cite{ABSCPI_data}, which provides quarterly Consumer Price Index (CPI) information for Australia from 1948 to 2023. These data are aggregated to an annual scale to align with the DFA model’s time resolution.

For projections, we obtain socio-economic variables from the IIASA SSP database \citep{RiVaKr17}, offering forecasts of GDP, population, and energy production at the five-year interval under multiple development paths at the country level. Spline interpolation \citep{Er13} is then used to convert these projections to annual data.

\paragraph{Financial data}

To calibrate the equity models described in Section \ref{Section:Equity}, we employ Australian gross corporate operating profits and total returns data for the All-Ordinaries share index. The corporate operating profits dataset, sourced from \cite{ABSBI_data}, covers the total industry on a quarterly basis from 1994 to 2023. The All-Ordinaries index data, obtained from \cite{FactSet}, provides daily-to-annual total returns from 1992 to 2023.

\paragraph{Insurance market statistics}

In addition to the data sources mentioned above, we draw on Australian insurance market statistics from the General Insurance Performance Statistics database \citep{APRAQuarterlyStats} and the General Insurance Institution-Level Statistics database \citep{APRAGIIL_data}. These data inform our market assumptions and support the calibration of the non-catastrophe loss model (see Section \ref{Section:InsuranceCost}). Because no direct non-catastrophe loss data are publicly available, we derive industry-level non-catastrophe losses by subtracting the ICA-recorded catastrophe losses from the total industry losses reported in the General Insurance Performance Statistics database \citep{APRAGIIL_data}, after joining those two datasets.

The General Insurance Performance Statistics database contains quarterly aggregate financial data on Australian general insurers from 2002 to 2023, while the General Insurance Institution-Level Statistics database provides annual, institution-level financial information from 2005 to 2023. Both databases include key metrics such as insurance losses, premiums, equity bases, and the number of underwritten risks.

\subsection{\rev{Main calibration procedures and results}} \label{Appendix:MainCalibration}

Due to the large number of parameters, the bias-correction and noise volatility parameters calibrated for selected climate variables in the climate module are presented in Online Appendix \ref{Appendix:CalibrationResultsClimate} for illustrative purposes.

Table \ref{Table:CalibrationResultsHazard} presents the calibrated parameters for the selected model of each hazard type\footnote{\footnotesize{Each coefficient $\beta$ links a climate variable (superscript) to a distribution parameter (subscript). For example, $\beta_{(\lambda)}^{\text{rx5day}}$ denotes the sensitivity of frequency ($\lambda$) to \texttt{rx5day}, while $\beta_{(\mu)}^{\text{rx5day}}$ refers to its effect on the severity location parameter.}}. The full set of candidate models is detailed in Online Appendix \ref{Appendix:CalibrationResultsHazards}. Final models were chosen based on both physical relevance (see Section \ref{Section:WeatherCovariatesSelection}) and statistical performance. For example, in Table \ref{Table:FloodFrequencyModels}, all flood frequency models show positive coefficients for precipitation, consistent with known flood mechanisms. Model 3, using \texttt{rx5day} as an extreme precipitation proxy, performed best by AIC and BIC and its covariate is statistically significant at the $5\%$ level. Thus, Model 3 is selected. As another case, although the atmospheric temperature coefficient is not statistically significant at the $5\%$ or $10\%$ levels, we retain it in the hailstorm model due to the theoretical importance of temperature effects (see Section \ref{Section:HailstormFormulation}).

\begin{table}[H]
\small
\centering
\begin{tabular}{lllll}
Hazards & Parameters    & Values    & Data (for calibration)  & Data (for projection) \\ \hline
Flood       & $\beta_{(\lambda)}^{\text{rx5day}}$      & 0.037**         &   ERA 5 re-analysis; ICA data             &  CMIP 6       \\ 
       & $\beta_{(\mu)}^{\text{rx5day}}$      & 0.035*        & ERA 5 re-analysis; ICA data                 & CMIP 6              \\ \hline
Bushfire       & $\beta_{(\lambda)}^{\text{mfwixx}}$      & 0.084**         &   ERA 5 re-analysis; ICA data             &  CMIP 6       \\ \hline

Cyclones       & $\beta_{(\lambda)}^{\text{SST}}$      & 1.213***         &   ERA 5 re-analysis; ICA data             &  CMIP 6       \\ \hline

Storms       & $\beta_{(\lambda)}^{\text{SST}}$      & 0.348**         &   ERA 5 re-analysis; ICA data             &  CMIP 6       \\
  & $\beta_{(\mu)}^{\text{SST}}$      & 0.239        &   ERA 5 re-analysis; ICA data             &  CMIP 6       \\ \hline
East Coast Low       & $\beta_{(\lambda)}^{\Delta \text{SST}}$      & 2.189*         &   ERA 5 re-analysis; ICA data             &  CMIP 6       \\ \hline
Hails       & $\beta_{(\lambda)}^{T^{\text{NS}}}$      & 0.211***         &   ERA 5 re-analysis; ICA data             &  CMIP 6       \\ \hline

  & $\beta_{(\lambda)}^{T^{\text{MT}}}$      & -0.079        &   ERA 5 re-analysis; ICA data             &  CMIP 6       \\ \hline
\bottomrule
\multicolumn{4}{l}{\rule{0pt}{1em}+ p $<$ 0.1, * p $<$ 0.05, ** p $<$ 0.01, *** p $<$ 0.001}\\
\end{tabular}
\caption{Calibration of key parameters: Hazard modules}
\label{Table:CalibrationResultsHazard}
\end{table}

The calibrated parameters for the real risk-free interest rate and inflation rate models are shown in Table \ref{Table:CalibrationResultsMEV}. The positive impact of potential GDP growth on real rates (as captured by the parameter $\beta_1$) aligns with the expectations discussed in Section \ref{Section:RiskfreeRates}. For temperature impacts on inflation, the coefficients $\alpha_{1+L}$ and $\beta_{1 + L}$ are taken from \cite{KoKuLiNi24}, who calibrated these parameters across 121 countries to evaluate climate-related effects on inflation.

\begin{table}[H]
\small
\centering
\begin{tabular}{lllll}
Model  &Parameters    & Values    & Data (for calibration)  & Data (for projection) \\ \hline
Real rates &$\beta_1$ \tiny{($\partial{r_t}/\partial{g_t}$)}  & 2.206** &RBA cash rate; & SSP database (IIASA) \\
 & & &World Bank Potential Growth Database &  \\
 & $\mu_r$ & -0.0005 &same as above. & \\
 & $\phi_r$ &0.478** &same as above. & \\
 & $\sigma_r$ & 0.025 &same as above. & \\
\\ \hline
Inflation  & $\alpha_{1+L}$, $\beta_{1 + L}$ & \textemdash &\citep{KoKuLiNi24} &CMIP 6  \\
&$\mu_i$ &0.0517*** &ABS CPI data &  \\
&$a_i$ &0.713*** &ABS CPI data &  \\
& $\sigma_i$ & 0.0309 &ABS CPI data &  \\ \hline
\bottomrule
\multicolumn{4}{l}{\rule{0pt}{1em}+ p $<$ 0.1, * p $<$ 0.05, ** p $<$ 0.01, *** p $<$ 0.001}\\
\end{tabular}
\caption{Calibration of key parameters: Macro-economic variables}
\label{Table:CalibrationResultsMEV}
\end{table}

Table \ref{Table:CalibrationResultsEquity} presents the calibrated equity model parameters. $\alpha_1$ (sensitivity of operating profit growth to consumption growth) and $\sigma_O$ (its standard deviation) are calibrated using Australian operating profit data \citep{ABSBI_data} and consumption growth data \citep{WorldBankIndicators_data}. $\beta_1$ (sensitivity of excess equity returns to operating profit growth) and $\sigma_X$ (standard deviation of excess equity returns) are calibrated using operating profit data \citep{ABSBI_data} and All-Ordinaries total returns \citep{FactSet}. Results indicate positive relationships between consumption and profit growth, and between profit growth and equity returns, consistent with expectations as discussed in Online Appendix \ref{Appendix:AnalyticalEquitySupplementary}.

To calibrate the sensitivity of operating profits in brown firms \footnote{\footnotesize{Here, we define brown firms as those operating within the oil and gas sector.}} to planned changes in production under various climate scenarios (sourced from the SSP database \citep{RiVaKr17}), we adopt a simplified approach akin to that used in \cite{GrMa20}. We estimate the impact of changes in output on operating profits for a representative firm in the Australian energy sector by considering its fixed and variable costs. These results are then extrapolated to other oil and gas firms within the sector. Woodside Energy Group Ltd (WDS) is selected as the representative firm, given its dominant market share of $66\%$ in the Australian energy sector as of March 2024 \citep{FactSet}. The impact of output changes is calculated as the average percentage change in operating profit per $1\%$ change in production, based on the historical financial statements available from 2014 to 2021 in FactSet. While this approach is intentionally simplified, general insurers may refine it to better align with their portfolio compositions. Enhancements could include integrating more granular transition risk metrics, potentially incorporating proprietary data sources.

\begin{table}[H]
\small
\centering
\begin{tabular}{llll}
Parameters    & Values    & Data (for calibration)  & Data (for projection) \\ \hline
$\alpha_1$ \tiny{($\partial{\Delta {OP}_t}/\partial{\Delta C_t}$)} &3.824*  
 &World Development Indicators (World Bank);  & SSP database (IIASA)   \\ 
 &  & ABS Business Indicators &CMIP 6  \\ 
$\sigma_{\text{O}}$ & 0.083 &same as above. & \\ 
$\beta_1$ \tiny{($\partial{x_t}/\partial{\Delta {OP}_t}$)}  &0.047  
 &ABS Business Indicators; All-Ordinaries index \tiny{(Factset)}  & SSP database (IIASA)  \\ 
 $\sigma_{\text{x}}$ & 0.103 &same as above. & \\ 
 $\beta^{\text{(B)}}$ \tiny{($\partial{\Delta {OP}_t^B}/\partial{\Delta Y_t^B}$)} &1.768  &Income statements of WDS (Factset) &SSP database (IIASA) \\ \hline
 \bottomrule
\multicolumn{4}{l}{\rule{0pt}{1em}+ p $<$ 0.1, * p $<$ 0.05, ** p $<$ 0.01, *** p $<$ 0.001}\\
\end{tabular}
\caption{Calibration of key parameters: Equity returns}
\label{Table:CalibrationResultsEquity}
\end{table}

Table \ref{Table:MarketAssumptions} presents the market assumptions underlying our simulations, derived from general insurance market statistics \citep{APRAGIperf_data} and financial disclosures from individual insurers. In the projections, both excess and limit levels are adjusted for changes in GDP and CPI across future periods and under different climate scenarios.

\begin{table}[H]
\small
\centering
\begin{tabular}{lllll}
Size of Insurers & Numbers & Market Shares & Excess (normalised)  & Limit (normalised) \\ \hline
Large            & 4       & 20\%          & \$1000 million                 & \$600 million                 \\ 
Medium           & 4       & 3\%           & \$150 million                  & \$90 million                  \\ 
Small            & 12      & 0.67\%        & \$33 million                   & \$20 million                  \\ \hline
\end{tabular}
\caption{Market assumptions}
\label{Table:MarketAssumptions}
\end{table}

Table \ref{Table:OtherAssumptions} summarises the additional assumptions used in the simulations. The target capital ratio is determined as the general insurance industry's average ratio of eligible equity to the Minimum Capital Requirement (MCR), based on data from the APRA General Insurance Institution-Level Statistics database \citep{APRAGIIL_data}. The uninsured-to-insured loss ratio is used to scale catastrophe insurance losses from the hazard module to uninsured economic damage, serving as an input to the equity model. This ratio is calculated as the average proportion of uninsured to insured losses over the historical period from 1985 to 2023, using data from the EM-DAT database \citep{EMDAT}, since the ICA dataset records only insurance losses. It should also be noted that this ratio represents only the proportion of uninsured to insured damages to assets, which are primarily properties and motor vehicles that tend to have relatively high insurance coverage due to regulatory and mortgage requirements \citep{MeLeRu24}, during natural disasters. It does not, however, reflect the overall insurance penetration of the broader economy. Moreover, this ratio may vary across countries and can be affected by differences in the quality and reporting protocols of hazard impact data across regions \citep{EMDAT2023Issues}.
\begin{table}[H]
\small
\centering
\begin{tabular}{llll}
 Parameters    & Type    & Values  & Source \\ \hline
$\rho$  &Risk-aversion &0.55 & \cite{KuMiRa11, PaBoAe13} \\ 
$\tau$  &Target capital ratio  &1.75  & \cite{APRAGIIL_data}
\\ 
$\eta$ &Uninsured-to-insured loss ratio  & 1.22 & \cite{EMDAT} \\ 
$w_{rf}$ &Allocation to risk-free assets & $60\%$ \footnote{\footnotesize{The assumption is that the majority of bonds are invested in risk-free government securities, as insurers typically prefer government bonds over corporate bonds \citep{OECDInsurMarket23}. Additionally, cash and deposits are considered nearly risk-free assets}} & \cite{OECDInsurMarket23}  \\ 
$w_{B}$ &Allocation to brown assets & $3\%$ & \cite{GaOz24} \\ \hline
\end{tabular}
\caption{Other assumptions} 
\label{Table:OtherAssumptions}
\end{table}

\begin{limitation} \label{Remark:ExposureAssumption}
   The data used to calibrate the hazard models are normalised to 2022 level for both CPI and exposure. For future projections, we adjust hazard losses using projected CPI and real GDP growth under each climate scenario. This differs from the normalisation technique in \cite{McSaCrMoMuPiGi19} (also adopted by ICA), which adjusts for dwelling numbers, values, sizes, and building standards based on historical data. Therefore, our approach assumes dwelling values across regions grow in line with nominal GDP, with no changes in dwelling size or building standards under different climate scenarios. We acknowledge this is a limitation, as climate change may affect dwelling numbers and values unevenly across regions, and adaptation measures could alter dwelling size and standards. These aspects are left for future research.
\end{limitation}

\section{Calibration results of the stationary DFA model} \label{Appendix:CalibrationStationaryDFA}

This section presents the calibrated parameters of the benchmark stationary DFA model specified in Section \ref{Section:StationaryDFA}. Table \ref{Table:CalibrationResultsHazardStationary} reports the calibrated parameters of the hazard loss distributions, while the calibrated parameters for the remaining model components are provided in Table \ref{Table:CalibrationResultsMEVStationary}. 

The stationary DFA model is calibrated using Australian data and, where applicable, consistent data sources as those employed for the climate-dependent DFA model. Data sources that are not required for the stationary specification (e.g., climate variables listed in Online Appendix \ref{Appendix:MainCalibration}) are excluded. 

To ensure comparability with the climate-dependent DFA results, we adopt the same market assumptions and decision variables as reported in Tables \ref{Table:MarketAssumptions} and \ref{Table:OtherAssumptions} in Online Appendix \ref{Appendix:MainCalibration}.

\begin{table}[H]
\small
\centering
\begin{tabular}{llll}
\toprule
Hazards & Parameters & Values & Data (for calibration) \\ 
\midrule
Flood & $\{\lambda^{\text{FL}}, \mu^{\text{FL}}, \sigma^{\text{FL}}\}$ 
& $\{0.857,\; 19.038,\; 1.420\}$ & ICA data \\ 

Bushfire & $\{\lambda^{\text{BF}}, \mu^{\text{BF}}, \sigma^{\text{BF}}\}$ 
& $\{0.661,\; 18.797,\; 1.406\}$ & ICA data \\ 

Tropical Cyclones & $\{\lambda^{\text{TC}}, \mu^{\text{TC}}, \sigma^{\text{TC}}\}$ 
& $\{0.679,\; 19.270,\; 1.601\}$ & ICA data \\ 

Storms & $\{\lambda^{\text{ST}}, \mu^{\text{ST}}, \sigma^{\text{ST}}\}$ 
& $\{0.554,\; 18.543,\; 1.230\}$ & ICA data \\ 

East Coast Low & $\{\lambda^{\text{ECL}}, \mu^{\text{ECL}}, \sigma^{\text{ECL}}\}$ 
& $\{0.125,\; 18.981,\; 1.391\}$ & ICA data \\ 

Hail & $\{\lambda^{\text{HS}}, \mu^{\text{HS}}, \sigma^{\text{HS}}\}$ 
& $\{0.268,\; 19.964,\; 1.307\}$ & ICA data \\ 
\bottomrule
\end{tabular}
\caption{\rev{Calibration of hazard models (stationary)}}
\label{Table:CalibrationResultsHazardStationary}
\end{table}

\begin{table}[H]
\small
\centering

\begin{tabular}{lllll}
\toprule
Variables  &Parameters    & Values    & Data (for calibration)   \\ \hline
Real rates &$\{g, m_r,\varphi_r, \sigma_r\}$  & $\{0,0.0249,0.637, 0.0246\}$ &RBA cash rate;  \\ 
Inflation  &$\{\mu_i, a_i, \sigma_i\}$ &$\{0.0517, 0.713, 0.0309\}$ &ABS CPI data \\ 
Excess equity returns &$\{\mu_{\text{x}}, \sigma_{\text{x}}\}$  &$\{0.0494, 0.0940\}$ &All-Ordinaries index \tiny{(Factset)} \\
Exposure growth &$\{\mu_{\omega},  \phi_{\omega}, \sigma_{\omega}\}$ \footnote{\footnotesize{Note that the volatility parameter is reported for completeness. In the projections, exposure growth is treated as deterministic, consistent with the climate-dependent DFA specification and the recommendations of \cite{KaGaKl01}.}} &$\{0.0380,0.374, 0.0168\}$ &APRA data \\ \hline
\end{tabular}

\caption{\rev{Calibration of macroeconomic, excess equity return, and exposure growth models (stationary)}}
\label{Table:CalibrationResultsMEVStationary}
\end{table}

\section{Model calibration results: supplementary details} \label{Appendix:CalibrationResultsClimateandHazard}

\subsection{Climate model supplementary calibration results} \label{Appendix:CalibrationResultsClimate}

The results presented here show the calibrated bias-correction and noise volatility parameters for selected climate variables within the climate module. The detailed calibration methodology is provided in Section \ref{Section:Climate} of the main paper.

\begin{table}[H]
\centering
\begin{tabular}{llll}
\toprule
Model names  & $\hat{\beta}_0^{(m)}$        & $\hat{\beta}_1^{(m)}$       & $\hat{\sigma}_{(m)}$        \\
\midrule
ACCESS-CM2    & 2.257       & 0.883       & 1.271     \\
CanESM5-CanOE & 2.484       & 0.889       & 1.354     \\
CESM2         & -0.065      & 0.987       & 1.256     \\
CMCC-CM2-SR5  & -2.18       & 1.085       & 1.405     \\
CNRM-CM6-1    & 0.735       & 0.974       & 1.321     \\
CNRM-ESM2-1   & 0.868       & 0.942       & 1.186     \\
FGOALS-f3-L   & 0.944       & 0.98        & 1.294     \\
FGOALS-g3     & -1.132      & 1.039       & 1.23      \\
INM-CM4-8     & 1.345       & 0.971       & 1.199     \\
INM-CM5-0     & 2.222       & 0.948       & 1.195     \\
IPSL-CM6A-LR  & -0.139      & 1.055       & 1.428     \\
MCM-UA-1-0    & 1.329       & 0.899       & 1.401     \\
MIROC-ES2L    & -2.734      & 1.045       & 1.255     \\
MIROC6        & -0.85       & 0.931       & 1.325     \\
MPI-ESM1-2-LR & 1.441       & 0.916       & 1.288     \\
MRI-ESM2-0    & -0.046      & 0.96        & 1.375     \\
NorESM2-MM    & -0.858      & 1.036       & 1.267  \\
\bottomrule
\end{tabular}
\caption{Near-surface temperature (monthly-average): Calibrated bias-correction coefficients ($\hat{\beta}_0^{(m)}$ and $\hat{\beta}_1^{(m)}$), and standard deviation of noise ($\hat{\sigma}_{(m)}$)}
\end{table}

\begin{table}[H]
\centering
\begin{tabular}{llll}
\toprule
Model names  & $\hat{\beta}_0^{(m)}$        & $\hat{\beta}_1^{(m)}$       & $\hat{\sigma}_{(m)}$        \\
\midrule
ACCESS-CM2    & 0.926       & 1.045       & 1.622     \\
CanESM5-CanOE & -4.122      & 0.924       & 1.32      \\
CESM2         & -4.897      & 0.881       & 1.491     \\
CNRM-CM6-1    & 6.783       & 1.04        & 1.412     \\
CNRM-ESM2-1   & 5.617       & 1.025       & 1.366     \\
INM-CM4-8     & 12.879      & 1.361       & 1.346     \\
INM-CM5-0     & 14.56       & 1.384       & 1.502     \\
IPSL-CM6A-LR  & 1.294       & 1.001       & 1.409     \\
MCM-UA-1-0    & -0.307      & 0.982       & 1.368     \\
MIROC-ES2L    & -10.139     & 0.675       & 1.488     \\
MIROC6        & -4.526      & 0.86        & 1.369     \\
MPI-ESM1-2-LR & 1.847       & 1.036       & 1.475     \\
MRI-ESM2-0    & 3.861       & 1.102       & 1.448     \\
NorESM2-MM    & -6.233      & 0.838       & 1.408     \\
UKESM1-0-LL   & 2.097       & 1.066       & 1.596    \\
\bottomrule
\end{tabular}
\caption{Air temperature (monthly-average): Calibrated bias-correction coefficients ($\hat{\beta}_0^{(m)}$ and $\hat{\beta}_1^{(m)}$), and standard deviation of noise ($\hat{\sigma}_{(m)}$)}
\end{table}

\begin{table}[H]
\centering
\begin{tabular}{llll}
\toprule
Model names  & $\hat{\beta}_0^{(m)}$        & $\hat{\beta}_1^{(m)}$       & $\hat{\sigma}_{(m)}$        \\
\midrule
ACCESS-CM2   & -2.418      & 0.996       & 0.316     \\
CNRM-CM6-1   & 4.446       & 0.801       & 0.263     \\
FGOALS-f3-L  & 2.266       & 0.924       & 0.35      \\
FGOALS-g3    & 0.792       & 0.955       & 0.279     \\
INM-CM4-8    & 1.414       & 0.944       & 0.239     \\
INM-CM5-0    & 0.636       & 0.994       & 0.255     \\
MCM-UA-1-0   & 2.26        & 0.881       & 0.378     \\
MIROC-ES2L   & -5.092      & 1.204       & 0.356     \\
UKESM1-0-LL  & 1.157       & 0.943       & 0.262 \\
\bottomrule
\end{tabular}
\caption{Sea-surface temperature (monthly-average): Calibrated bias-correction coefficients ($\hat{\beta}_0^{(m)}$ and $\hat{\beta}_1^{(m)}$), and standard deviation of noise ($\hat{\sigma}_{(m)}$)}
\end{table}

\begin{table}[H]
\centering
\begin{tabular}{llll}
\toprule
Model names  & $\hat{\beta}_0^{(m)}$        & $\hat{\beta}_1^{(m)}$       & $\hat{\sigma}_{(m)}$        \\
\midrule
ACCESS-CM2    & -17.133     & 1.306       & 8.675     \\
ACCESS-ESM1-5 & -4.806      & 1.239       & 8.346     \\
CanESM5       & -0.279      & 1.234       & 9.263     \\
CMCC-CM2-SR5  & 57.288      & 1.052       & 7.644     \\
CMCC-ESM2     & 53.705      & 0.959       & 10.141    \\
EC-Earth3     & -45.561     & 1.726       & 8.791     \\
FGOALS-g3     & 8.191       & 1.323       & 8.697     \\
GFDL-ESM4     & 5.471       & 1.261       & 8.534     \\
INM-CM4-8     & 17.222      & 1.582       & 9.137     \\
INM-CM5-0     & -9.668      & 1.898       & 8.932     \\
IPSL-CM6A-LR  & 42.053      & 1.074       & 8.376     \\
KACE-1-0-G    & -35.045     & 1.209       & 8.592     \\
MIROC6        & 30.162      & 0.911       & 9.741     \\
MPI-ESM1-2-HR & -42.055     & 1.476       & 8.781     \\
MPI-ESM1-2-LR & -23.581     & 1.359       & 8.883     \\
MRI-ESM2-0    & -20.99      & 1.108       & 8.844     \\
NorESM2-MM    & 40.105      & 0.992       & 8.963     \\
TaiESM1       & 50.755      & 1.001       & 9.536  \\
\bottomrule
\end{tabular}
\caption{Fire Weather Index: Calibrated bias-correction coefficients ($\hat{\beta}_0^{(m)}$ and $\hat{\beta}_1^{(m)}$), and standard deviation of noise ($\hat{\sigma}_{(m)}$)}
\end{table}

\begin{table}[H]
\centering
\begin{tabular}{llll}
\toprule
Model names  & $\hat{\beta}_0^{(m)}$        & $\hat{\beta}_1^{(m)}$       & $\hat{\sigma}_{(m)}$        \\
\midrule
ACCESS-CM2    & 2471.323    & 0.975       & 146.064   \\
CanESM5-CanOE & 18345.06    & 0.819       & 136.658   \\
CESM2         & 14433.794   & 0.857       & 154.145   \\
CMCC-CM2-SR5  & 14248.639   & 0.859       & 144.904   \\
CNRM-ESM2-1   & 2981.637    & 0.97        & 144.572   \\
INM-CM4-8     & -12552.876  & 1.125       & 133.775   \\
INM-CM5-0     & -25169.454  & 1.25        & 145.112   \\
MCM-UA-1-0    & 37431.853   & 0.629       & 125.137   \\
MIROC-ES2L    & 6250.218    & 0.942       & 179.028   \\
MIROC6        & 624.259     & 0.998       & 169.48    \\
MPI-ESM1-2-LR & -8807.205   & 1.087       & 170.452   \\
MRI-ESM2-0    & -23938.494  & 1.234       & 185.582   \\
NorESM2-MM    & 15219.903   & 0.85        & 155.486   \\
UKESM1-0-LL   & 10695.377   & 0.894       & 142.076  \\
\bottomrule
\end{tabular}
\caption{Mean sea-level pressure (MSLP): Calibrated bias-correction coefficients ($\hat{\beta}_0^{(m)}$ and $\hat{\beta}_1^{(m)}$), and standard deviation of noise ($\hat{\sigma}_{(m)}$)}
\end{table}

\begin{table}[H]
\centering
\begin{tabular}{llll}
\toprule
Model names  & $\hat{\beta}_0^{(m)}$        & $\hat{\beta}_1^{(m)}$       & $\hat{\sigma}_{(m)}$        \\
\midrule
access-cm2    & 2.811       & 1.156       & 18.276    \\
access-esm1-5 & 4.2         & 1.051       & 17.659    \\
bcc-csm2-mr   & -22.791     & 1.417       & 20.799    \\
canesm5       & -12.108     & 1.517       & 18.868    \\
cmcc-esm2     & 6.167       & 1.007       & 19.607    \\
ec-earth3     & -12.368     & 1.2         & 19.394    \\
gfdl-esm4     & -14.868     & 1.321       & 16.222    \\
inm-cm4-8     & 11.73       & 0.985       & 18.008    \\
inm-cm5-0     & 3.857       & 1.115       & 18.983    \\
ipsl-cm6a-lr  & -5.697      & 1.305       & 17.26     \\
kace-1-0-g    & 16.459      & 0.953       & 16.631    \\
miroc6        & -45.993     & 1.728       & 19.952    \\
mpi-esm1-2-hr & 14.079      & 0.875       & 17.57     \\
mpi-esm1-2-lr & -3.133      & 1.185       & 17.843    \\
mri-esm2-0    & 20.245      & 0.835       & 16.37     \\
noresm2-lm    & -19.198     & 1.364       & 18.09     \\
noresm2-mm    & -17.776     & 1.263       & 18.665    \\
\bottomrule
\end{tabular}
\caption{Largest five-day cumulative precipitation (rx5day): Calibrated bias-correction coefficients ($\hat{\beta}_0^{(m)}$ and $\hat{\beta}_1^{(m)}$), and standard deviation of noise ($\hat{\sigma}_{(m)}$)}
\end{table}

\begin{table}[H]
\centering
\begin{tabular}{llll}
\toprule
Model names  & $\hat{\beta}_0^{(m)}$        & $\hat{\beta}_1^{(m)}$       & $\hat{\sigma}_{(m)}$        \\
\midrule
ACCESS-CM2   & -0.199      & 1.31        & 0.498     \\
CNRM-CM6-1   & 0.399       & 1.053       & 0.485     \\
FGOALS-f3-L  & -1.589      & 0.722       & 0.513     \\
FGOALS-g3    & -1.713      & 1.047       & 0.529     \\
INM-CM4-8    & 0.63        & 0.765       & 0.501     \\
INM-CM5-0    & 0.074       & 0.71        & 0.481     \\
MCM-UA-1-0   & -0.229      & 1.044       & 0.47      \\
MIROC-ES2L   & -0.264      & 0.774       & 0.437     \\
UKESM1-0-LL  & -0.021      & 0.636       & 0.462    \\
\bottomrule
\end{tabular}
\caption{Sea-surface temperature gradients: Calibrated bias-correction coefficients ($\hat{\beta}_0^{(m)}$ and $\hat{\beta}_1^{(m)}$), and standard deviation of noise ($\hat{\sigma}_{(m)}$)}
\end{table}

\subsection{Normality tests of climate model residuals} \label{Appendix:NormalityTest}

The Shapiro–Wilk test is applied to assess the normality of the climate model residuals. The resulting p-values are reported below. For most ensemble members, the p-values exceed the 5\% threshold, indicating insufficient statistical evidence to reject the null hypothesis of normality at the conventional 5\% significance level.

\begin{figure}[H]
\begin{minipage}{\textwidth}
\begin{minipage}[t]{0.33\textwidth}
    \centering
    \includegraphics[width=\textwidth]{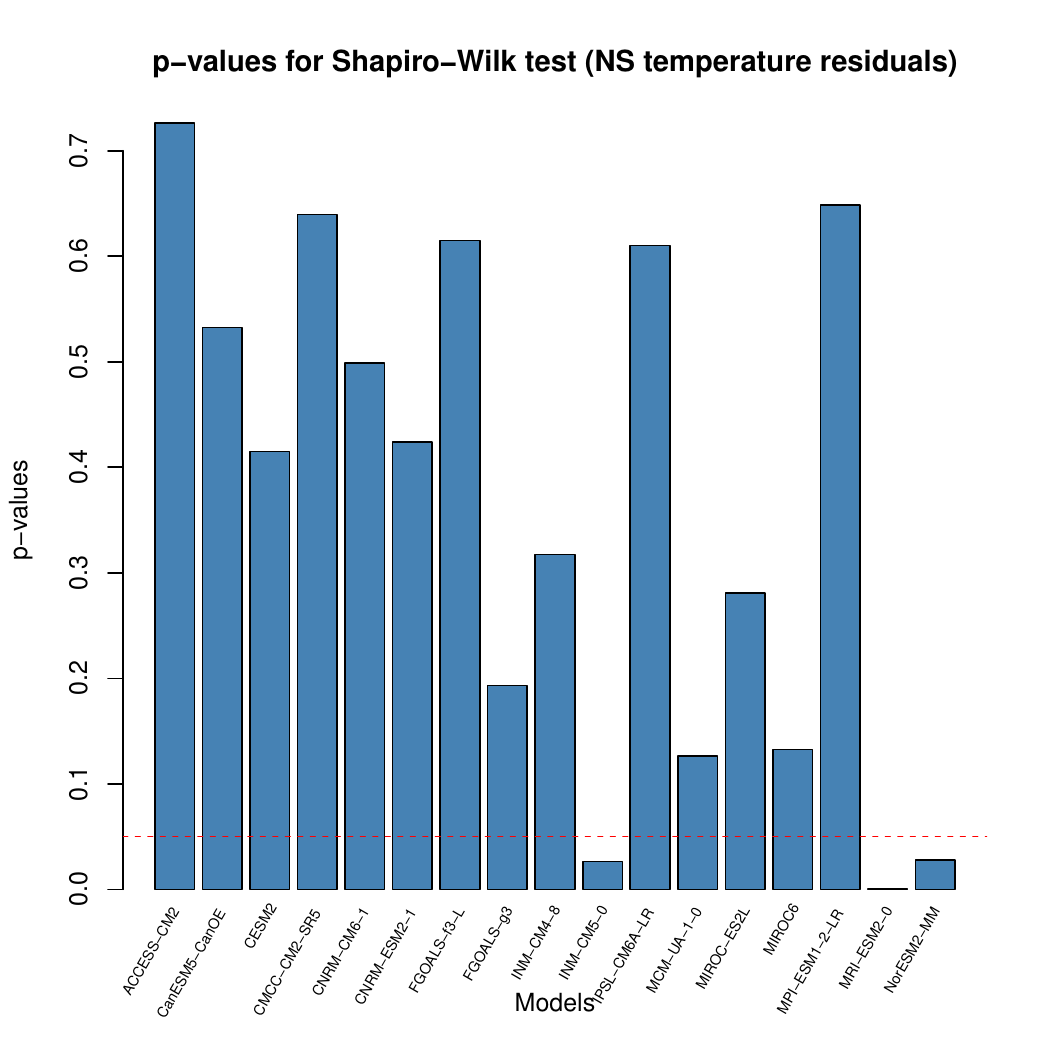}
    \label{Fig:pValuesNS} 
    \end{minipage}
  \begin{minipage}[t]{0.33\textwidth}
    \centering
    \includegraphics[width=\textwidth]{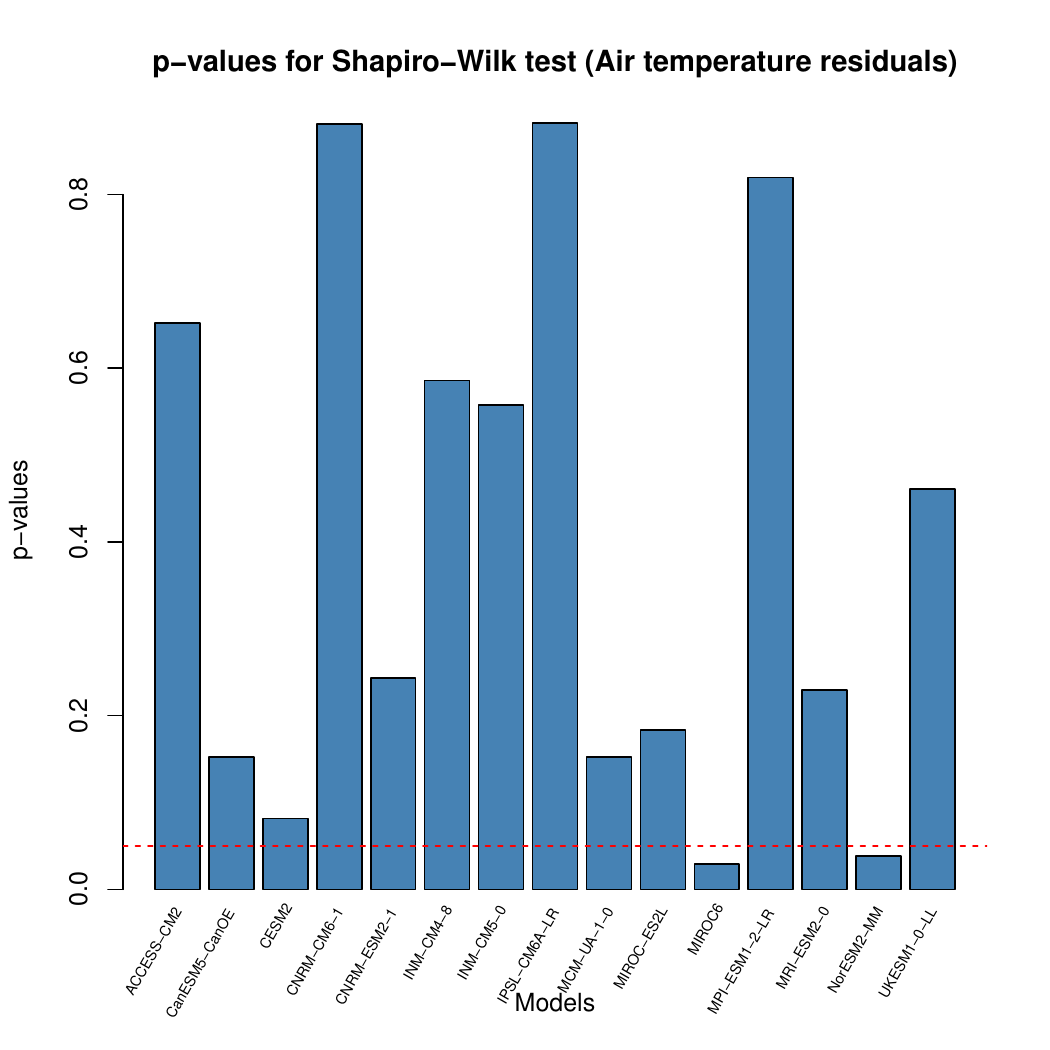}
    \label{Fig:pValuesAir} 
  \end{minipage}
  \begin{minipage}[t]{0.33\textwidth}
    \centering
    \includegraphics[width=\textwidth]{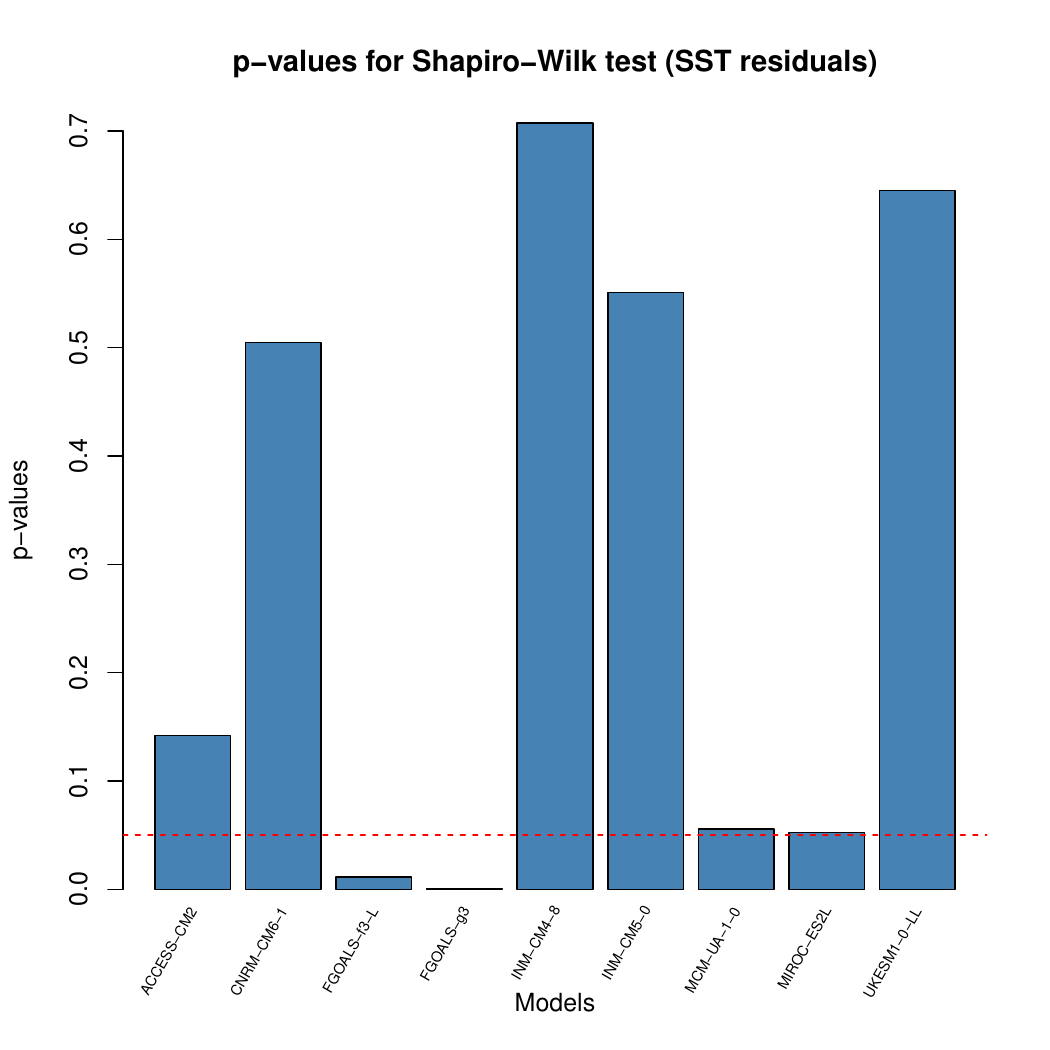}
  \end{minipage}
  \hfill
  \end{minipage}
\captionof{figure}{p-values from Shapiro-Wilk test of climate model residuals: Near-surface temperature (left), air temperature (middle), and sea-surface temperature (right)}
    \label{Fig:pValuesAir} 
\end{figure}

\begin{figure}[H]
\begin{minipage}{\textwidth}
\begin{minipage}[t]{0.5\textwidth}
    \centering
    \includegraphics[width=0.8\textwidth]{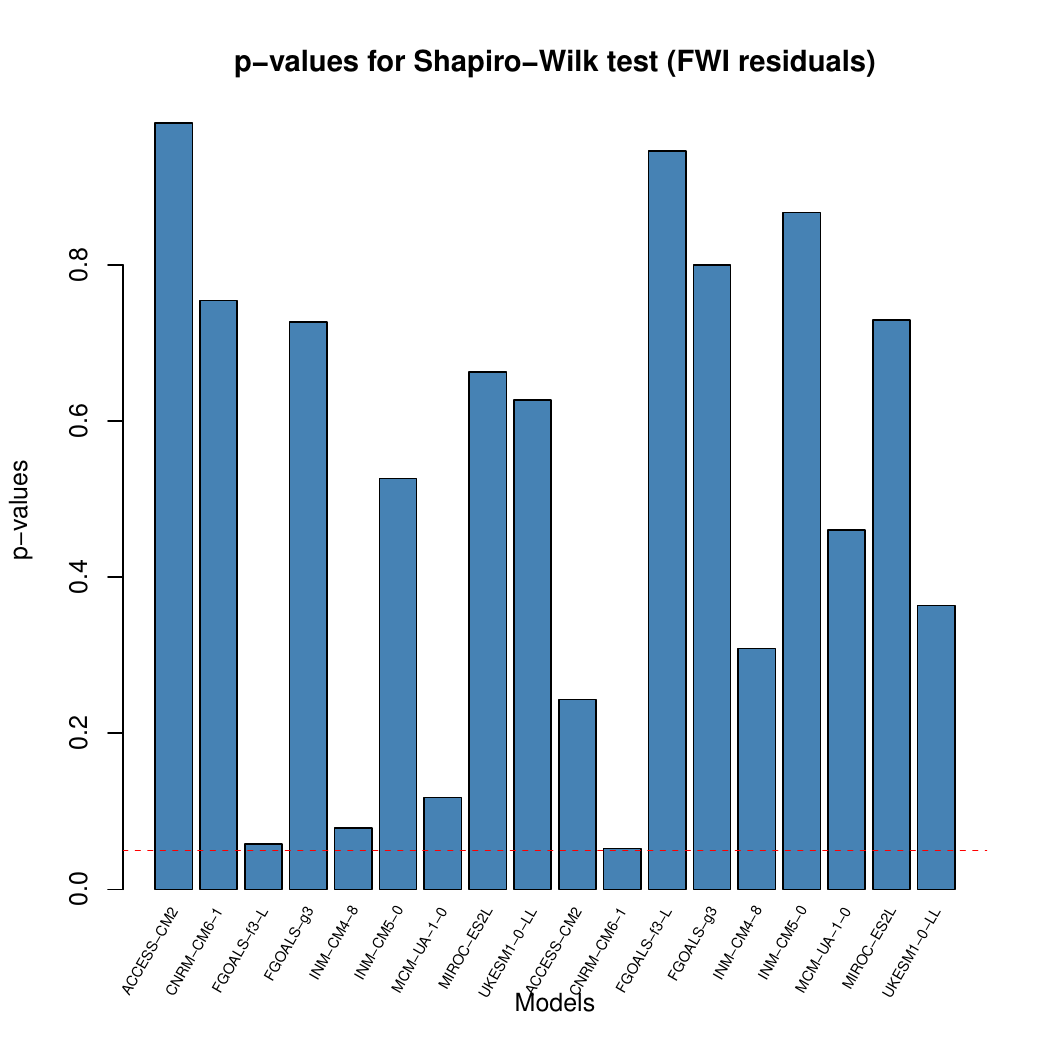}
    \label{Fig:pValuesFWI} 
    \end{minipage}
  \begin{minipage}[t]{0.5\textwidth}
    \centering
    \includegraphics[width=0.8\textwidth]{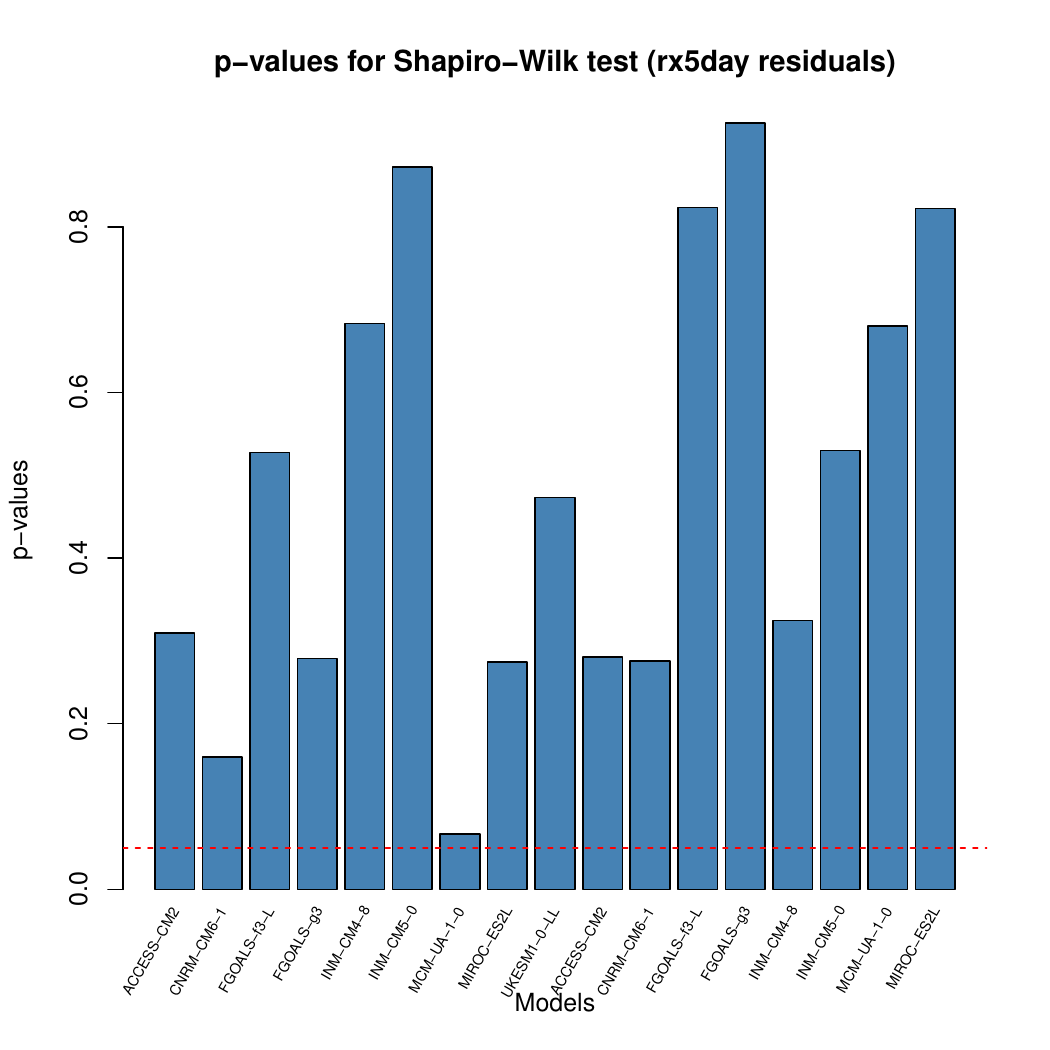}
    \label{Fig:pValuesRx5day} 
  \end{minipage}
  \end{minipage}
   \captionof{figure}{p-values from Shapiro-Wilk test of climate model residuals: Fire Weather Index (left) and extreme precipitation (rx5day) (right)}
\end{figure}

\subsection{Hazard model supplementary calibration results} \label{Appendix:CalibrationResultsHazards}

The results presented here show the full calibration outcomes for the hazard models, including all candidate models considered for each hazard type. The methodology for calibrating these parameters is provided in Section \ref{Section:Hazards} \rev{and \ref{Section:WeatherCovariatesSelection}} of the main paper, \rev{as well as in Online Appendix~\ref{Appendix:DataSourcesandMainCalibration}.}

\subsubsection{Selection of catastrophe loss distributions} \label{Section:CompareDistFit}

Table~\ref{Table:GoodnessofFit} reports the AIC and BIC values for the candidate heavy-tailed severity distributions \citep{EmKlMi13}. The Log-normal distribution achieves the best fit according to both criteria.

\begin{table}[H]
\centering
\begin{tabular}{lllll} 
\toprule
Distributions & AIC              & BIC              & Rank (AIC) & Rank (BIC) \\
\midrule
Log-Normal    & 9366.825 & 9366.825 & 1         & 1         \\
Weibull       & 9405.851 & 9412.683 & 3         & 3         \\
Pareto        & 9379.475 & 9386.307  & 2         & 2         \\
Cauchy        & 9560.989 & 9567.821 & 4         & 4       \\
\bottomrule
\end{tabular}
\caption{Goodness-of-fit for severity distributions: AIC and BIC} 
\label{Table:GoodnessofFit}
\end{table}

As for the frequency of catastrophe events, we assess the suitability of the Poisson distribution assumption using a two-sided dispersion test \citep{CaTr90}. Overdispersion or underdispersion would indicate that an alternative distribution (e.g., Negative Binomial) may be more appropriate. The estimated dispersion parameters across all hazard types are close to one, and the p-values show no statistically significant evidence of over- or under-dispersion at the 5\% level. Hence, the Poisson distribution is deemed appropriate for modelling event frequency in the current application.

\begin{table}[H]
\centering
\small
\setlength{\tabcolsep}{10pt}
\begin{tabular}{lccc}
\toprule
Hazard type & Dispersion estimates & z-statistic & p-value \\
\midrule
Flood & 1.061 & 0.376 & 0.707 \\
Bushfire & 1.117 & 0.608 & 0.543 \\
Tropical Cyclones & 0.968 & -1.693 & 0.090 \\
ECL & 0.990 & -1.333 & 0.183 \\
Hailstorms & 1.144 & 1.112 & 0.266 \\
\bottomrule
\end{tabular}
\caption{Dispersion test results for Poisson frequency models across different hazard types.}
\label{Table:DispersionTests}
\end{table}

\subsubsection{Flood}

\begin{table}[H]
\centering
\begin{tabular}[t]{lcccccc}
\toprule
  & Model 1 & Model 2 & Model 3* & Model 4 & Model 5 & Model 6\\
\midrule
(Intercept) & \num{-2.177}** & \num{-4.295}** & \num{-3.714}** & \num{-4.326}** & \num{-3.666}** & \num{-4.191}**\\
 & (\num{0.004}) & (\num{0.002}) & (\num{0.001}) & (\num{0.005}) & (\num{0.002}) & (\num{0.002})\\
Precipitation & \num{0.004}** &  &  & \num{0.000} & \num{0.000} & \\
 & (\num{0.005}) &  &  & (\num{0.967}) & (\num{0.902}) & \\
rx1day &  & \num{0.081}** &  & \num{0.083}+ &  & \num{0.041}\\
 &  & (\num{0.002}) &  & (\num{0.098}) &  & (\num{0.515})\\
rx5day &  &  & \num{0.037}** &  & \num{0.035}+ & \num{0.020}\\
 &  &  & (\num{0.001}) &  & (\num{0.099}) & (\num{0.487})\\
\midrule
AIC & \num{138.6} & \num{136.1} & \num{136.0} & \num{138.1} & \num{138.0} & \num{137.6}\\
BIC & \num{142.7} & \num{140.1} & \num{140.0} & \num{144.1} & \num{144.1} & \num{143.7}\\
Log.Lik. & \num{-67.306} & \num{-66.026} & \num{-65.998} & \num{-66.025} & \num{-65.991} & \num{-65.788}\\
F & \num{7.838} & \num{9.906} & \num{10.511} & \num{4.948} & \num{5.285} & \num{5.269}\\
\bottomrule
\multicolumn{7}{l}{\rule{0pt}{1em}+ p $<$ 0.1, * p $<$ 0.05, ** p $<$ 0.01, *** p $<$ 0.001}\\
\end{tabular}
\caption{Flood: candidate frequency models}
\label{Table:FloodFrequencyModels}
\end{table}

\begin{table}[H]
\centering
\begin{tabular}[t]{lcccccc}
\toprule
  & Model 1 & Model 2 & Model 3* & Model 4 & Model 5 & Model 6\\
\midrule
Precipitation & \num{0.003}+ &  &  & \num{0.000} & \num{-0.001} & \\
 & (\num{0.086}) &  &  & (\num{0.942}) & (\num{0.826}) & \\
rx1day &  & \num{0.071}* &  & \num{0.076} &  & \num{0.004}\\
 &  & (\num{0.049}) &  & (\num{0.335}) &  & (\num{0.963})\\
rx5day &  &  & \num{0.035}* &  & \num{0.040} & \num{0.034}\\
 &  &  & (\num{0.029}) &  & (\num{0.170}) & (\num{0.339})\\
\midrule
AIC & \num{2000.5} & \num{1999.6} & \num{1998.7} & \num{2001.6} & \num{2000.6} & \num{2000.6}\\
BIC & \num{2006.1} & \num{2005.2} & \num{2004.3} & \num{2009.1} & \num{2008.1} & \num{2008.1}\\
\bottomrule
\multicolumn{7}{l}{\rule{0pt}{1em}+ p $<$ 0.1, * p $<$ 0.05, ** p $<$ 0.01, *** p $<$ 0.001}\\
\end{tabular}
\caption{Flood: candidate severity models}
\end{table}

\subsubsection{Bushfire}

Here, \texttt{mfwixx} represents the annual maximum of the Fire Weather Index averaged across all gridded cells in Australia, while \texttt{xfwixx} denotes the highest annual maximum of the Fire Weather Index across all gridded cells. Similarly, \texttt{mfwixd} represents the average duration of extreme fire weather across all gridded cells, whereas \texttt{xfwixd} denotes the longest duration of extreme fire weather observed across all gridded cells in Australia.

\begin{table}[H]
\centering
\begin{tabular}[t]{lcccccc}
\toprule
  & Model 1* & Model 2 & Model 3 & Model 4 & Model 5 & Model 6\\
\midrule
mfwixx & \num{0.084}** &  & \num{0.092}** &  &  & \\
 & (\num{0.002}) &  & (\num{0.001}) &  &  & \\
xfwixx &  & \num{0.004} & \num{-0.005} &  &  & \\
 &  & (\num{0.418}) & (\num{0.482}) &  &  & \\
mfwixd &  &  &  & \num{0.072}** &  & \num{0.109}**\\
 &  &  &  & (\num{0.001}) &  & (\num{0.007})\\
xfwixd &  &  &  &  & \num{0.012} & \num{-0.014}\\
 &  &  &  &  & (\num{0.111}) & (\num{0.262})\\
\midrule
AIC & \num{118.3} & \num{127.9} & \num{119.7} & \num{120.1} & \num{126.3} & \num{120.8}\\
BIC & \num{122.3} & \num{132.0} & \num{125.8} & \num{124.2} & \num{130.3} & \num{126.9}\\
Log.Lik. & \num{-57.130} & \num{-61.968} & \num{-56.858} & \num{-58.053} & \num{-61.137} & \num{-57.394}\\
F & \num{9.889} & \num{0.657} & \num{5.433} & \num{10.195} & \num{2.547} & \num{5.470}\\
\bottomrule
\multicolumn{7}{l}{\rule{0pt}{1em}+ p $<$ 0.1, * p $<$ 0.05, ** p $<$ 0.01, *** p $<$ 0.001}\\
\end{tabular}
\caption{Bushfire: candidate frequency models}
\end{table}

As all coefficients in the bushfire severity model are statistically insignificant and the stationary Log-Normal distribution yields the lowest AIC and BIC values, we adopt the stationary Log-Normal distribution to model bushfire severity.

\begin{table}[H]
\centering
\begin{tabular}[t]{lcccccc}
\toprule
  & Model 1 & Model 2 & Model 3 & Model 4 & Model 5 & Model 6\\
\midrule
mfwixx & \num{0.016} &  & \num{0.008} &  &  & \\
 & (\num{0.652}) &  & (\num{0.822}) &  &  & \\
xfwixx &  & \num{0.008} & \num{0.007} &  &  & \\
 &  & (\num{0.461}) & (\num{0.532}) &  &  & \\
mfwixd &  &  &  & \num{-0.003} &  & \num{0.029}\\
 &  &  &  & (\num{0.931}) &  & (\num{0.603})\\
xfwixd &  &  &  &  & \num{-0.004} & \num{-0.012}\\
 &  &  &  &  & (\num{0.652}) & (\num{0.495})\\
\midrule
Num.Obs. & \num{37} & \num{37} & \num{37} & \num{37} & \num{37} & \num{37}\\
AIC & \num{1527.0} & \num{1526.6} & \num{1528.6} & \num{1527.2} & \num{1527.0} & \num{1528.7}\\
BIC & \num{1531.8} & \num{1531.5} & \num{1535.0} & \num{1532.0} & \num{1531.8} & \num{1535.1}\\
\bottomrule
\multicolumn{7}{l}{\rule{0pt}{1em}+ p $<$ 0.1, * p $<$ 0.05, ** p $<$ 0.01, *** p $<$ 0.001}\\
\end{tabular}
\caption{Bushfire: candidate severity models}
\end{table}

\subsubsection{Tropical cyclones}

\begin{table}[H]
\centering
\begin{tabular}[t]{lccc}
\toprule
  & Model 1* & Model 2 & Model 3\\
\midrule
Intercept & \num{-33.005}*** & \num{545.650}*** & \num{367.462}***\\
 & ($<$0.001) & ($<$0.001) & ($<$0.001)\\
SST & \num{1.213}*** &  & \num{0.989}***\\
 & ($<$0.001) &  & ($<$0.001)\\
MSLP &  & \num{-0.005}*** & \num{-0.004}***\\
 &  & ($<$0.001) & ($<$0.001)\\
\midrule
AIC & \num{257.3} & \num{258.0} & \num{244.1}\\
BIC & \num{266.3} & \num{267.0} & \num{257.7}\\
Log.Lik. & \num{-126.625} & \num{-127.005} & \num{-119.067}\\
F & \num{27.719} & \num{36.066} & \num{16.122}\\
\bottomrule
\multicolumn{4}{l}{\rule{0pt}{1em}+ p $<$ 0.1, * p $<$ 0.05, ** p $<$ 0.01, *** p $<$ 0.001}\\
\end{tabular}
\caption{Tropical cyclones: candidate frequency models}
\label{Table:TCFrequencyModels}
\end{table}

\begin{table}[H]
\centering
\begin{tabular}[t]{lcccc}
\toprule
  & Model 1 & Model 2 & Model 3 & Model 4\\
\midrule
SST & \num{-79.405} &  &  & \\
 & (\num{0.189}) &  &  & \\
$\text{SST}^2$ & \num{1.560} &  &  & \\
 & (\num{0.194}) &  &  & \\
MSLP &  & \num{-2.022} &  & \\
 &  & (\num{0.342}) &  & \\
$\text{MSLP}^2$ &  & \num{0.000} &  & \\
 &  & (\num{0.343}) &  & \\
cs(SST) &  &  & \num{-0.886} & \num{-1.096}*\\
 &  &  & (\num{0.129}) & (\num{0.048})\\
cs(MSLP) &  &  &  & \num{-0.001}\\
 &  &  &  & (\num{0.346})\\
\midrule
AIC & \num{1612.2} & \num{1613.0} & \num{1614.5} & \num{1613.1}\\
BIC & \num{1618.8} & \num{1619.6} & \num{1624.3} & \num{1629.5}\\
\bottomrule
\multicolumn{5}{l}{\rule{0pt}{1em}+ p $<$ 0.1, * p $<$ 0.05, ** p $<$ 0.01, *** p $<$ 0.001}\\
\end{tabular}
\caption{Tropical cyclones: candidate severity models}
\end{table}

\subsubsection{Storms}

\begin{table}[H]
\centering
\begin{tabular}[t]{lccc}
\toprule
  & Model 1* & Model 2 & Model 3\\
\midrule
Intercept & \num{-11.573}*** & \num{161.061}** & \num{71.386}\\
 & ($<$0.001) & (\num{0.008}) & (\num{0.364})\\
SST & \num{0.348}** &  & \num{0.260}+\\
 & (\num{0.003}) &  & (\num{0.073})\\
MSLP &  & \num{-0.002}** & \num{-0.001}\\
 &  & (\num{0.007}) & (\num{0.291})\\
\midrule
AIC & \num{516.1} & \num{518.2} & \num{517.0}\\
BIC & \num{526.5} & \num{528.6} & \num{532.6}\\
Log.Lik. & \num{-256.058} & \num{-257.101} & \num{-255.499}\\
F & \num{8.800} & \num{7.240} & \num{4.888}\\
\bottomrule
\multicolumn{4}{l}{\rule{0pt}{1em}+ p $<$ 0.1, * p $<$ 0.05, ** p $<$ 0.01, *** p $<$ 0.001}\\
\end{tabular}
\caption{Storms: candidate frequency models}
\end{table}

\begin{table}[H]
\centering
\begin{tabular}[t]{lccc}
\toprule
  & Model 1* & Model 2 & Model 3\\
\midrule
SST & \num{0.239} &  & \num{0.176}\\
 & (\num{0.103}) &  & (\num{0.296})\\
MSLP &  & \num{-0.001} & \num{-0.001}\\
 &  & (\num{0.149}) & (\num{0.471})\\
\midrule
AIC & \num{2504.3} & \num{2504.9} & \num{2505.8}\\
BIC & \num{2510.7} & \num{2511.3} & \num{2514.3}\\
\bottomrule
\multicolumn{4}{l}{\rule{0pt}{1em}+ p $<$ 0.1, * p $<$ 0.05, ** p $<$ 0.01, *** p $<$ 0.001}\\
\end{tabular}
\caption{Storms: candidate severity models}
\end{table}

\subsubsection{East Coast Low}

\begin{table}[H]
\centering
\begin{tabular}[t]{lcc}
\toprule
  & Model 1 & Model 2*\\
\midrule
Intercept & \num{-2.593} & \num{-2.260}*\\
 & (\num{0.518}) & (\num{0.041})\\
SST & \num{-0.098} & \\
 & (\num{0.625}) & \\
SST gradients &  & \num{2.189}+\\
 &  & (\num{0.053})\\
\midrule
AIC & \num{81.7} & \num{77.8}\\
BIC & \num{90.7} & \num{86.8}\\
Log.Lik. & \num{-38.828} & \num{-36.888}\\
F & \num{0.239} & \num{3.732}\\
\bottomrule
\multicolumn{3}{l}{\rule{0pt}{1em}+ p $<$ 0.1, * p $<$ 0.05, ** p $<$ 0.01, *** p $<$ 0.001}\\
\end{tabular}
\caption{East Coast Low: candidate frequency models}
\end{table}

Given the limited data and high uncertainty in projected East Coast Low intensities under climate change \citep{PeDiJiAlEvSh16}, we assume a stationary distribution for the normalized severity of East Coast Lows.

\subsubsection{Hailstorms}

\begin{table}[H]
\centering
\begin{tabular}[t]{lc}
\toprule
  & Model 1\\
\midrule
Intercept & \num{-11.589}*\\
 & (\num{0.014})\\
Atmospheric temperature & \num{-0.079}\\
 & (\num{0.452})\\
Near-surface temperature & \num{0.211}***\\
 & ($<$0.001)\\
\midrule
AIC & \num{281.3}\\
BIC & \num{296.9}\\
Log.Lik. & \num{-137.654}\\
F & \num{6.437}\\
\bottomrule
\multicolumn{2}{l}{\rule{0pt}{1em}+ p $<$ 0.1, * p $<$ 0.05, ** p $<$ 0.01, *** p $<$ 0.001}\\
\end{tabular}
\caption{Hailstorm: frequency model}
\label{Table:HailstormFrequencyModels}
\end{table}

As all coefficients are statistically insignificant and their signs do not align with expectations (e.g., a positive relationship between near-surface temperature and hailstorm intensity is anticipated), we adopt a stationary Log-Normal distribution to model hailstorm severity.

\begin{table}[H]
\centering
\begin{tabular}[t]{lc}
\toprule
  & Model 1\\
\midrule
Atmospheric temperature & \num{-0.049}\\
 & (\num{0.724})\\
Near-surface temperature & \num{-0.083}\\
 & (\num{0.394})\\
\midrule
AIC & \num{1304.8}\\
BIC & \num{1310.4}\\
\bottomrule
\multicolumn{2}{l}{\rule{0pt}{1em}+ p $<$ 0.1, * p $<$ 0.05, ** p $<$ 0.01, *** p $<$ 0.001}\\
\end{tabular}
\caption{Hailstorm: severity model}
\end{table}

\section{Model simulation results: supplementary details} \label{Appendix:SupplementaryModelSimulationResults}

\subsection{Flood Loss: Investigation of simulation peaks} \label{Appendix:FloodLossInvestigation}

The simulated flood losses presented in Section~\ref{Section: ResultsClimateHazards} of the main paper indicate a pronounced peak under the SSP 8.5 scenario, particularly at the $95^{\text{th}}$ percentile. As shown in Figure \ref{Fig:FloodLoss95thquantileSSP85}, the peak in simulated flood losses around the year 2047 is primarily driven by the increase in projected extreme precipitation (the climate covariate used to model flood frequency and severity), which reaches a record-high value during this period. This surge in projected extreme precipitation reflects the inter-annual variability within the CMIP6 ensemble simulations \citep{JaSc23}, as well as the projected rise in the frequency of extreme ENSO events under high-emission scenarios reported in the literature \citep[see, e.g.,][]{CaGuMc15, HeFe23}.

\begin{figure}[H]

\begin{minipage}[t]{0.49\textwidth}
    \centering
    \includegraphics[width=0.95\textwidth]{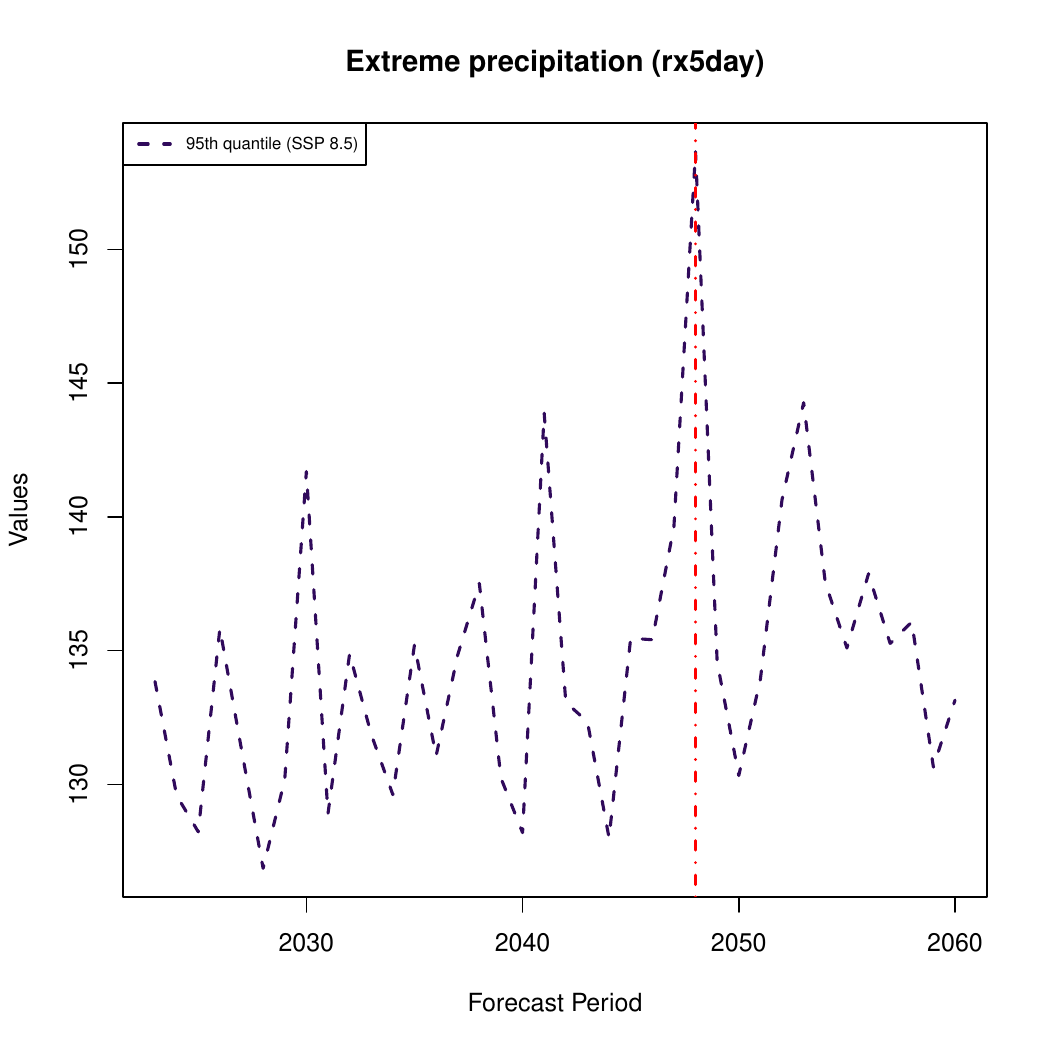}
    \end{minipage}
  \begin{minipage}[t]{0.49\textwidth}
    \centering
    \includegraphics[width=0.95\textwidth]{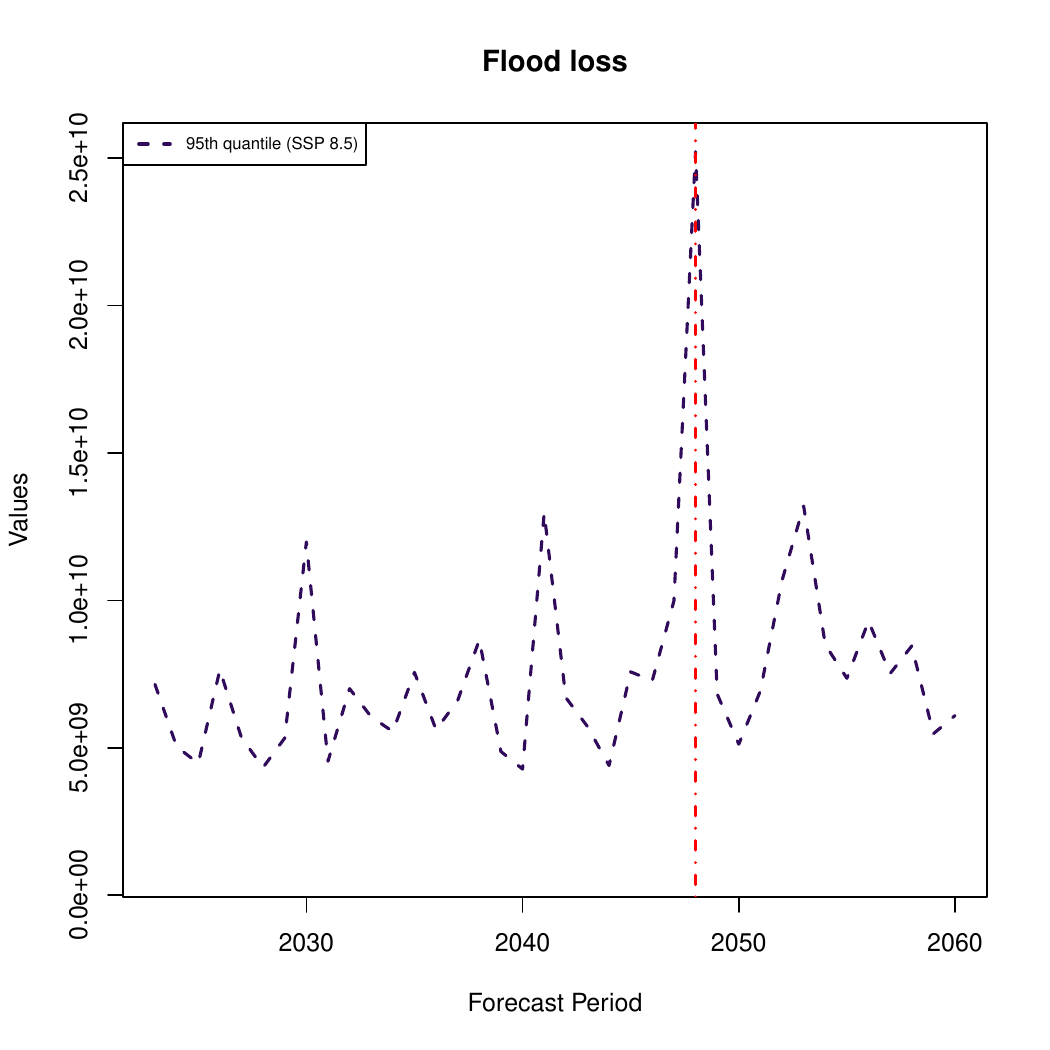}
    \label{Fig:FloodLoss95thquantileSSP85} 
  \end{minipage}
  \hfill
  \captionof{figure}{Simulated extreme precipitation (rx5day) at the 95th percentile under the SSP 8.5 scenario (left), and simulated flood losses at the 95th percentile under the same scenario (right). The red dashed line marks the peak in the simulated flood loss.}
  \label{Fig:FloodLoss95thquantileSSP85} 
\end{figure}

\subsection{Bushfire Loss: Simulation with extended projection horizon} \label{Appendix:BushfireLossInvestigation}

To better illustrate the long-term trend in bushfire losses, we extend the simulation horizon of normalised bushfire losses to 2100, as shown in Figure~\ref{fig:BushfireLossExtendedHorizon}. The results reveal a clear upward trend under the high-emission scenarios SSP~8.5 and SSP~7.0 across both the mean and higher quantiles. Specifically, the mean bushfire loss under SSP~8.5 is projected to increase by $91\%$ by 2100 relative to its initial value, and to be $95\%$ higher than the corresponding mean loss under SSP~2.6 at that time. This increasing trend aligns with the projected rise in fire susceptibility in Australia \citep[pp.~32--37]{NCRAHazards25}.

\begin{figure}[H]
\begin{minipage}{0.7\textwidth}
\includegraphics[width=0.8\linewidth]{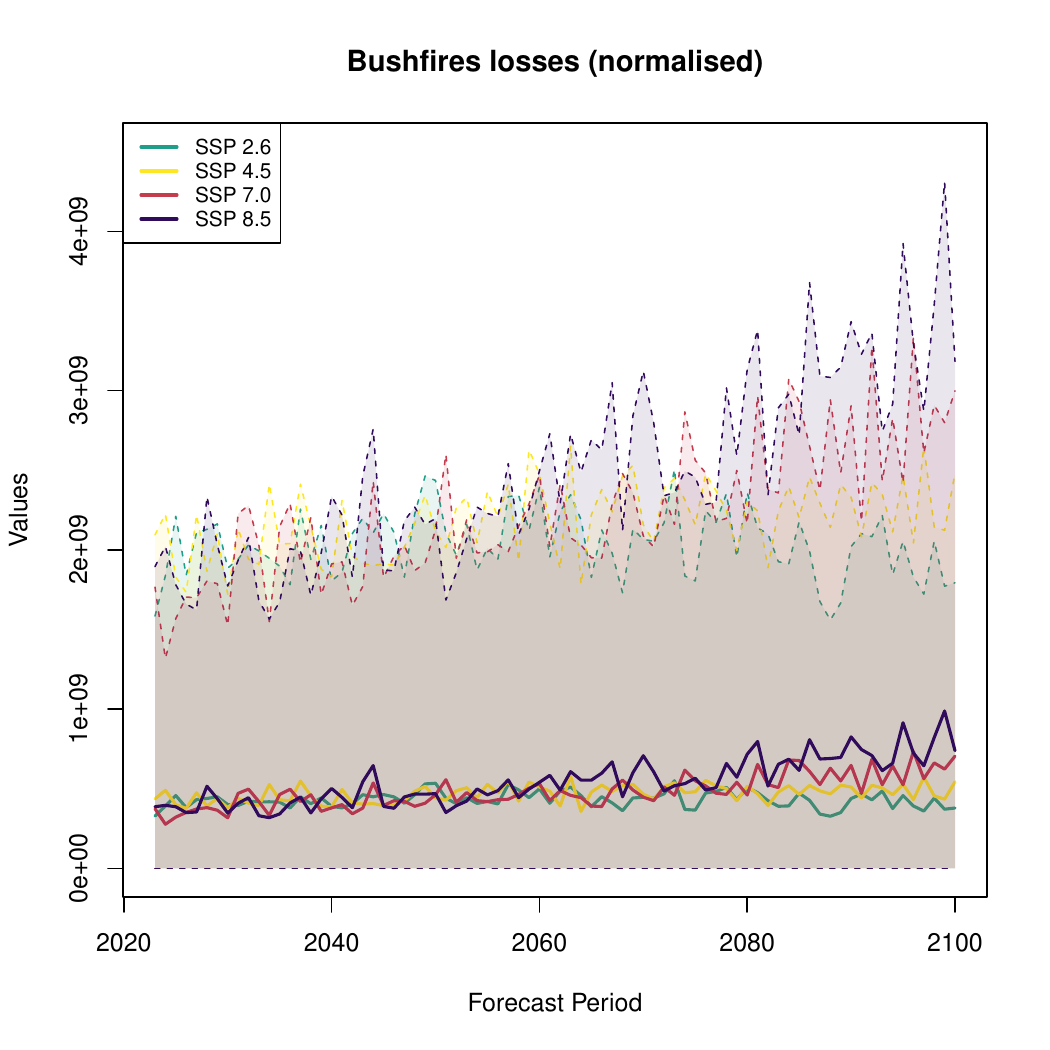} 
\end{minipage}
    \begin{minipage}{0.3\textwidth}
        \captionof{figure}{Simulation results of normalised bushfire losses extended to 2100. Solid lines represent the average simulation paths under different climate scenarios, while dashed lines indicate the $5^{\text{th}}$ and $95^{\text{th}}$ percentiles.}
\label{fig:BushfireLossExtendedHorizon}
\end{minipage}
\end{figure}

\section{Equity models: supplementary details} \label{Appendix:EquitySupplementary}

\subsection{Modelling assumptions of the equity model} \label{Appendix:AnalyticalEquitySupplementary}

In Section \ref{Section:Equity} of the main paper, equity returns are modelled as a function of operating profit growth, which in turn is modelled as a function of consumption growth. In this section, we examine the validity of this modelling assumption.

Dividend growth is a critical driver of total equity returns \citep[see, e.g.,][]{LeLu05}. The total dividends paid by corporations can be expressed as:
\begin{equation} \label{Eq:DividendsDecomposition}
    D_t = (\beta_t (1-\text{Tax}_t)\zeta_t) C_t,
\end{equation}
where $\beta_t$ is the share of corporate's operating profits in the total economy at time $t$, $\text{Tax}_t$ is the tax rate, and $\zeta_t$ is the dividend payout ratio at time $t$.  Consequently, dividend growth is given by:
\begin{equation} \label{Eq:DividendsGrowth}
    \Delta d_t  = \frac{(\beta_t (1-\text{Tax}_t)\zeta_t) C_t}{(\beta_{t-1} (1-\text{Tax}_{t-1})\zeta_{t-1}) C_{t-1}}-1.
\end{equation}
If the share of corporate operating profits, tax rates, and dividend payout ratios remain constant over time, dividend growth would be equal to consumption growth. 

However, this assumption may be too restrictive. Instead of assuming a constant dividend payout ratio, we allow its change to be relatively constant over time (i.e., $1+\Delta \zeta_t = \zeta_1 + \epsilon$, with $\epsilon \sim N(0, \sigma^2)$), then the change in dividend growth could be written as a linear function of the operating profit growth:
\begin{equation} \label{Eq:ChangeinDiv}
    \Delta d_t = \zeta_0 + \zeta_1 \Delta \text{OP}_t + \tilde{\epsilon}_t,
\end{equation}
where $\Delta \text{OP}_t$ is the change in operating profits defined as $\text{OP}_t = \beta_t (1-\text{Tax}_t) C_t$.  This formulation is analogous to our empirical model shown in the main paper. 

Furthermore, the change in operating profit can be expressed as a function of consumption growth:
\begin{equation} \label{Eq:OpinConsumption}
    \Delta \text{OP}_t = (1 + \Delta \gamma_t) \Delta C_t + \Delta \gamma_t,
\end{equation}
where $1 + \Delta \gamma_t = (1 + \Delta \beta_t)(1+\Delta (1-\text{Tax}_t))$. Assuming that the changes in the proportion of corporate profits to total consumption (i.e., $\Delta \beta_t$) and in tax rates (i.e., $\Delta (1-\text{Tax}_t)$) are relatively constant over time, the change in operating profit can be modelled as a linear function of consumption growth:
\begin{equation} \label{Eq:ChangeinOP}
    \Delta \text{OP}_t = \alpha_0 + \alpha_1 \Delta C_t + \tilde{\epsilon}_t,
\end{equation}
which is consistent with our empirical model shown in the main paper. As shown in Figure \ref{Fig:AuCorporateTaxRates}, the assumption of a constant change in tax rates is supported by historical data, with Australian corporate tax rates remaining at $30\%$ since 2002. Moreover, the proportion of corporate earnings is relatively stable over time, as demonstrated in Figure \ref{Fig:AuCorporateProfitsProp}. 

\begin{figure}[H]
\begin{minipage}{\textwidth}
\begin{minipage}[t]{0.49\textwidth}
    \centering
    \includegraphics[width=\textwidth]{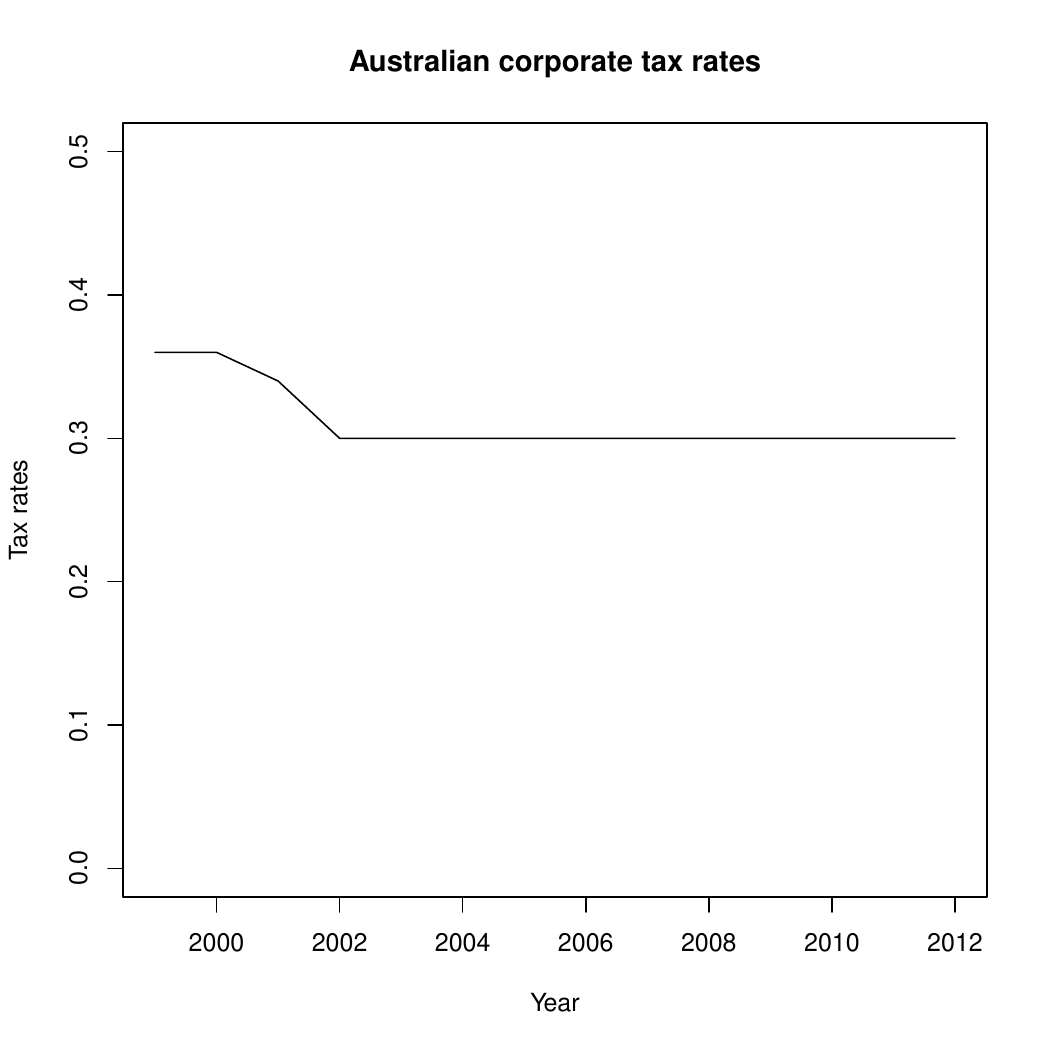}
    \captionof{figure}{Australian corporate tax rates (Data source: \cite{CorporateTaxRate_data}}
    \label{Fig:AuCorporateTaxRates} 
    \end{minipage}
  \begin{minipage}[t]{0.49\textwidth}
    \centering
    \includegraphics[width=\textwidth]{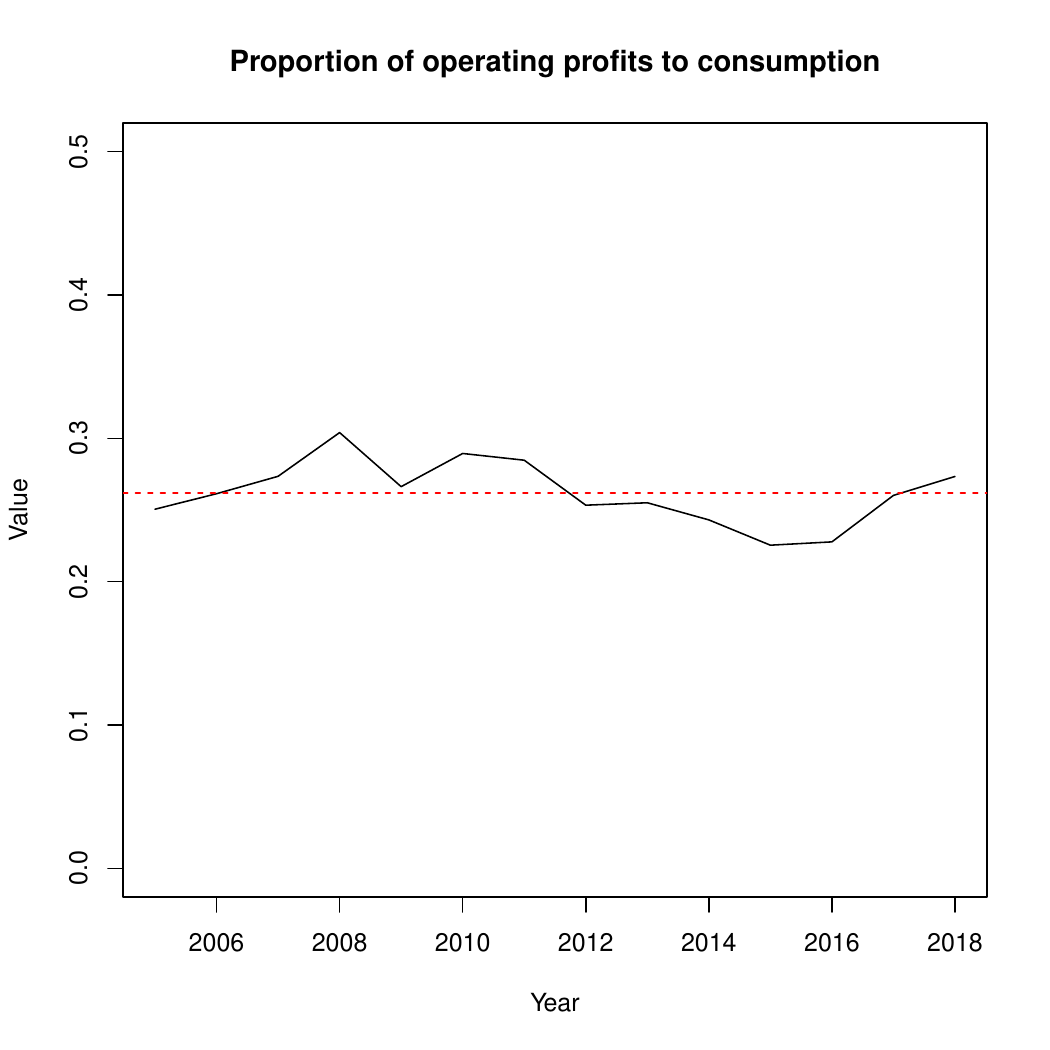}
    \captionof{figure}{Proportion of aggregate Australian corporate operating profits to total consumption in Australia (Data source: \cite{ABSBI_data})}
    \label{Fig:AuCorporateProfitsProp} 
  \end{minipage}
  \hfill
  \end{minipage}
\end{figure}

\subsection{Simulation results for general and brown investment portfolios} \label{Appendix:BrownPortfolioResults}

The cumulative equity returns for the general and brown portfolios are presented in Figure \ref{Fig:InvestmentReturnsGeneralBrowns}, showing trends broadly consistent with those of the total investment portfolios discussed in Section \ref{Section: ResultsInvestmentReturns}.

To examine the difference in investment performance between the general and brown portfolios, Figure \ref{fig:RelativeDifferenceBrownvsGeneral} plots the relative difference in excess equity returns (i.e., the percentage deviation of the brown portfolio’s excess return from that of the general portfolio). The results are generated under the assumption that, in the absence of transition effects, the brown portfolio yields the same returns as the general portfolio, consistent with the assumption in \cite{GaOz24}. 

A positive deviation is projected under the SSP 8.5 scenario throughout the projection horizon, reflecting its assumption of increasing reliance on fossil fuels, which supports operating profit growth in brown energy firms. Under SSP 4.5, positive deviations appear early in the projection period, driven by temporary increases in fossil fuel production, but turn negative later as dependence on fossil fuels slowly declines \citep{OnKrEbKeRiRo17}. A similar pattern is observed under SSP~7.0, albeit with smaller excess returns; however, the lower fossil fuel consumption in OECD countries is attributed to the declining population in developed economies assumed in this scenario, rather than to a reduction in fossil fuel dependence \citep{RiVaKr17}.

Under the low-emission scenario (i.e., SSP~2.6), a negative deviation from the general portfolio return is projected, driven by lower operating profit growth resulting from declining fossil fuel production, which in turn reflects the reduced reliance on fossil energy \citep{OnKrEbKeRiRo17, RiVaKr17}. The projected stress magnitude on equity returns is broadly consistent with that under the ``Below~2$^\circ\mathrm{C}$'' scenario of the NGFS framework projected by the REgional Model of Investment and Development (REMIND) model \citep{BeCoGuMcMe23}. However, the transition stress on the brown portfolio under SSP~2.6 projected here is relatively mild compared with that projected under the ``Delayed Transition'' scenario in the NGFS framework \citep{BeCoGuMcMe23}, which may be explained by the assumption of a gradual and orderly transition to a sustainable economy in SSP~2.6. \citep{OnKrEbKeRiRo17}. 

\begin{figure}[H]
\begin{minipage}{\textwidth}
\begin{minipage}[t]{0.49\textwidth}
    \centering
    \includegraphics[width=\textwidth]{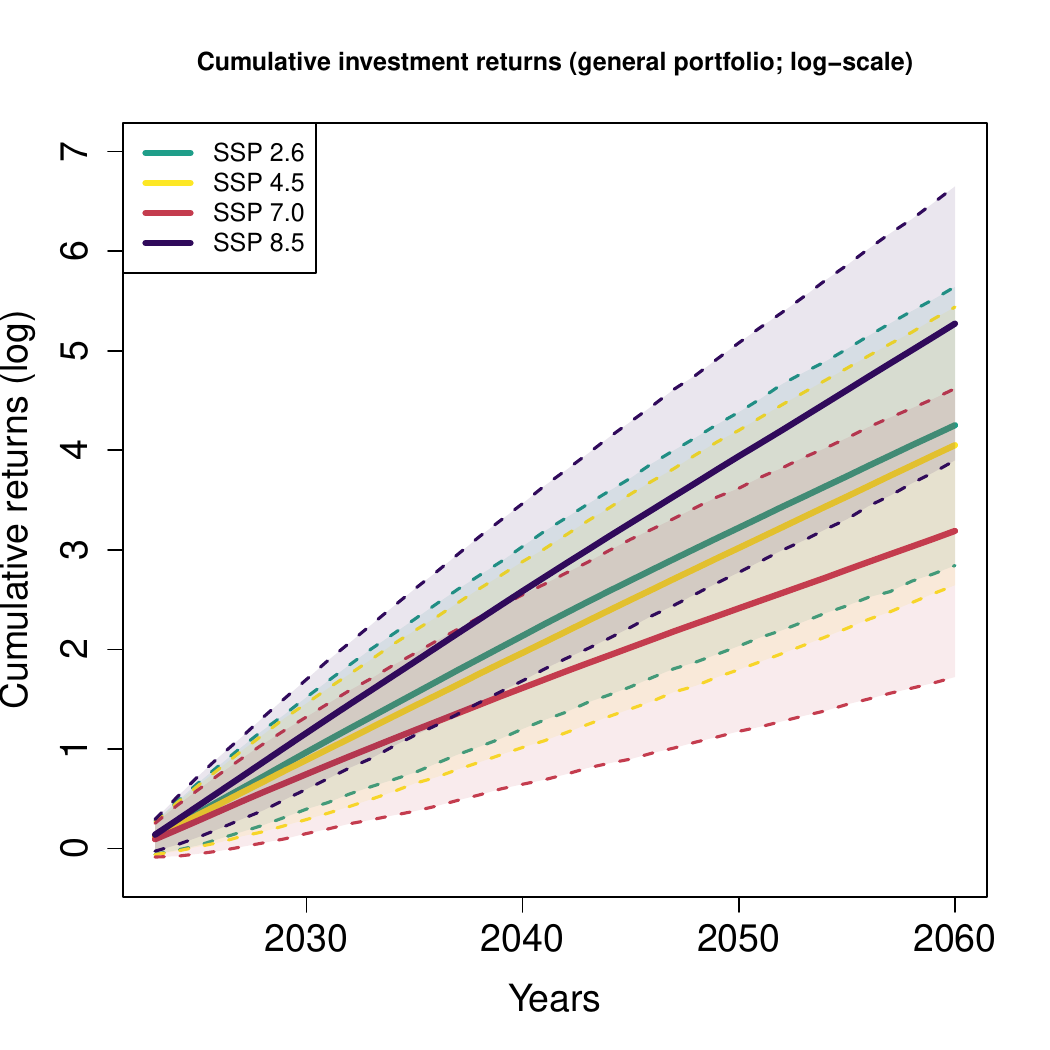}
    \end{minipage}
  \begin{minipage}[t]{0.49\textwidth}
    \centering
    \includegraphics[width=\textwidth]{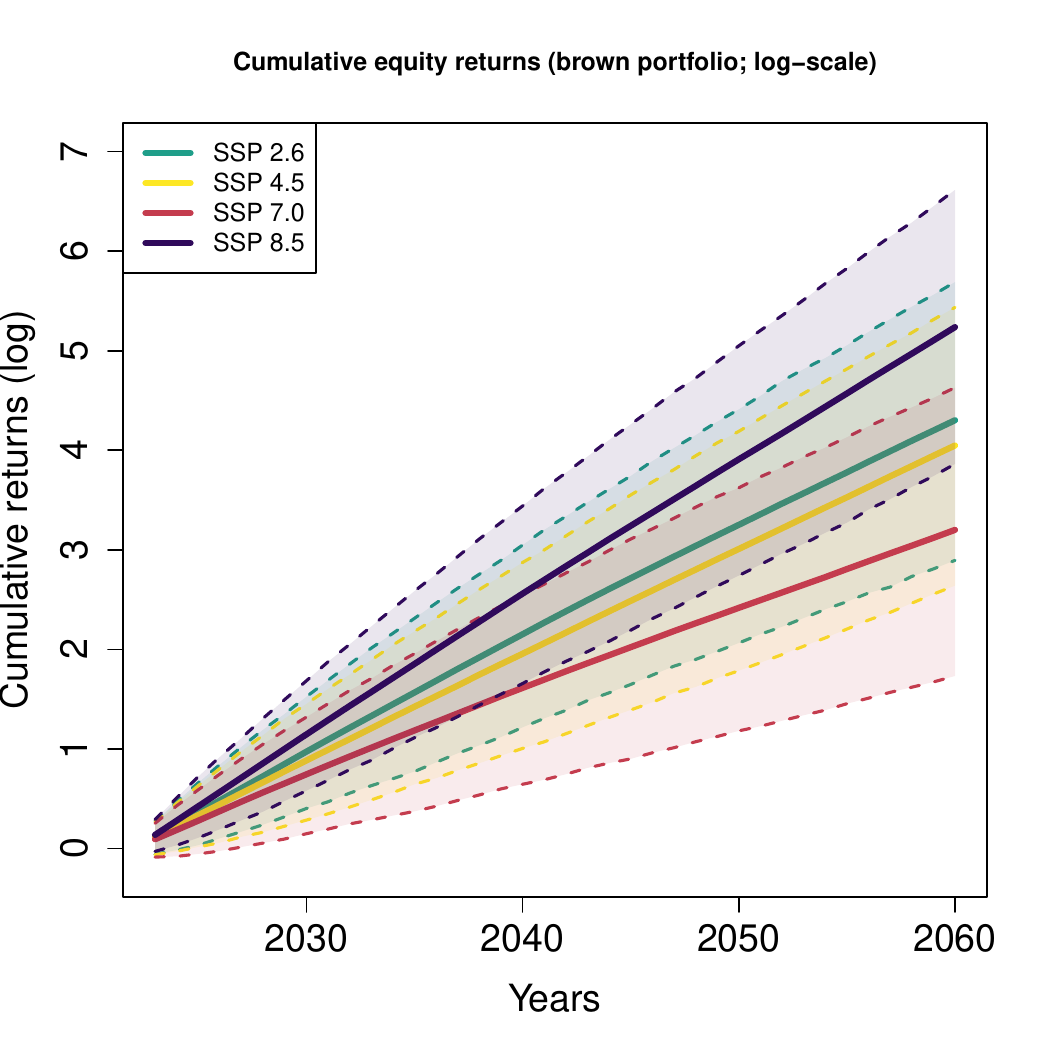}
  \end{minipage}
  \hfill
  \end{minipage}
  \captionof{figure}{Simulated (log) compounded equity returns on the general portfolio (a) and the brown portfolio (b). }
  \label{Fig:InvestmentReturnsGeneralBrowns} 
\end{figure}

\begin{figure}[H]
    \centering
    \includegraphics[width=0.5\linewidth]{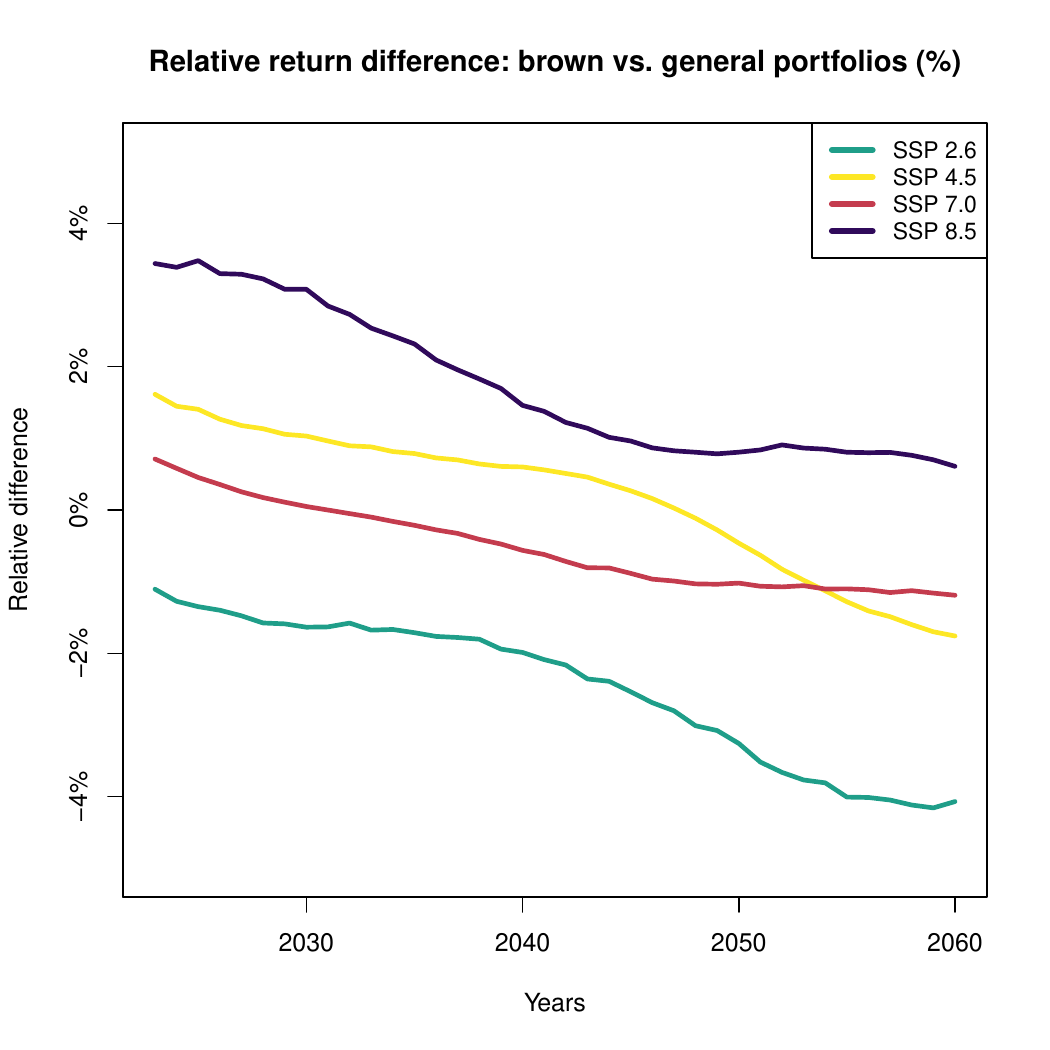}
    \caption{Relative difference (in \%) between excess equity returns of the brown portfolio and the general portfolio, averaged across all simulation paths.}
    \label{fig:RelativeDifferenceBrownvsGeneral}
\end{figure}

\section{Supplementary benchmarking analysis: Comparison of projected damage ratios in literature} \label{Appendix: ProjectedDamageRatiosComparison}

The projected economic damage ratios from our climate-dependent DFA model (as presented in Section \ref{Section: ResultsInvestmentReturns} of the main paper) are compared with estimates from the Dynamic Integrated Model of Climate and the Economy (DICE), originally proposed by \cite{Nord92}. The DICE model has been widely adopted in climate economics to estimate the social cost of carbon and evaluate climate policies. For this comparison, we use damage ratio estimates from the two latest versions of the DICE model: DICE-2023 \citep{BaNo24} and DICE-2016 \citep{Nord18}.

For benchmarking, we focus on three scenarios from the DICE model: baseline, cost-benefit optimal, and 2 $^{\circ}\text{C}$ target. The baseline scenario in DICE assumes a 3.6 $^{\circ}\text{C}$ global temperature increase by 2100 \citep{BaNo24}, which aligns broadly with the SSP 7.0 scenario (4.1 $^{\circ}\text{C}$ by 2100). The cost-benefit optimal scenario projects a 2.6 $^{\circ}\text{C}$ rise by 2100 \citep{BaNo24}, corresponding to SSP 4.5 (2.63 $^{\circ}\text{C}$ by 2100). Lastly, the 2 $^{\circ}\text{C}$ target scenario assumes a 2 $^{\circ}\text{C}$ limit on warming by 2100 \citep{BaNo24}, which is comparable to SSP 2.6 (1.76 $^{\circ}\text{C}$ by 2100) \citep{RiVaKr17}. However, caution is warranted, as the narratives underlying the DICE scenarios differ from those of the SSP framework, and here we only focus on temperature alignment. For a detailed discussion of these narratives, see \cite{BaNo24}.

Figure \ref{fig:DamageRatiosComparison} presents a comparison between our simulated damage ratios (Section \ref{Section: ResultsInvestmentReturns}) and the DICE estimates. Since DICE projections are provided at five-year intervals, we apply spline interpolation to convert them to annual values for benchmarking. Our mean projected damage ratios align more closely with DICE-2016 estimates in the early projection horizon but fall below them in later years. Across the entire horizon, our projections are also lower than those from DICE-2023. However, the estimated damage ratios from both DICE-2016 and DICE-2023 generally fall within the 5th–95th percentile prediction intervals of our simulations in the corresponding SSP scenarios, suggesting that our projections encompass their estimated damage paths.

However, caution is warranted when interpreting these benchmarking results. Firstly, the DICE model estimates global climate impacts, whereas our analysis focuses on localised impacts (specifically, Australia). Given the regional heterogeneity in climate risks, economic structures, and infrastructure profiles, differences between the DICE projections and our results are anticipated. Secondly, the DICE model captures the cumulative, long-term economic impacts of global warming using a quadratic damage function \citep{BaNo24}, whereas our simulations focus on the instantaneous impacts of natural catastrophe events, which are more relevant for general insurance applications. We do not account for chronic climate change effects. This distinction may explain the higher damage ratios projected by the DICE model, particularly in the later projection horizon, where cumulative climate impacts on economic outputs become more pronounced. 


\begin{figure}[H]
    \centering
    \includegraphics[width=0.6\textwidth]{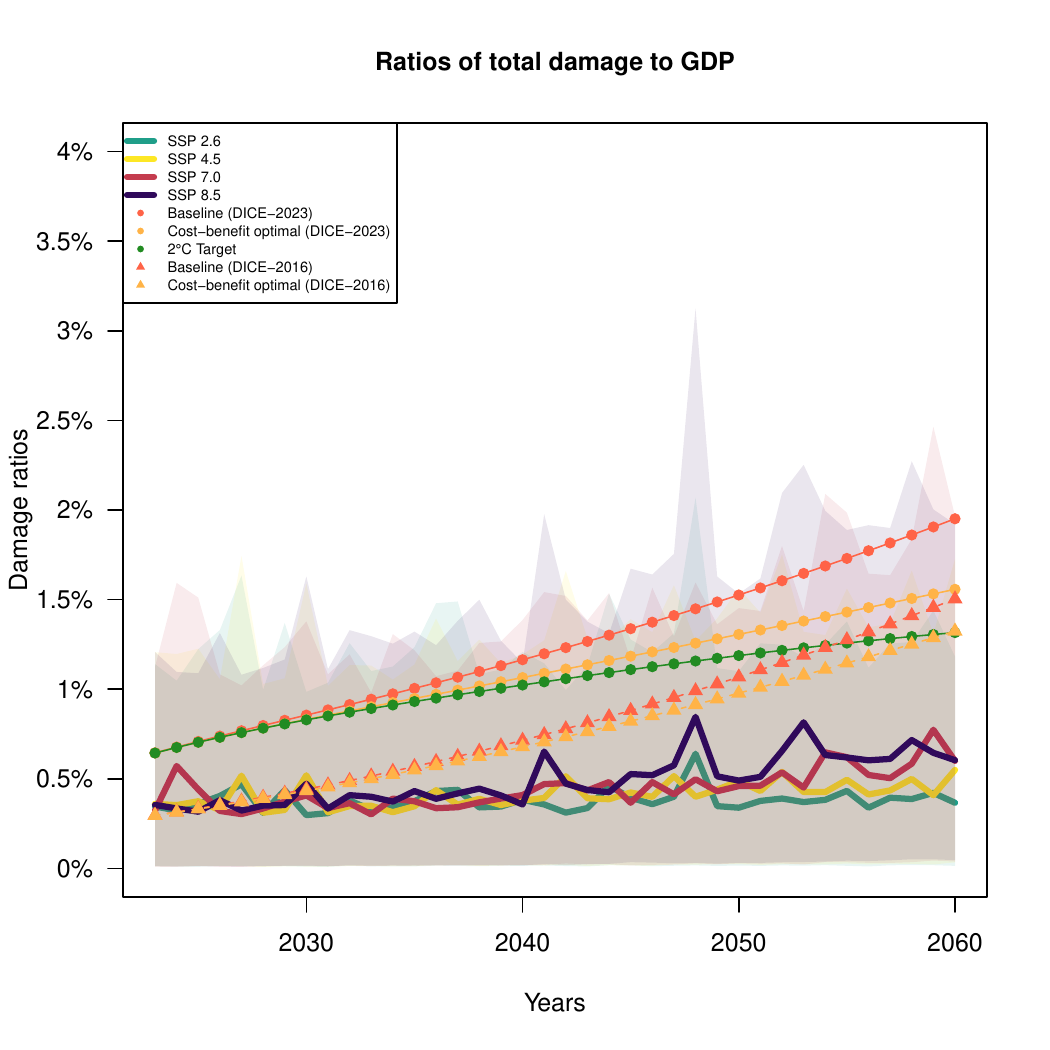}
    \caption{Comparison of projected economic damage ratios: Climate-dependent DFA (solid lines), DICE-2023 (circled markers), and DICE-2016 (triangular markers)}
    \label{fig:DamageRatiosComparison}
\end{figure}

\section{Model extensions} \label{Appendix:ModelExtension}

\subsection{Multi-regional Collective Risk Models and portfolio optimisation} \label{Appendix:MultiRegionalCollectiveRiskModels}

The Collective Risk Model used for modelling hazard events, as defined in \eqref{Eq:AggregatLoss} in Section \ref{Section:Hazards} of the main paper, can be extended to a multi-regional form:
\begin{equation} \label{Eq:AggregatLossMultiRegions}
    \tilde{X}_t = \sum_{r=1}^R \sum_{i=1}^{I} \sum_{m=1}^{M_t^{(i, r)}} \tilde{X}_t^{(i), m, r},
\end{equation}
where $M_t^{(i, r)}$ denotes the number of events of hazard type $i$ in year $t$ occurring in region $r$, and $\tilde{X}_t^{(i), m, r}$ represents the insurance loss associated with the $m^{\text{th}}$ event of hazard type $i$ in year $t$ in region $r$. Region-specific climate covariates can then be incorporated into the modelling of hazard frequency and severity.


The aggregated simulated catastrophe (CAT) losses across regions, as defined in \eqref{Eq:AggregatLossMultiRegions}, can be incorporated into the climate-dependent DFA framework to simulate other modelled variables. In particular, the gross catastrophe loss incurred by insurer $j$ can be expressed as:
\begin{equation} \label{Eq:GrossCATLossMultiRegions}
    \tilde{X}_{t,(j)} = \sum_{r=1}^R w_j^{(r)} \tilde{X}_t^{(r)}, 
    \quad \text{where} \quad 
    \tilde{X}_t^{(r)} = \sum_{i=1}^{I} \sum_{m=1}^{M_t^{(i, r)}} \tilde{X}_t^{(i), m, r}.
\end{equation}
Here, $w_j^{(r)}$ denotes the exposure of insurer $j$ to region $r$. This allocation can be treated as a decision variable, determined by optimising underwriting performance over a planning horizon under each climate scenario, thereby accounting for the heterogeneous impacts of climate change across regions \citep[Chapter~12 of][pp.~1767--1876]{IPCC2021Ch12}. An empirical application of the multi-regional hazard model and portfolio optimisation under the impact of climate change is left for future research.

\subsection{Integration of CAT modelling outputs} \label{Appendix:IntegrationofCATModels}

The modelling framework proposed in this paper is intentionally broad yet simplified. Future research may incorporate more sophisticated component models to enhance precision, such as proprietary catastrophe (CAT) models.

To enable integration into the climate-dependent DFA framework, information on projected hazard frequency and severity under different climate scenarios need to be obtained from catastrophe (CAT) model providers. As leading CAT modelling firms begin to incorporate the impacts of climate change into their models \citep[see, e.g.,][]{RMS2022}, such integration is becoming increasingly feasible.

CAT model outputs, such as those produced by Risk Management Solutions (RMS), typically provide event-level projections of annual hazard frequency and severity. These outputs can be aggregated to parametrise a Collective Risk Model within the DFA framework. The resulting hazard loss projections can then be incorporated into the broader climate-dependent DFA framework to assess the impacts of climate change on general insurers.

\section{Sensitivity testing of dividend payments}\label{Appendix:SensitivityDividendsPaids}

\subsection{Modelling of dividend payments} \label{Appendix:ModellingDividentPayments}

The results on surplus and insolvency probability presented in Section \ref{Section:Returns} are based on the assumption that no dividends are paid from the surplus, consistent with the paper's scope of providing a baseline model. As a supplementary analysis, this section examines how varying dividend payout ratios, which represent different management strategies, affect the surplus and insolvency probability of general insurers.

Following \citet{Ta12}, we assume that dividends are distributed as a fixed proportion of the surplus above the target solvency level. In addition, we impose a ceiling on the maximum dividend payment that may be distributed. Under these assumptions, the dividend payment for insurer $j$ is given by:
\begin{equation} \label{Eq:Dividend}
     D_{t}^{(j)} = \min \left[\max \left(0, \varphi \cdot (K_t^{(j)} - K_t^{(j),*})\right), \max(0, \Upsilon \cdot \Pi_t^{(j)}) \right],
\end{equation}
where $K_t^{(j),*}$ denotes the target capital level, $\varphi$ is the dividend payout ratio, $\Pi_t^{(j)}$ denotes the net profit (i.e., $\Pi_t^{(j)} = (1 + \tilde{r}_t^{(I)}) \cdot \tilde{\pi}_t^{(j)} + \tilde{r}_t^{(I)} \cdot K_{t-1}^{(j)} - (\tilde{X}_{t,(j)}^{\text{net}} + \tilde{X}_{t,(j)}^{\text{NC}})$), and $\Upsilon$ denotes the cap on dividend payments, expressed as a proportion of the net profit. The target capital level is defined as a multiple of the gross written premium: $K_t^{(j),*} = S^* \pi_{t,(j)}$, where $S^*$ represents the target solvency ratio. This formulation ensures that no dividends are paid when the capital level falls below the target solvency threshold, thereby triggering a recovery plan to restore solvency. In addition, the dividend payment is capped at a proportion of net profit (i.e., $\Upsilon$), which is typically determined by the external regulatory environment.

For the sensitivity analysis, we consider dividend payout ratios $\varphi \in \{0.1, 0.3, 0.5, 0.7\}$. The target solvency ratio is set to 3, guided by the average ratio of general insurers' capital base to gross written premium over a 15-year period (see Figure \ref{fig:RatiosofCapitaltoPremiums}) derived from the APRA statistics \citep{APRAGIIL_data}. The cap on dividend payments is set at 50\% of net profit (i.e., $\Upsilon = 50\%$), consistent with guidance previously issued by APRA during periods of financial stress \citep{APRA2020}. We also repeated the sensitivity analysis under a higher ceiling on dividend payments, set at 100\% of net profit (i.e., $\Upsilon = 100\%$). The results, presented in Section \ref{Appendix:ResultsSensitivityDividentsHigherCap}, are generally consistent with those obtained under the default ceiling of 50\%.

\begin{figure}
    \centering
    \includegraphics[width=0.5\linewidth]{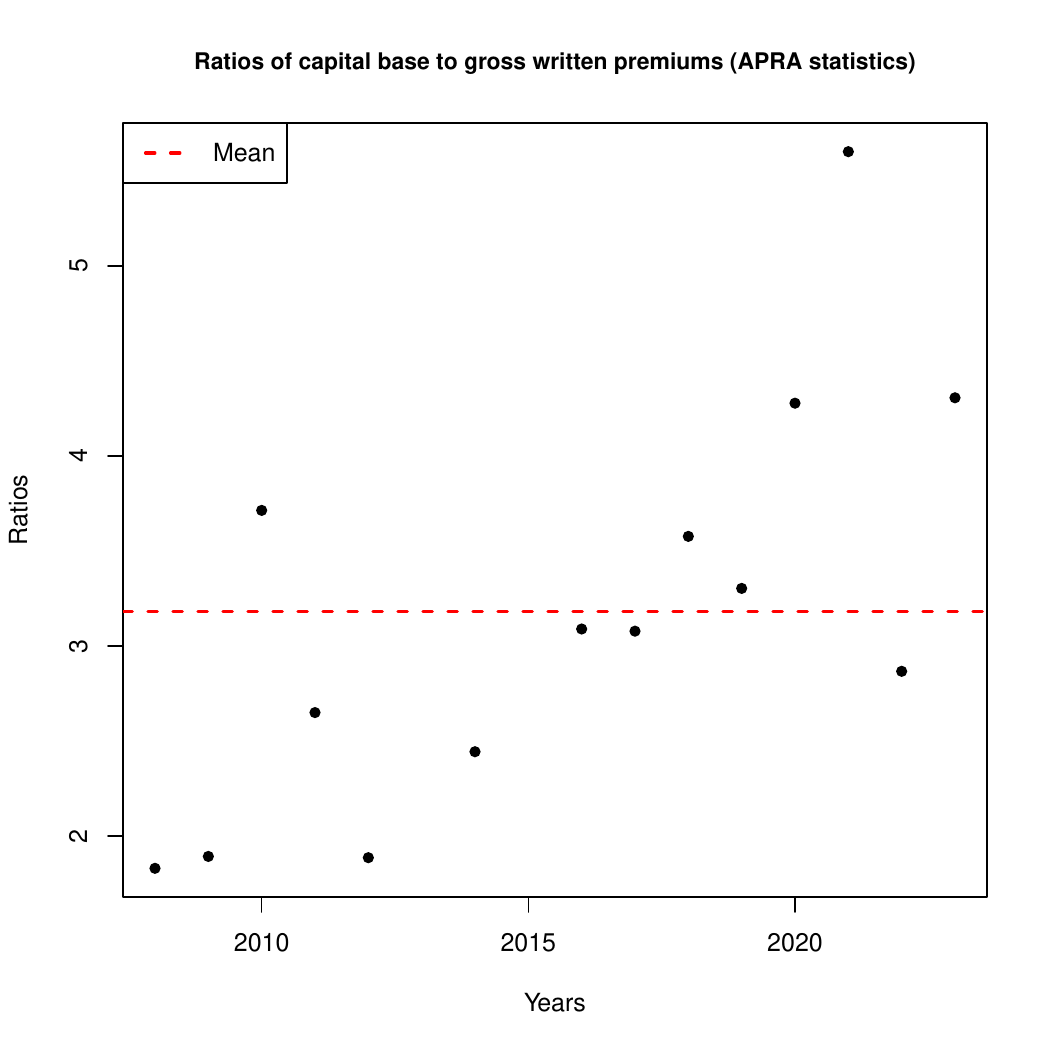}
    \caption{Ratios of capital base to gross written premium for the Australian general insurance market, averaged across market participants (APRA statistics)}
    \label{fig:RatiosofCapitaltoPremiums}
\end{figure}

\subsection{Results of sensitivity testing on dividend payments} \label{Appendix:ResultsSensitivityDividents}

The impacts of different dividend policies on the mean surplus and on the differences in mean surplus across scenarios are illustrated in Figure \ref{fig:MeanSurplusbyPayoutRatiosScenarios} and Figure \ref{fig:MeanSurplusbyScenariosPayoutRatios}, respectively. As shown in Figure~\ref{fig:MeanSurplusbyPayoutRatiosScenarios}, the mean surplus decreases as the dividend payout ratio increases, which aligns with expectations since higher dividend payouts slow capital accumulation. Figure \ref{fig:MeanSurplusbyScenariosPayoutRatios} compares the mean surplus across SSP scenarios under each dividend policy. The highest mean surplus remains under the SSP 8.5 scenario, followed by SSP 2.6, SSP 4.5, and SSP 7.0, consistent with the findings in the main paper. However, the difference in mean surplus between SSP 2.6 and SSP 8.5 narrows as the dividend payout ratio increases, which is further supported by the relative differences shown in Table \ref{tab:Mean_diff_surplus_dividends}. Since dividend payments are only triggered when the surplus exceeds the target level, their effects are more pronounced in scenarios with higher surpluses (i.e., SSP~8.5).

The impacts of different dividend policies on insolvency probability and on the differences in insolvency probability across scenarios are shown in Figure \ref{fig:MeanInsolvencyProbsbyPayoutRatiosScenarios} and Figure \ref{fig:MeanInsolvencyProbsScenariosPayoutRatios}, respectively. As shown in Figure~\ref{fig:MeanInsolvencyProbsbyPayoutRatiosScenarios}, insolvency probability increases with higher dividend payout ratios across all scenarios, as expected, since slower capital accumulation reduces the buffer available to absorb future losses. The largest increase is observed under the SSP 8.5 scenario, particularly in the later projection years. This supports the argument in Section \ref{Section:Returns} that the low insolvency probability under SSP 8.5, despite its higher physical risk, is primarily driven by faster capital accumulation and stronger investment returns earlier in the projection horizon, which provide greater resilience against later losses. As dividend payouts increase, this capital buffer is diminished, leaving less available capital to absorb future losses. Given that the SSP 8.5 scenario involves the highest physical risk, this reduction in capital resilience leads to the largest rise in insolvency probability among the four scenarios. Figure \ref{fig:MeanInsolvencyProbsScenariosPayoutRatios} compares insolvency probabilities across SSP scenarios, showing that the ranking identified in the main paper (i.e., highest under SSP 7.0 and lowest under SSP 8.5) remains consistent. However, the gap in insolvency probability between SSP 2.6 and SSP 8.5 narrows, reflecting the relatively greater increase in insolvency risk under SSP 8.5, as discussed above.

\begin{figure}[H]
    \centering
    \includegraphics[width=0.7\linewidth]{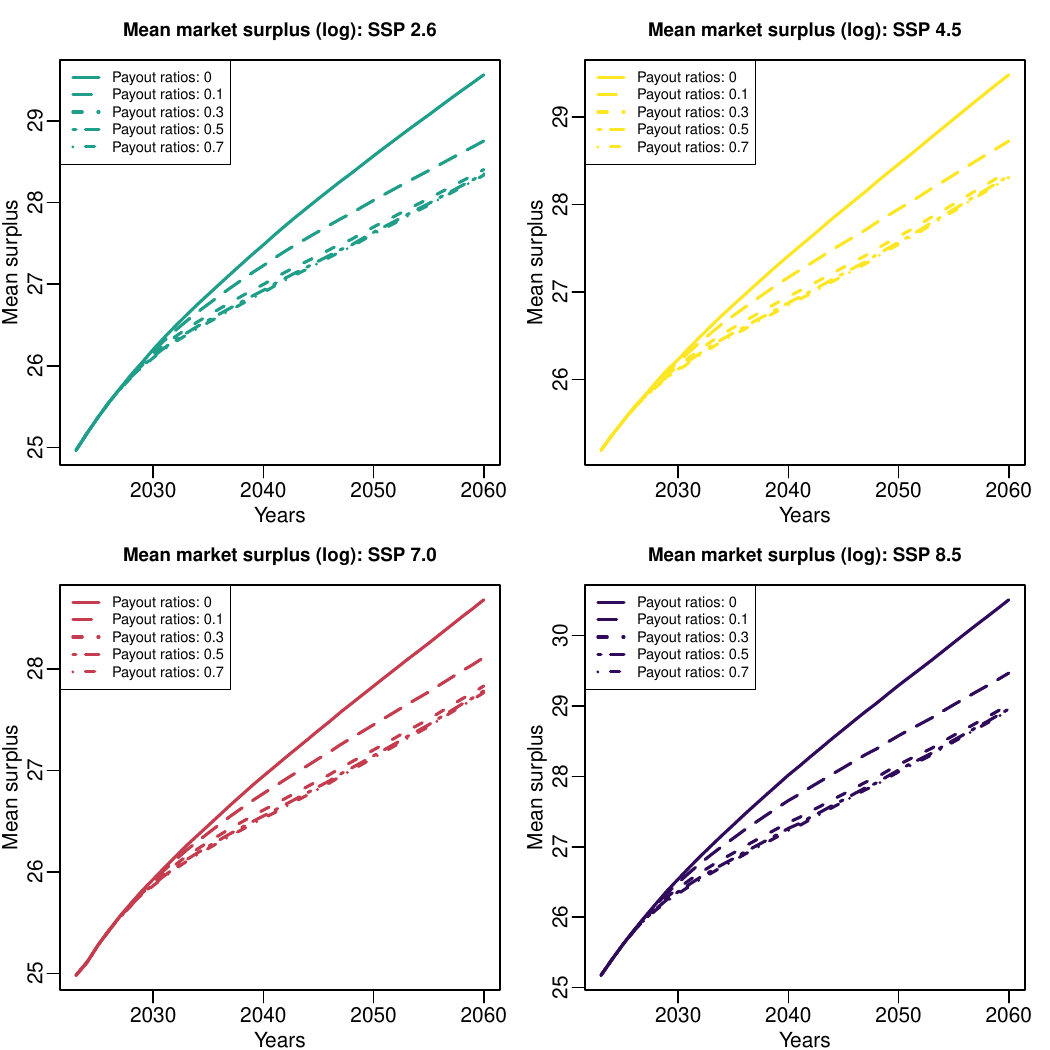}
   \caption{Comparison of mean surplus under different dividend payout ratio assumptions within each SSP scenario.}
    \label{fig:MeanSurplusbyPayoutRatiosScenarios}
\end{figure}

\begin{figure}[H]
    \centering
    \includegraphics[width=0.7\linewidth]{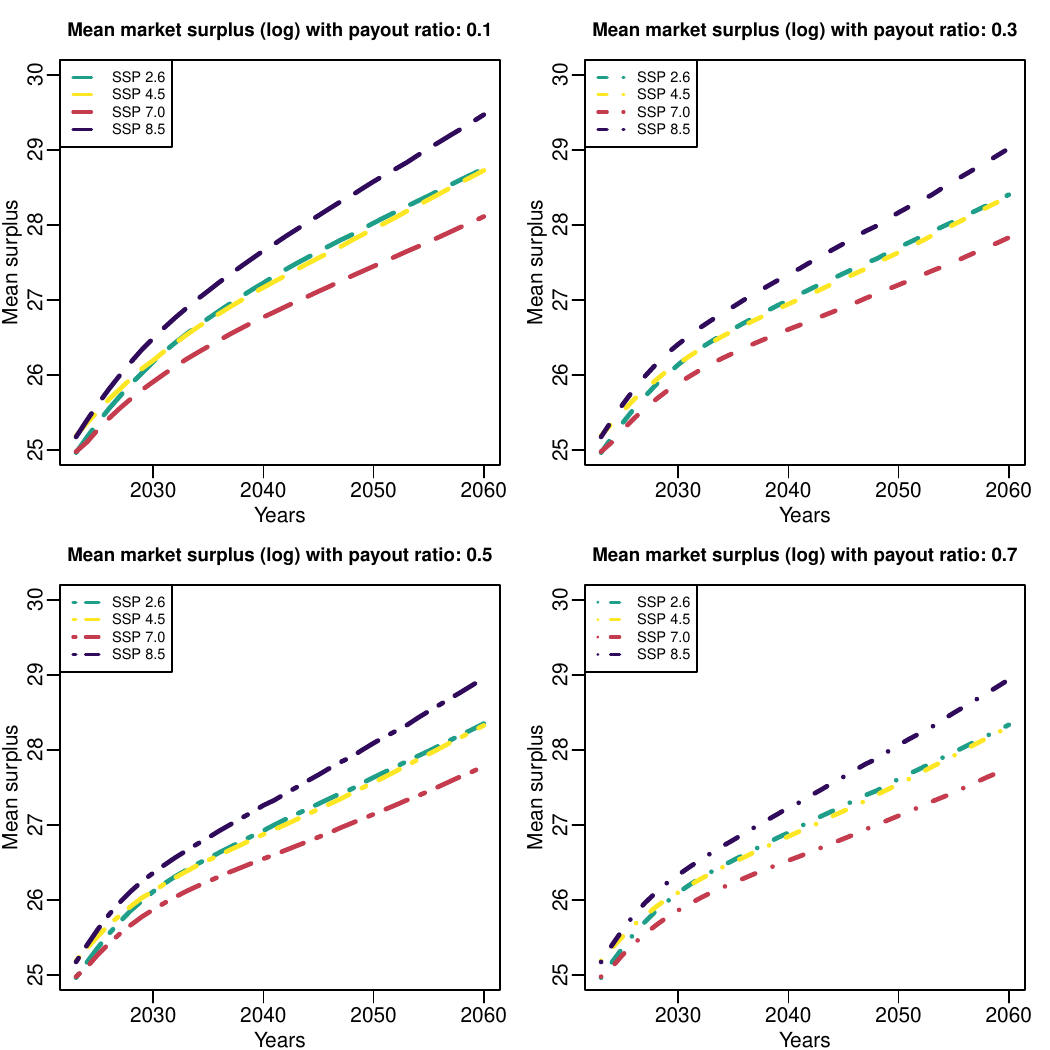}
    \caption{Comparison of mean surplus across SSP scenarios for each dividend payout ratio assumption.}
    \label{fig:MeanSurplusbyScenariosPayoutRatios}
\end{figure}

\begin{table}[H]
\centering
\caption{Average relative difference in mean surplus between SSP 2.6 and SSP 8.5 scenarios under different dividend payout ratios}
\label{tab:Mean_diff_surplus_dividends}
\small
\begin{tabular}{lc}
\toprule
\textbf{Dividend payout ratio} & \textbf{Average relative difference in mean surplus (SSP 2.6 v.s. SSP 8.5)} \\
\midrule
0.0 & \grad{100}{-79.7\%} \\  
0.1 & \grad{38}{-58.5\%} \\
0.3 & \grad{6}{-47.6\%} \\
0.5 & \grad{2}{-45.9\%} \\
0.7 & \grad{0}{-45.2\%} \\    
\bottomrule
\end{tabular}
\end{table}

\begin{figure}[H]
    \centering
    \includegraphics[width=0.7\linewidth]{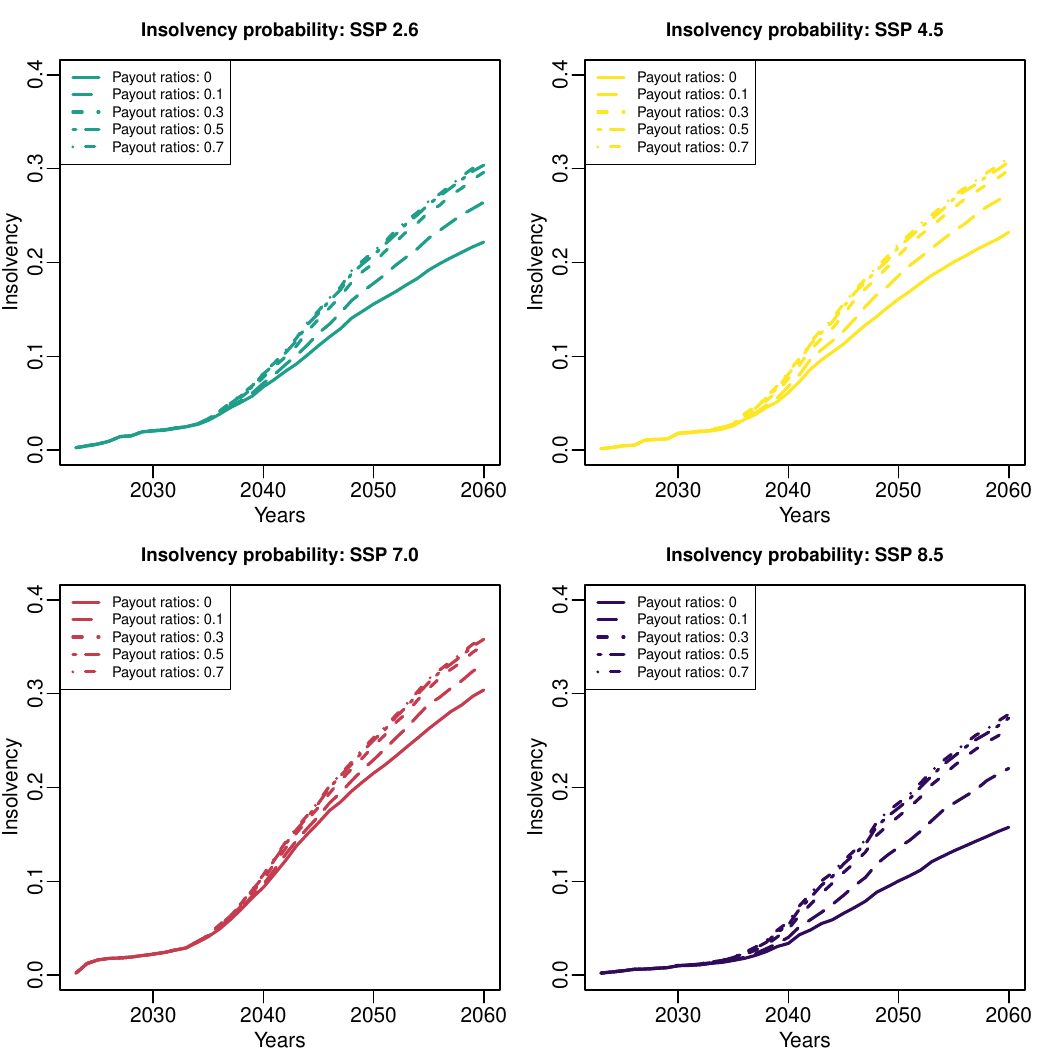}
   \caption{Comparison of insolvency probability under different dividend payout ratio assumptions within each SSP scenario.}
    \label{fig:MeanInsolvencyProbsbyPayoutRatiosScenarios}
\end{figure}

\begin{figure}[H]
    \centering
    \includegraphics[width=0.7\linewidth]{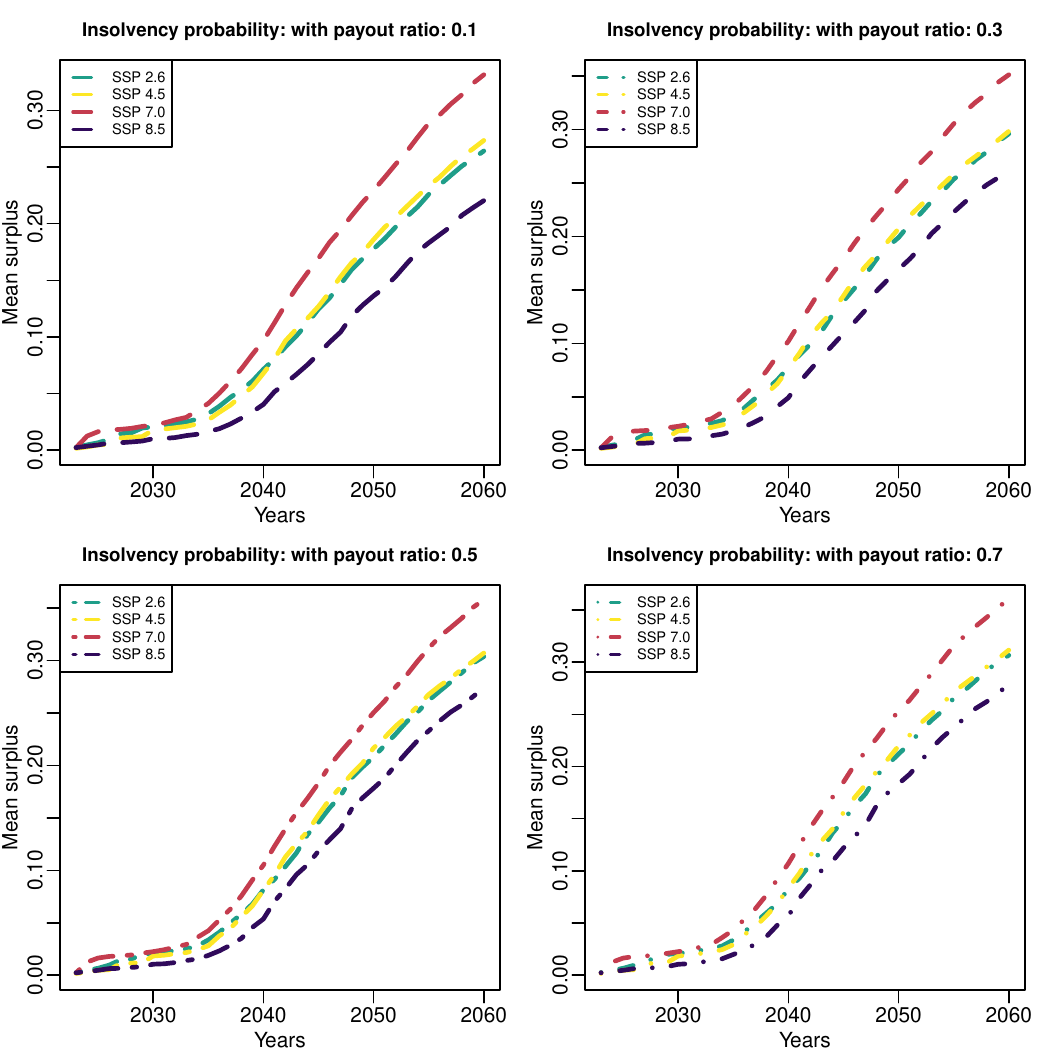}
    \caption{Comparison of insolvency probability across SSP scenarios for each dividend payout ratio assumption.}
    \label{fig:MeanInsolvencyProbsScenariosPayoutRatios}
\end{figure}

\begin{table}[H]
\centering
\caption{Mean difference in insolvency probability between SSP 2.6 and SSP 8.5 scenarios under different dividend payout ratios}
\label{tab:Mean_diff_insolvency_dividends}
\small
\begin{tabular}{lc}
\toprule
\textbf{Dividend payout ratios} & \textbf{Mean difference in insolvency probability (SSP 2.6 v.s. SSP 8.5)} \\
\midrule
0.0 & \grad{100}{0.0341} \\
0.1 & \grad{54}{0.0274} \\
0.3 & \grad{15}{0.0217} \\
0.5 & \grad{7}{0.0206} \\
0.7 & \grad{0}{0.0196} \\
\bottomrule
\end{tabular}

\end{table}

\subsection{Results of sensitivity testing on dividend payments under a higher cap on maximum dividend payments} \label{Appendix:ResultsSensitivityDividentsHigherCap}

The following figures present the mean surplus and insolvency probability under different dividend payout assumptions with a 100\% net profit cap. Overall, the results are consistent with those reported in the previous section based on the 50\% net profit cap.

\begin{figure}[H]
    \centering
    \includegraphics[width=0.7\linewidth]{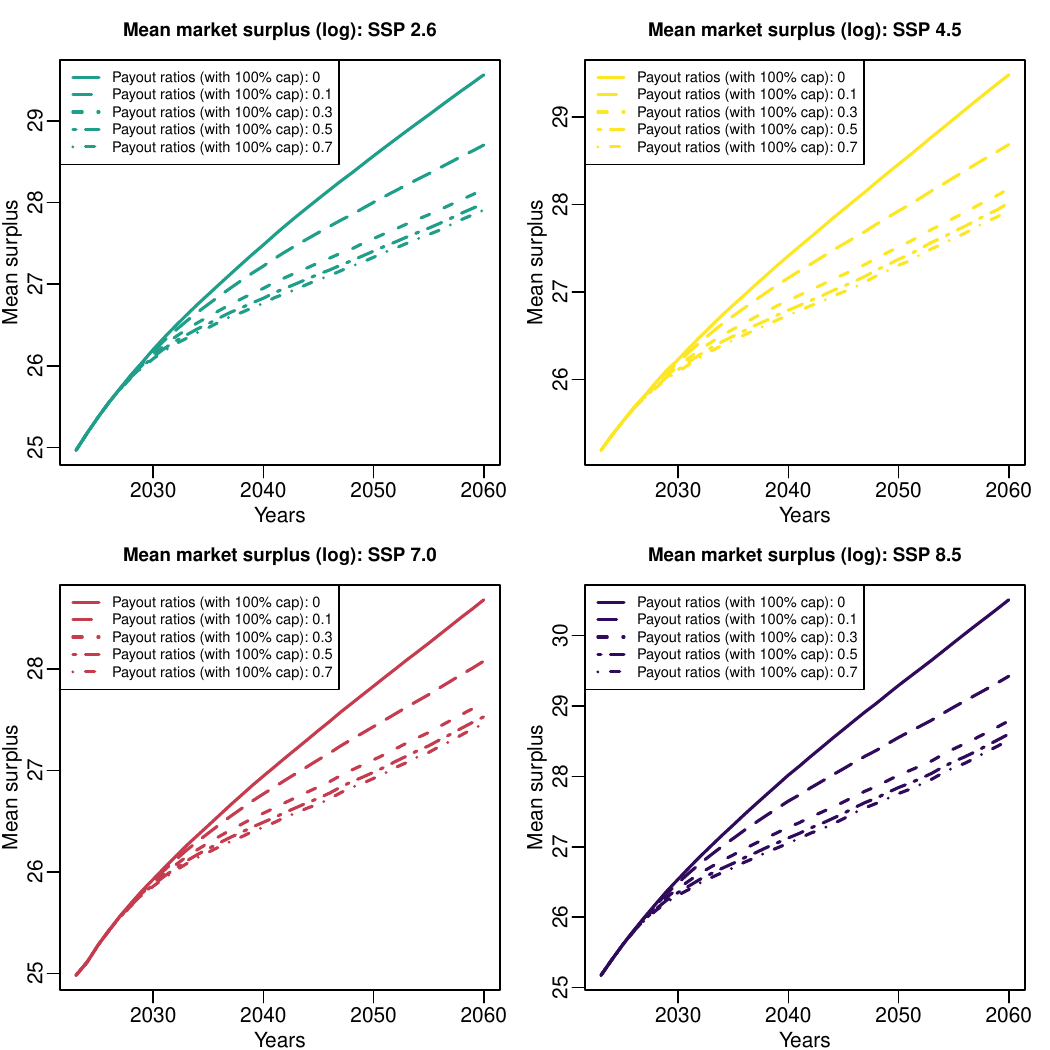}
   \caption{Comparison of mean surplus under different dividend payout ratio assumptions (under 100\% net profit cap) within each SSP scenario.}
    \label{fig:MeanSurplusbyPayoutRatiosScenariosHigherCap}
\end{figure}

\begin{figure}[H]
    \centering
    \includegraphics[width=0.7\linewidth]{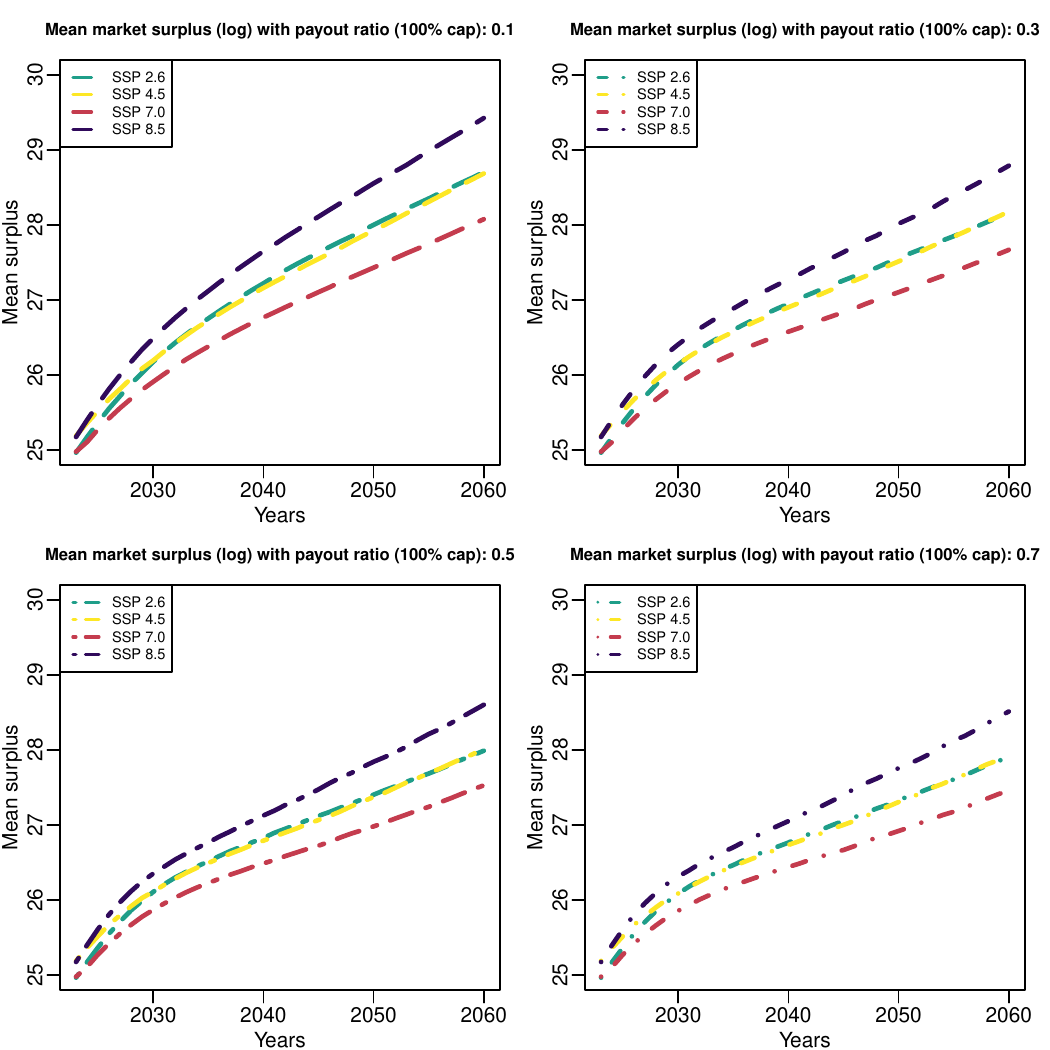}
    \caption{Comparison of mean surplus across SSP scenarios for each dividend payout ratio assumption (under 100\% net profit cap).}
    \label{fig:MeanSurplusbyScenariosPayoutRatiosHigherCap}
\end{figure}

\begin{figure}[H]
    \centering
    \includegraphics[width=0.7\linewidth]{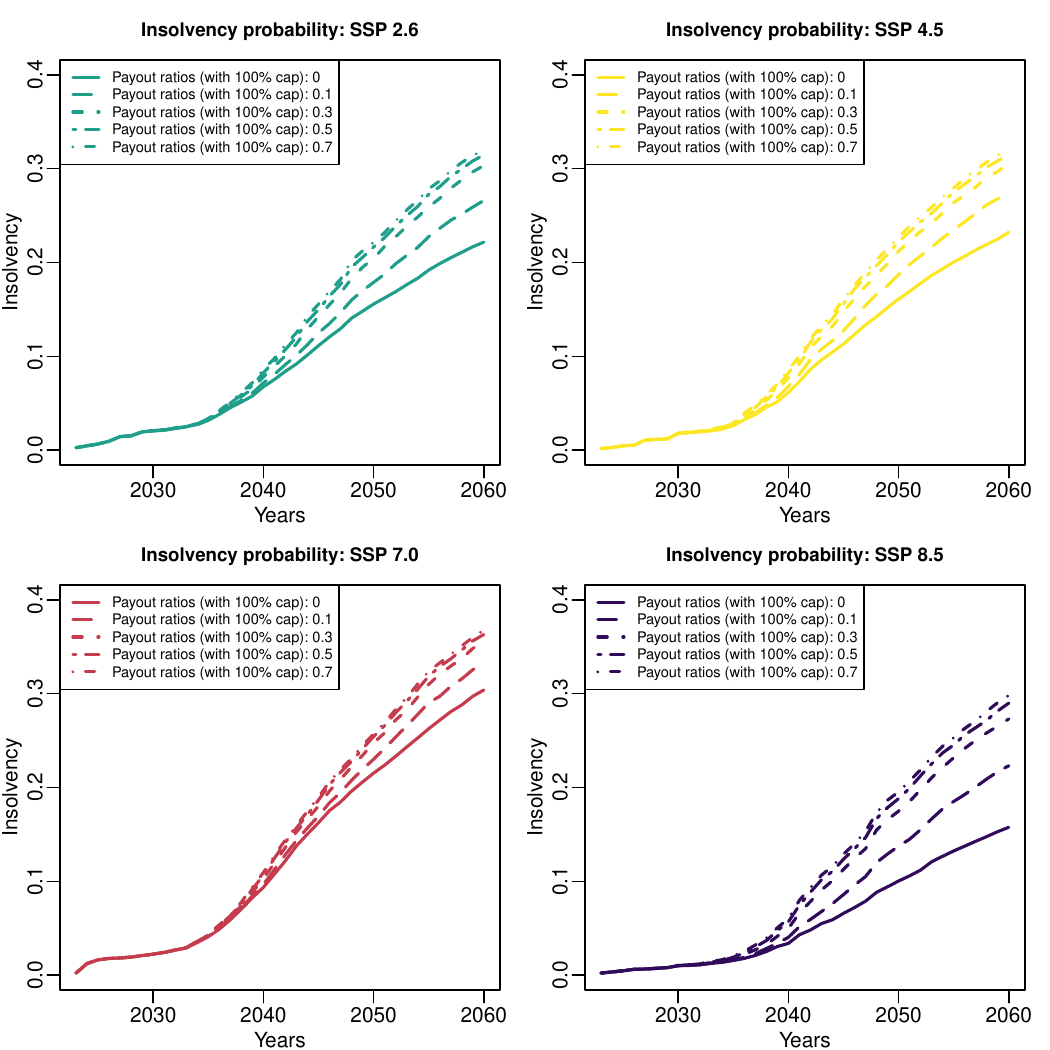}
   \caption{Comparison of insolvency probability under different dividend payout ratio assumptions (under 100\% net profit cap) within each SSP scenario.}
    \label{fig:MeanInsolvencyProbsbyPayoutRatiosScenariosHigherCap}
\end{figure}

\begin{figure}[H]
    \centering
    \includegraphics[width=0.7\linewidth]{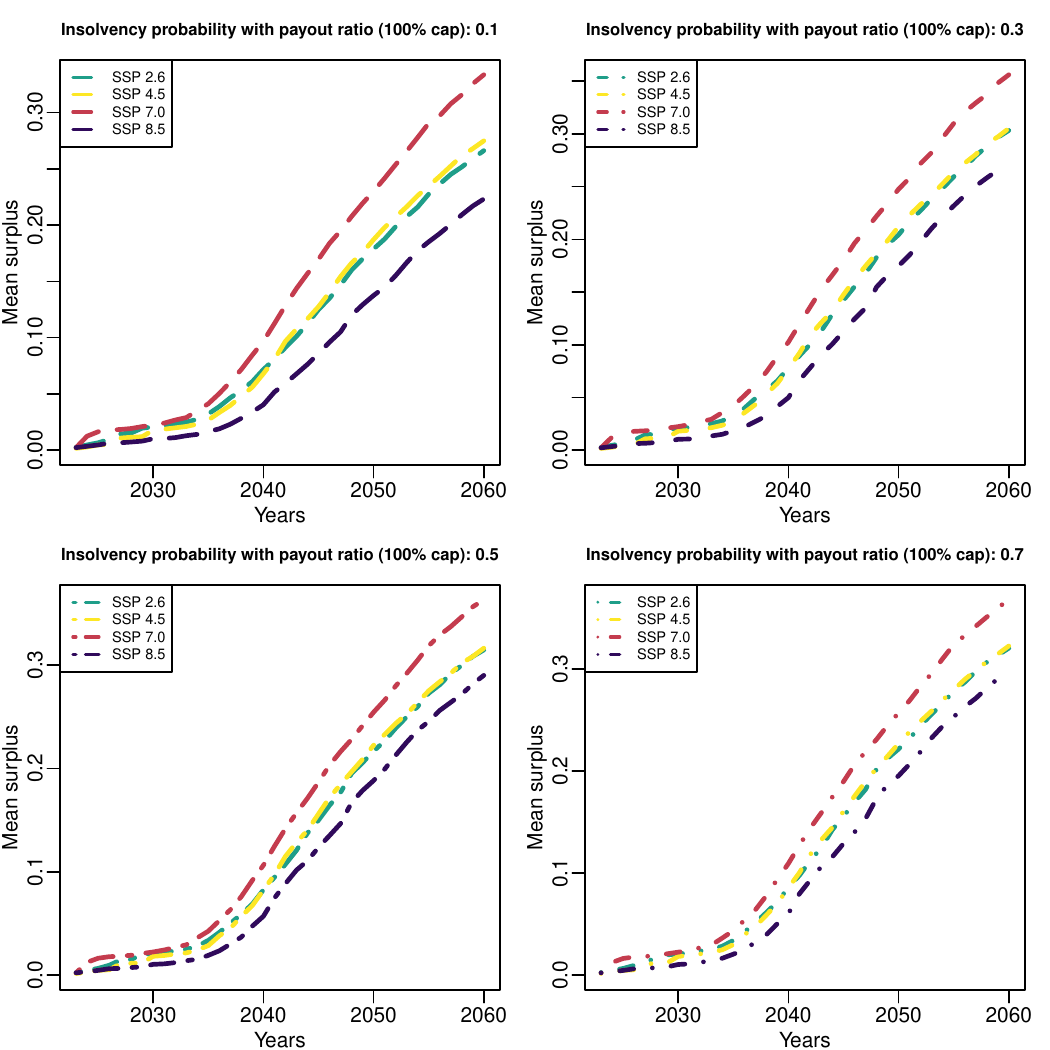}
    \caption{Comparison of insolvency probability across SSP scenarios for each dividend payout ratio assumption (under 100\% net profit cap).}
    \label{fig:MeanInsolvencyProbsScenariosPayoutRatiosHigherCap}
\end{figure}

\section{Results of sensitivity testing on economic damages}\label{Appendix:SensitivityEconomicDamages}

This section investigates the impact of alternative economic damage assumptions on simulated investment returns by calibrating the projected economic damages in this study to those reported in the literature. In particular, we adopt the damage function used in the Dynamic Integrated model of Climate and the Economy (DICE), which has been widely applied in the climate–economy literature since its introduction by \citet{Nord92}. Specifically, we use the functional form implemented in the latest version, DICE-2023 \citep{BaNo24}:
\begin{equation} \label{Eq:DICEDamagefunction}
    \Omega^*(t) = \gamma (\Delta T)^2,
\end{equation}
where $\Omega^*(t)$ denotes the ratio of total climate damage to GDP (i.e., the damage ratio), and $\Delta T$ represents the increase in Global Mean Surface Temperature (GMST) relative to the pre-industrial level. The damage parameter $\gamma$ is set to 0.003467 in DICE-2023 \citep{BaNo24}, compared with 0.00227 in the earlier DICE-2016R2 version \citep{Nord18DICE26R2} and 0.001478 in the original DICE-1992 version \citep{No92}. However, as noted in Online Appendix~\ref{Appendix: ProjectedDamageRatiosComparison}, caution is warranted since the DICE damage function is calibrated at the global scale, whereas our simulated damages are specific to Australia. Therefore, the benchmarking results presented in this section should be interpreted solely as sensitivity tests illustrating the implications of alternative economic damage assumptions, rather than as predictive estimates.

To derive the damage ratios implied by the DICE damage function under the SSP scenarios, we substitute the projected GMST increases for each SSP pathway into the functional form defined in \eqref{Eq:DICEDamagefunction}. The resulting damage ratios from the DICE-1992, DICE-2016R2 and DICE-2023 versions, together with the damage-adjusted real GDP based on the DICE-2023 damage function, are presented in Figure~\ref{Fig:DamageRatiosComparisonandGDP}. The DICE-2023 parameters are adopted as the benchmark for economic damage, producing more conservative estimates of damage ratios than those under DICE-2016R2 and DICE-1992. For simplicity, we refer to ``\textit{DICE-implied damage ratios}'' as those derived from the functional form and parameters of the \textbf{DICE-2023} damage function in the subsequent sections. As shown in Figure~\ref{Fig:DamageRatiosComparisonandGDP}, although SSP~8.5 yields the highest projected damage ratio, its damage-adjusted real GDP remains the highest across scenarios, reflecting the strong potential economic growth driven by competitive markets and innovation assumed in this pathway \citep{OnKrEbKeRiRo17}.

\begin{figure}[H]
\begin{minipage}{\textwidth}
\begin{minipage}[t]{0.49\textwidth}
    \centering
    \includegraphics[width=\textwidth]{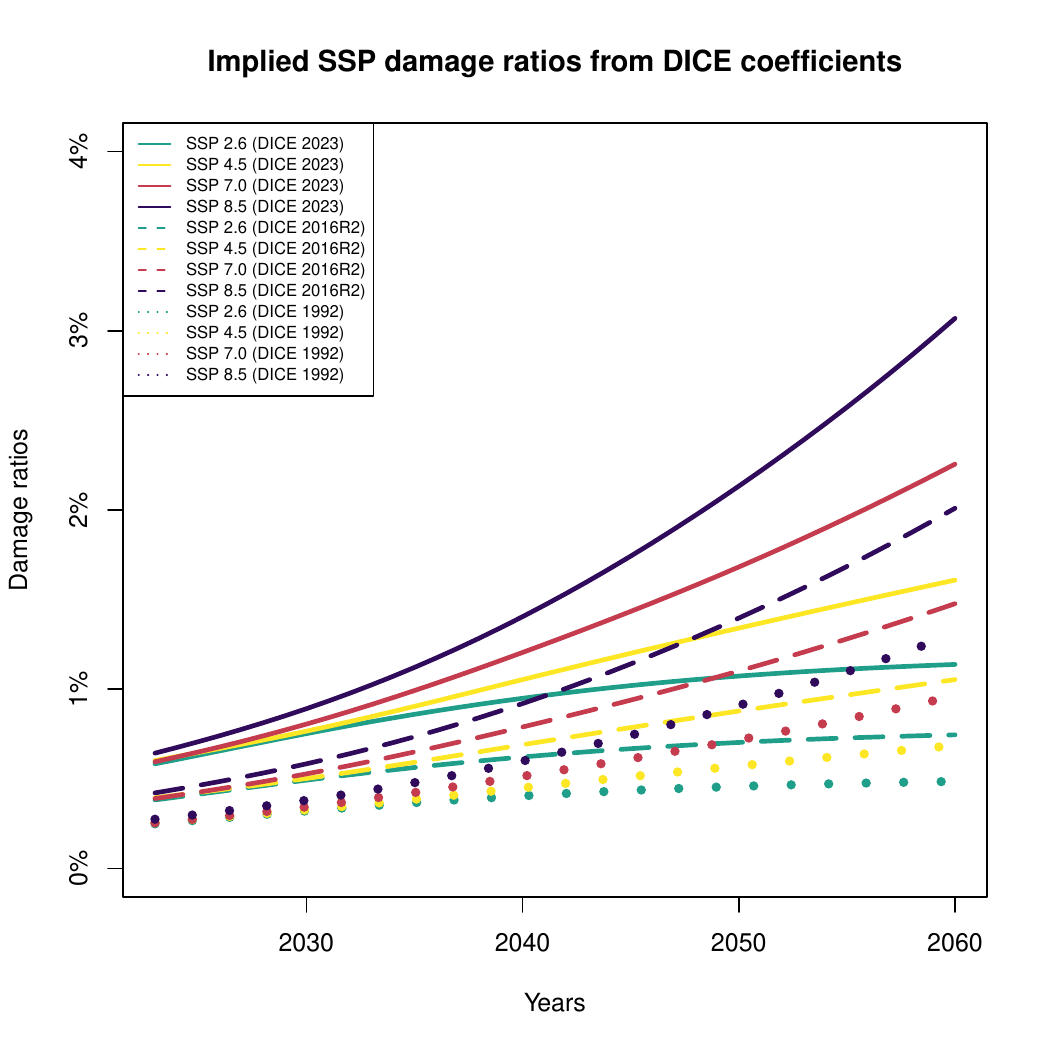}
    \end{minipage}
  \begin{minipage}[t]{0.49\textwidth}
    \centering
    \includegraphics[width=\textwidth]{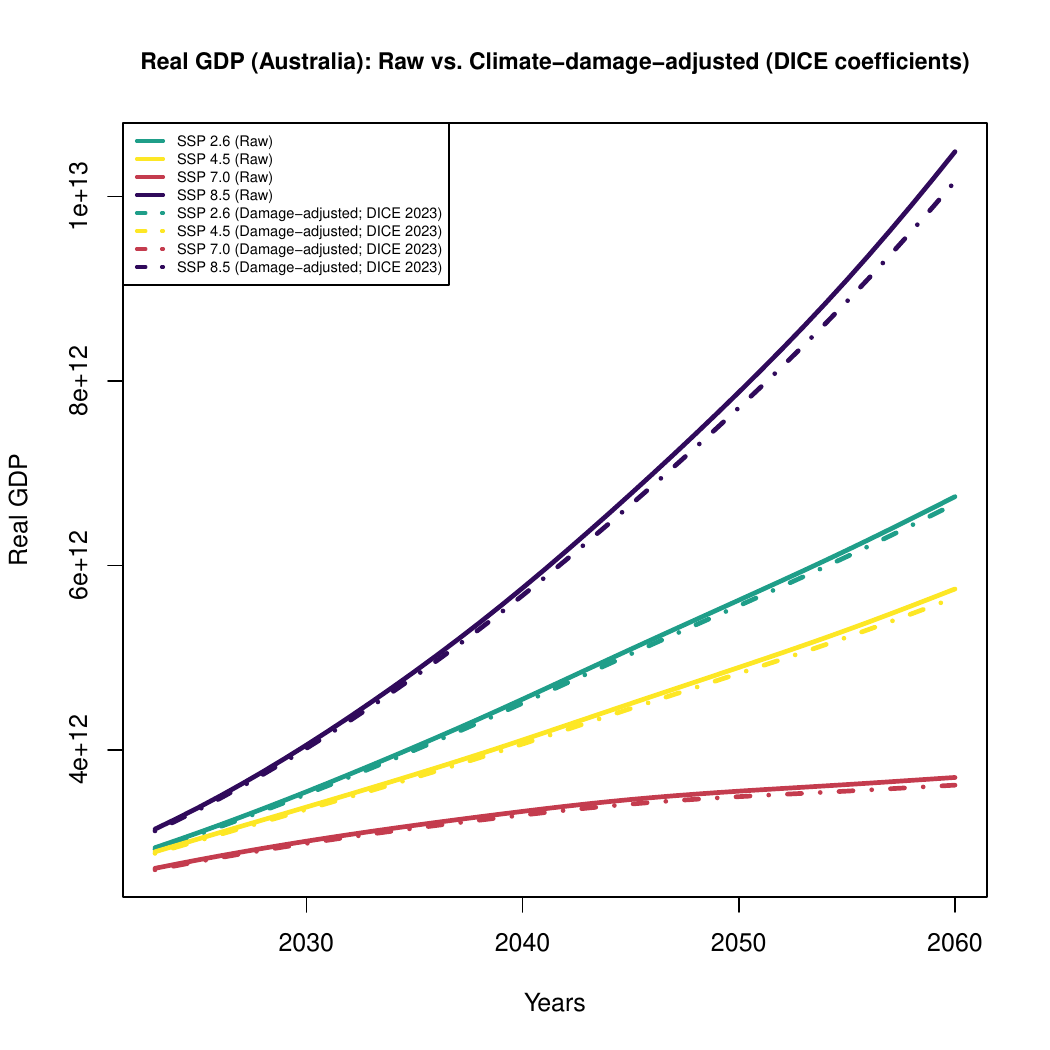}
  \end{minipage}
  \hfill
  \end{minipage}
  \captionof{figure}{Left panel (a) shows the implied damage ratios under SSP scenarios based on the DICE damage functions and parameters from the DICE-2023 (solid lines) and DICE-2016R2 (dashed lines) versions; right panel (b) presents the real GDP projections for Australia under SSP scenarios and the damage-adjusted values based on the DICE-2023 damage function.}
  \label{Fig:DamageRatiosComparisonandGDP} 
\end{figure}

To calibrate our projected economic damages with those implied by the DICE-2023 damage function specified in \eqref{Eq:DICEDamagefunction}, we adopt a quantile-mapping approach with linear transformation function \citep{PiWeBeGoViHaHa10, QiCh21} rather than applying a constant multiplier to scale insured losses to economic damages, as in Section~\ref{Section:Equity} of the main paper. Specifically, we scale the simulated insurance losses by matching the quantiles of the mean projected total economic damages to those of the DICE-implied damages over the projection horizon:
\begin{equation} \label{Eq:EconomicDamageQuantilesMapping}
   Q_{\bm{\Psi^*}}(q) = \hat{\alpha}_q + \hat{\beta}_q Q_{\bm{\Xi}}(q),
\end{equation}
where $\bm{\Psi}^* = \{\psi_1^*, \ldots, \psi_t^*, \ldots, \psi_T^*\}$ denotes the set of DICE-implied economic damages over the projection horizon $T$, with $\psi_t^* = Y_t \Omega^*(t)$; and $\bm{\Xi} = \{\mathbb{E}(\tilde{X}_1), \ldots, \mathbb{E}(\tilde{X}_t), \ldots, \mathbb{E}(\tilde{X}_T)\}$ denotes the expected insurance damages derived from the hazard module over the same horizon. Here, $Q_{\bm{\Xi}}(q)$ and $Q_{\bm{\Psi^*}}(q)$ denote the $q^{\text{th}}$ quantiles of $\bm{\Xi}$ and $\bm{\Psi^*}$, respectively, and the parameters $\hat{\alpha}_q$ and $\hat{\beta}_q$ are estimated via regression. The adjusted simulated economic damage is then given by:
\begin{equation} \label{Eq:AdjustedEconomicDamage}
    \psi_t = \hat{\alpha}_q + \hat{\beta}_q \tilde{X}_t.
\end{equation}
This adjustment ensures that the magnitudes and ranges of the mean simulated economic damages are broadly consistent with those of the DICE-implied economic damages over the projection horizon. In the following sections, we refer to ``\textit{DICE-adjusted damage ratios}'' as the simulated economic damage ratios calibrated to the \textit{DICE-implied damage ratios} using \eqref{Eq:AdjustedEconomicDamage}.

The mean \textit{DICE-adjusted damage ratios} based on this approach are illustrated in Figure~\ref{Fig:DamageRatiosQuantileMapping}. The quantiles of the \textit{DICE-adjusted damage ratios} are calibrated to the corresponding quantiles of the \textit{DICE-implied damage ratios} over the defined projection horizon. Despite inter-annual fluctuations in the simulated mean \textit{DICE-adjusted damage ratios}, driven by the variability in hazard losses discussed in Section \ref{Section: ResultsClimateHazards} of the main paper, the overall trends remain broadly consistent with the \textit{DICE-implied damage ratios} across scenarios.

\begin{figure}[H]
\begin{minipage}{\textwidth}
\begin{minipage}[t]{0.49\textwidth}
    \centering
    \includegraphics[width=\textwidth]{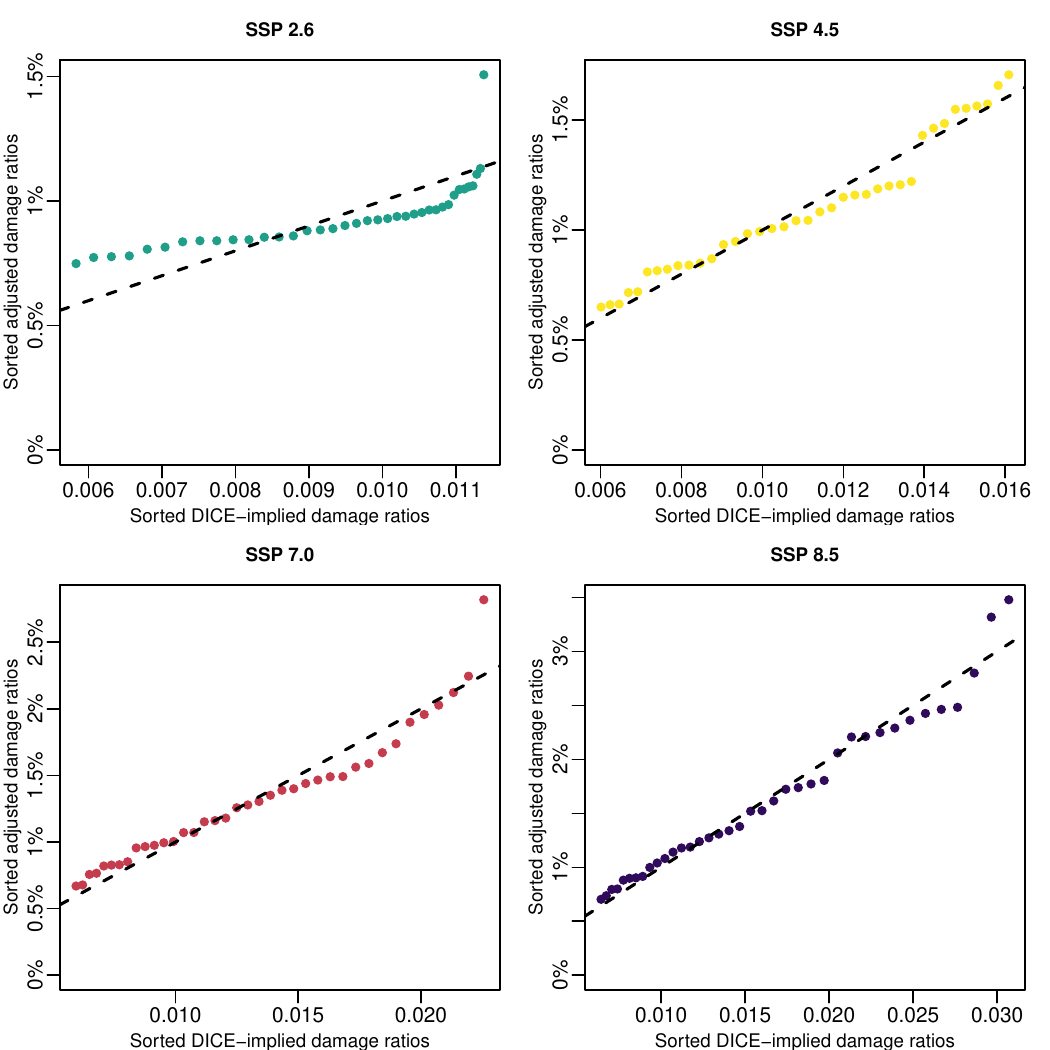}
    \end{minipage}
  \begin{minipage}[t]{0.49\textwidth}
    \centering
    \includegraphics[width=\textwidth]{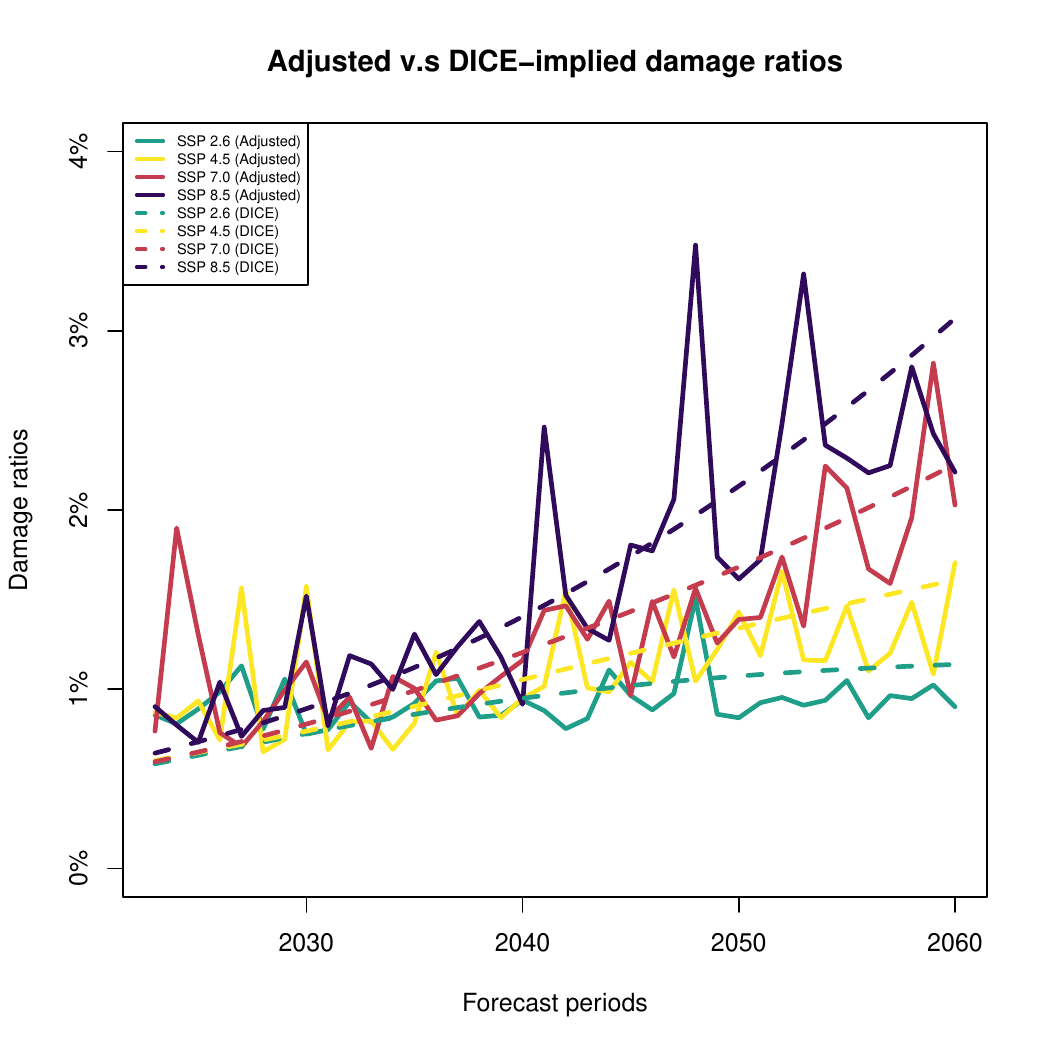}
  \end{minipage}
  \hfill
  \end{minipage}
  \captionof{figure}{Left panel (a) plots the sorted mean \textit{DICE-adjusted damage ratios} from \eqref{Eq:AdjustedEconomicDamage} against the sorted \textit{DICE-implied damage ratios} over the projection horizon to 2060; right panel (b) shows the mean \textit{DICE-adjusted damage ratios} together with the \textit{DICE-implied damage ratios} over the same period.}
  \label{Fig:DamageRatiosQuantileMapping} 
\end{figure}

The left panel of Figure \ref{Fig:DamageRatiosBenchmarkingResults} compares the simulated damage-adjusted cumulative economic growth under both the original simulated damage ratios and the \textit{DICE-adjusted damage ratios} with the damage-unadjusted cumulative growth. While the \textit{DICE-adjusted damage ratios} lead to a greater reduction in real GDP, the ranking of cumulative economic growth remains consistent across all cases -- raw, damage-adjusted, and \textit{DICE-adjusted} -- with SSP~8.5 exhibiting the highest growth, followed by SSP~2.6, SSP~4.5, and SSP~7.0. This ranking aligns with the scenario narratives in \citet{OnKrEbKeRiRo17}. Although the projected damage ratio is highest under SSP~8.5, the corresponding damage-adjusted GDP remains the largest due to its stronger baseline growth assumptions, which are consistent with the narrative of ``the attainment of human development goals, robust economic growth, and highly engineered infrastructure resulting in relatively low challenges to adaptation'' under this scenario \citep{OnKrEbKeRiRo17}.

The right panel of Figure~\ref{Fig:DamageRatiosBenchmarkingResults} shows the cumulative investment returns based on the \textit{DICE-adjusted damage ratios}, with trends and rankings across scenarios reflecting the corresponding patterns in cumulative economic growth. These results are consistent with those reported in Section \ref{Section: ResultsInvestmentReturns}, suggesting that investment returns are primarily driven by the underlying potential growth assumptions of each SSP scenario.

\begin{figure}[H]
\begin{minipage}{\textwidth}
\begin{minipage}[t]{0.49\textwidth}
    \centering
    \includegraphics[width=\textwidth]{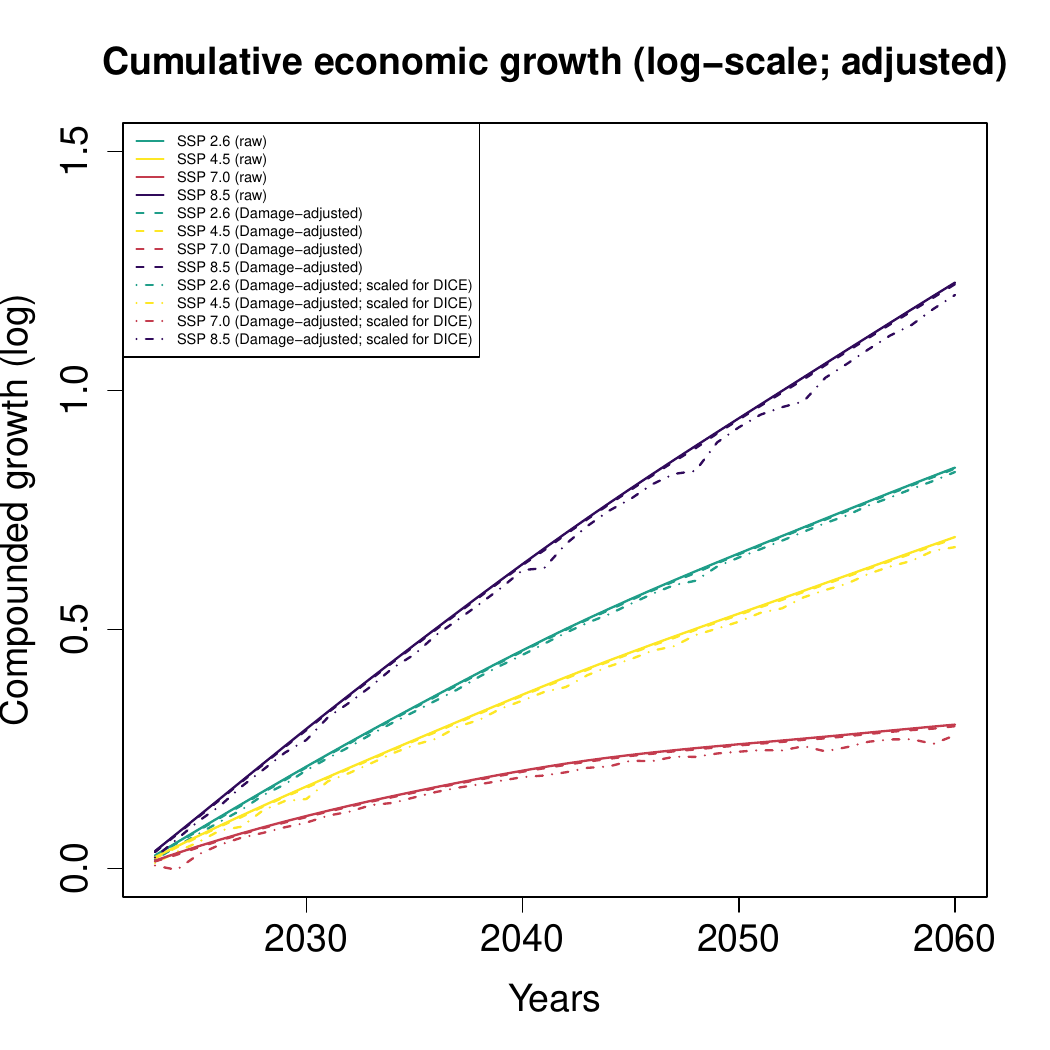}
    \end{minipage}
  \begin{minipage}[t]{0.49\textwidth}
    \centering
    \includegraphics[width=\textwidth]{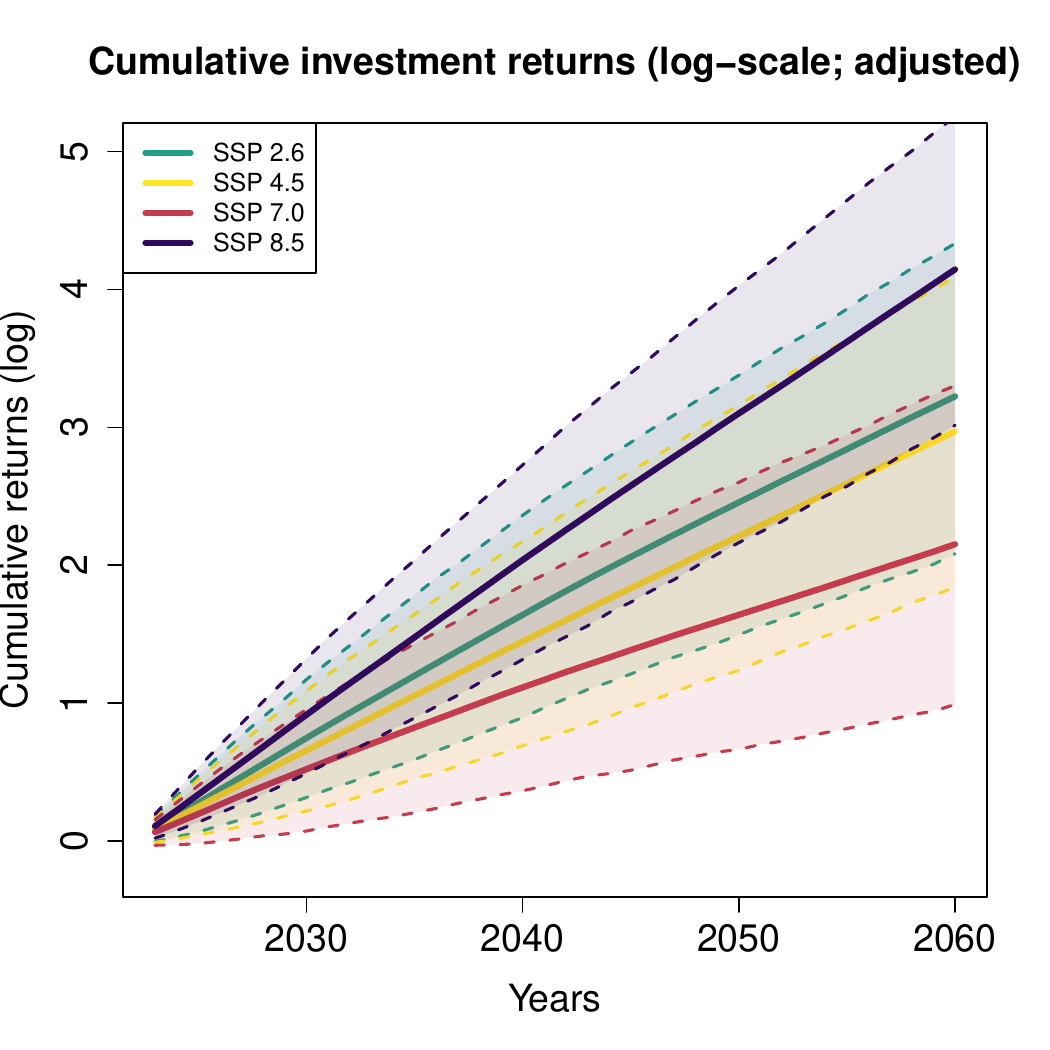}
  \end{minipage}
  \hfill
  \end{minipage}
  \captionof{figure}{Left panel (a) shows the raw (log) cumulative economic growth based on real GDP pathways (i.e., unadjusted for climate damages) underlying the SSP scenarios \citep{RiVaKr17} (solid lines), the damage-adjusted (log) cumulative economic growth based on the original simulated damages from the hazard module (dashed lines), and the damage-adjusted (log) cumulative economic growth based on the \textit{DICE-adjusted damage ratios} (dash-dotted lines). Right panel (b) shows the (log) cumulative total investment returns based on the \textit{DICE-adjusted damage ratios}, with solid lines representing the mean simulation paths and dashed lines indicating the $5^{\text{th}}$ and $95^{\text{th}}$ percentiles.}
  \label{Fig:DamageRatiosBenchmarkingResults} 
\end{figure}


It is noted that the DICE damage function represents only one of several approaches to estimating the economic impacts of rising temperatures, and that damage estimates vary substantially across studies, as shown in the left panel of Figure~\ref{Fig:DamageRatiosBenchmarkingResults2060}. The right panel of Figure~\ref{Fig:DamageRatiosBenchmarkingResults2060} compares the damage-adjusted GDP under SSP~2.6 and SSP~8.5 in~2060 with the corresponding baseline GDP assumptions. Despite higher estimated damage ratios under SSP~8.5, the resulting damage-adjusted GDP remains above that under SSP~2.6 across all studies by 2060, owing to the stronger baseline (damage-unadjusted) economic growth assumptions under SSP~8.5. Accordingly, while we do not perform a detailed sensitivity analysis of investment returns to alternative damage specifications, it might be reasonable to expect that the relative trends and ranking of investment returns across scenarios (at least at the mean and median levels) -- based on these damage estimates up to~2060 -- would remain broadly consistent with the results presented in Section~\ref{Section: ResultsInvestmentReturns} of the main paper.

\begin{figure}[H]
\begin{minipage}{\textwidth}
\begin{minipage}[t]{0.49\textwidth}
    \centering
    \includegraphics[width=\textwidth]{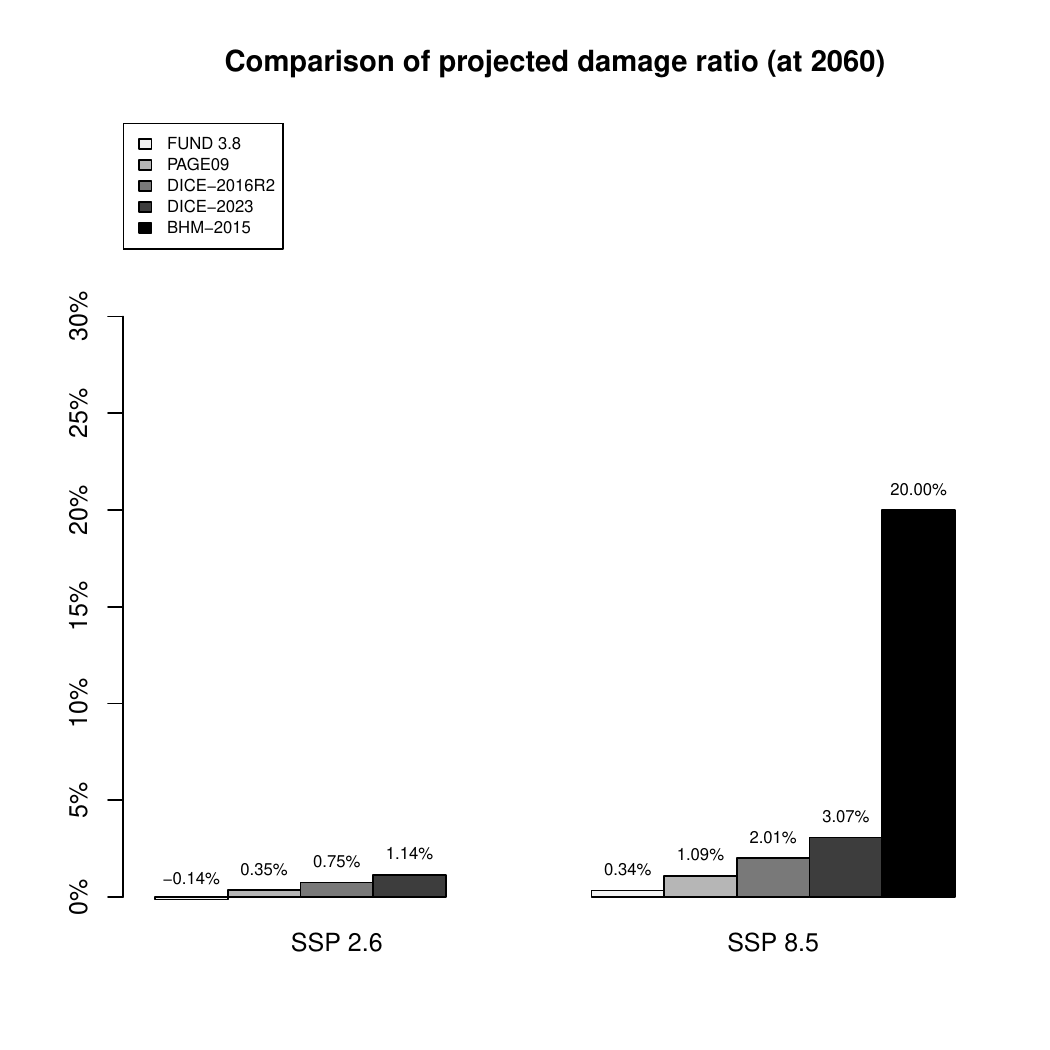}
    \end{minipage}
  \begin{minipage}[t]{0.49\textwidth}
    \centering
    \includegraphics[width=\textwidth]{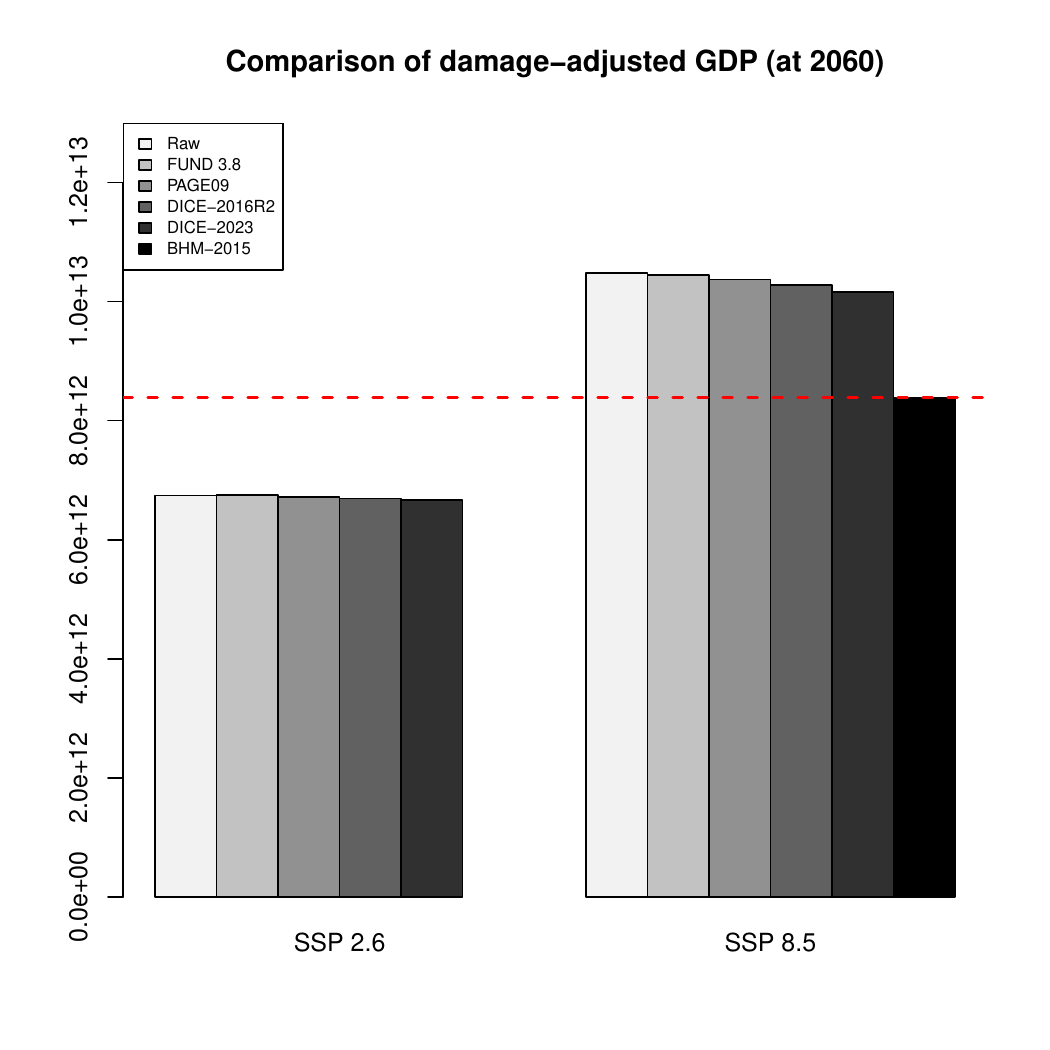}
  \end{minipage}
  \hfill
  \end{minipage}
  \captionof{figure}{Left panel shows the projected damage ratios at 2060 under SSP 2.6 and SSP 8.5 from FUND3.8 \citep{FUND38}, PAGE09 \citep{PAGE09}, DICE-2016R2 \citep{Nord18DICE26R2}, DICE-2023 \citep{BaNo24}, and BHM-2015 \citep{BuHsMi15}. Damage ratios can be interpreted as the percentage decline in GDP relative to the baseline GDP assumptions underlying each climate scenarios, and they \textbf{do not} represent the decline relative to the current GDP level. The right panel presents the corresponding damage-unadjusted and damage-adjusted real GDP at 2060 under SSP 2.6 and SSP 8.5, with adjustments based on the same model-specific damage ratios. The SSP 2.6 projection is not available in BHM-2015 \citep{BuHsMi15} and therefore left blank in the figure.}
  \label{Fig:DamageRatiosBenchmarkingResults2060} 
\end{figure}

\end{document}